\documentclass[12pt,english]{article}

\usepackage[T1]{fontenc}
\usepackage[utf8]{inputenc}
\usepackage{geometry}
\usepackage{pgfplots}
\pgfplotsset{compat=1.17}
\usepackage{amsmath}
\usepackage{amsthm}
\usepackage{amssymb}
\usepackage{bbm}
\usepackage{graphicx}
\usepackage{datetime}
\usepackage{bbm}
\usepackage{dsfont}
\usepackage{comment}
\usepackage[format=plain, labelfont=it, textfont=it]{caption}
\usepackage{atbegshi}
\usepackage{setspace}
\usepackage{esint}
\usepackage{enumitem}
\usepackage[unicode=true]{hyperref}
\usepackage[dvipsnames]{xcolor}
\usepackage{lipsum} 
\usepackage{rotating}
\usepackage{lipsum}
\usepackage{float}
\restylefloat{table}
\usepackage{rotating}
\usepackage[longnamesfirst]{natbib}
\usepackage{xcolor}
\usepackage{soul}
\usepackage{algorithm}

\usepackage{cleveref}
\usepackage{caption}
\usepackage{subcaption}

\usetikzlibrary{arrows.meta}
\hypersetup{
  colorlinks,
  linkcolor={red!44!black},
  citecolor={blue!44!black},
  urlcolor={blue}
}

\graphicspath{{images/}}

\usepackage{tikz}
\usetikzlibrary{fit}
\usetikzlibrary{calc}
\usetikzlibrary{arrows}
\usetikzlibrary{decorations.pathreplacing,angles,quotes}

\theoremstyle{plain}

\newtheorem{corollary}{Corollary}
\newtheorem{assumption}{Assumption}
\newtheorem*{assumption*}{Assumption}

\newtheorem{example}{Example}
\newtheorem*{example*}{Example}
\newtheorem{lemma}{Lemma}

\newtheorem{proposition}{Proposition}

\newtheorem*{theorem*}{Theorem}
\newtheorem*{prop*}{Proposition}
\newtheorem*{lemma*}{Lemma}

\newtheorem*{definition*}{Definition}

\newtheorem{ex}{Example}

\newtheorem*{continuancex}{Example \continuanceref \hspace{0.15cm}Continued}
\newenvironment{continuance}[1]
{\newcommand\continuanceref{\ref{#1}}\continuancex}{\endcontinuancex}

\makeatletter
\newcommand{\pushright}[1]{\ifmeasuring@#1\else\omit\hfill$\displaystyle#1$\fi\ignorespaces}
\newcommand{\pushleft}[1]{\ifmeasuring@#1\else\omit$\displaystyle#1$\hfill\fi\ignorespaces}
\makeatother

\newdateformat{monthdayyear}{%
 \monthname[\THEMONTH] \THEDAY, \THEYEAR}

\begin{document}

\author{Marc Claveria-Mayol\thanks{Universitat de les Illes Balears (\href{mailto:mclaveriamayol@gmail.com}{mclaveriamayol@gmail.com})} \hspace{0.2cm} \and \hspace{0.2cm} Pau Mil\'{a}n\thanks{UAB-Barcelona School of Economics (\textit{corresponding author:} \href{mailto:pau.milan@gmail.com}{pau.milan@gmail.com})} \hspace{0.2cm} \and \hspace{0.2cm} Nicol\'{a}s Oviedo-D\'{a}vila\thanks{UAB-Barcelona School of Economics (\href{mailto:nicolasoviedod@gmail.com}{nicolasoviedod@gmail.com})}} 

\title{Incentive Contracts and Peer Effects in the Workplace\thanks{We thank Yann Bramoull\'{e},  Hector Chade, Fred Deroian, Pradeep Dubey, Jan Eeckhout, Andrea Galeotti, Ben Golub, Sanjeev Goyal, Ting Liu, In\'{e}s Macho-Stadler, Mihai Manea, David Perez-Castrillo, Eran Shmaya, Ran Shorrer, participants at the 9th European Conference on Networks (Essex), 27th CTN Conference (Budapest), 2024 BSE Summer Forum (Barcelona), 2024 CMID (Budapest), 2024 EWMES and 49th SAEe (Palma), 2025 BiNoMa (Granada), 12th ERMAS (Cluj), and seminar participants at Cambridge, Columbia University, Duke University, Northwestern University, Stanford University, University of Michigan, Boston University, Johns Hopkins University, Aix-Marseille, ECARES, UAB, UCL, UPF and TSE for useful comments. Claveria-Mayol acknowledges financial support from grant PID2024-158079NB-I00, funded by MICIU/AEI/10.13039/501100011033 and ERDF/EU. Mil\'{a}n acknowledges financial support from the Spanish Ministry of Economy and Competitiveness, through grant ECO2017-83534-P and grant PID2020-116771GB-I00, and from the Severo Ochoa Program for Centres of Excellence in R\&D (CEX2019-000915-S). Oviedo-D\'{a}vila acknowledges financial support from the Spanish Ministry of Science and Innovation, through grants PID2021-122403NB-I00 and PID2022-140014NB-I00, funded by MCIN/AEI. \href{www.Refine.ink}{Refine.ink} was used to check multiple drafts of this paper for clarity and consistency. All errors are ours. Declarations of interests: none.}}

\date{\monthdayyear\today}
\maketitle
\enlargethispage{1\baselineskip}
\thispagestyle{empty}
\vspace{-12mm}
\begin{abstract}
 We analyze how firms should design wage contracts when workers collaborate in teams and effort costs depend on colleagues through a peer network. Performance-based compensation generates incentives that cascade through the organization, which firms target to boost profits. We analyze optimal incentive design if firms can---and can't---fully discriminate across workers, and when the production technology is separable or complementary across divisions. When workers' effort is substitutable, the most central workers---those who influence most colleagues directly and indirectly---receive the steepest incentives only when output risk is sufficiently large; otherwise firms prioritize workers who are closer to those they influence. When production technology exhibits complementarity across teams, stronger incentives are assigned to workers who influence colleagues in small teams that receive little influence from others. For high levels of output risk, we derive a sufficient network statistic that approximates the profit loss when firms must compensate workers of varying centrality equally. Finally, we apply our findings to organizational design questions, such as optimal firm structure and workforce investments.
\end{abstract}
\medskip{}
\textbf{\phantom{lks}JEL Codes:} D21, D23, D85, D86, L14, L22

\newpage
\pagenumbering{arabic}
\newpage
\section{Introduction}

Earnings inequality within firms has increased significantly in recent decades, with top earners capturing disproportionately large shares of total compensation. Consequently, wage distributions have become highly skewed, with increasingly pronounced right tails. One key driver of this trend is the heightened sensitivity of top earners' salaries to firm performance  through bonuses, stock options, and other forms of \textit{variable pay}: recent micro-level evidence shows that earnings of the top 1\% of U.S. employees respond four times more to firm performance than earnings at the bottom 1\%. At the same time, organizational structures have flattened, with fewer hierarchical layers and larger spans of control at the top. Empirical findings indicate that greater spans of control correlate with steeper earnings profiles within firms and across occupations, suggesting that an employee's position in an organization shapes compensation.\footnote{For a survey of earnings inequality in firms and its theoretical explanations, see \cite{neal2000theories}. \cite{jensen2004remuneration} report that CEO pay in S\&P 500 firms rose from 850,000 in 1970 to over 14 million in 2000, with stock options driving over half the increase. \cite{wallskog2024within} attribute 40\% of the CEO-median worker pay gap growth (1980–2013) to the differing pay sensitivity of the top and bottom 1\%. See \cite{bertrand2009ceos} for a survey on CEO pay.

On firm flattening, \cite{rajan2006flattening} show that the number of managers reporting to the CEO rose from about four in 1986 to over seven by 1999. Possible drivers include knowledge hierarchies \citep{garicano2000hierarchies} and trade liberalization eroding tall corporate structures \citep{guadalupe2010flattening}.

\cite{fox2009firm} finds span-of-control wage gaps widen up the hierarchy: for sales workers, rank-4 supervisors managing three times as many workers earn 1\% more, while rank-7 supervisors earn 3.4\% more. \cite{smeets2008too} confirm this, showing that managers overseeing twice the average team size earn 2.8\% more, with middle managers indirectly responsible for twice as many workers earning 4.1\% more.}

This paper ties salaries to organizational structure by analyzing how firms design wage contracts when workers influence each other's productivity. We extend a standard moral hazard model by including peer effects. In our model, motivating one worker improves the performance of others throughout the organization. 
Such spillovers appear in many workplace settings: doctors treat patients more efficiently when nurses monitor symptoms diligently; nurses work harder when matched with diligent peers; and plant managers who streamline protocols reduce subordinates' workloads. As these examples suggest, peer effects reflect assistance, knowledge, delegation, or peer pressure, and they can flow across hierarchical levels (vertical spillovers) or between co-workers at similar ranks (horizontal spillovers). Evidence shows that peer effects significantly impact productivity, contributing nearly half the total effect of performance pay in some settings \citep{ashraf2018review}.\footnote{\cite{mas2009peers} finds that the pattern of peer effects reflects firms' physical layout, while \cite{bandiera2005social} detects informal friendship bonds. Workers' effort has been found to respond to co-workers' effort even when remuneration is independent of output \citep{falk2006clean}.}

How should firms design monetary incentives to exploit productivity networks?\footnote{Previous work has focused on optimizing team composition to leverage social incentives. For instance, \cite{mas2009peers} find that by maximizing skill diversity in each shift, a supermarket chain could save up to 123,529 hours worked per year which, in 2009 wage costs, amounted to \$2.5 million per year.} Our framework integrates contract theory with network games in order to characterize optimal compensation.\footnote{Network games describe environments where agents best-reply to a ``local'' subset of other players and the overlapping sets of players define a graph, or network. \cite{ballester2006} pioneered the case of strategic complements and \cite{bramoulle2007public} the case of strategic substitutes.} In our model, risk-averse workers collaborate in a team and each worker's effort cost depends on colleagues' effort levels. The firm cannot observe individual effort but can condition wages on total output, which combines workers' efforts plus a random component. Workers receive both fixed and performance-based wages. Variable payments motivate effort but also introduce risk, requiring the firm to provide risk compensation, which reduces profits. Firms must therefore balance incentive provision against the cost of exposing workers to risk—a fundamental trade-off in classical incentive design \citep{holmstrom1987aggregation, holmstrom1991multitask, bolton2004contract}. Our theoretical contribution highlights how firms can leverage the structure of peer complementarities to distribute incentives across the organization while optimally managing workers' exposure to risk.

We analyze optimal incentive design while varying two key aspects of our environment: institutional constraints and the production technology of the firm. Specifically, we characterize incentive rules for three related cases. We first assume that the firm can write personalized wage contracts and we design incentives for two very different production technologies: a linear technology and a ``modular'' technology in which workers are divided into teams that are complementary in production. We then consider a constrained environment in which the firm cannot write personalized wage contracts and must offer the same contract to entire categories of workers.\footnote{There is a fourth possible case: that the firm's technology is modular and the firm cannot write personalized contracts. This complicates the analysis without providing additional insights, so we focus on the other three.}

Our first result (Proposition \ref{Optimal Contracts}) characterizes incentives for the simplest case (linear technology and personalized contracts). We show that each worker's performance-pay is determined as a linear combination of everyone's \textit{Bonacich centrality}---a network statistic that accumulates direct and indirect paths along the organization. This structure reflects how the firm uses incentives as an instrument to both raise output and reduce costs. On the one hand, incentives \textit{amplify} effort by triggering network spillovers, so workers with greater Bonacich centrality generate larger marginal benefits. On the other hand, incentivizing one worker makes others more tolerant to stronger incentives, lowering the marginal cost of incentivizing them. Crucially, this cost-reduction channel depends on how workers’ spheres of influence \textit{overlap}—that is, on the extent to which two workers affect the same downstream individuals, directly or through longer paths.  As a result, a worker’s incentive depends not only on her own centrality, but also on the centrality of others, with greater weight placed on those with more shared downstream influence. We show how these countervailing forces enter the optimal incentive allocation rule as a function of the firm’s level of output risk.

\begin{figure}
 \hspace{-35mm}
    \includegraphics[width=0.45\linewidth]{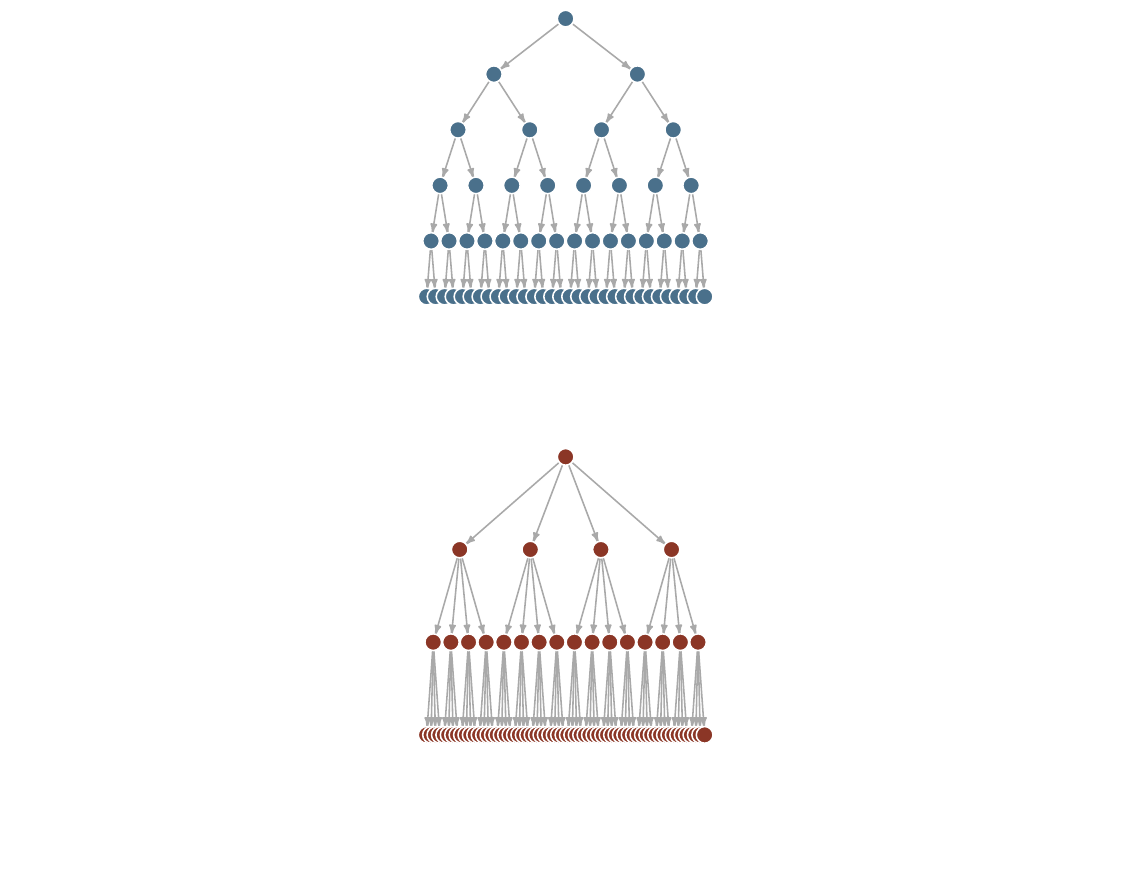}\hspace{-25mm}
    \includegraphics[width=0.49\linewidth]{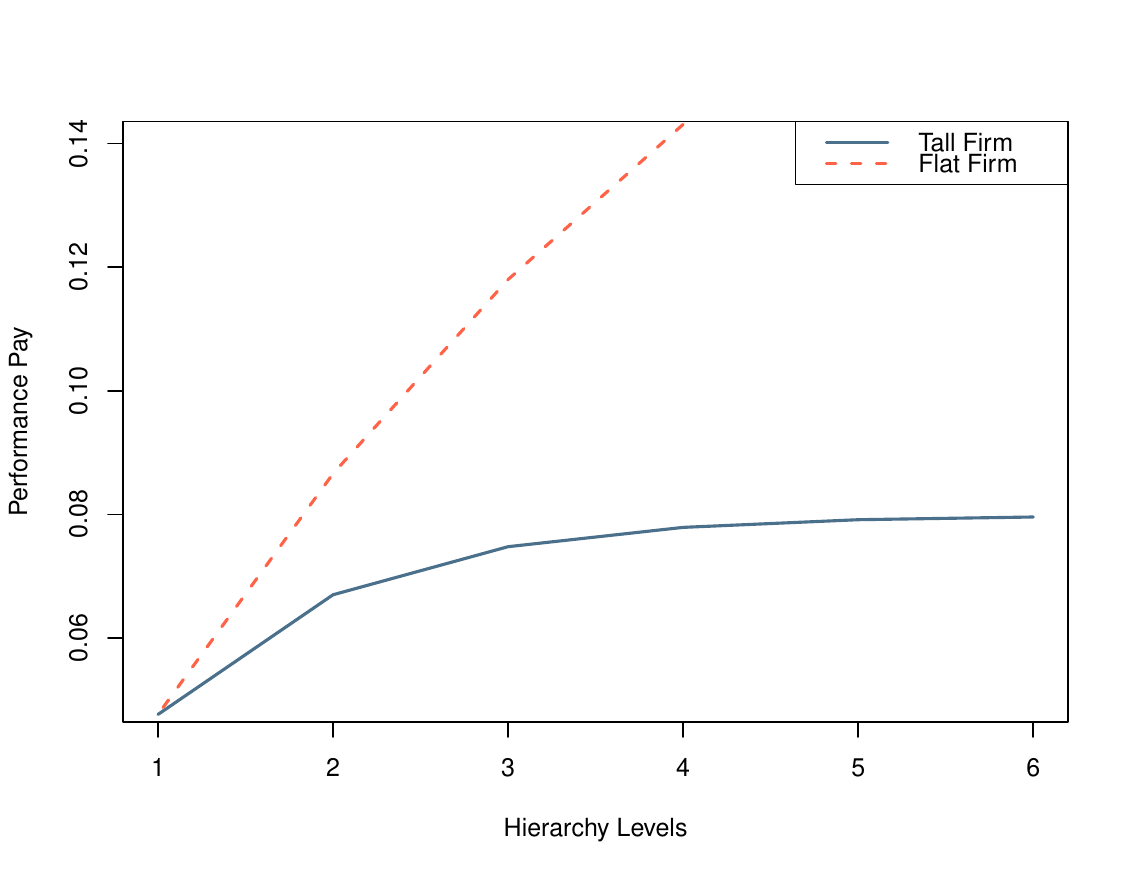}
    \hspace{-5mm}
      \includegraphics[width=0.49\linewidth]{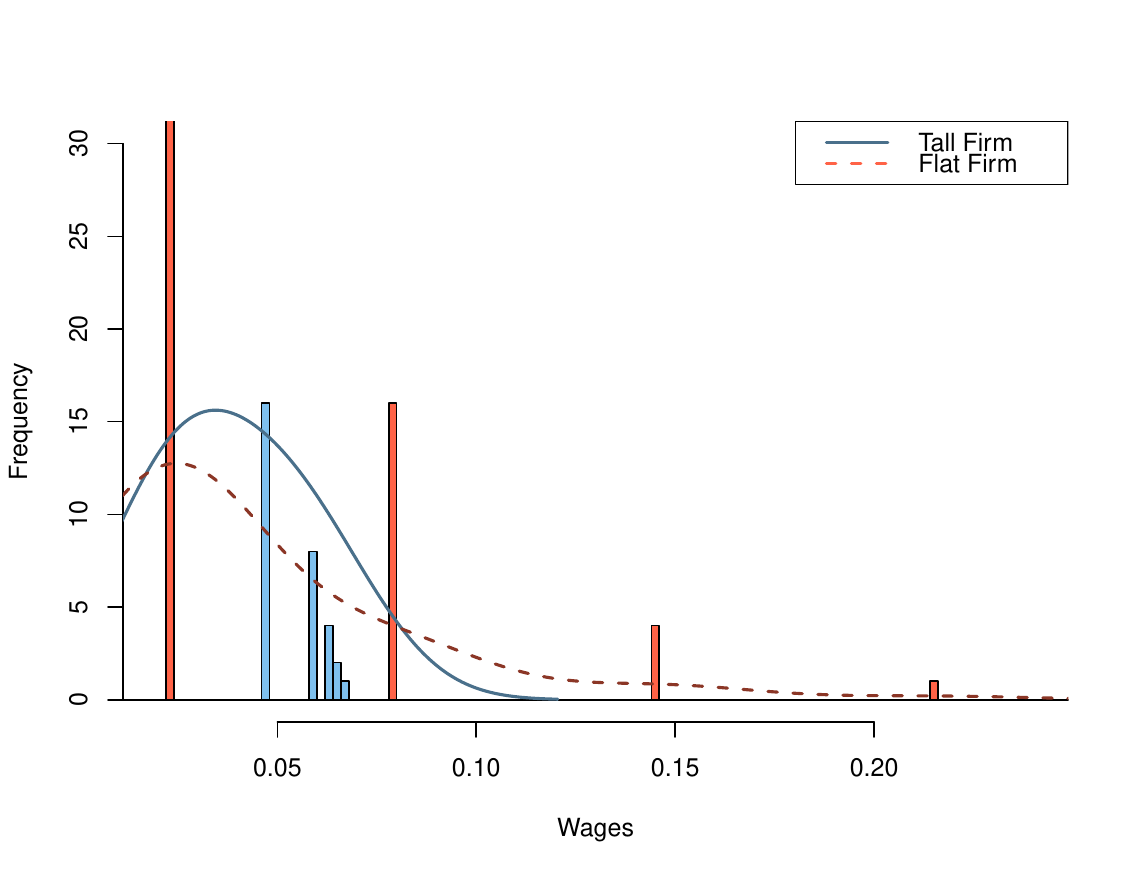}
    \caption{Performance pay and overall wage distribution for two different organizational structures: a tall (blue) firm and a flat (red) firm.}
    \label{hierarchies}
\end{figure}

Figure \ref{hierarchies} illustrates how our framework can explain the relationship between wage dispersion and organizational structure by plotting the optimal allocation rule for a simple example with two hierarchical firms. To keep things simple, assume that a worker's effort cost is affected only by her direct supervisor. This is drawn as an arrow from a worker to her subordinates.\footnote{In reality, and in our model, peer effects need not coincide with the firm's formal chain of command. We do this here to link earnings distribution with flat and tall firms.} We simulate wages for a \textit{tall firm} with 6 levels and a span of control of 2, and a \textit{flat firm} with 4 levels and a span of control of 4. The left panel shows optimal performance-pay across different hierarchy levels. Consistent with empirical findings described above, performance-pay increases as we move up the organization and, more importantly, the pay-profile is steeper for the flat firm. The right panel shows the distribution of total earnings, which sums performance-pay and fixed salaries. As in the data, the flatter firm has a longer right tail because the wage gap between top earners is much larger. In other words, optimal wage contracts with peer effects can replicate the stylized facts described above because in flatter firms upper management has shorter paths to the rest of the workforce, which makes concentrating incentives on the top more profitable when output risk is significant.\footnote{Alternative explanations for within-firm earnings inequality typically assume an initial distribution of managerial talent which is sorted across job levels and firms over time in order to produce the observed earnings distribution \citep {roy1951some, heckman1990empirical,gabaix2008has,garicano2006organization}. We complement this perspective by emphasizing that firms tie wages not only to workers' attributes but also to job positions, as evidenced by the fact that talent does not change on promotion day. Having said this, our model is flexible enough to allow for variation in individual skills.}

Second, we generalize the production function of the firm. Our analysis so far assumed that workers are substitutable in output. To address this limitation, we extend our analysis to modular production where output depends on the minimum performance across essential components. By ``modular'', we mean production processes where the final output requires successful completion of multiple distinct tasks or modules---like an assembly line where a failure at any stage compromises the entire product.\footnote{\cite{kremer1993ring} named the O-ring theory after the fatal Challenger spacecraft incident in 1986, which malfunctioned due to the failure of one small metal gasket. See also \cite{matouschek2023organizing} for a recent analysis on how to organize communication networks in a firm with modular production.} To model this, we divide the workforce into modules, and we assume the production function is substitutable within modules but perfectly complementary across modules. Since we allow for any partition of workers into modules, modular production nests our baseline environment because if the entire workforce belongs to a single module, output is linear.   

With modular production, incentives are allocated very differently (Proposition \ref{Modular Production}). A worker's performance pay depends on the entire organization---even on workers she is not connected to who are in different modules, since all modules depend on each other to generate output. Although the allocation rule is computed analogously to before, the relevant centrality measure now weighs paths by module-specific factors, which reflect how relatively costly it is for modules to match each other's performance.
We highlight two special cases. First, we turn off peer effects and show how incentives depend on relative module size: managers who validate the work of many subordinates become critical choke points, similar to senior engineers who must approve all code before deployment. We show that managerial bottlenecks create large pay disparities across hierarchical levels. Second, when every worker is essential, firms prioritize incentivizing workers with fewer incoming links rather than those with greater outgoing influence, since these workers face higher costs and are more tempted to reduce effort. More generally, we show that wage profiles exhibit sharp discontinuities between modules---contrary to the smoother profiles seen in non-modular production---and that these jumps can be much larger than the variation in pay within modules.

Third, we consider how institutional constraints affect our findings if firms are forced to offer identical contracts to multiple workers  (Proposition \ref{Coarse Contracts}). In the environments we have in mind, workers grouped under the same contract are performing the same formal role (e.g., junior associates, nurses, sales agents), and are therefore compensated according to a common wage schedule. What differs across these workers is not their job description, but their position in the informal network of peer complementarities—i.e., who influences whom, and how strongly effort propagates through the
organization.\footnote{This \textit{wage benchmarking} scenario is motivated by recent policy responses to the negative effects associated with large wage disparities. Significant disparities in peers' salaries lead to job dissatisfaction and higher quit rates \citep{card2012inequality, breza2018morale}. As a result, governments and agencies have encouraged wage transparency \citep{mas2017does,obloj2022influence, cullen2024pay}, salary benchmarking \citep{cullen2025}, and national wage setting \citep{hazell2022national}, all of which compress wages within occupational categories, especially at lower skill levels.} How should firms design coarse contracts? In this case, incentives are allocated according to a group-wide network measure, which aggregates network paths across all workers in the same job category. More importantly, forcing a common contract on workers with different network positions prevents the firm from tailoring incentives and diminishes profits. We show that, for high enough values of risk, the profit loss from these constraints can be approximated by within-group centrality variance (Proposition~\ref{profits_variance}): uniform wages impose no profit loss when all members of the same category are equally central (such as in the simplified example in Figure~\ref{hierarchies}), but in more realistic scenarios losses increase linearly with within-group variability in centrality. This finding has significant practical implications, suggesting that standardized pay structures, while equitable, can be economically costly.

Finally, while our primary focus is deriving incentive rules across different production technologies and institutional constraints, our model has broader implications for \textit{organizational design} questions---such as how firms should structure internally, or how they should invest in their workforce. To address these questions, we derive a \textit{spectral decomposition} of profits based on eigenvalues and eigenvectors of the associated network, reducing organizational complexity to its principal components. In Section \ref{orgdesign}, we leverage well-known spectral properties to establish how profits depend on the fundamental structure of the organization. Among other things, we find that profits are greater when centrality is evenly distributed, everything else equal. We also evaluate investment strategies by establishing an average connectivity threshold that determines when firms should prioritize ``team-building exercises'' that strengthen peer effects over comparable investments in individual human capital.

\bigskip
\noindent \textbf{Related Literature -} We contribute to the contract design literature with multiple agents \citep{mookherjee1984optimal, machostadler1993moral, bolton2004contract, winter2010contract}, tracing back to \cite{holmstrom1982moral} on moral hazard in teams. We link this framework to firms' organizational structure by analyzing pay-for-output incentives when workers interact via productivity spillovers. Under certain conditions, worker behavior aligns with a modified linear-quadratic network game \citep{ballester2006}. Our key contribution is extending the classic moral hazard problem with multiple agents using network-based peer effects. Unlike prior work on network-targeted incentives, we focus on classical performance-pay schemes \citep{demange2017optimal,belhaj2018targeting,galeotti2020targeting,parise2023graphon}.

The main applied contribution is to use our model to speak to questions at the heart of organizational economics. This allows us to explore adjacent questions, such as how optimal incentive design depends on the granularity of the contract and the modular structure of production. There has been significant recent interest connecting contracts to peer networks, focusing on very different topics, from relative performance compensation schemes \citep{demarzo2023finance} to the psychological costs of status concerns and the allocation of psycho-therapeutic resources \citep{sun2024contract}. In recent work \cite{shi2022optimal} studies a riskless team-production environment in which agents choose effort and bilateral helping along network links under a Cobb–Douglas technology. Unlike our analysis, there is no incentive–insurance trade-off and the principal retains no surplus. Because incentives operate through equilibrium helping, the characterization of optimal shares is implicit.

Our paper is most similar to work by \cite{dasaratha2024equity}. Relative to \cite{dasaratha2024equity}, we study linear contracts under a CARA–normal environment, which yields a certainty-equivalent problem and closed-form incentive rules. By contrast, they allow general outcome-contingent transfers over a finite state space under limited liability and describe optimal contracts through a set of necessary conditions. The treatment of spillovers also differs. In our framework peer effects operate through spillovers in workers’ costs of effort and their strength is independent of the contract, whereas in \cite{dasaratha2024equity} the contract itself affects the strength of spillovers. As a result, the two approaches emphasize different objects for empirical implementation: ours delivers closed-forms and comparative statics directly in terms of network structure and risk parameters, while they provide a balance condition that links incentives to deeper technological primitives such as the curvature of the production technology, and state-contingent marginal utilities.

Recent work on optimal price discrimination with local network effects shares many conceptual similarities with our approach. In these studies, a monopolist designs a menu of prices or discounts to leverage consumer externalities within an existing network. For example, \cite{bloch2013pricing} identify market conditions under which price discrimination may not be optimal, as the incentive to subsidize central consumers is counterbalanced by their higher willingness to pay. Similarly, \cite{fainmesser2016pricing} analyze how varying levels of information about network effects can lead to increased price discrimination and even improve overall welfare when price effects are strong. \cite{candogan2012optimal} is most similar to our analysis because they relate prices (in our case wages) to Bonacich centrality and they also consider a restricted scenario where the firm cannot fully price discriminate.

Finally, our theory provides a new set of testable predictions that speak to a strand of empirical work estimating peer effects, group composition, and team incentives in organizations \citep{hamilton2003team, cornelissen2017peer, calvo2009peer, amodio2018input}. Most of these papers emphasize how different remuneration schemes or other aspects of the contract---like employment termination decisions---affect productivity spillovers across workers. Our framework takes the peer effect structure as given and solves for the optimal contract. 

\noindent \textbf{Roadmap -} The rest of the paper is organized as follows. Section \ref{baseline} presents the baseline model (linear technology and personalized contracts) and solves the optimal contract. We provide comparative statics results and discuss negative spillovers and incomplete information. In Section \ref{sec:Modular} we generalize the firm's technology and derive optimal incentive allocation rules for modular production. In Section~\ref{sec: coarse contracts} we consider wage benchmarking and derive its impact on profits. In Section \ref{orgdesign} we summarize our results on organizational design, which we relegate to Supplementary Appendix \ref{Design} to keep the paper concise. We conclude in Section \ref{sec: conclusion} with a discussion of future lines of research. All proofs are in the Appendix.

\section{The Model}\label{baseline}

\subsection{Basic Setup}
Consider a risk-neutral firm that hires $n$ workers $N = \{1, 2, \ldots, n \}$ to conduct a joint production process.\footnote{We consider individual production in Supplementary Appendix \ref{SupApp_IndividualProduction} and show that the optimal structure of incentives is equivalent under individual and joint production.} Each worker chooses individual effort $e_i \in \mathbb{R}_+$ and the firm's production is given by 
\begin{equation*}
X(\mathbf{e}) = \sum_{i\in N} e_i + \varepsilon,
\end{equation*}
where $\mathbf{e}$ is the vector of workers' efforts and $\varepsilon \sim \mathcal{N}(0, \sigma^2)$ is an unobserved random shock to output. Because individual effort is not observable, contractual wage agreements must be based on observable (and verifiable) outcomes, such as output. We focus on the case in which the firm offers linear wage schemes of the form
\[\omega_i(X) = \beta_i + \alpha_i X,\] 
where $\beta_i$ is a fixed payment and $\alpha_i$ captures the contract's variable payment or performance-based compensation.\footnote{Although $\alpha_i$ can be thought of as a form of equity compensation whereby a share of the firm is transferred to the worker, one can also consider cases where $\sum_i \alpha_i > 1$ and $\beta_i < 0$, in which case the contract corresponds to a franchise contractual arrangement.} Although linear contracts may seem restrictive, they are parsimonious and resemble most equity payments and bonus schemes typically offered in corporate wage contracts. They are also optimal in some circumstances.\footnote{\cite{holmstrom1987aggregation} show that with continuous efforts in a dynamic setting the optimal contract is linear in the final outcome. \cite{carroll2015robustness} also demonstrates that linear contracts are optimal with limited liability and risk neutrality, particularly when the principal is uncertain about the agent's available technology.} 

A worker's cost of effort depends on the effort that is exerted by her co-workers. Following the peer effects literature, we assume a linear-quadratic function of own and neighbors' efforts: 
\begin{equation}\label{psi}
    \psi_i(\mathbf{e};\mathbf{G})= \frac{1}{2} e_i^2-\lambda e_i \sum_{j\in N} g_{ij} e_j,
\end{equation}
where $g_{ij}>0$ if $j$ is $i$'s \textit{co-worker} and $0$ otherwise.\footnote{The network $\mathbf{G}$ is allowed to be undirected or directed, and either unweighted or weighted.  Among other things, a directed and weighted network allows us to capture \textit{congestion effects}, whereby a supervisor with many subordinates might have less influence on each of them. This could be captured by defining a new network $\mathbf{G}_{D}$ where $(\mathbf{G}_{D})_{ij} = g_{ij}/(\mathbf{G}'\mathbf{1})_{j}$ if $(\mathbf{G}'\mathbf{1})_{j}>0$ and $(\mathbf{G}_{D})_{ij} = 0$ if $(\mathbf{G}'\mathbf{1})_{j}=0$.} In other words, we define $i$'s co-workers as those members of the firm that can influence $i$'s costs by exerting more or less effort. These co-workers can (but need not) reflect the formal organizational structure of the firm: they may be $i$'s subordinates, managers, or just someone that sits next to $i$. We allow for any co-worker structure and define it by a fixed and exogenous network $\mathbf{G}$. The parameter $\lambda$ captures the strength of peer effects.  When $\lambda =0$ the model reduces to the classical textbook model in \cite{bolton2004contract}. For clarity, we focus on nonnegative peer effects, $\lambda \geq 0$ in the main text, and show how our results extend to negative externalities in Appendix~\ref{sec: App Negative Spillovers}.

Workers are risk averse with constant absolute risk aversion (CARA) parameter $r$: 
\[u_i(\mathbf{e},\mathbf{G}, X; \alpha_i, \beta_i) = -\exp\left[-r\left(\omega_i(X; \alpha_i, \beta_i) - \psi_i(\mathbf{e},\mathbf{G})\right)\right]. \]
Since wages are linear and output is normally distributed, expected utility takes a tractable form as 
\begin{equation*}
 \mathbb{E}[u_i(\mathbf{e},\mathbf{G}; \alpha_i, \beta_i)] \equiv -\exp\left[-r  \: \operatorname{CE}_i(\mathbf{e},\mathbf{G}; \alpha_i, \beta_i)\right],
\end{equation*}
where the certainty equivalent of worker $i$, $\operatorname{CE}_i$, is defined as:
\begin{equation}\label{CE}
\operatorname{CE}_i(\mathbf{e},\mathbf{G}; \alpha_i, \beta_i) = \beta_i+\alpha_i\sum_{j\in N} e_j - \frac{1}{2}  e_i^2 + \lambda e_i \sum_{j\in N} g_{ij} e_j - \alpha_i^2 \frac{r\sigma^2}{2}.
\end{equation}
The above functional form is conveniently analogous to the utility functions proposed by \cite{ballester2006} and \cite{calvo2009peer}, with an additional term correcting for risk. The last term captures how adding risk into workers' compensation (through $\alpha_i$) decreases individual welfare. 

If a contract ($\alpha_i$, $\beta_i$) is acceptable, worker $i$ will optimally choose the effort level that maximizes expected utility, taking all other workers' equilibrium effort levels as given, 
\[e^{*}_i \in \arg \max_{\hat{e}_i \in \mathbb{R}_+} \operatorname{CE}_i(\hat{e}_i, \mathbf{e}^{*}_{-i}).\]
A worker accepts the contract only if the certainty equivalent in equilibrium is greater than or equal to her reservation utility, $U_i$. We consider therefore a situation in which the firm has all bargaining power and essentially makes a take-it-or-leave-it offer to the worker.\footnote{A natural extension considers what the optimal contract looks like when firms compete for workers in different industrial structures. We leave this for future work.}  To shorten the exposition and save on notation, we assume for now that $U_i=0$ for all $i$,  and we analyze the full model in Supplementary Appendix \ref{SuppApp_General}. Therefore, we can write the IR constraint as
\[\operatorname{CE}_i(\mathbf{e}) \geq 0.\]

The firm will select a contract profile (represented by vectors $\boldsymbol{\alpha}$ and $\boldsymbol{\beta}$) that maximizes expected profits subject to these constraints:
\begin{align*}
 \max_{\boldsymbol{\alpha}, \boldsymbol{\beta}} \: \mathbb{E}[\pi(\mathbf{e}\mid \boldsymbol{\alpha}, \boldsymbol{\beta})]&= \sum_{i\in N} e_{i} - \sum_{i\in N}\left(\beta_i + \alpha_i \sum_{j \in N} e_j\right) \\
 \text{subject to:} \phantom{d}  \operatorname{CE}_i(\boldsymbol{\alpha}, \boldsymbol{\beta}, \mathbf{G}) &\geq 0, \, \forall i \in N\quad \tag{IR} \\
  \phantom{dd} e_i \in \arg \max_{\hat{e}_i \in \mathbb{R}_+} & \, \operatorname{CE}_i(\hat{e}_i, \mathbf{e}_{-i}), \, \forall i \in N.\quad \tag{IC}
\end{align*} 

    \subsubsection{Main Modeling Assumption}

Before solving the model, we briefly discuss why we choose to model peer effects through workers' cost structure, as defined in equation \eqref{psi}. There are two natural ways to introduce interactions across workers. One is through the production function (a technology channel), where effort complementarities affect output. The other is through the cost of effort (a preference-based channel), where coworkers affect the marginal dis-utility of effort. Both channels may operate simultaneously, while in some environments one effect may dominate.

This paper focuses on the preference-based channel in order to isolate a mechanism that has been documented repeatedly in settings where individual production is technologically separable.\footnote{For example, \cite{bandiera2005social} study a fruit-picking setting in which workers are assigned separate rows and individual productivity depends exclusively on their own effort and the fruit available on those rows. Despite the absence of technological complementarities, they find strong peer effects.} We interpret this evidence as suggesting that coworkers can affect how effort is subjectively experienced, not just how each unit of effort maps into output. This can happen by raising the social cost of shirking or creating a workplace culture in which sustained effort feels lighter. For instance, this is how \cite{mas2009peers} model peer effects in their theoretical approach. We view our approach as building on this by introducing rich heterogeneity through the interaction structure.

At the same time, we recognize that technological complementarities are natural. Our approach is not meant to rule out this channel, and a more complete model would incorporate both. Incorporating bilateral effort complementarities through production, however, complicates the analysis by coupling the firm's contracting problem with the strength of strategic interactions: changing incentives affects how strongly wages respond to output, which in turn alters workers' responses to their coworkers' effort. This feedback makes explicit solutions to the contracting problem more difficult to obtain. \cite{dasaratha2024equity} make progress by characterizing relative incentives conditional on the active set in a special case, while acknowledging that an explicit characterization with general networks and heterogeneous standalone productivities remains unavailable. Conversely, when peer effects enter through effort costs, the coefficients governing strategic interactions are independent of the contract. This separation keeps our contract-design problem tractable and allows us to derive optimal incentive levels explicitly in terms of network structure and risk parameters.

Finally, modeling peer effects through effort costs does not mean utility levels and effort incentives move in the same direction, nor does it require production to be additively separable. Appendix~\ref{SuppApp:MoreFlexibleExternalities} allows for more flexible preference interactions, including peer effects that encourage effort while lowering utility. Section~\ref{sec:Modular} introduces technological complementarities through modular production while preserving tractability. These extensions show that the framework accommodates richer preference and production structures while retaining closed-form incentive rules.

\subsection{Optimal Incentive Contracts}

We now characterize the optimal contract.  Consider first the optimal effort decision of the worker for any contract $\boldsymbol{\alpha}, \boldsymbol{\beta}\in \mathbb{R}^{n}$. 
Workers play a non-cooperative game similar to that in \cite{ballester2006}. The best-reply function of worker $i$ is given by 
\begin{equation}
e_i^{*}(\mathbf{e}_{-i}) = \max\left\{ 0, \alpha_i + \lambda \sum_{j\in N} g_{ij} e_j \right\}, \quad \forall i\in N.
\label{eq:BestResponse}
\end{equation}
Notice that the contract's fixed payment $\beta_i$ has no effect on workers' effort incentives. A worker is motivated to work only through performance-based compensation $\alpha_i$, and by the actions of peers. 

If the firm were to select a contract $\boldsymbol{\alpha}\in \mathbb{R}^n$ such that worker $i$ remains inactive ($e_i^*=0$), it must be that
$$
\alpha_i \leq-\lambda \sum g_{i j} e_j \leq 0 .
$$
In this case, the firm could raise $\alpha_i$ to the boundary value $-\lambda \sum_j g_{ij}e_j$ and adjust $\beta_i$ in order to keep the participation constraint binding. This preserves worker $i$'s zero-effort best reply while weakly reducing the risk premium and hence the firm's expected wage cost. The firm can therefore restrict attention, without loss of generality, to contracts for which equilibrium effort satisfies
\begin{equation}\label{FOCs}
(\mathbf{I}-\lambda \mathbf{G}) \mathbf{e}^*=\boldsymbol{\alpha} .
\end{equation}
For these contracts, the linear equations, together with $\mathbf{e}^* \geq \mathbf{0}$, characterize workers' best responses, including at zero effort. We impose the following standard restriction.

\begin{assumption}\label{assumption1}
The spectral radius of $\lambda\mathbf{G}$ is less than 1.
\end{assumption}
Under Assumption \ref{assumption1}, the equilibrium is unique and can be written as
$$
\mathbf{e}^*=\mathbf{C} \boldsymbol{\alpha}, \quad \mathbf{C}:=(\mathbf{I}-\lambda \mathbf{G})^{-1},
$$
where $\mathbf{C}$ is entrywise nonnegative. Finally, the firm sets fixed payments $\beta_i$ in order to extract all surplus from workers, such that $\operatorname{CE}_i(\mathbf{e})=0$.  We can therefore reformulate the firm's problem as:\footnote{We use the first-order approach and replace IC constraints by workers' first-order conditions. The worker’s objective is strictly concave in own effort: given others’ efforts, the certainty equivalent is quadratic with negative curvature and strategic interactions enter linearly. Hence the FOC yields the unique global maximizer and fully characterizes IC.}
\begin{align*}
 \max_{\boldsymbol{\alpha}, \boldsymbol{\beta}} \: \mathbb{E}[\pi(\mathbf{e}\mid \boldsymbol{\alpha}, \boldsymbol{\beta})] \\
 \text{subject to: }\phantom{d} \operatorname{CE}_i(\boldsymbol{\alpha}, \boldsymbol{\beta}, \mathbf{G}) &= 0, \, \forall i \in N\quad \tag{IR} \\
 \mathbf{e}^{*} = \mathbf{C}\boldsymbol{\alpha},   \, \, \mathbf{e}^{*}&\geq \mathbf{0} .\tag{IC}
\end{align*} 
As with the worker's problem, the firm's problem is concave provided peer effects are sufficiently weak. However, concavity of the firm's problem requires a further restriction on $\lambda$ than Assumption 1:

\begin{assumption}\label{spectralradius}
     The spectral radius of $\lambda^{2}/(1+r\sigma^{2})(\mathbf{GC})'\mathbf{GC}$ is less than 1.
 \end{assumption}

We are now ready to characterize the optimal incentive rule. By Assumption \ref{assumption1}, $\mathbf{C}:=(\mathbf{I}-\lambda \mathbf{G})^{-1}=\sum_{q=0}^{\infty} \lambda^q \mathbf{G}^q$ so $\mathbf{C}^{\prime} \mathbf{1}$ is the vector of \textit{Bonacich centralities}. Its  $i$-th component, $b_i(\lambda)$, aggregates all incentive paths (of all lengths) emanating from worker $i$.\footnote{We use column sums of $\mathbf{C}$, rather than the usual row sums, to capture outgoing paths from each worker to those they influence. See \cite{ballester2006} and \cite{jackson2008social} for further details on Katz-Bonacich and related centrality measures.}  The following result shows that the optimal incentive rule, $\boldsymbol{\alpha}^{*}$, is a linear transformation of this vector.\footnote{The fixed payments $\boldsymbol{\beta}$ ensure that workers receive their reservation utilities. Their expression, which depends on outside options and $\boldsymbol{\alpha}$, is uninformative and we provide it in the Appendix.} 
\begin{proposition}[Optimal Incentives Rule]\label{Optimal Contracts}
        Under Assumptions $\ref{assumption1}$ and $\ref{spectralradius}$, there exists a unique profit-maximizing incentive rule for any peer network $\mathbf{G}$ given by
        \begin{equation}\label{alpha}
            \boldsymbol{\alpha}^{*} = \mathbf{W}\mathbf{C'1},
        \end{equation}     
 where $\mathbf{W}=\left[(1+r\sigma^2) \mathbf{I}- (\lambda\mathbf{GC})'\lambda\mathbf{GC} \right]^{-1}$ and $\mathbf{C}=(\mathbf{I}-\lambda\mathbf{G})^{-1}$.
   \end{proposition} 

To understand Proposition \ref{Optimal Contracts}, consider the marginal benefit of increasing worker $i$'s performance pay. At the margin, worker $s$ will respond to $i$'s incentives in proportion to how many \textit{incentive paths} (of any length) lead to $s$ starting from $i$:\footnote{To see this take a derivative of the IC constraint.} 
\begin{equation}\label{incentivepaths}
  \frac{\partial e_s}{\partial \alpha_i}=\sum_{q=0}^\infty \lambda^q \mathbf{G}^q_{si}\geq 0  .
\end{equation}
If $\frac{\partial e_s}{\partial \alpha_i}>0$, we say that $s$ is an \textit{incentive target} of $i$. The total marginal benefit of $\alpha_i$ is therefore obtained by summing  across all of $i$'s targets. This equals  the \textit{Bonacich centrality measure}, which captures the total \textit{amplification potential} of worker $i$:
$$MB_{\alpha_i}= \sum_s\frac{\partial e_s}{\partial \alpha_i}  =\sum_s \sum_{q=0}^\infty\lambda^q \mathbf{G}^q_{si}:=b_i(\lambda).$$
In other words, workers with larger centrality generate larger downstream responses from others, raising the \textit{marginal benefit} from additional incentives.

We now turn to the \textit{marginal cost} of incentives: the change in the firm's expected wage bill from raising incentives. Because the firm sets fixed payments $\beta_s$ so that workers' participation constraints bind, raising $\alpha_i$ affects the firm's expected wage bill both by changing workers' equilibrium effort costs and by exposing worker $i$ to additional wage risk. Formally, the marginal cost associated to $\alpha_i$ is:
\begin{equation}\label{MCalpha}
  MC_{\alpha_i}=\underbrace{\sum_s e_s \frac{\partial e_s}{\partial \alpha_i}-\lambda \sum_s \sum_\ell g_{s \ell}\left(e_s \frac{\partial e_\ell}{\partial \alpha_i}+e_\ell \frac{\partial e_s}{\partial \alpha_i}\right)}_{\text{effort costs}}+ \underbrace{\: \phantom{\sum_i}r\sigma^2 \alpha_i \phantom{\sum_i} }_{\text{risk exposure}}.
\end{equation}
At the same time, $e_s$ depends on many workers' incentives since $e_s=\sum_j \alpha_j \frac{\partial e_s}{\partial \alpha_j}$ (as seen from the IC constraint). Substituting this in equation \eqref{MCalpha}, we obtain: 
\begin{equation}\label{MCalphalong}
MC_{\alpha_i}= \sum_s \sum_j \alpha_j \frac{\partial e_s}{\partial \alpha_j} \frac{\partial e_s}{\partial \alpha_i}-\lambda \sum_s \sum_\ell g_{s \ell} \sum_j \alpha_j\left(\frac{\partial e_s}{\partial \alpha_j}\frac{\partial e_\ell}{\partial \alpha_i} +\frac{\partial e_\ell}{\partial \alpha_j}\frac{\partial e_s}{\partial \alpha_i} \right) + r\sigma^2\alpha_i.
\end{equation}
Equation \eqref{MCalphalong} shows how one worker's incentives affect the expected compensation cost of incentivizing another worker. It operates through two opposing channels: 

\begin{enumerate}
    \item The first term on the right hand side shows that incentivizing any worker $j$ who shares incentive targets with $i$ raises the marginal cost of $\alpha_i$. This is because incentivizing $j$ pushes $j$'s targets to exert more effort, which means that they become more cost-sensitive (i.e., less tolerant) to additional incentives from $i$.  As a result, $\alpha_j$ and $\alpha_i$ are \textit{substitutes} (see Figure \ref{fig: substitutes}). 

    \item The second term shows that incentivizing a worker $j$ who influences the \textit{neighbors} of $i$'s targets (rather than the targets themselves) lowers $\alpha_i$'s marginal costs. This is because incentivizing $j$ again pushes $j$'s targets to exert more effort, which through complementarity links makes their neighbors more tolerant to additional incentives; this includes $i$'s target so, as a result, $\alpha_i$ and $\alpha_j$ are \textit{complements} (see Figure \ref{fig:complements}). 

\end{enumerate}

 \begin{figure}
     \begin{minipage}{0.4\textwidth}
     \centering
\begin{tikzpicture}
    \node[circle,draw=blue!50 ,minimum width=6mm,minimum height=6mm, fill=blue!10] (i) at (0,0) {\small $s$};
    \node[circle,draw=red!50, fill=red!10,minimum width=6mm,minimum height=6mm,inner sep=0pt] (j) at (-1.5,2) {\small $i$};
    \node[circle,draw=red!50, fill=red!10, minimum width=6mm,minimum height=6mm,inner sep=0pt] (k) at (1.5,2) {\small $j$};

    \draw[dashed, thick] (i) -- (-1.1, 1.1);
    \draw[solid, thick] (i) -- (-0.6, 0.6);
    \draw[dashed, thick] (i) -- (1.1,1.1);
     \draw[solid, thick] (i) -- (0.6, 0.6);
    \draw[dashed, thick] (i) -- (1,-1.5);
     \draw[solid, thick] (i) -- (0.5, -0.75);
     
    \draw[->, -{Triangle}, decorate, decoration={snake, amplitude=0.7mm, segment length=5.5mm},  thick, color= red!50] (j) to[bend right=20] (-0.5,0.1);
    \draw[->, -{Triangle}, decorate, decoration={snake, amplitude=0.7mm, segment length=5.5mm},  thick, color= red!50] (k) to[bend left=20] (0.5,0.1);
    
    \node at (-2.3,0.7) {\small $\frac{\partial e_s}{\partial \alpha_i} > 0$};
    \node at (2.3,0.7) {\small $\frac{\partial e_s}{\partial \alpha_j} > 0$};

\end{tikzpicture}
     \caption{\textbf{Substitutes:} Workers $i$ and $j$ both have incentive paths leading to worker $s$. This means that raising $\alpha_i$ increases the marginal cost of $\alpha_j$. }\label{fig: substitutes}
     \end{minipage}
     \hfill
     \begin{minipage}{0.55\textwidth}
     \centering
\begin{tikzpicture}
    \node[circle,draw=blue!50, minimum width=6mm,minimum height=6mm, fill=blue!10] (i) at (0,0) {\small $s$};
    \node[circle,draw=red!50, fill=red!10,minimum width=6mm,minimum height=6mm,inner sep=0pt] (k) at (-1.7,2) {\small $i$};
    \node[circle,draw=blue!50,minimum width=6mm,minimum height=6mm, fill=blue!10,inner sep=3pt] (j) at (3,0.8) {\small $\ell$};;
\node[circle,draw=red!50, fill=red!10,minimum width=6mm,minimum height=6mm,inner sep=0pt] (s) at (4.6,-1.3) {\small $j$};

    \draw[solid, thick] (i) -- (j);
    \draw[dashed, thick] (i) -- (-1.1,1.1);
     \draw[solid, thick] (i) -- (-0.6, 0.6);
    \draw[dashed, thick] (i) -- (-1,-1.5);
     \draw[solid, thick] (i) -- (-0.5, -0.75);
      \draw[dashed, thick] (j) -- (4,-1);
     \draw[solid, thick] (j) -- (3.61, -0.3);
     
    \draw[->, -{Triangle}, decorate, decoration={snake, amplitude=0.7mm, segment length=5.5mm},  thick, color= red!50] (k) to[bend right=20] (-0.5,0.1);
    \draw[->, -{Triangle}, decorate, decoration={snake, amplitude=0.7mm, segment length=5.5mm},  thick, color= red!50] (s) to[bend right=20] (3.4,0.4);
    
    \node at (-2.3,0.7) {\small $\frac{\partial e_s}{\partial \alpha_i} > 0$};
    \node at (5.3,-0.3) {\small $\frac{\partial e_\ell}{\partial \alpha_j} > 0$};

\end{tikzpicture}
      \caption{\textbf{Complements:} Worker $i$ has an incentive path leading to worker $s$, who is a neighbor of $j$'s incentive target, $\ell$. This means that raising $\alpha_i$ decreases the marginal cost of $\alpha_j$. }  \label{fig:complements}
      \end{minipage}
   
 \end{figure}

Incentives can operate through both substitution and complementarity channels simultaneously. Every complementarity effect is accompanied by a substitution effect along a path with one additional step. For instance, workers $i$ and $j$ in Figure \ref{fig:complements} also share worker $s$ as a common target, so they must also be related as in Figure \ref{fig: substitutes}. This implies that we can collect terms in equation \eqref{MCalphalong} and reduce it.  After collecting such terms, the firm's profit-maximizing conditions, $MB_{\alpha_i}=MC_{\alpha_i}$, can be written as follows:
\begin{equation}\label{systemprofits}
b_i(\lambda)=\left(1+r \sigma^2\right) \alpha_i-\sum_j \alpha_j \sum_s\left(\lambda\sum_{\ell} \frac{\partial e_\ell}{\partial \alpha_{j}} g_{s\ell}\right)\left(\lambda\sum_m \frac{\partial e_m}{\partial \alpha_{i}} g_{sm}\right) \quad \quad \text{for all }\: i\in N.
\end{equation}
Equation \eqref{systemprofits} gives the economic meaning of \textit{common influence}. For $\alpha_j$ to lower the marginal compensation cost of $\alpha_i$, workers $i$ and $j$ must commonly influence some worker $s$ through incentive paths reaching $s$'s neighbors. Thus, common influence is not simply a graph pattern: it identifies precisely the network configurations through which one worker's incentives reduce the expected wage cost of incentivizing another.

 \begin{figure}
    \centering

    \begin{tikzpicture}[scale=0.7]
        \tikzset{
            blueNode/.style={circle, draw=blue!50, fill=blue!10, thick, minimum size=3.5mm, inner sep=0.5mm},
            redNode/.style={circle, draw=red!50, fill=red!10, thick, minimum size=3.5mm, inner sep=0.5mm},
        }

        \node[redNode] (R1) at (1,2) {1};
        \node[redNode] (R3) at (5,2) {3};
        \node[redNode] (R5) at (9,2) {5};
        
        \node[blueNode] (B2) at (3,0) {2};
        \node[blueNode] (B4) at (7,0) {4};

        \draw[->,-{Triangle}] (R1) -- (B2);
        \draw[->,-{Triangle}] (R3) -- (B2);
        \draw[->,-{Triangle}] (R3) -- (B4);
        \draw[->,-{Triangle}] (R5) -- (B4);

    \end{tikzpicture}
    
    \caption{A directed network with 5 workers.}
    \label{fig:directed_graph}
\end{figure}
 
\begin{ex}[Common Influence] \label{example1}
Consider the network in Figure \ref{fig:directed_graph}. Workers $1$ and $3$ commonly influence worker $2$, while workers $3$ and $5$ commonly influence worker $4$. These pairs interact directly through the cost-reduction term in equation \eqref{systemprofits}. Workers $1$ and $5$, by contrast, do not directly share a target. Nevertheless, as we show below, solving the system of optimality conditions links them indirectly because both share common influence with worker $3$.
   \end{ex}

Finally, stacking the $n$ conditions in equation \eqref{systemprofits} gives
$
\mathbf{C}^{\prime} \mathbf{1}=\mathbf{W}^{-1} \boldsymbol{\alpha} .
$
Inverting this system returns the optimal incentive rule in Proposition \ref{Optimal Contracts}. Hence, for each worker $i$,
$$
\alpha_i^*=\sum_{j \in N} w_{i j}\left(\lambda, \sigma^2\right) b_j(\lambda), \quad \forall i \in N,
$$
where $w_{i j}$ is the $(i, j)$ element of the symmetric weighting matrix $\mathbf{W}$. The weight $w_{i j}\geq 0$ captures the interaction between workers $i$ and $j$ in the firm's marginal wage cost, as summarized by the degree of common influence between them---i.e., not only when $i$ and $j$ directly share common influence, but also when connected through longer chains of common influence.\footnote{To see this, note that $\lambda(\mathbf{GC})_{ij}$ contains the same information as $\mathbf{C}_{ij}$ but ignores walks of length zero. Thus, $(\lambda\mathbf{GC})'\lambda\mathbf{GC}$ forms a symmetric matrix where each $(i,j)$ element sums over workers indirectly influenced by both $i$ and $j$. Under Assumption \ref{spectralradius}, the weight matrix $\mathbf{W}$ can be expressed as a geometric series of these common-influence matrices:  
\[
\mathbf{W} = \frac{1}{1+r\sigma^{2}}\sum_{q=0}^{\infty} \left(\frac{\lambda^2}{1+r\sigma^{2}}\right)^{q}\left((\mathbf{GC})'\mathbf{GC}\right)^{q}.
\]
Each power $((\mathbf{GC})'\mathbf{GC})^q$ tracks the (discounted) common influence of two workers through $q$ intermediaries.}

\begin{continuance}{example1}[Chains of Common Influence]
Consider worker $1$'s incentives in Figure \ref{fig:directed_graph}. Worker $1$ only shares direct common influence with worker $3$. Applying equation \eqref{systemprofits} we have that 
$b_1 = (1+r\sigma^2 -\lambda^{2}) \alpha_1 - \lambda^2 \alpha_3$.
However, worker $5$ has a similar expression that also depends on $\alpha_3$. Solving the system \eqref{systemprofits} reveals that $\alpha_1$ indeed depends on $b_5$, because $w_{1,5}= \frac{\lambda^4}{\left(1+r \sigma^2\right)(1+ r\sigma^{2}-\lambda^{2})(1+r\sigma^{2}-3\lambda^{2})}>0$. In other words, since workers $1$ and $5$ both share common influence with $3$, they are indirectly linked by a chain of common influence.
\end{continuance}

\begin{figure}[t]
\centering

        \includegraphics[scale=0.5]{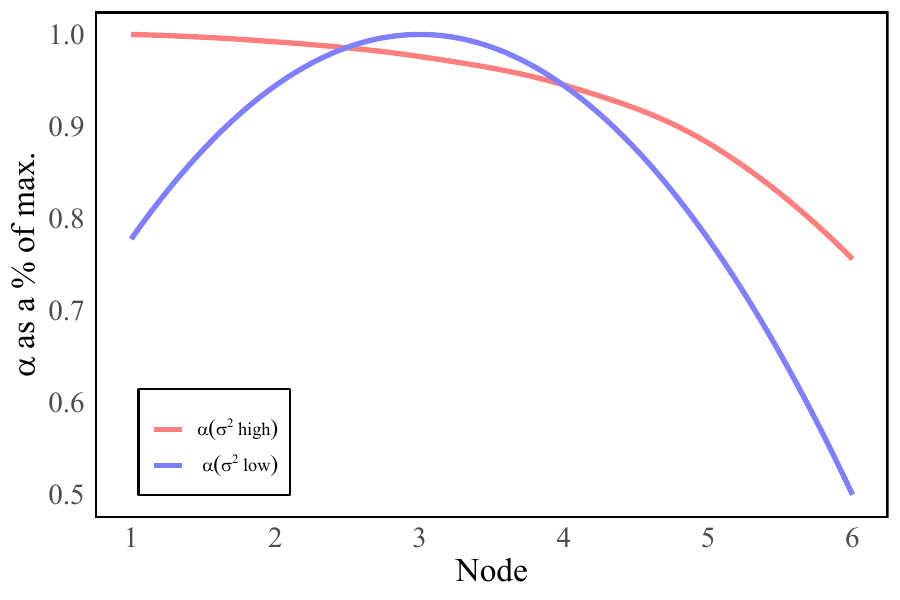}
   
     \begin{tikzpicture}[node distance=2cm, every node/.style={circle, draw, minimum size=6mm}]
    \tikzset{
            blueNode/.style={circle, draw=blue!50, fill=blue!10, thick},
            redNode/.style={circle, draw=red!50, fill=red!10, thick},
        }
        \node[blueNode] (1) {1};
        \node[blueNode] (2) [right of=1] {2};
        \node[blueNode] (3) [right of=2] {3};
        \node[blueNode] (4) [right of=3] {4};
        \node[blueNode] (N) [right of=4, xshift=2cm] {$n$};
        
        \draw[->, -{Triangle},thick] (1) -- (2);
        \draw[->, -{Triangle},thick] (2) -- (3);
        \draw[->, -{Triangle},thick] (3) -- (4);
        \draw[-, ,thick] (4) -- (7.1,0);
        \draw[dashed,,thick] (7.2,0) -- (8.5,0);
        \draw[->, -{Triangle}, thick] (8.5,0) -- (N);
    \end{tikzpicture}
    \caption{In the supermarket chain network from \cite{mas2009peers} it can pay off (if risk is low) to incentivize nodes in the middle more than nodes on the left, even though nodes on the left are more central.}
    \label{fig:chain}
\end{figure}

To sum up, incentives affect the firm's profits through two channels: they amplify output by stimulating effort throughout the network, and they reduce the marginal compensation cost of providing incentives to other workers. The optimal contract must account for both effects. Consequently, a worker with lower Bonacich centrality may receive stronger incentives than a more central worker if increasing the former's incentives generates sufficiently large cost reductions elsewhere in the organization. Figure \ref{fig:chain} illustrates this possibility: when risk is low, a worker in the center of a line can receive stronger incentives despite having lower centrality. As output risk increases, however, the direct risk-premium cost of incentives becomes relatively more important, weakening the influence of cross-worker wage bill reductions on the optimal allocation. When risk is sufficiently high, incentives are ordered according to Bonacich centrality (see the red line in Figure \ref{fig:chain}). This leads to the following allocation rule.

\begin{proposition}[Monotonicity]\label{monotonic}
   There exists a value $\bar{\sigma}^2(\mathbf{G})$ such that if $\sigma^2>\bar{\sigma}^2$ optimal incentives preserve the strict ordering of Bonacich centrality: $b_i(\lambda)>b_j(\lambda) \implies \alpha_i^{*} >\alpha_j^{*}$. 
\end{proposition}

Before moving on, it is useful to consider our incentive rule when there are no spillovers. We can see this result directly from equation \eqref{systemprofits}. Since no incentive paths exist when $\lambda=0$ (i.e., $\partial e_s / \partial \alpha_i = 0$ for all $i\neq s$), then no amplification is possible and $b_i(0)=1\: \forall i$. Equation \eqref{systemprofits} boils down to $1=(1+r\sigma^2)\alpha_i$, which yields the textbook result. 
\begin{corollary}[No Peer Effects]\label{no peer effects}
    In the absence of peer effects (i.e., $\lambda=0$) incentives are constant across workers and equal to
    \[\alpha_i^{*} = \frac{1}{1+r\sigma^2}, \quad \forall i\in N.\] 
\end{corollary}
Moreover, when risk is also zero (i.e., $\sigma^2=0$), the principal effectively makes each worker a residual claimant on output (i.e., sets $\alpha_i=1$ for all $i$). As we discuss below, this induces the first-best level of effort.

\subsubsection*{Implications for Effort Allocation} 

Having characterized the optimal contract, it is natural to ask what
pattern of effort it induces in equilibrium. Do more central workers
exert greater effort? How does the resulting effort profile compare with the social optimum?

As shown in the previous section, when risk is sufficiently large, the
optimal incentive vector $\boldsymbol{\alpha}$ is ordered according to
workers' outward centrality. However, even in this case more central
workers need not exert more effort. Effort depends on \emph{inward}
exposure-- notice that $e_i$ is determined by $(\mathbf C\boldsymbol{\alpha})_i$ and not by $(\mathbf C^\prime \boldsymbol{\alpha})_i$-- so a worker who is not very influential may nonetheless exert high effort if she is
strongly influenced by highly central co-workers upstream in the
network. In other words, while centrality governs how incentives
propagate through the network, equilibrium effort reflects the
centrality of one's neighbors rather than one's own amplification potential.

It is also useful to compare the induced effort allocation to the
social optimum. In Supplementary Appendix~\ref{Sec: First-Best} we characterize the socially optimal  profile, maximizing workers' and firms' joint payoff,
\[
\mathbf{e}^{0}
=
\left(\mathbf{I}-\lambda(\mathbf{G}+\mathbf{G}')\right)^{-1}\mathbf{1},
\]
and show that when $\sigma^2=0$ the optimal contract in Proposition~\ref{Optimal Contracts} implements this profile as a Nash equilibrium of the workers' game. In other words, surplus is lost in our model because the provision of incentives shifts risk from the risk-neutral principal to risk-averse agents, rather than because individual effort is unobservable per se. In fact, when $\sigma^2 = 0$, the firm perfectly observes group output but cannot contract on individual effort. This corresponds to what \citet{holmstrom1982moral} calls ``control deficiency'': agents can cover improper actions behind the uncertainty of who was at fault. \citet{holmstrom1982moral} analyzes a version of our model with $\lambda=\sigma^2=0$ and argues that, in order to penalize workers sufficiently for a deviation, they would have to be made residual claimants on output (i.e., $\alpha_i=1$) which would break the budget: $\sum_i \alpha_i=1$. 
In our setting, the presence of a principal allows this constraint to be relaxed: the firm can use fixed transfers to implement efficient effort (recall Corollary \ref{no peer effects}) while extracting residual surplus. More importantly, we show in Supplementary Appendix~\ref{Sec: First-Best} that Holmstrom's logic extends to environments with network spillovers (i.e., $\lambda \neq 0$).

\subsubsection*{Comparative Statics: The Strength of Connections}

How should firms react to changes in the structure of spillovers? One might think that when worker $j$'s influence over $i$ intensifies, the firm should decrease incentives for other workers and concentrate them on $i$ and $j$'s strengthened relationship. We show below that this is not optimal.

\begin{proposition}[An Increase in Link Strength]\label{CS_Link}

Every worker's incentive pay {\normalfont weakly} increases in any link's strength (i.e., $\partial \alpha_s / \partial g_{ij}\geq 0$ for all $i,j,s\in N$). Moreover, an increase in $g_{ij}$  {\normalfont 
 strictly} increases the incentive pay of worker $j$ and any worker $s$ who has a common influence with $j$ (i.e., $w_{j,s}>0$). 
\end{proposition}

Proposition \ref{CS_Link} states that incentives and efforts do not decrease as \emph{any} link is strengthened. It turns out that the firm recognizes that incentive spillovers travel in all directions, even if the network is directed, and thus allocates more incentives to all influential workers that share influence with $j$.

\subsubsection{Heterogeneous Workers}

We have assumed for simplicity that workers only differ in their connections, but are otherwise identical in  productivity, risk aversion, and outside options. In Supplementary Appendix \ref{SuppApp_General} we provide a general characterization of optimal contracts for arbitrary heterogeneity in these parameters.  One particularly interesting case considers differences in productivity, with output defined as $X(\mathbf{e})=\sum_i \theta_i e_i + \varepsilon$. In this environment, incentives propagate through the network exactly as in the baseline model, but their impact is scaled by $\boldsymbol{\theta}=(\theta_1,\ldots,\theta_n)^\prime$. More productive workers respond more strongly to incentives: at the level of the best-response (holding others' effort fixed), the direct effect is proportional to $\theta_i$. In equilibrium, this response is amplified through the network via $\mathbf{C}$, so $\frac{\partial e_j}{\partial \alpha_i}=C_{ji}\theta_i$.  Moreover, the effort they induce in others generates more output when those targets are themselves productive $\frac{\partial X(\mathbf{e})}{\partial e_j}=\theta_j$. As a result, productivity enters the marginal benefit of incentives twice: through the responsiveness of the worker receiving incentives and through the productivity of her incentive targets. Formally, $M B_{\alpha_i}= \sum_j \theta_j \frac{\partial e_j}{\partial \alpha_i} =\sum_j \theta_j C_{j i} \theta_i$. This leads to a modified incentive rule: 
$$\boldsymbol{\alpha}^{*}= \mathbf{W_{\theta}\boldsymbol{\Theta}C^\prime}\boldsymbol{\Theta}\mathbf{1},$$ where $\boldsymbol{\Theta} = \operatorname{diag}(\boldsymbol{\theta})$, and  $\mathbf{W}_\theta$ is a modified weighting matrix of common influence paths, which are similarly weighted by the productivity parameters of source and target.

We also show in Supplementary Appendix \ref{SuppApp_General} that heterogeneity in risk aversion and outside options does not alter workers' incentives to exert effort, but instead modifies the cost the firm must incur to implement a given contract, via the participation constraints that determine fixed payments $\boldsymbol{\beta}$. With risk aversion, this cost effect depends on the incentive scheme--steeper incentives increase risk exposure and require higher compensation--so it affects the optimal incentive rule through $\mathbf{W}$ only. Outside options, instead, shift fixed payments uniformly across incentive schemes and therefore have no effect at all on the optimal incentive rule. Finally, heterogeneity in spillovers, e.g., $\frac{1}{2} e_i^2-e_i \sum_j g_{i j} \lambda_{i j} e_j$, is already encompassed by our baseline model since $\mathbf{G}$ is allowed to be weighted.

\subsubsection{Optimal Incentives with Negative Spillovers}

We have focused thus far on peer-to-peer complementarities ($\lambda>0$) but many workplaces experience negative spillovers ($\lambda<0$), such as free-riding documented by \cite{amodio2018input} and \cite{bandiera2005social}. With negative spillovers, efforts are strategic substitutes. A worker receiving low incentives may optimally exert zero effort if she is surrounded by sufficiently active neighbors. In other words, very unequal incentives (i.e., low incentives to a worker and high incentives to her connections) can generate free-riding and push effort towards corners. We show in Supplementary Appendix~\ref{sec: App Negative Spillovers} that when $\lambda<0$ is sufficiently close to zero, the firm will never find it optimal to set incentives that generate corner solutions. Therefore, Propositions \ref{Optimal Contracts} and \ref{monotonic}  continue to apply, though the identity of workers receiving incentives changes with $\lambda$. For example, in Figure \ref{signcentral}, nodes most central under strategic complements are least central under strategic substitutes, flipping incentive allocations.

Recent work by \cite{galeotti2020targeting} shows that when $\lambda<0$, optimal interventions alternate incentives between adjacent individuals due to convex costs. Their model assumes quadratic costs for adjusting incentives from some baseline level $\hat{\alpha}_i$, $\sum_{i \in N}(\alpha_i-\hat{\alpha}_i)^2.$
In contrast, our model's costs depend on the wage risk, expressed as $\frac{r\sigma^2}{2}\sum_{i \in N} (\alpha_i^2 - \hat{\alpha}_i^2).$ 
This difference is significant: lowering incentives in our model reduces costs by decreasing risk exposure, making it unprofitable to exploit alternating patterns as in their convex cost specification. Consequently, the firm's optimal strategy diverges from the alternating structure in \cite{galeotti2020targeting}.

\begin{figure}[t]
\begin{tikzpicture}[scale=0.8, every node/.style={circle, draw=blue, fill=lightblue, thick, minimum size=5mm, inner sep=0pt}, 
                    every edge/.style={draw, thick}]

    \definecolor{lightblue}{rgb}{0.68, 0.85, 0.9}

     \node (1) at (-1,1) [draw=blue!50, fill=blue!20, minimum size=0.3cm] {};
    \node (2) at (-1,-1) [draw=blue!50, fill=blue!20, minimum size=0.3cm] {};
    \node (3) at (.5,1) [draw=red!50, fill=red!20, minimum size=0.6cm] {};
    \node (4) at (.5,-1) [draw=red!50, fill=red!20, minimum size=0.6cm] {};
    \node (5) at (2,0) [draw=gray!50, fill=gray!20, minimum size=0.45cm]{};
    \node (6) at (3.5,0) [draw=gray!50, fill=gray!20, minimum size=0.45cm] {};
    \node (7) at (5,1) [draw=red!50, fill=red!20, minimum size=0.6cm]{};
    \node (8) at (5,-1) [draw=red!50, fill=red!20, minimum size=0.6cm]{};
    \node (9) at (6.5,1) [draw=blue!50, fill=blue!20, minimum size=0.3cm]{};
    \node (10) at (6.5,-1) [draw=blue!50, fill=blue!20, minimum size=0.3cm]{};

    \path (1) edge (3);
    \path (1) edge (4);
    \path (2) edge (3);
    \path (2) edge (4);
    \path (3) edge (4);
    \path (3) edge (5);
    \path (4) edge (5);
    \path (6) edge (5);
    \path (6) edge (7);
    \path (6) edge (8);
    \path (7) edge (8);
    \path (7) edge (9);
    \path (7) edge (10);
    \path (8) edge (9);
    \path (8) edge (10);

\end{tikzpicture}
\hfill
\begin{tikzpicture}[scale=0.8, every node/.style={circle, fill=lightblue, thick, minimum size=5mm, inner sep=0pt}, 
                    every edge/.style={draw, thick}]

    \definecolor{lightblue}{rgb}{0.68, 0.85, 0.9}

     \node (1) at (-1,1) [draw=red!50, fill=red!20, minimum size=0.6cm] {};
    \node (2) at (-1,-1) [draw=red!50, fill=red!20, minimum size=0.6cm] {};
    \node (3) at (.5,1) [draw=blue!50, fill=blue!20, minimum size=0.3cm] {};
    \node (4) at (.5,-1) [draw=blue!50, fill=blue!20, minimum size=0.3cm] {};
    \node (5) at (2,0) [draw=gray!50, fill=gray!20, minimum size=0.4cm]{};
    \node (6) at (3.5,0) [draw=gray!50, fill=gray!20, minimum size=0.4cm] {};
    \node (7) at (5,1) [draw=blue!50, fill=blue!20, minimum size=0.3cm]{};
    \node (8) at (5,-1) [draw=blue!50, fill=blue!20, minimum size=0.3cm]{};
    \node (9) at (6.5,1) [draw=red!50, fill=red!20, minimum size=0.6cm]{};
    \node (10) at (6.5,-1) [draw=red!50, fill=red!20, minimum size=0.6cm]{};

    \path (1) edge (3);
    \path (1) edge (4);
    \path (2) edge (3);
    \path (2) edge (4);
    \path (3) edge (4);
    \path (3) edge (5);
    \path (4) edge (5);
    \path (6) edge (5);
    \path (6) edge (7);
    \path (6) edge (8);
    \path (7) edge (8);
    \path (7) edge (9);
    \path (7) edge (10);
    \path (8) edge (9);
    \path (8) edge (10);

\end{tikzpicture}
\caption{ \textbf{Panel A:} Positive peer effects (i.e., $\lambda>0$); \textbf{Panel B:} Negative peer effects (i.e., $\lambda<0$). }
 \label{signcentral}
\end{figure}

\subsubsection{More Flexible Externalities}

Our analysis so far has focused on the simplest form of externalities that admit a linear–quadratic game. One can imagine richer settings where the overall utility effect runs counter to the strategic effect---for instance, I may exert more effort because a hardworking neighbor observes me, even though this lowers my overall welfare.\footnote{We thank an anonymous referee for suggesting this alternative formulation.} We can capture this margin by adding a level effect (modulated by $\omega$) into the worker's cost structure: 
$$\psi_i(\mathbf{e} ; \mathbf{G})=\frac{1}{2} e_i^2-\lambda e_i \sum_{j \in N} g_{i j} e_j+\omega \sum_{j\in N} g_{i j} e_j.$$
In Supplementary Appendix~\ref{SuppApp:MoreFlexibleExternalities} we derive the optimal contract rule under these more flexible externalities. We show that the optimal incentive rule retains the same structure as in our baseline model but with a modified notion of network centrality that reflects the net value of incentive spillovers. Since spillovers also raise workers’ baseline costs, the amplification generated by outgoing incentive paths becomes less valuable for the firm,
and the effective centrality of workers correspondingly declines:
$$
\tilde{b}_i(\lambda, \omega)=1+\left(1-\frac{\omega}{\lambda}\right) \sum_s \sum_{q=1}^{\infty} \lambda^q \mathbf{G}_{s i}^q .
$$
If the level effect exactly offsets the marginal benefit of spillovers ($\omega=\lambda$), network amplification becomes neutral from the firm's perspective, and the optimal contract treats all workers as if they were equally central:
$$
\boldsymbol{\alpha}=\mathbf{W1}.
$$
In Supplementary Appendix~\ref{SuppApp:MoreFlexibleExternalities} we also show how this alternative form of externalities affects firm's profits in equilibrium. 

\subsubsection{Optimal Incentives with Partial Information}

We have assumed that the firm knows the entire network structure and can fully condition on it when designing optimal contracts. There are two main directions in which to relax this assumption. The first is that the firm may know the entire network but may be unable to write contracts that fully discriminate across workers. This is the approach we take in Section \ref{sec: coarse contracts} where we force firms to offer the same contract to entire sections of the workforce. The second approach is to assume that the firm may have only partial information about the relevant peer-to-peer network. 

To address this, in Supplementary Appendix~\ref{sec: random_networks}, we modify the firm's problem and derive the optimal incentive rule as a function of the parameters in a simple random-graph model, rather than a realized network $\mathbf{G}$. Specifically, we assume that, when designing the contract, the firm knows the linking probabilities between each pair of workers, but not the realized linking structure. Using a mean-field approximation, we provide a closed-form representation of the incentive rule in this case and show that worker $i$'s incentives depend on her expected connections, but not on the likelihood that $i$'s connections are themselves connected.

\section{Modular Production}\label{sec:Modular}

So far, we have assumed output is a linear function of effort, implying workers' efforts are substitutable despite occupying different positions in the peer network. This allowed us to isolate the impact of peer effects on wages and profits but overlooked how a firm's production function shapes its organization. We now examine how incentive contracts with peer effects are structured when firms have fragmented organizational structures.

We modify our production function to incorporate \textit{modules}, assuming final output depends on the weakest-performing module. Formally, $n$ workers are assigned to $K$ teams, $k_1, k_2, \dots, k_K$, each responsible for a separate module.\footnote{We use $K$ to denote both the number of teams and the set of teams $k\in K$. Moreover, we use $k(i)$ to denote worker $i$'s team.} Within teams, performance is substitutable, but across teams, it is perfectly complementary.\footnote{This assumption simplifies the model while remaining flexible, as we impose no restrictions on module size or composition. Unlike \cite{matouschek2023organizing}, who model modular production as a network of interdependent decisions, our approach avoids introducing an additional ``modular production network'' alongside the peer effects network. We just have to partition workers into modules.} Firm output is now:
\begin{equation}\label{modular}
    X(\mathbf{e}) = \min \left\{\sum_{i\in k_1} e_i, \sum_{i\in k_2} e_i, \ldots, \sum_{i\in k_K} e_i\right\}+\varepsilon.
\end{equation}
This technology is extremely versatile. It nests our original production function as a special case in which all workers belong to the same module.  However, we can also capture \textit{managerial bottlenecks} within organizations, like when a senior software engineer must approve all code before deployment.  Equation \eqref{modular} can also capture \textit{interdependent production teams}, whereby a small failure in a single critical component of an airplane wing, for instance, compromises the final product.\footnote{In an extreme and popular example, the \textit{Challenger} NASA spacecraft disaster occurred because a single small metal gasket called the O-ring failed. See \cite{kremer1993ring, garud2009managing, baldwin2003managing} for more details on this and other examples of modular production.}

To find the optimal contract, we must reconsider workers' equilibrium effort profile. In any equilibrium, only worker(s) in the weakest module(s) determine output. Therefore, a worker in an overproducing module can always reduce effort without affecting her salary. She will do so, unless she receives strong enough benefits from her peers such that her marginal effort costs become negative: $e_i - \lambda \sum_j g_{ij}e_j<0$. This happens only if $\lambda$ is large enough that workers not only exert effort without incentive pay, they actually pay to work in the firm (i.e., $\beta_i <0$).   We rule out these counterintuitive cases here and consider them in detail in Supplementary Appendix~\ref{App:Modular_Production_Uneq}. It follows that all modules must contribute the same total effort in equilibrium: $\sum_{i\in k}e_i=\bar{e}$ for all $k\in K$, and some $\bar{e} \geq 0$.  Decreasing effort is only profitable if the marginal cost ($\alpha_i$) is less than the marginal benefit ($e_i-\lambda \sum_{j\in N} g_{i j} e_j$). Putting everything together, any equilibrium must satisfy: 
\begin{equation}\label{inequality} \alpha_i \geq e_i-\lambda \sum_{j\in N} g_{i j} e_j\geq 0, \quad \forall i \in N. 
\end{equation}

The question now is which values of $\bar{e}$ constitute a Nash equilibrium. A contract proposed by the firm may admit multiple equilibria. For instance, $e_i=0$ for all $i\in N$ (and thus $\bar{e}=0$) is always an equilibrium.\footnote{Given others’ effort, a unilateral increase in $e_i$ never pays off.} We focus on the \textit{maximal equilibrium} $\hat{e}(\boldsymbol{\alpha})$,  such that any $\bar{e}>\hat{e}(\boldsymbol{\alpha})$ is not a Nash equilibrium. Still, given a contract $(\boldsymbol{\alpha},\boldsymbol{\beta})$, many equilibrium vectors $\mathbf{e}$ can be consistent with the same $\hat{e}$.\footnote{For instance, with modules $\{1,2\}$ and $\{3\}$, and $\lambda=0$, if the firm were to choose $\boldsymbol{\alpha}=(\hat{e},\hat{e},\hat{e})$ then any vector $\mathbf{e}=(t, \hat{e}-t, \hat{e})$ with $t\in [0,\hat{e}]$ is a NE. We thank a referee for this observation.} However, the firm will never offer a contract $(\boldsymbol{\alpha},\boldsymbol{\beta})$ that induces a maximal equilibrium $\hat{e}(\boldsymbol{\alpha})$ in which the first inequality  in \eqref{inequality} does not bind. If it did, the firm could lower $\alpha_i$ to the threshold level, $ e_i-\lambda\sum_j g_{ij} e_j$, and induce $\hat{e}$ more cheaply. Moreover, this cheaper contract is associated with a unique effort vector $\mathbf{e}$, and thus a unique wage cost for the firm; so profits are well defined. It follows that any optimal wage contract with modular production must satisfy the following two restrictions:
\begin{enumerate}
    \item $\alpha_i = e_i-\lambda \sum_{j\in N} g_{i j} e_j\geq 0, \:\: \forall i\in N$. 
    \item $\sum_{i\in k}e_i= \hat{e}, \:\: \forall k\in K.$ 
\end{enumerate}
Define the \textit{module assignment matrix} $\mathbf{M}_{K \times n}$, such that $\mathbf{M}_{ki}=1$ if worker $i\in N$ is assigned to module $k\in K$, and zero otherwise.\footnote{
        For example, in a firm with workers $1$, $2$, and $3$ in module $k_1$ and workers $4$ and $5$ in module $k_2$, the module assignment matrix is $\mathbf{M}_{2\times 5} = \begin{bmatrix}
            1 & 1 & 1 & 0 & 0 \\
            0 & 0 & 0 & 1 & 1\\
            \end{bmatrix}$, and $\mathbf{M}\mathbf{e} = \begin{bmatrix}
            \sum_{i \in k_1} e_i  \\
            \sum_{i \in k_2} e_i  \\
            \end{bmatrix} = \begin{bmatrix}
            \hat{e} \\
            \hat{e}\\
            \end{bmatrix}=\hat{e}\boldsymbol{1}_2.$}
We can now use matrix $\mathbf{M}$ in order to write down the firm's problem as optimizing profits subject to restrictions 1 and 2 above.
        \begin{align*}
        \max_{\boldsymbol{\alpha}, \hat{e}} \: \mathbb{E}[\pi(\mathbf{e}\mid \boldsymbol{\alpha}, \boldsymbol{\beta})]&= \left( \hat{e} - \frac{1}{2} \mathbf{e}' (\mathbf{I}-2\lambda \mathbf{G})\mathbf{e} - \frac{\sigma^2 r}{2} \boldsymbol{\alpha}' \boldsymbol{\alpha} \right) \\  \text{subject to: } 
       \boldsymbol{\alpha}&=\left(\mathbf{I}-\lambda \mathbf{G}\right)\mathbf{e}\geq \mathbf{0}, \quad \mathbf{e}\geq 0 \\ \quad \quad \quad \quad \quad \mathbf{M}\mathbf{e}&=\hat{e}\boldsymbol{1}_K.
        \end{align*}
When peer effects are not too strong, the firm's preferred contract satisfies $\left(\mathbf{I}-\lambda \mathbf{G}\right)\mathbf{e}\geq \mathbf{0}$, so these constraints do not bind. We therefore omit them when deriving the explicit, closed-form contract in Proposition \ref{Modular Production} below. Supplementary Appendix~\ref{App:Modular_Production_eq} derives the more general solution with non-negativity multipliers and discusses how they modify the incentive rule in equation \eqref{modular_alpha}.

   \begin{proposition}[Modular Production]\label{Modular Production}
        The optimal allocation of incentives under modular production with module assignment $\mathbf{M}$ is given by:
        \begin{equation}\label{modular_alpha}
           \boldsymbol{\alpha}^{*}= \mathbf{WC^\prime} \boldsymbol{\mu},
        \end{equation}
where $\boldsymbol{\mu}= \mathbf{M}' \mathbf{H}^{-1}\mathbf{1}/ (\mathbf{1}' \mathbf{H}^{-1} \mathbf{1})$ is an $n \times 1$ vector of $K$ module-specific weights $(\mu_{k(1)},\ldots, \mu_{k(n)})^\prime$ and $\mathbf{H=(MC)W(MC)^\prime}$.
    \end{proposition}
Proposition \ref{Modular Production} characterizes optimal incentive allocations under modular production, for any assignment of workers to modules, as captured by matrix $\mathbf{M}$ and any peer-effects network $\mathbf{G}$. In other words, equation \eqref{modular_alpha} generalizes Proposition \ref{Optimal Contracts} to any number of modules with arbitrary sizes, and for any structure of incentive links within and across modules. Notice that incentives are aggregated using the same weights $\mathbf{W}$ as before, but the relevant centrality measure now is $\mathbf{C}^\prime\boldsymbol{\mu}$ rather than $\mathbf{C^\prime1}$.\footnote{If there is only one module (i.e., if $\mathbf{M}=\mathbf{1}_n^\prime$) then $\boldsymbol{\mu}=\mathbf{1}$ and equation \eqref{modular_alpha} corresponds to the incentive rule from Proposition \ref{Optimal Contracts}, as expected.}

The intuition behind this incentive rule builds on our earlier discussion surrounding equation \eqref{systemprofits}. Notice that, regardless of how workers are partitioned into modules, the marginal cost $MC_{\alpha_i}$ remains unchanged; it depends only on the network structure through $\mathbf{W}^{-1}$.
The marginal benefit $MB_{\alpha_i}$, however, changes because the firm’s revenue is determined by the weakest-performing module. In equilibrium, all modules are induced to produce the same level $\hat e$, and the marginal contribution of increasing effort in module $k$ is given by its shadow value $\mu_k$. Therefore, a worker’s effort no longer increases revenue one-for-one, but instead in proportion to the marginal value of her module. Specifically, worker $j$’s effort increases revenue by $\mu_{k(j)}$, where $\mu_{k(j)}$ is a \textit{modular factor} associated with her module. As a result, $MB_{\alpha_i}$ weights each of $i$’s incentive paths by the module of the target:
$$MB_{\alpha_i}= \sum_j \frac{\partial e_j}{\partial \alpha_i} \mu_{k(j)}.$$
Stacking across all workers yields the vector of marginal benefits, which corresponds to the centrality measure $\mathbf{C}'\boldsymbol{\mu}$ in equation \eqref{modular_alpha}.

The values of $\mu_k$ (which sum to one) are determined by module size and link structure, as captured by $\mathbf{H}^{-1}$ in Proposition \ref{Modular Production}. As shown in the Appendix, $\mu_k$ is the shadow value (Lagrange multiplier) on the constraint that module $k$ delivers $\hat e$ in equilibrium. It reflects the marginal cost of raising module $k$’s performance relative to other modules.
Modules where effort can be generated more cheaply—due, for instance, to stronger peer effects or more favorable network positions—have lower shadow values $\mu_k$. Conversely, modules where effort is more costly to sustain receive higher weights.

For example, with two equally-sized and symmetrically-linked modules, we obtain $\mu_{k_1}=\mu_{k_2}=0.5$, so incentives are uniformly scaled down by a factor of one half relative to Proposition \ref{Optimal Contracts}. This reflects that, under modular production, output depends on the joint performance of all modules: at the symmetric equilibrium, each module contributes equally to relaxing the bottleneck, so the marginal value of increasing effort in any given module is only a fraction of total output. As a result, each worker’s marginal benefit is proportionally reduced, leading to uniformly lower incentives.

The following example shows that, for asymmetric cases, $\mu_k$ varies inversely with module $k$'s aggregate centrality.

\begin{figure}
    \centering
\begin{tikzpicture}[scale=1, every node/.style={circle, draw=blue, fill=lightblue, thick, minimum size=6mm, inner sep=0.2pt}, every edge/.style={draw, thick}]

    \definecolor{lightblue}{rgb}{0.68, 0.85, 0.9}

     \node (11) at (0,0) [draw=blue!50, fill=blue!10] {};
    \node (12) at (0,-1) [draw=blue!50, fill=blue!10] {};
    \node (13) at (1,0) [draw=blue!50, fill=blue!10] {$\scriptscriptstyle .30$};
    \node (14) at (1,-1) [draw=blue!50, fill=blue!10] {};
    \node (15) at (2.5,0) [draw=blue!50, fill=blue!10]{$\scriptscriptstyle .20$};
    \node (16) at (2.5,-1) [draw=blue!50, fill=blue!10] {};
    \node (17) at (3.5,0) [draw=blue!50, fill=blue!10]{};
    \node (18) at (3.5,-1) [draw=blue!50, fill=blue!10]{};
    \node (19) at (4.55,0) [draw=blue!50, fill=blue!10]{};
    \node (20) at (4.5,-1) [draw=blue!50, fill=blue!10]{};

    \fill[gray,opacity=0.1, rounded corners=0.4cm] (6,.5) rectangle (8,-1.5);
    \draw[thick, gray!50, rounded corners=0.4cm] (6,.5) rectangle (8,-1.5);
    
    \fill[gray,opacity=0.1, rounded corners=0.4cm] (8.5,.5) rectangle (11.5,-1.5);
    \draw[thick, gray!50, rounded corners=0.4cm] (8.5,.5) rectangle (11.5,-1.5);
    
    \node at (7, 0.9) [draw=none, fill=none] {$ \mu_{k_1}=0.60$} ;
	\node at (10, 0.9) [draw=none, fill=none]{$\mu_{k_2}=0.40$};
    

    \draw[line width=1.2pt,dashed, gray] (5.5,-1.7) -- (5.5,1); 

%
     \node (1) at (6.5,0) [draw=blue!50, fill=blue!10] {};
    \node (2) at (6.5,-1) [draw=blue!50, fill=blue!10] {};
    \node (3) at (7.5,0) [draw=blue!50, fill=blue!10] {$\scriptscriptstyle .47$};
    \node (4) at (7.5,-1) [draw=blue!50, fill=blue!10] {};
    \node (5) at (9,0) [draw=blue!50, fill=blue!10]{$\scriptscriptstyle .31$};
    \node (6) at (9,-1) [draw=blue!50, fill=blue!10] {};
    \node (7) at (10,0) [draw=blue!50, fill=blue!10]{};
    \node (8) at (10,-1) [draw=blue!50, fill=blue!10]{};
    \node (9) at (11,0) [draw=blue!50, fill=blue!10]{};
    \node (10) at (11,-1) [draw=blue!50, fill=blue!10]{};

    \fill[gray,opacity=0.1, rounded corners=0.4cm] (-.5,.5) rectangle (1.5,-1.5);
    \draw[thick, gray!50, rounded corners=0.4cm] (-.5,.5) rectangle (1.5,-1.5);
    
    \fill[gray,opacity=0.1, rounded corners=0.4cm] (2,.5) rectangle (5,-1.5);
    \draw[thick, gray!50, rounded corners=0.4cm] (2,.5) rectangle (5,-1.5);
    
    \node at (0.5, 0.9) [draw=none, fill=none] {$ \mu_{k_1}=0.60$} ;
	\node at (3.5, 0.9) [draw=none, fill=none]{$\mu_{k_2}=0.40$};
    
    \path (1) edge (3);
    \path (3) edge (4);
    \path (2) edge (4);
    \path (2) edge (1);
    \path (5) edge (7);
    \path (7) edge (9);
    \path (9) edge (10);
    \path (10) edge (8);
    \path (8) edge (6);
    \path (6) edge (5);

    \draw[line width=1.2pt,dashed, gray] (-0.5,-1.8) -- (11.5,-1.8); 

    \begin{scope}[yshift=-3.2cm] 
    
    \node (11) at (0,0) [draw=blue!50, fill=blue!10] {};
    \node (12) at (0,-1) [draw=blue!50, fill=blue!10] {};
    \node (13) at (1,0) [draw=blue!50, fill=blue!10] {$\scriptscriptstyle .55$};
    \node (14) at (1,-1) [draw=blue!50, fill=blue!10] {};
    \node (15) at (2.5,0) [draw=blue!50, fill=blue!10]{$\scriptscriptstyle .47$};
    \node (16) at (2.5,-1) [draw=blue!50, fill=blue!10] {};
    \node (17) at (3.5,0) [draw=blue!50, fill=blue!10]{};
    \node (18) at (3.5,-1) [draw=blue!50, fill=blue!10]{};
    \node (19) at (4.55,0) [draw=blue!50, fill=blue!10]{};
    \node (20) at (4.5,-1) [draw=blue!50, fill=blue!10]{};

    \fill[gray,opacity=0.1, rounded corners=0.4cm] (6,.5) rectangle (8,-1.5);
    \draw[thick, gray!50, rounded corners=0.4cm] (6,.5) rectangle (8,-1.5);
    
    \fill[gray,opacity=0.1, rounded corners=0.4cm] (8.5,.5) rectangle (11.5,-1.5);
    \draw[thick, gray!50, rounded corners=0.4cm] (8.5,.5) rectangle (11.5,-1.5);
    
    \node at (7, 0.9) [draw=none, fill=none] {$ \mu_{k_1}=0.37$} ;
	\node at (10, 0.9) [draw=none, fill=none]{$\mu_{k_2}=0.63$};
    
    \path (11) edge (12);
    \path (12) edge (14);
    \path (12) edge (13);
    \path (11) edge (13);
    \path (11) edge (14);
    \path (13) edge (14);
    \path (15) edge (16);
    \path (15) edge (17);
    \path (17) edge (19);
    \path (16) edge (18);
    \path (19) edge (20);
    \path (18) edge (20);

    \draw[line width=1.2pt,dashed, gray] (5.5,-1.5) -- (5.5,1.4); 

%
     \node (1) at (6.5,0) [draw=blue!50, fill=blue!10] {};
    \node (2) at (6.5,-1) [draw=blue!50, fill=blue!10] {};
    \node (3) at (7.5,0) [draw=blue!50, fill=blue!10] {$\scriptscriptstyle .62$};
    \node (4) at (7.5,-1) [draw=blue!50, fill=blue!10] {$\scriptscriptstyle .73$};
    \node (5) at (9,0) [draw=blue!50, fill=blue!10]{$\scriptscriptstyle .55$};
    \node (6) at (9,-1) [draw=blue!50, fill=blue!10] {$\scriptscriptstyle .68$};
    \node (7) at (10,0) [draw=blue!50, fill=blue!10]{$\scriptscriptstyle .51$};
    \node (8) at (10,-1) [draw=blue!50, fill=blue!10]{$\scriptscriptstyle .55$};
    \node (9) at (11,0) [draw=blue!50, fill=blue!10]{$\scriptscriptstyle .50$};
    \node (10) at (11,-1) [draw=blue!50, fill=blue!10]{$\scriptscriptstyle .51$};

    \fill[gray,opacity=0.1, rounded corners=0.4cm] (-.5,.5) rectangle (1.5,-1.5);
    \draw[thick, gray!50, rounded corners=0.4cm] (-.5,.5) rectangle (1.5,-1.5);
    
    \fill[gray,opacity=0.1, rounded corners=0.4cm] (2,.5) rectangle (5,-1.5);
    \draw[thick, gray!50, rounded corners=0.4cm] (2,.5) rectangle (5,-1.5);
    
    \node at (0.5, 0.9) [draw=none, fill=none] {$ \mu_{k_1}=0.40$} ;
	\node at (3.5, 0.9) [draw=none, fill=none]{$\mu_{k_2}=0.60$};
    
    \path (1) edge (2);
    \path (2) edge (4);
    \path (3) edge (4);
    \path (1) edge (3);
    \path (1) edge (4);
    \path (2) edge (3);
    \path (4) edge (6);
    \path (5) edge (6);
    \path (5) edge (7);
    \path (7) edge (9);
    \path (6) edge (8);
    \path (9) edge (10);
    \path (8) edge (10);
\end{scope}

\end{tikzpicture}

\caption{Modular shares, $\mu_k$, and incentives, $\alpha_i$, in four different firm configurations. Each firm has a 4-worker and 6-worker module.  (Parameters: $\lambda=0.15$ and $r\sigma^2=1$).}
\label{fig:mus}
\end{figure}

\begin{ex} [Modular Shares] \label{example2}
     Figure \ref{fig:mus} reports $(\mu_{k_1}, \mu_{k_2})$ for four firms with two modules each. The top-left panel shows that, without links, size differences drive $\mu_k$: workers in smaller modules require stronger incentives to achieve a given $\hat{e}$, a result formalized in Corollary \ref{modules_nopeers} below. The top-right panel confirms that symmetric connections preserve $\mu_k$'s since the (relative) incentives required to achieve $\hat{e}$ have not changed. However, relative to the previous case, $\alpha_i$'s increase by 56\% for everyone, due to uniform centrality gains.\footnote{This is computed as the top right $\alpha_i$'s over the top left $\alpha_i$'s,  $0.47/0.3=0.31/0.2=1.56$}  The bottom-left panel shows that workers in small but well-connected modules receive lower $\mu_k$ than those in larger but poorly connected ones, as it becomes relatively easier for them to generate $\hat{e}$. Consequently, incentives rise less (17\%) for these workers than for workers in the larger module (52\%).  Lastly, the bottom-right panel shows that cross-module links have asymmetric effects, boosting incentives more for loosely connected groups. 
\end{ex}

This example highlights another key feature of modular production: whereas in the linear technology of Section \ref{baseline}, network components were independent, here incentives depend on the entire network, including links in separate components. To see this, note that incentives for the large module in Figure \ref{fig:mus} increase from $0.31$ (top-right) to $0.47$ (bottom-left), even though the only additional links are formed in a separate component of the peer network. Before proceeding, we examine the incentive rule in detail for two special cases that illustrate how modular technology \eqref{modular} can represent distinct organizational environments.

\subsubsection{Modular Production without Peer Effects}

The case without peer effects (i.e., $\lambda=0$) teaches us a lot already about how modular production drives earnings disparities, and connects with some of the lessons from the literature on knowledge hierarchies \citep{garicano2000hierarchies, garicano2006organization}. More specifically, a technology like \eqref{modular} can capture managerial bottlenecks and sequential hierarchies, such as when a set of $100$ medical officers at the FDA (module 1) prepare reports that must be processed and approved by a senior official (module 2) before drugs can go to market. Even without peer-to-peer complementarities, this 100-to-1 span-of-control ratio in the firm's production function yields large differences in incentive pay. 

Recall that, without peer effects, there are no incentive paths (i.e., $\partial e_i /\partial \alpha_j=0$ for $i\neq j$ and $\partial e_i /\partial \alpha_i=1$). Therefore, the marginal benefit from incentivizing worker $i$ simplifies to her own module's productivity weight: i.e., $MB_{\alpha_i}=\mu_{k(i)}$. Following equation \eqref{systemprofits} the profit-maximizing incentive rules must  satisfy $\mu_{k(i)}=(1+r\sigma^2)\alpha_i$, for some constants $\mu_{k(i)}$ such that $\sum_k \mu_k=1$.  Since the manager alone must validate $100$ workers' worth of output, the firm must weigh the manager with a productivity share, $\mu_k$, $100$ times larger than that of workers. Thus, relative to Corollary \ref{no peer effects}, incentive rules are now scaled to capture differences in module size. The following corollary generalizes this intuition.

\begin{corollary}[No Peer Effects]\label{modules_nopeers}
   Take $\lambda=0$. Consider a firm with $K$ modules with sizes $n_1,n_2,\ldots,n_K$ and let $k(i)$ represent worker $i$'s module. Incentives are allocated according to the following rule: 
   \begin{equation}\label{eq: modules_nopeers}
       \alpha_i^{*} = \frac{1}{1+r\sigma^2}\: \frac{1}{n_{k(i)}} \left( \sum_{s\in K} \frac{1}{n_s}\right)^{-1},\quad \forall\: i\in N.
   \end{equation}
   If modules are of equal size, $\alpha_i^{*}= \frac{1}{1+r\sigma^2}\frac{1}{K}$, and if all workers form part of one single module, $\alpha_i$ simplifies to that in Corollary $\ref{no peer effects}$.
\end{corollary}
Corollary \ref{modules_nopeers} implies that workers in small modules are compensated more in performance-pay when they collaborate with larger modules, and that the ratio in incentives is inversely proportional to the ratio in module size: $\alpha_i/\alpha_j = n_{k(j)}/n_{k(i)}$. This hitherto unknown result in the theory of optimal contracts sheds light on how non-linear technologies can drive large wage disparities (even in the absence of peer effects). It captures the simple fact that workers in small modules are individually valuable, since they are essential in order to validate the output of larger groups.  

\subsubsection{When Every Worker is Essential}

Now imagine a ``weakest-link'' production function, where every worker is \textit{essential} (i.e., each worker belongs to a separate module). This captures extremely precise production processes that consist of many indispensable steps, whereby a single error halts the entire process. For instance, a small defect rate in extreme ultraviolet lithography (EUV) can cause massive yield losses in semiconductor fabrication---one worker miscalibrating a machine in a cleanroom can disrupt production for days. 

Although all modules have the same size (i.e., size of 1), workers differ in how they are connected to each other. Since everyone contributes $\hat{e}$ in equilibrium, differences in the length and targets of their influence paths determine the productivity factor $\mu_k$ to weigh each module. It turns out that $\mathbf{H}=\mathbf{CWC^\prime}$ and therefore, following equation \eqref{modular_alpha},  $\boldsymbol{\alpha}^{*} \propto \mathbf{C}^{-1}\mathbf{1}$. We thus find that incentives are allocated very differently. Firms no longer concentrate incentives on workers with more outgoing paths (i.e., higher Bonacich centrality). Rather, firms prioritize workers with fewer incoming links.

\begin{corollary}\label{single_module}
    When every worker is essential (i.e., every module is of size $1$), incentives are allocated inversely to workers' in-degree. Formally, optimal incentives are allocated as follows:$$
\alpha_i^{*}=\frac{1-\lambda d_i}{\xi}, \quad \forall i\in N,
$$
where $d_i$ is worker $i$'s in-degree and $\xi=\sum_{j\in N} \left( 1-2 \lambda d_j +r\sigma^2 \left(1-\lambda d_j\right)^2 \right)$ is common across all workers.
\end{corollary}

Intuitively, all workers exert the same effort in equilibrium so paying central workers a large $\alpha_i$ won't raise others' contributions (beyond their own value of $\alpha$). On the other hand, those with few incoming links have large effort costs and therefore stand more to gain by lowering their effort, unless variable pay is set sufficiently high.\footnote{Recall that with modular technology it never pays to increase effort, given that everyone is doing the same effort. It only (sometimes) pays to lower effort.} These two forces imply that firms ensure a sufficiently large level of output by disregarding Bonacich centrality and allocating incentives based on incoming links instead.\footnote{On some networks (see Figure \ref{hierarchies}) the worker with the least incoming links might also be the most central, but this is generally not the case.}

\begin{figure}
    \centering
    \includegraphics[width=0.45\linewidth]{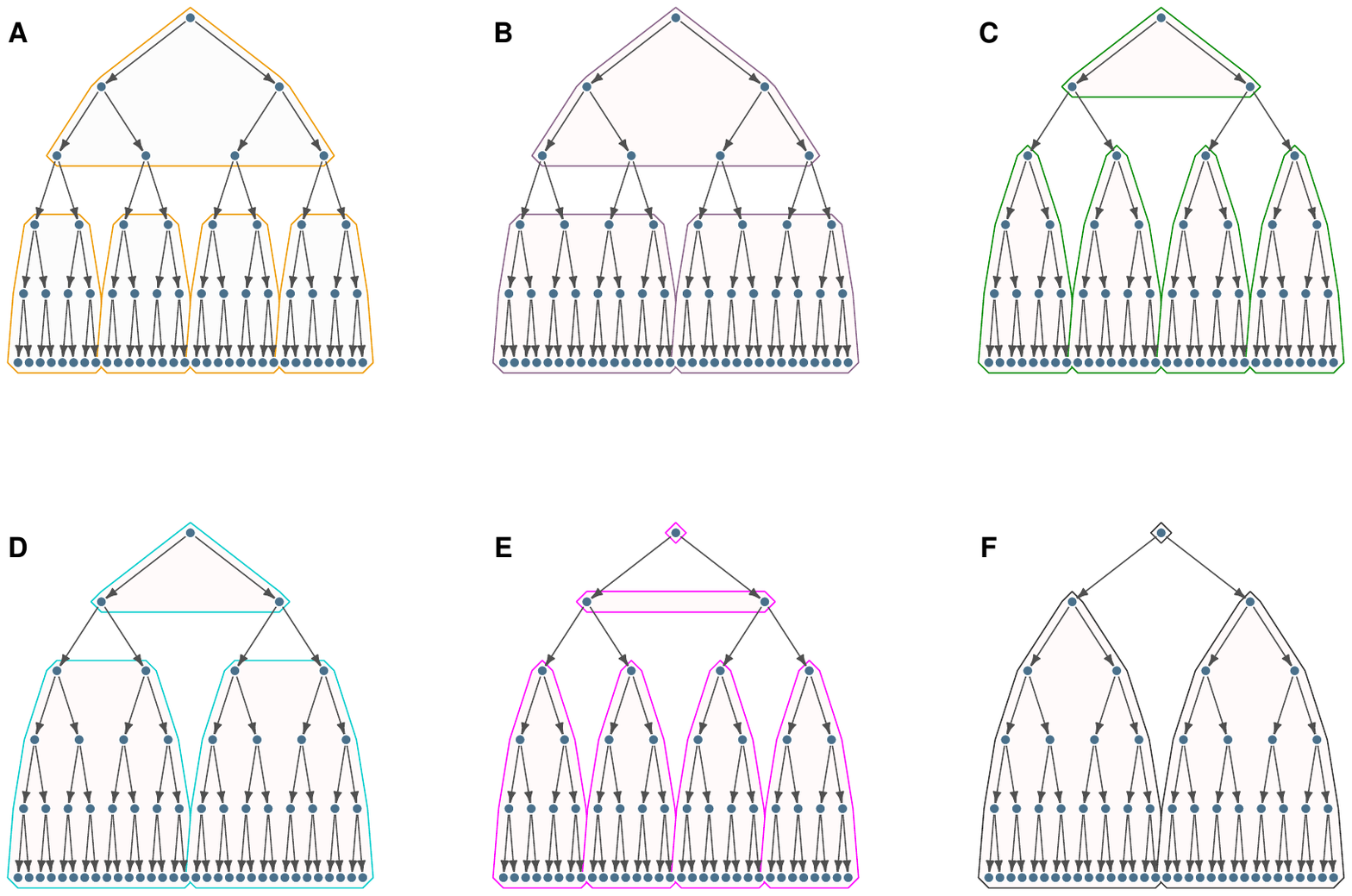}\hfill
    \includegraphics[width=0.55\linewidth]{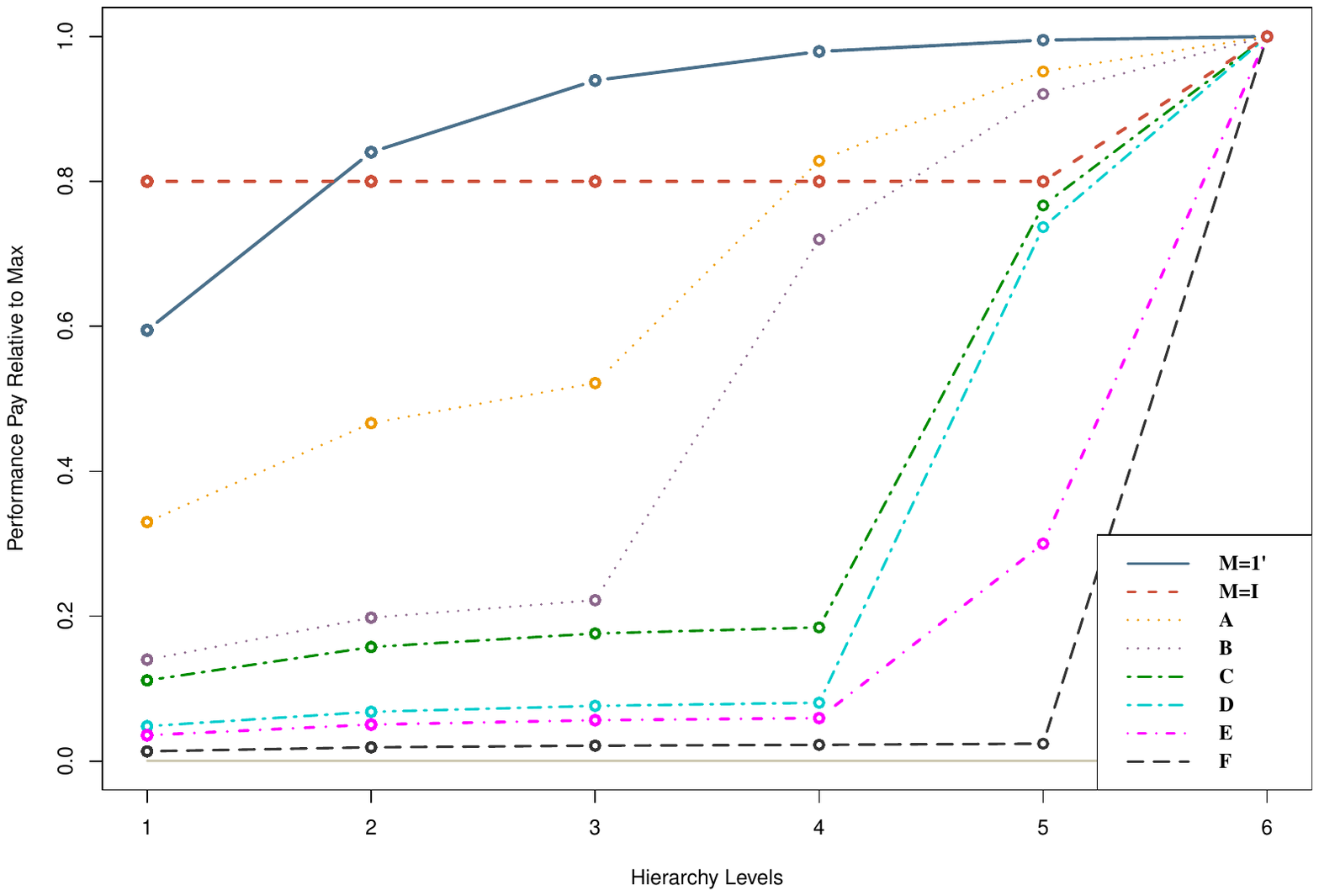}
    \caption{Performance-pay for Different Modular Configurations. Simulations are run for $\lambda=0.2$, $\sigma^2 = 2$ and $r=5$.}
    \label{fig:modularprofiles}
\end{figure}

\subsubsection{Modules in Hierarchies}

Before moving on, let's consider how modular structure impacts earnings profiles by revisiting the hierarchy example from Figure \ref{hierarchies}. The right panel of Figure \ref{fig:modularprofiles} tracks performance-pay along the firm's hierarchy for different module configurations. There are various things to note from this figure. First, the $\mathbf{M=1}^\prime$ case replicates the curve in Figure \ref{hierarchies}, as expected. Second, the $\mathbf{M=I}$ case shows everyone receiving $80\%$ of the CEO's performance compensation. This is in line with Corollary \ref{single_module} since the CEO has an in-degree of $0$ while everyone else has an in-degree of $1$, and $1-0.2*1=0.8$. Now consider other modular configurations, which are drawn on the left panel. Notice that performance-pay jumps up as we cross to a higher module, which occurs, for instance, in level $3$ for firms $A$ and $B$, and level $4$ for firms $C$ and $D$.  Notice also that the jump is more pronounced the larger are the lower modules. Within a module, performance pay tracks network centrality (as in the $\mathbf{M=1}^\prime$ curve), but the network effect is dampened by the large jump across modules (for instance in firm $F$). In fact, we decompose total variance in performance pay and find that \textit{within}-module variation only accounts for somewhere from 3\% (firm $D$) to 35\% (firm $A$) of the total variation: the rest is driven by variation in pay \textit{between} modules. In fact, firm $E$ shows that with multiple module jumps performance-pay profile can even look convex.

\section{Wage Benchmarking}\label{sec: coarse contracts}

In this section, we consider allocating incentives when firms cannot write fully personalized contracts. We divide the workforce into occupational categories and assume that the firm must offer the same linear wage contract to every member of a given category, while everything else remains as in Section \ref{baseline}.

This restriction reflects a common feature of compensation systems: firms typically set pay at the level of job titles or grades rather than tailoring contracts worker-by-worker.\footnote{Recent evidence shows that salary benchmarking is widely used and reduces within-job wage dispersion \citep{cullen2024pay}, and that increased pay transparency further compresses wages within firms \citep{card2012inequality, cullen2025}.} Workers within the same category perform similar formal roles, but differ in their position within the informal network of peer complementarities—i.e., in who influences whom and how effort propagates through the organization.\footnote{Even within narrowly defined roles, a large empirical literature documents substantial heterogeneity in peer effects depending on local interactions (e.g., \cite{mas2009peers, bandiera2005social}).} Our framework captures the implications of this mismatch between contractible job categories and non-contractible interaction structures for optimal incentive design. As the partition becomes finer, this distortion vanishes and the model converges to the benchmark with fully personalized contracts.

 Since contracts can't perfectly discriminate, the firm no longer extracts all surplus from its workforce. More importantly, the peer-effects network will determine exactly how surplus is distributed, with more ``central'' workers extracting larger rents from the employer. To see this, we assign $n$ workers into $K\leq n$ groups (or occupational categories) and we assume that the firm must offer the same wage contract to all workers in category $k\in K$:
 \begin{equation*}
     \omega_i = \beta_k + \alpha_k X, \quad \forall i \in k.
 \end{equation*}
We allow for any level of granularity in the contract because we make no restrictions on how to partition workers into categories.\footnote{As a partition, each worker must be assigned to one and only one category.}  Define the \textit{group assignment matrix} $\mathbf{T}_{K \times n}$ such that $\mathbf{T}_{ki}=1$ if worker $i\in N$ is assigned to occupational category $k\in K$, and zero otherwise.\footnote{For example, if in a law firm workers $1$, $2$, and $3$ are in one occupational category (e.g. paralegals) and workers $4$ and $5$ are in another (e.g. senior partners), the group assignment matrix is $\mathbf{T}_{2\times 5} = \begin{bmatrix}
            1 & 1 & 1 & 0 & 0 \\
            0 & 0 & 0 & 1 & 1\\
            \end{bmatrix}$. We choose to use different notation across Sections~\ref{sec:Modular} and~\ref{sec: coarse contracts} to emphasize that the $K$ occupational categories in this section have no relation to the $K$ modules of Section \ref{sec:Modular}. Analyzing modular production and wage benchmarking simultaneously is beyond the scope of this paper, and we leave it for future research.} 
Let $n_{k}:=\sum_{i}T_{ki} $ denote the size of group $k$. Define $\hat{\boldsymbol{\alpha}}$ as the $K \times 1$ vector of incentives chosen by the firm: the $k$-th term, $\hat{\alpha}_k$, corresponds to the value of $\alpha$ offered to all workers in group $k$. We can now relate $\hat{\boldsymbol{\alpha}}$ to the full $n$-vector of incentives by the following simple relation: $\boldsymbol{\alpha}= \mathbf{T}^\prime \hat{\boldsymbol{\alpha}}$ (and similarly for $\hat{\boldsymbol{\beta}}$). It follows that the Incentive Compatibility (IC) constraint can be obtained from the $K\leq n$ contracts as $\mathbf{e}^{*}=\mathbf{CT}^\prime\hat{\boldsymbol{\alpha}}$ together with $\mathbf{e}\geq 0$.

When contracts are coarse ($K<n$) the Individual Rationality (IR) constraints look different.\footnote{Notice that if $K=n$ (i.e., if each worker is allowed to have a different job-title/contract) then $\mathbf{T}=\mathbf{I}$ and we get the original setup of Section \ref{baseline}.} Firms no longer extract full rents from all workers because if a group contains multiple workers occupying different network positions, then a single contract cannot simultaneously guarantee that everyone is exactly compensated at their reservation utility. We assume that the firm sets $\beta_k$ such that no one in group $k$ rejects the contract. This means that every worker will extract (weakly) positive rents from their contract, and only the ``highest-cost worker'' will receive her reservation utility. 
Let $\psi_i=\frac{1}{2} e_i^2-\lambda e_i \sum_j g_{i j} e_j$ denote worker $i$'s effort cost, and let $\bar{i}(k)$ represent a worker with highest effort cost in group $k$.\footnote{There may be multiple workers with highest cost in group $k$. It is not important which of these is identified by $\bar{i}(k)$.}
The common fixed payment for group $k$ is given by:
\[\beta_k=\frac{r \sigma^2}{2} \alpha_k^2-\alpha_k \sum_{j\in N} e_j+\psi_i + (\psi_{\bar{i}(k)} - \psi_i). \]
Notice that $\beta_k$ looks just like in our baseline model, but has an additional term---in parentheses---that captures the rents extracted by those members of the team with lower effort costs. Following equation \eqref{CE}, every worker $i\in k$ satisfies $\operatorname{CE}(\hat{\boldsymbol{\alpha}}, \hat{\boldsymbol{\beta}}; \mathbf{G}, \mathbf{T})_i\geq 0$, while the worker(s) with the highest effort cost have binding (IR), $\operatorname{CE}(\hat{\boldsymbol{\alpha}}, \hat{\boldsymbol{\beta}}; \mathbf{G}, \mathbf{T})_{\bar{i}(k)}=0$. 
We can now re-write the firm's problem under coarse contracts using modified (IR) and (IC) constraints as: 
        \begin{align*}
        \max_{\hat{\boldsymbol{\alpha}}, \hat{\boldsymbol{\beta}}} \mathbb{E}[\pi(\mathbf{e}\mid \hat{\boldsymbol{\alpha}}, \hat{\boldsymbol{\beta}})] 
        \\ \text{subject to: }        \operatorname{CE}_i(\mathbf{\hat{\boldsymbol{\alpha}}, \hat{\boldsymbol{\beta}}; G, T}) &\geq 0, \quad \forall i\in N \hfill \tag{IR} \\
        \mathbf{e} = \mathbf{C}\mathbf{T}^\prime\boldsymbol{\hat{\alpha}}, \,\,\, \mathbf{e} &\geq 0.\hfill \tag{IC}
        \end{align*}

To ensure that the firm's problem remains concave under wage benchmarking, we strengthen Assumption~\ref{spectralradius} to account for group size. Define $\bar{n}:= \max_{k} n_{k}$.
\renewcommand{\theassumption}{2-WB}
\begin{assumption}\label{spectralradiusWB}
     The spectral radius of $\bar{n}\lambda^{2}/(1+r\sigma^{2})(\mathbf{GC})'\mathbf{GC}$ is less than 1.
 \end{assumption}
 
We can now characterize optimal incentives under wage benchmarking.

\begin{proposition}[Wage Benchmarking]\label{Coarse Contracts}
        Under Assumptions~\ref{assumption1} and \ref{spectralradiusWB},  the optimal allocation of incentives under wage benchmarking with group assignment $\mathbf{T}$ is implicitly defined by:
        \begin{equation}\label{coarse_alpha}
            \boldsymbol{\hat{\alpha}}^{*} = \mathbf{W_{M}}\mathbf{TC}^\prime\mathbf{1}, 
        \end{equation}
        where $\mathbf{W_{M}} =  [\mathbf{T}((1+r\sigma^2) \mathbf{I}- (\lambda \mathbf{GC})'\mathbf{M}(\lambda\mathbf{GC}) )\mathbf{T}']^{-1}$ and $\mathbf{M}=\operatorname{diag}(\boldsymbol{\mu})$ is a diagonal matrix of multipliers associated to the IR constraints, with $\sum_{i\in k}\mu_{i} = n_{k}$ for each group $k$. 
    \end{proposition}

Equation \eqref{coarse_alpha} differs from our original incentive rule in Proposition \ref{Optimal Contracts} in a very natural way. Incentives are now allocated at the group-level according to the aggregated centrality measure, $\mathbf{TC}^\prime\mathbf{1}$, which simply sums the centrality of all members in each group. The weighting matrix $\mathbf{W_M}$ also changes because, under wage benchmarking, only workers with binding IR constraints affect the firm's marginal cost of raising incentives---all other workers receive strictly positive surplus, so any marginal change in their costs affects their certainty equivalent without changing the firm’s profits. The matrix $\mathbf{M}$ essentially identifies which workers are relevant at the margin: only binding workers with $\mu_i>0$ contribute to marginal incentive costs, while non-binding workers with $\mu_i=0$ do not. Because salaries are set at the group level, these shadow values aggregate within each group, so that $\sum_{i \in k} \mu_i=n_k$.\footnote{We show in the appendix that the constants $\mu_i$ are jointly determined by this equation and \eqref{coarse_alpha}. In some situations the $\mu$'s are unique, but if multiple binders in the same team have equal network position vis-a-vis all groups---i.e. the rows of $\lambda \mathbf{GCT}'$ are equal---then any vector satisfying $\sum_{i\in k}\mu_i=n_k$ yields the same contract $\boldsymbol{\alpha}$.} 
Putting everything together, the terms in $\mathbf{W}_{\mathbf{M}}$ effectively re-weight the network because incentives are shaped not by all influence paths in the network, but only by those that target workers whose participation constraints bind.\footnote{See Supplementary Appendix~\ref{App: WageBenchmarking} for more details on how equation \eqref{coarse_alpha} relates to equation \eqref{systemprofits}, and a simple example that shows how $\mathbf{W_M}$ determines group-level incentives.}

When the firm can write personalized wages (i.e., if $\mathbf{T}=\mathbf{I}$), all IR constraints bind. Therefore, $\boldsymbol{\mu}=\mathbf{1}$ and $\mathbf{M}=\mathbf{I}$, and the characterization collapses to the one in Proposition \ref{Optimal Contracts}, as expected. Finally, notice that in the absence of peer effects coarse contracts also coincide with personalized contracts. Without peer effects we have $\lambda=0$ and $\mathbf{C}=\mathbf{I}$. This implies workers are identical and so all IR constraints bind, i.e., $\mathbf{M}=\mathbf{I}$. Thus we have $\hat{\boldsymbol{\alpha}}^{*} = \left[(1+r\sigma^{2})\mathbf{TT}'\right]^{-1}\mathbf{T}\mathbf{1}$, and $\alpha_{i}^{*} = 1/(1+r\sigma^{2})$ (Corollary \ref{no peer effects}).

\subsection{Loss in Profits due to Wage Benchmarking}

Wage benchmarking restricts the firm to offering the same contract to all workers within a group. This prevents the firm from fully tailoring incentives to workers' positions in the peer network and forbids full surplus extraction by the firm. In this section, we develop an intuitive way of measuring the profit loss generated by any group assignment $\mathbf{T}$ on any peer network $\mathbf{G}$. Before that, we derive a simple expression for profits under the optimal contract.\footnote{The profit expression also holds in the textbook case and in our baseline model of Section~\ref{baseline}, as we show in Supplementary Appendix~\ref{Design}.}

\begin{lemma}[Profits with Wage Benchmarking]\label{profits_coarse}
With wage benchmarking, a firm's profits in expectation are maximized at one-half of equilibrium output:
\[\mathbb{E}(\pi^{*}(\mathbf{e}^{*}|\hat{\boldsymbol{\alpha}}^{*}, \hat{\boldsymbol{\beta}}^{*})) = \frac{1}{2}\sum_i e_i^{*}. \]
\end{lemma}
For a given group assignment $\mathbf{T}$, let  $\mathbf{e}^{*}(\mathbf{T}):=\mathbf{C}\mathbf{T}^\prime\boldsymbol{\hat{\alpha}}^{*}(\mathbf{T})$
denote the vector of equilibrium effort contributions under the optimal contract. Then, the loss in profits between group assignment $\mathbf{T}$ and group assignment $\widetilde{\mathbf{T}}$ is given by:
\[ \Delta \Pi_{\widetilde{\mathbf{T}}-\mathbf{T}} := \mathbb{E} \big(\pi^{*}\big(\mathbf{e}^* (\widetilde{\mathbf{T}})\mid \hat{\boldsymbol{\alpha}}^*(\widetilde{\mathbf{T}}), \hat{\boldsymbol{\beta}}^*(\widetilde{\mathbf{T}}) \big) \big) -  \mathbb{E} \big(\pi^{*} \big( \mathbf{e}^* (\mathbf{T})\mid \hat{\boldsymbol{\alpha}}^*(\mathbf{T}),  \hat{\boldsymbol{\beta}}^*(\mathbf{T}) \big)\big)  = \frac{1}{2}\mathbf{1}'(\mathbf{e}^{*}(\widetilde{\mathbf{T}}) - \mathbf{e}^{*}(\mathbf{T})). \]
The following result focuses on $\Delta \Pi_{\mathbf{I-T}}$, which corresponds to the loss of profits when groups are assigned by $\mathbf{T}$, relative to fully personalized contracts, $\mathbf{I}$. We show that, for high enough values of risk, profit loss can be approximated easily by computing within-group dispersion in centrality.

\begin{figure} [t]
\begin{tikzpicture}[scale=1, every node/.style={circle, draw=blue, fill=lightblue, thick, minimum size=5mm, inner sep=0pt}, 
                    every edge/.style={draw, thick}]

    \definecolor{lightblue}{rgb}{0.68, 0.85, 0.9}

     \node (1) at (-1,0) [draw=blue!50, fill=blue!10] {};
    \node (2) at (-1,-1) [draw=blue!50, fill=blue!10] {};
    \node (3) at (-1,-2) [draw=blue!50, fill=blue!10] {};
    \node (4) at (0.5,-1) [draw=red, fill=red!10] {};
    \node (5) at (2.5,-1) [draw=red, fill=red!10]{};
    \node (6) at (4,0) [draw=blue!50, fill=blue!10] {};
    \node (7) at (4,-1) [draw=blue!50, fill=blue!10]{};
    \node (8) at (4,-2) [draw=blue!50, fill=blue!10]{};
    
    \fill[gray,opacity=0.1, rounded corners=0.4cm] (0,-1.5) rectangle (3,-0.5);
    \draw[thick, gray!50, rounded corners=0.4cm] (0,-1.5) rectangle (3,-0.5);
    
    \fill[gray,opacity=0.1, rounded corners=0.4cm] (-1.5,-2.5) rectangle (-0.5,0.5);
    \draw[thick, gray!50, rounded corners=0.4cm] (-1.5,-2.5) rectangle (-0.5,0.5);
    
    \fill[gray,opacity=0.1, rounded corners=0.4cm] (3.5,-2.5) rectangle (4.5,0.5);
    \draw[thick, gray!50, rounded corners=0.4cm] (3.5,-2.5) rectangle (4.5,0.5);
    
    \path (1) edge (4);
    \path (2) edge (4);
    \path (3) edge (4);
    \path (4) edge (5);
    \path (6) edge (5);
    \path (7) edge (5);
    \path (8) edge (5);

\end{tikzpicture}
\hfill
\begin{tikzpicture}[scale=1, every node/.style={circle, draw=blue, fill=lightblue, thick, minimum size=5mm, inner sep=0pt}, 
                    every edge/.style={draw, thick}]

    \definecolor{lightblue}{rgb}{0.68, 0.85, 0.9}

     \node (1) at (-1,0) [draw=blue!50, fill=blue!10] {};
    \node (2) at (-1,-1) [draw=blue!50, fill=blue!10] {};
    \node (3) at (-1,-2) [draw=blue!50, fill=blue!10] {};
    \node (4) at (0.5,-1) [draw=red!50, fill=red!10] {};
    \node (5) at (2.5,-1) [draw=red!50, fill=red!10]{};
    \node (6) at (4,0) [draw=blue!50, fill=blue!10] {};
    \node (7) at (4,-1) [draw=blue!50, fill=blue!10]{};
    \node (8) at (4,-2) [draw=blue!50, fill=blue!10]{};

 \fill[gray,opacity=0.1, rounded corners=0.4cm] (-1.5,-2.5) rectangle (1,0.5);
    \draw[thick, gray!50, rounded corners=0.4cm] (-1.5,-2.5) rectangle (1,0.5);

    \fill[gray,opacity=0.1, rounded corners=0.4cm] (2,-2.5) rectangle (4.5,0.5);
    \draw[thick, gray!50, rounded corners=0.4cm] (2,-2.5) rectangle (4.5,0.5);

    \path (1) edge (4);
    \path (2) edge (4);
    \path (3) edge (4);
    \path (4) edge (5);
    \path (6) edge (5);
    \path (7) edge (5);
    \path (8) edge (5);

\end{tikzpicture}
\caption{ \textbf{Panel A:} Within-group variance is zero. No profit loss. \textbf{Panel B:} Within-group variance is 0.53. Profit loss is about 1.97 ($\lambda=0.25, \sigma^{2}=4$ and $r=1$).}
\label{fig: surplus}
\end{figure}

\begin{proposition}[Loss in Profits]\label{profits_variance}
As $\sigma^2$ grows, the profit loss due to wage benchmarking with group assignment $\mathbf{T}$ approaches the sum of within-group variances in Bonacich centrality, weighted by group size. In other words, as $\sigma^2 \rightarrow \infty$:
\[
 \left(1+r \sigma^2\right) \Delta \Pi_{\mathbf{I}-\mathbf{T}} \longrightarrow\frac{1}{2}\sum_{k \in K} n_k \operatorname{Var}\left(\mathbf{b}_k\right),
\]
where $\mathbf{b}_{k}$ is the (sub)vector of Bonacich centralities for workers in group $k$. 
\end{proposition}

Intuitively, if $\operatorname{Var}\left(\mathbf{b}_k\right)=0$ for all $k \in K$, then all workers with the same job title are equally central. Setting uniform contracts within each group only matters to the firm if equally-central workers affect incentive costs differently, but this cost-saving feature is negligible to the firm when risk is high (see Proposition \ref{monotonic}). More importantly, Proposition \ref{profits_variance} allows for arbitrary differences in centrality across groups---so long as within-group variability is zero, profits remain unaffected by wage benchmarking. This underscores that the interaction between the peer structure $\mathbf{G}$ and group assignment $\mathbf{T}$ is what truly matters. For instance, Figure~\ref{fig: surplus} illustrates how the same network under different assignments $\mathbf{T}$ leads to starkly different values of $\Delta \Pi_{\mathbf{I-T}}$.

\begin{figure}[t]
    \centering
   \includegraphics[scale=0.25]{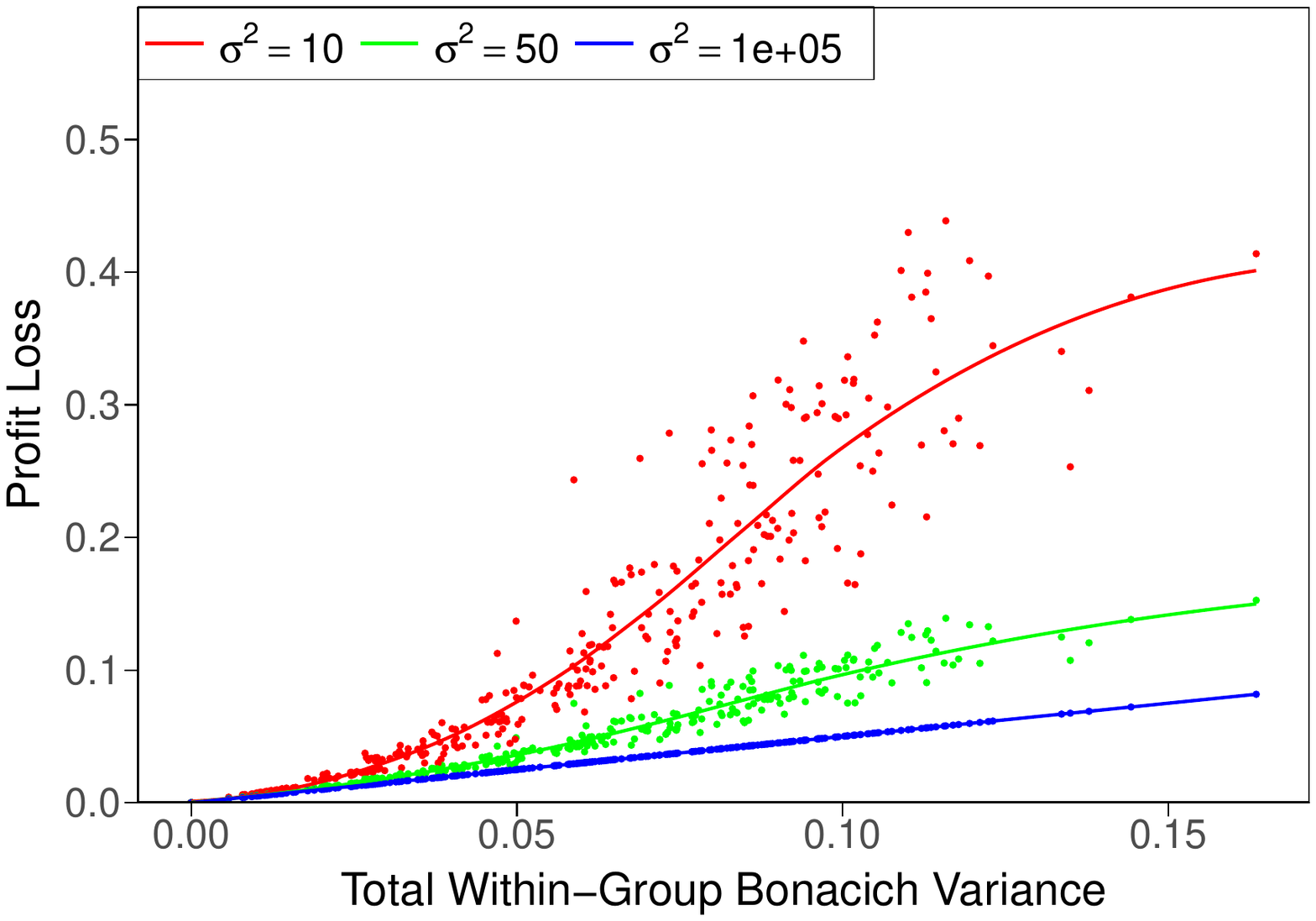}
    \caption{The profit loss $ \left(1+r \sigma^2\right) \Delta \Pi_{\mathbf{I}-\mathbf{T}}$ is increasing in within-group variance in centrality. The relationship is approximately linear and becomes deterministic as $\sigma^2$ increases. }
    \label{variance_plot}
\end{figure}

Proposition \ref{profits_variance}  establishes a limiting result for large $\sigma^2$.  One might still ask whether profit loss is  determined by the sum of within-group variances at finite levels of risk. For instance, in \textit{regular networks}---where all individuals are identical in every respect--- the profit gap is zero for any $\sigma^2\geq 0$.  Figure \ref{variance_plot} presents simulation results from Erdős-Rényi random graphs, showing that $\Delta \Pi_{\mathbf{I-T}}$ tends to rise with $\sum_k n_k \operatorname{Var}\left(\mathbf{b}_k\right)$ and converges to an exact linear relationship as $\sigma^2$ grows. Thus, while proportionality holds strictly only in the limit, the profit loss due to wage benchmarking is well captured by this simple statistic.

\section{Implications for Organizational Design}\label{orgdesign}

While our focus throughout has been on deriving incentive rules across different production technologies and institutional constraints, our model also has broad implications for \textit{organizational design}. We can use our framework to ask how firms should structure their organizations---or how they should invest in their workforce---based on the relationship between wages and the underlying network of bilateral ties. We explore these questions in detail in Supplementary Appendix~\ref{Design} and provide some intuition for our findings below. 

We first show that the firm's overall profits in equilibrium are proportional to aggregate output, a result that extends the canonical intuition on team production to our setting with a peer-to-peer network of complementarities. More importantly,  using a decomposition method as in \cite{galeotti2020targeting}, Proposition~\ref{Profits and Spectrum} derives an explicit formula linking expected profits to network structure through its \textit{principal components}, for both undirected and directed (but normal) networks. The general expressions are in Supplementary Appendix~\ref{Design}, but a lot of intuition can be obtained by focusing on undirected  networks and taking a first-order approximation.\footnote{In Supplementary Appendix~\ref{Design} we argue that this approximation, akin to principal component analysis, is particularly good for networks with large spectral gaps.} Letting $\mu_1$ and $\mathbf{u}_1$ represent the leading eigenvalue and unit-eigenvector of $\mathbf{G}$ respectively, we can write profits as:
\begin{equation}\label{profits_approx}
\mathbb{E}\left(\pi^{*}\right) \approx \frac{n}{2} \frac{1-n \operatorname{Var}\left(\mathbf{u}_1\right)}{\left(1+r \sigma^2\right)\left(1-\lambda \mu_1\right)^2-\left(\lambda \mu_1\right)^2}.
\end{equation}

This expression highlights four key features. First, profits are decreasing in $\sigma^2$ because higher risk makes incentive provision more costly, inducing the firm to flatten incentives.
Second, profits are increasing in $\mu_1$ (if $\lambda>0$), meaning that networks with larger leading eigenvalues---i.e., denser and more expansive graphs---generate larger profits. This makes sense since these networks are better at amplifying incentives (notice the effect is reversed if $\lambda<0$). Third, and perhaps most interestingly, profits decrease with $\operatorname{Var}(\mathbf{u}_1)$. In other words, profits are larger when eigenvector centrality is evenly distributed across workers.\footnote{For instance, an undirected ring and star for $n=5$ have the same leading eigenvalue, but because the ring has a lower centrality variance, it generates larger profits according to \eqref{profits_approx}.} Finally, we show that the numerator $n\left(1-n\operatorname{Var}(\mathbf{u}_1)\right)$ captures the \textit{number of effective workers} $N_{\mathrm{eff}}$: the size of a uniform network that would generate the same aggregate influence as the observed network.\footnote{For instance, if $\mathbf{u}_1$ is uniform, $\mathbf{u}_1=\frac{1}{\sqrt{n}} 1$, then
$\operatorname{Var}\left(\mathbf{u}_1\right)=0$ and $N_{\text {eff }}=n$, while if $\mathbf{u}_1$ is concentrated on $k$ nodes, then $\operatorname{Var}\left(\mathbf{u}_1\right) \approx \frac{1}{n}-\frac{k}{n^2} $ and $ N_{\mathrm{eff}} =k$.} As firms grow, profits increase only if this effective size grows with $n$; otherwise, adding workers has little impact on performance.

To showcase these effects, we compare networks whose degree distribution follows a power law, $P(d) \sim d^{-\gamma}$. As $\gamma$ decreases, both $\mu_1$ and $\operatorname{Var}\left(\mathbf{u}_1\right)$ increase because more weight is placed on the tail of the degree distribution, creating dominant hubs with many links. These two changes have opposing effects on profits. A larger $\mu_1$ increases network amplification and therefore profits, while a larger $\operatorname{Var}\left(\mathbf{u}_1\right)$ concentrates centrality and reduces profits. Consequently, the net effect of $\gamma$ is generally ambiguous and depends on parameter values. Supplementary Appendix~\ref{Design} illustrates this trade-off using standard configuration-type approximations.
    

Our profit decomposition result also enables an exact comparison of different organizational structures using well-known spectral properties for different families of networks. For instance, we can show that, among all complete bipartite networks, profits are maximized when divisions are of equal size (Corollary \ref{cor: bipartite}). Similarly, in networks where each worker has the same number of connections, expected profits are independent of how the organization is partitioned (Corollary \ref{cor: regular}). We also consider how community structure might influence profits, and find that it doesn't (Corollary \ref{RandomNetworks}). 

Our analysis also highlights how firms should allocate resources to maximize profits.  Imagine firms must decide whether to enhance workers' human capital by investing in a general training program or whether to strengthen peer effects by investing in team-building exercises. If network connectivity is sufficiently sparse, investing in human capital is (obviously) preferable, but the advantage shifts towards strengthening peer effects if workers are sufficiently connected. Proposition \ref{ER_model} provides a simple sufficient condition 
in Erd\H{o}s--R\'enyi random graphs, showing that, once each worker has at least one expected connection, investments in team strength yield higher profits than investments in individual skills.\footnote{This sufficient condition coincides with the threshold for the emergence of a giant component in Erd\H{o}s--R\'enyi random graphs.}

These findings underscore a broader lesson: organizational design decisions should be guided by workers' interactions structure. While standard economic models often focus on individual incentives, this analysis demonstrates that network structure fundamentally shapes firm performance. This first set of results suggests that this is a very promising line of future research.

\section{Concluding Remarks}\label{sec: conclusion}
This paper examines optimal wage contracts in firms with productivity spillovers. We show that firms can leverage network-based incentives to boost productivity, with optimal contract design depending on co-worker externalities, production technology, and institutional constraints.
Our framework provides a rationale for observed trends in the steepening earnings profile within firms that doesn't rely on assuming different endowments of managerial talent or intricate market forces that elevate talent up the firm's hierarchy, to larger teams, or to more valuable organizations. \cite{neal2000theories}, for instance, note that the shape of the earnings distribution cannot be explained by the ``super-star'' CEO phenomenon because scale economies imply there are not enough of them to make a dent in the upper tail of the earnings distribution. By introducing peer effects into standard contract theory, we connect salaries to workers' span of control and rationalize why variable-pay varies so much within organizations.

Empirical validation using co-worker network data could test whether performance-based
pay tracks worker centrality and identify untapped productivity gains. Future research could also explore firm competition \citep{chade2023managers} and the impact of common ownership on managerial incentives and firm strategies \citep{garud2009managing}. The model could also be used to study the excess burden associated to income taxation using a modified labor wedge that accounts for peer interaction \citep{kotlikoff1987tax}. We can also provide quantitative measures of the welfare effects associated to benchmarking salaries when peer effects are taken into account. We expect the impact to depend on organizational structure in interesting ways. These extensions would allow us to understand how variations in network structures across firms might lead to different competitive outcomes and the sorting patterns that ensue \citep{gabaix2008has}.

Finally, our framework offers a foundation to study how artificial intelligence (AI) and automation may disrupt organizational structures. As AI systems automate cognitive tasks and relocate humans to routine knowledge work, the dynamics of employee interactions and organizational structure are likely to shift \citep{ide2024artificial}. The integration of AI is poised to transform production processes, potentially leading to more modular configurations and certainly disrupting the structure of peer-to-peer complementarities in the workplace. Firms will need to reevaluate incentive structures and contract designs moving forward.

\newpage
\setlength{\bibsep}{1.3pt}
\bibliographystyle{ecta}
\bibliography{contracts}

\appendix

\newpage
\section{Proofs}

\subsection{Proof of Proposition \ref{Optimal Contracts}}

For any fixed $\boldsymbol{\alpha}$, the fixed payment $\beta_{i}$ does not affect worker $i$'s first-order condition. That is, varying $\beta_{i}$ leaves effort and the (IC) constraints unchanged. Since firm's profits are strictly decreasing in $\beta_{i}$, if $CE_{i}>0$ for some $i$, the firm can profitably reduce $\beta_{i}$ while holding $\boldsymbol{\alpha}$ fixed. It follows that, at an optimum, the (IR) constraints bind. Thus, we may impose $CE_{i}=0$ for all $i\in N$ to write the firm's problem in matrix form as follows:\footnote{We proceed using the unconstrained best response $\mathbf{e}=\mathbf{C}\boldsymbol{\alpha}$ and verify below that efforts are indeed interior at $\boldsymbol{\alpha}^{*}$.}
\begin{align*}
\max_{\boldsymbol{\alpha},\boldsymbol{\beta}} \:  \mathbb{E}[\pi(\mathbf{e}\mid \boldsymbol{\alpha}, \boldsymbol{\beta})]  & = \mathbf{1}'\mathbf{e} - \mathbf{1}'\mathbf{w} = \mathbf{1}'\mathbf{e} - \mathbf{1}'\boldsymbol{\alpha}(\mathbf{1}'\mathbf{e}) - \mathbf{1}'\boldsymbol{\beta}\\
 \text{subject to }\phantom{d} & \\
\mathbf{1}' \boldsymbol{\beta} & = \frac{1}{2}\mathbf{e}'\mathbf{e} -\lambda \mathbf{e}'\mathbf{Ge} - \mathbf{1}'\boldsymbol{\alpha}(\mathbf{1}'\mathbf{e}) +\frac{1}{2}r\sigma^{2}\boldsymbol{\alpha}'\boldsymbol{\alpha}  \quad &\text{(IRs)} \\
\mathbf{e} &= \mathbf{C}\boldsymbol{\alpha}  \quad &\text{(IC)}
\end{align*}
Substituting in the (IRs) and (IC) constraints, the firm's problem as a function of $\boldsymbol{\alpha}$ is given by: 
\begin{align*}
\max_{\boldsymbol{\alpha}} \: \mathbb{E}[\pi(\mathbf{e}\mid \boldsymbol{\alpha}, \boldsymbol{\beta})]  = \boldsymbol{\alpha}' \mathbf{C}'\mathbf{1} - \frac{1}{2}\boldsymbol{\alpha}' \mathbf{C}' \mathbf{C}\boldsymbol{\alpha} + \lambda \boldsymbol{\alpha}' \mathbf{C}'\mathbf{G} \mathbf{C}\boldsymbol{\alpha} - \frac{1}{2}\sigma^2 r\boldsymbol{\alpha}'\boldsymbol{\alpha}.
\end{align*}
Taking the first order condition and solving for $\boldsymbol{\alpha}$ we get:
\begin{align*}
 \mathbf{C}'\boldsymbol{1} - \mathbf{C}'\mathbf{C}\boldsymbol{\alpha}^{*} + \lambda \mathbf{C}'(\mathbf{G}+\mathbf{G}')\mathbf{C}\boldsymbol{\alpha}^{*} - \sigma^2 r\boldsymbol{\alpha}^{*} = 0 
\implies \mathbf{C}'\mathbf{1}  = \left[ \mathbf{C} '(\mathbf{I} - \lambda (\mathbf{G}+\mathbf{G}') )\mathbf{C}+ \sigma^2 r\mathbf{I} \right]\boldsymbol{\alpha}^{*}
\end{align*}
Under Assumption 2, we can solve for the optimal incentive rule as:
\begin{equation*}
    \boldsymbol{\alpha}^{*} = \left( \mathbf{C} '(\mathbf{I} - \lambda (\mathbf{G}+\mathbf{G}') )\mathbf{C}+ \sigma^2 r\mathbf{I} \right)^{-1} \mathbf{C}'\mathbf{1} = \mathbf{W}\mathbf{C}'\mathbf{1}
\end{equation*}

To get the form of the symmetric matrix $\mathbf{W}$ given in the proposition, we can first use the fact that $\mathbf{C}'=(\mathbf{I}-\lambda\mathbf{G}')^{-1}$ to see that $\mathbf{C} '(\mathbf{I} - \lambda (\mathbf{G}+\mathbf{G}') )\mathbf{C}= 
\mathbf{C}'(\mathbf{I} - \lambda\mathbf{G}')\mathbf{C} 
- \lambda\mathbf{C}' \mathbf{G}\mathbf{C} =  \mathbf{C} 
- \lambda\mathbf{C}' \mathbf{G}\mathbf{C}$. Thus, we have that
\begin{align*}
\mathbf{W} &= \left( \mathbf{C} - \lambda\mathbf{C}' \mathbf{G}\mathbf{C} + r\sigma^{2}\mathbf{I} \right)^{-1} =  \left( \mathbf{C} -\mathbf{I} - \lambda\mathbf{C}'\mathbf{G}\mathbf{C} + (1+r\sigma^{2})\mathbf{I} \right)^{-1}.
\end{align*}
Next, we use the facts that $\lambda\mathbf{G}\mathbf{C}=\lambda\mathbf{C}\mathbf{G}=\lambda\mathbf{G}+\lambda^{2}\mathbf{G}^2+...=(\mathbf{C}-\mathbf{I})$ and $\lambda(\mathbf{G}\mathbf{C})'=(\mathbf{C}'-\mathbf{I})$ to obtain:
\begin{align*}
\mathbf{W} &= \left( \mathbf{C} -\mathbf{I} - \lambda\mathbf{C}' \mathbf{G}\mathbf{C} + (1+r\sigma^{2})\mathbf{I} \right)^{-1} =  \left(\lambda \mathbf{C}\mathbf{G}  -\lambda\mathbf{C}'\mathbf{C}\mathbf{G}+ (1+\sigma^2 r)\mathbf{I} \right)^{-1}  \\
& = \left((1+\sigma^2 r)\mathbf{I}-\lambda(\mathbf{C}'-\mathbf{I}) \mathbf{C}\mathbf{G} \right)^{-1} = \left((1+\sigma^2 r)\mathbf{I}-\lambda^2(\mathbf{C}\mathbf{G})' \mathbf{C}\mathbf{G} \right)^{-1}.
\end{align*}

 Next, notice that $\alpha_{i}^{*}> 0$ for all $i\in N$ and, thus, workers' efforts are interior. Under Assumptions~\ref{assumption1} and~\ref{spectralradius}, $\mathbf{C}$ and $\mathbf{W}$ are entrywise nonnegative. Thus, we have $\mathbf{C}'\mathbf{1}\gg \mathbf{0}$ and so $\boldsymbol{\alpha}^{*}=\mathbf{W}\mathbf{C}'\mathbf{1}\gg 0$. It follows that $\mathbf{e}^{*}=\mathbf{C}\boldsymbol{\alpha}^{*}\gg 0$. Because the workers' problem is strictly concave in $\mathbf{e}$, the constrained best response coincides with the FOC solution $\mathbf{e}=\mathbf{C}\boldsymbol{\alpha}$ at $\boldsymbol{\alpha}^{*}$.

For uniqueness, notice that the firm's objective is strictly concave as the Hessian matrix $\mathbf{H} = -\left[ \mathbf{C} '(\mathbf{I} - \lambda (\mathbf{G}+\mathbf{G}') )\mathbf{C}+ \sigma^2 r\mathbf{I}\right] = -\left[(1+r\sigma^{2})\mathbf{I}-\lambda^2 (\mathbf{GC})'\mathbf{GC} \right]$ is negative definite since Assumption 2 guarantees that $\mathbf{W}^{-1} \succ  0$, and thus $\mathbf{H}$ is negative definite.\footnote{To see this, 
notice that for any nonzero vector $\mathbf{x}$ we have that $\mathbf{x}'((1+r\sigma^{2})\mathbf{I} - \lambda^2(\mathbf{GC})'\mathbf{GC})\mathbf{x} = (1+r\sigma^{2})\|\mathbf{x}\|^{2} - \lambda^2\mathbf{x}'(\mathbf{GC})'\mathbf{GC}\mathbf{x} = \|\mathbf{x}\|^{2}(1+r\sigma^{2}- \lambda^2\frac{\mathbf{x}'(\mathbf{GC})'\mathbf{GC}\mathbf{x}}{\|x\|^{2}} )$, where $\|\mathbf{x\|}$ is the Euclidean norm. Moreover, for a symmetric matrix, the Rayleigh quotient $\lambda^2\frac{\mathbf{x}'(\mathbf{GC})'\mathbf{GC}\mathbf{x}}{\|x\|^{2}}$ is less than or equal than its spectral radius, which by assumption~\ref{spectralradius} is less than $1+r\sigma^{2}$. Thus, $\mathbf{x}'((1+r\sigma^{2})\mathbf{I} - \lambda^2(\mathbf{GC})'\mathbf{GC})\mathbf{x}> 0$.}

Finally, using $\mathbf{e}^{*}$ and $\boldsymbol{\alpha}^{*}$ we can write the (IR) constraints in matrix form and solve for $\boldsymbol{\beta}^{*}$ to obtain:
\[
\boldsymbol{\beta}^{*} (\boldsymbol{\alpha}^{*}, \mathbf{e}^{*}) = \frac{1}{2} (\mathbf{e}^{*} \circ \mathbf{e}^{*}) - (\mathbf{e}^{*} \circ \lambda \mathbf{G}\mathbf{e}^{*}) + \frac{1}{2}(r\sigma^2) (\boldsymbol{\alpha}^{*} \circ \boldsymbol{\alpha}^{*}) - \boldsymbol{\alpha}^{*} \circ \mathbf{1}(\mathbf{1}'\mathbf{e}^{*}),  
\]
where $\circ$ denotes the Hadamard product. Recalling that $\mathbf{e}^{*} = \mathbf{C}\boldsymbol{\alpha}^{*}$ completes the proof. \qed

\subsection{Proof of Proposition \ref{monotonic}}

By Proposition~\ref{Optimal Contracts}, \(\boldsymbol{\alpha}^\ast=\mathbf{W}\mathbf{C}'\mathbf{1}\). Factoring out $1+r\sigma^{2}$, we can write $\mathbf{W}= \frac{1}{1+r\sigma^2}
\left[ \mathbf{I}-\frac{(\lambda \mathbf{GC})^{\prime}\lambda \mathbf{GC}}{1+r\sigma^2} \right]^{-1}.$ Under Assumption~\ref{spectralradius}, we can use the Neumann Expansion to write:
\begin{equation*}
    \mathbf{W} = \frac{1}{1+r\sigma^2}\left[ \mathbf{I} + \frac{(\lambda \mathbf{GC})^{\prime}\lambda \mathbf{GC}}{1+r\sigma^2} + \left(\frac{(\lambda \mathbf{GC})^{\prime}\lambda \mathbf{GC}}{1+r\sigma^2}\right)^{2}+ \cdots \right]  = \frac{1}{1+r\sigma^2}\mathbf{I}+\mathbf{B}(\sigma^2),
\end{equation*}
where each entry of $\mathbf{B}(\sigma^2)$ is of order $O((1+r\sigma^2)^{-2})$. Therefore, $\boldsymbol{\alpha}^\ast =  \frac{1}{1+r\sigma^2}\mathbf{C}'\mathbf{1} +\mathbf{B}(\sigma^2)\mathbf{C}'\mathbf{1}$. As $\sigma^{2}\to\infty$, we have $(1+r\sigma^{2})\boldsymbol{\alpha}^{*} \to \mathbf{b}=\mathbf{C}'\mathbf{1}$.
It follows that there exists a finite \(\bar{\sigma}^2(G)\) such that if $b_{i}>b_{j}$ then $\alpha_i^\ast>\alpha_j^\ast$ for all \(\sigma^2\ge \bar{\sigma}^2(\mathbf{G})\). That is, for sufficiently high risk, the ranking of optimal incentives preserves the strict ordering of Bonacich centralities. \hfill\qed

\subsection{Proof of Proposition \ref{CS_Link}}

Recall that by Assumptions~\ref{assumption1} and~\ref{spectralradius} both $\mathbf{C}$ and $\mathbf{W}$ are entrywise nonnegative. Moreover, recall that $\mathbf{b}:=\mathbf{C}'\mathbf{1}$. 

Fix $i,j\in N$. We have that: 
\begin{equation*}
	\frac{\partial \mathbf{C}}{\partial g_{ij}} = \lambda \mathbf{C}\mathbf{E}_{ij}\mathbf{C}\geq 0,
\end{equation*}
where $\mathbf{E}_{ij}$ is the matrix with a one in position $(i,j)$ and zeros elsewhere. Therefore, we have that: 
\begin{equation*}
	\frac{\partial \mathbf{C}'\mathbf{1}}{\partial g_{ij}} = \lambda \mathbf{C}'\mathbf{E}_{ji}\mathbf{C}'\mathbf{1}=\lambda b_{i}\mathbf{C}_{j\cdot }'\geq 0. 
\end{equation*}
In particular, $\partial b_{j}/\partial g_{ij} = \lambda b_{i}c_{jj}>0$ since $b_{i}\geq 1$ and $c_{jj}\geq1$. Next, differentiating $\mathbf{W}$ we obtain: 
\begin{equation*}
	\frac{\partial \mathbf{W} }{\partial g_{ij}} = - \mathbf{W} \frac{\partial [(1+r\sigma^{2})\mathbf{I}-(\lambda\mathbf{GC})'(\lambda\mathbf{G}\mathbf{C})]}{\partial g_{ij}}\mathbf{W} = \mathbf{W} \left[ \left(\frac{ \partial\lambda\mathbf{GC}}{\partial g_{ij}}\right)'\lambda\mathbf{GC} + (\lambda\mathbf{GC})'\frac{\partial \lambda \mathbf{GC}}{\partial g_{ij}} \right] \mathbf{W}.
\end{equation*}
Since $\lambda\mathbf{GC} = \mathbf{C}-\mathbf{I}$, we have that: 
\begin{equation*}
	\frac{\partial \lambda \mathbf{GC}}{\partial g_{ij}} = \frac{\partial (\mathbf{C}-\mathbf{I}) }{\partial g_{ij}} = \frac{\partial \mathbf{C}}{\partial g_{ij}} = \lambda \mathbf{C}\mathbf{E}_{ij}\mathbf{C}\geq 0. 
\end{equation*} 
It follows that $\partial\mathbf{W}/\partial g_{ij}\geq 0$. Therefore, 
\begin{equation*}
	\frac{\partial \boldsymbol{\alpha}^{*}}{\partial g_{ij}} = \frac{\partial \mathbf{W}}{\partial g_{ij}}\mathbf{b} + \mathbf{W}\frac{\partial \mathbf{b}}{\partial g_{ij}}\geq 0. 
\end{equation*}

For the strict part, let $s$ be such that $w_{sj}>0$.  Then
\[
\frac{\partial \alpha_s^\ast}{\partial g_{ij}}
=
\sum_r \frac{\partial w_{sr}}{\partial g_{ij}}\, b_r
+
\sum_r w_{sr}\frac{\partial b_r}{\partial g_{ij}}
\ge
w_{sj}\frac{\partial b_j}{\partial g_{ij}}
>0.
\]
Thus, \(\partial \alpha_s^\ast/\partial g_{ij}>0\) for every worker \(s\) with \(w_{sj}>0\), i.e., for every worker who shares common influence with \(j\). Taking \(s=j\), and recalling that \(w_{jj}>0\), also gives $\partial \alpha_j^\ast/\partial g_{ij}>0$. \hfill \qed

\subsection{Proof of Proposition \ref{Modular Production}}


Notice that we can replace $\boldsymbol{\alpha}=(\mathbf{I}-\lambda \mathbf{G})\mathbf{e}$ and, using the auxiliary variable $\hat{e}$, solve the equivalent dual problem
        \begin{align*}
        &\max_{(\hat{e}, \mathbf{e})} \left( \hat{e} - \frac{1}{2} \mathbf{e}' [(\mathbf{I}-2\lambda \mathbf{G}) + \sigma^2 r(\mathbf{I}-\lambda \mathbf{G}')(\mathbf{I}-\lambda \mathbf{G})]\mathbf{e} \right) \\ \text{subject to} \\
        &\mathbf{M}\mathbf{e}=\hat{e}\boldsymbol{1}_K,
        \end{align*}
and retrieve $\boldsymbol{\alpha}^{*}$ using $\boldsymbol{\alpha}^{*}=(\mathbf{I}-\lambda \mathbf{G}) \mathbf{e}^{*}$. Let $\breve{\boldsymbol{\Sigma}} \equiv (\mathbf{I}-2\lambda \mathbf{G}) + \sigma^2 r(\mathbf{I}-\lambda \mathbf{G}^\prime)(\mathbf{I}-\lambda \mathbf{G})$. Considering the $K \times 1$ vector of Lagrangian multipliers $\boldsymbol{\mu}_{K}$, the Lagrangian of the above problem can be expressed as:
\[
\mathcal{L}(\mathbf{e}, \hat{e}, \boldsymbol{\mu}) = \hat{e} - \frac{1}{2} \mathbf{e}' \breve{\boldsymbol{\Sigma}} \mathbf{e} - \boldsymbol{\mu}_{K}' (\hat{e}\boldsymbol{1_K} - \mathbf{M}\mathbf{e}).
\]
We have the following system of first order conditions:
\begin{align}
\frac{\partial \mathcal{L}}{\partial \mathbf{e}} &= -\frac{1}{2}\left(\breve{\boldsymbol{\Sigma}}+\breve{\boldsymbol{\Sigma}}^\prime\right) \mathbf{e} + \mathbf{M}' \boldsymbol{\mu}_{K} = -\boldsymbol{\Sigma} \mathbf{e} + \mathbf{M}' \boldsymbol{\mu_{K}} = \boldsymbol{0_n}  \label{A_ModularFOC1} \\
\frac{\partial \mathcal{L}}{\partial \boldsymbol{\mu}} &= \mathbf{M}\mathbf{e} - \hat{e}\boldsymbol{1_K} = \boldsymbol{0_K} \label{A_ModularFOC2} \\
\frac{\partial \mathcal{L}}{\partial \hat{e}} &= 1-\boldsymbol{1_K}' \boldsymbol{\mu_{K}} = 0 \label{A_ModularFOC3} 
\end{align}
Let $\boldsymbol{\Sigma} := \frac{1}{2}\left(\breve{\boldsymbol{\Sigma}}+\breve{\boldsymbol{\Sigma}}^\prime\right) = (1+\sigma^2 r)(\mathbf{I}-\lambda (\mathbf{G}+\mathbf{G}')) + \sigma^2 r(\lambda \mathbf{G})^\prime(\lambda \mathbf{G})$.\footnote{Notice that $\breve{\boldsymbol{\Sigma}}^{\prime} = (\mathbf{I}-2\lambda \mathbf{G})^{\prime} + \sigma^2 r(\mathbf{I}-\lambda \mathbf{G}^\prime- \lambda\mathbf{G}+\lambda^{2} \mathbf{G}^{\prime}\mathbf{G})^{\prime} = \mathbf{I}-2 \lambda \mathbf{G}'+\sigma^2 r\left(\mathbf{I}-\lambda \mathbf{G} - \lambda \mathbf{G}' + \lambda^{2} \mathbf{G}'\mathbf{G}\right)$ since $(\mathbf{G}^{\prime}\mathbf{G})^{\prime}=\mathbf{G}^{\prime}(\mathbf{G}^{\prime})^{\prime}=\mathbf{G}^{\prime}\mathbf{G}$, i.e., $\mathbf{G}^{\prime}\mathbf{G}$ is a symmetric matrix. In the case of undirected graphs, $\mathbf{G}=\mathbf{G}'$, and $\mathbf{\Sigma}$ simplifies to   $(\mathbf{I}-2\lambda \mathbf{G})(1 + \sigma^2 r)+\sigma^2 r(\lambda \mathbf{G})^2$ and $\boldsymbol{\Sigma}=\breve{\boldsymbol{\Sigma}}$.} The Hessian with respect to $\mathbf{e}$ is $-\boldsymbol{\Sigma}$. From Proposition \ref{Optimal Contracts} we see that $\mathbf{W}=(\mathbf{C}'\boldsymbol{\Sigma}\mathbf{C})^{-1}$, which is well-defined and positive definite by Assumption \ref{spectralradius}. That is, $\mathbf{C}'\mathbf{\Sigma C}\succ 0$. Because $\mathbf{C}$ is invertible, this implies $\boldsymbol{\Sigma}\succ 0$. Therefore, the firm's problem is strictly concave in $\mathbf{e}$. 

From \eqref{A_ModularFOC1} we obtain: 
\begin{equation*}
    \mathbf{e} = \boldsymbol{\Sigma}^{-1} \mathbf{M}' \boldsymbol{\mu}_{K}.
\end{equation*}
Let $\mathbf{H} := \mathbf{M} \boldsymbol{\Sigma}^{-1} \mathbf{M}'$. Because $\boldsymbol{\Sigma}$ is positive definite and $\mathbf{M}$ has full row rank $K$ by construction, we have that $\mathbf{H}^{-1}$ exists. Plugging $\mathbf{e}$ into \eqref{A_ModularFOC2}, we can solve for $\boldsymbol{\mu}_{K}$:
\begin{equation*}
    \mathbf{M} \boldsymbol{\Sigma}^{-1} \mathbf{M}' \boldsymbol{\mu}_{K} = \hat{e} \mathbf{1}_{K}  \quad \implies \quad  \boldsymbol{\mu}_{K} = \hat{e} \mathbf{H}^{-1} \mathbf{1}_{K},
\end{equation*}
and use $\boldsymbol{\mu}_{K}$ in \eqref{A_ModularFOC3} to solve for $\hat{e}$:
\begin{equation*}
    \mathbf{1}_{K}' \boldsymbol{\mu}_{K} = \hat{e} \mathbf{1}_{K}' \mathbf{H}^{-1} \mathbf{1}_{K} = 1 \quad \Rightarrow \quad \hat{e} = \frac{1}{\mathbf{1}_{K}' \mathbf{H}^{-1} \mathbf{1}_{K}}.
\end{equation*}
Finally, putting everything together and using the fact that $\boldsymbol{\alpha} = (\mathbf{I} - \lambda \mathbf{G}) \mathbf{e}$, and defining the vector of worker-specific Lagrangian multipliers $\boldsymbol{\mu}=\mathbf{M} '\boldsymbol{\mu}_{K}$, we get:
\begin{align*}
\mathbf{e} =  \frac{1}{\mathbf{1}_{K}' \mathbf{H}^{-1} \mathbf{1}_{K}} \boldsymbol{\Sigma}^{-1} \mathbf{M}' \mathbf{H}^{-1} \mathbf{1}_{K} \quad \implies \quad \boldsymbol{\alpha} &= \frac{1}{\mathbf{1}_{K}' \mathbf{H}^{-1} \mathbf{1}_{K}} (\mathbf{I} - \lambda \mathbf{G}) \boldsymbol{\Sigma}^{-1} \mathbf{M}' \mathbf{H}^{-1} \mathbf{1}_{K}=(\boldsymbol{\Sigma} \mathbf{C})^{-1}\boldsymbol{\mu}. 
\end{align*}
Thus, $\boldsymbol{\mu}= \mathbf{M}' \mathbf{H}^{-1}\mathbf{1}/ (\mathbf{1}' \mathbf{H}^{-1} \mathbf{1})$. Notice that, since $\lambda \mathbf{G}\mathbf{C}= \mathbf{C}-\mathbf{I}$:
\begin{align*}
\mathbf{C}'\boldsymbol{\Sigma}\mathbf{C} 
&= (1+r\sigma^2)\mathbf{C}'\Bigl(\mathbf{I}-\lambda(\mathbf{G}'+\mathbf{G})\Bigr)\mathbf{C} + r\sigma^2\Bigl(\lambda\mathbf{G}\mathbf{C}\Bigr)'\lambda\mathbf{G}\mathbf{C} \\
&= (1+r\sigma^2)\Bigl(\mathbf{C}'\mathbf{C}-\mathbf{C}'(\lambda\mathbf{G}'+\lambda\mathbf{G})\mathbf{C}\Bigr)
+ r\sigma^2\Bigl(\mathbf{C}'\mathbf{C}-\mathbf{C}'-\mathbf{C}+\mathbf{I}\Bigr) \\
&= \mathbf{C}'\Bigl(\mathbf{I}-\lambda(\mathbf{G}'+\mathbf{G})\Bigr)\mathbf{C} + r\sigma^2\mathbf{I} 
= \mathbf{W}^{-1},
\end{align*}
where the last equality follows directly from Proposition \ref{Optimal Contracts}. As $ \mathbf{W}=(\mathbf{C}'\boldsymbol{\Sigma}\mathbf{C})^{-1}=(\boldsymbol{\Sigma}\mathbf{C})^{-1}\mathbf{C}'^{-1}$, $ \mathbf{W}\mathbf{C}'=(\boldsymbol{\Sigma}\mathbf{C})^{-1}$. Moreover, $\mathbf{H} = \mathbf{M} \boldsymbol{\Sigma}^{-1} \mathbf{M}'=(\mathbf{M}\mathbf{C})\mathbf{W}(\mathbf{M}\mathbf{C})'$ and  $\boldsymbol{\alpha}^{*}=\mathbf{WC}'\boldsymbol{\mu}$.\qed

\subsection{Proof of Corollary \ref{modules_nopeers}}

If $\lambda=0$ then $\boldsymbol{\Sigma}=(1+r\sigma^2)\mathbf{I}$ and $\mathbf{H}=\frac{1}{1+r\sigma^2}\mathbf{MM}^\prime$. Substituting into equation \eqref{modular_alpha}, we get
\[\boldsymbol{\alpha}^{*}=\frac{1}{1+r\sigma^2}\frac{\mathbf{M}^\prime(\mathbf{MM}^\prime)^{-1}\mathbf{1}}{\mathbf{1}^\prime (\mathbf{MM}^\prime)^{-1} \mathbf{1}}.\]
Notice that $\mathbf{MM}^\prime=\operatorname{diag}(n_1,n_2,\ldots,n_K)$ so the denominator of the second term above can be written as: 
\[\mathbf{1}^\prime (\mathbf{MM}^\prime)^{-1} \mathbf{1}=\sum_{k=1}^K \frac{1}{n_k}=\frac{\sum_r \prod_{k\in K\setminus r}\: n_k}{\prod_{k\in K}\: n_k}.\] Notice that the numerator is simply an $n\times 1$ vector where the $i$-th position is $1/n_{k(i)}$. Putting everything together we get the desired expression. Finally, if $n_1=n_2=\ldots=n_K=\tilde{n}$ then the expression in Corollary \ref{modules_nopeers} becomes 
\[\alpha_i^{*}= \frac{1}{1+r\sigma^2}\frac{1/\tilde{n}}{K\tilde{n}^{K-1}/\tilde{n}^K}=\frac{1}{1+r\sigma^2} \frac{1}{K}. \tag*{\qed}
\]

\subsection{Proof of Corollary \ref{single_module}}

Notice from Proposition \ref{Modular Production} that for $\mathbf{M} = \mathbf{I}$, we get $\mathbf{H} = \mathbf{M} \boldsymbol{\Sigma}^{-1} \mathbf{M}' = \boldsymbol{\Sigma}^{-1}$, and:
\begin{align*}
    \boldsymbol{\alpha}^* &= (\mathbf{I} - \lambda \mathbf{G}) \boldsymbol{\Sigma}^{-1} \frac{\mathbf{M}' \mathbf{H}^{-1} \mathbf{1}}{\mathbf{1}' \mathbf{H}^{-1} \mathbf{1}} = (\mathbf{I} - \lambda \mathbf{G}) \boldsymbol{\Sigma}^{-1} \boldsymbol{\Sigma} \mathbf{1}\frac{1}{\mathbf{1}' \boldsymbol{\Sigma} \mathbf{1}} = \frac{1}{\mathbf{1}' \boldsymbol{\Sigma} \mathbf{1}} (\mathbf{I} - \lambda \mathbf{G}) \mathbf{1}.
\end{align*}
Next, notice that $\mathbf{G} \mathbf{1}=\mathbf{d}$ where $\mathbf{d}:=(d_1,d_2,\ldots, d_n)^\prime$. Then,
\begin{align*}
    \mathbf{1}' \boldsymbol{\Sigma} \mathbf{1} &= \mathbf{1}' \left[(1+\sigma^2 r)(\mathbf{I}-\lambda (\mathbf{G}+\mathbf{G}')) + \sigma^2 r(\lambda \mathbf{G})^\prime(\lambda \mathbf{G})\right] \mathbf{1} \\
    &= (1 + r\sigma^2) \mathbf{1}^\prime (\mathbf{I} - \lambda (\mathbf{G}+\mathbf{G}')) \mathbf{1} + r\sigma^2 ( \lambda \mathbf{G}\mathbf{1})^\prime(\lambda \mathbf{G} \mathbf{1}) \\
     &= \sum_{j \in N} \left[(1 + r\sigma^2)(1 - 2\lambda d_j) + \lambda^2 r\sigma^2 d_j^2 \right]
\end{align*}
Therefore, $\boldsymbol{\alpha}^* = \frac{1}{\mathbf{1}' \boldsymbol{\Sigma} \mathbf{1}} (\mathbf{I} - \lambda \mathbf{G}) \mathbf{1} = \frac{1}{\xi} (\mathbf{1}- \lambda \mathbf{d})$. And thus, \(\alpha_i^* = \frac{1 - \lambda d_i}{\xi}\), as desired.
 \qed

 \subsection{Proof of Proposition \ref{Coarse Contracts}}
The firm's problem is given by:\footnote{As we did for Proposition~\ref{Optimal Contracts}, we proceed using the unconstrained best response only and verify below that efforts are indeed interior at $\hat{\boldsymbol{\alpha}}^{*}$.}
\begin{align*}
\max_{\hat{\boldsymbol{\alpha}},\hat{\boldsymbol{\beta}}}\quad 
& \mathbf{1}'\mathbf{C}\mathbf{T}'\hat{\boldsymbol{\alpha}}
-\big(\mathbf{1}'\mathbf{C}\mathbf{T}'\hat{\boldsymbol{\alpha}}\big)\big(\mathbf{1}'\mathbf{T}'\hat{\boldsymbol{\alpha}}\big)
-\mathbf{1}'\mathbf{T}'\hat{\boldsymbol{\beta}} \\[2mm]
\text{s.t.}\quad 
& \mathbf{T}'\hat{\boldsymbol{\beta}}
+\big(\mathbf{1}'\mathbf{C}\mathbf{T}'\hat{\boldsymbol{\alpha}}\big)\mathbf{T}'\hat{\boldsymbol{\alpha}}
-\boldsymbol{\psi}\!\big(\mathbf{C}\mathbf{T}'\hat{\boldsymbol{\alpha}}\big)
-\frac{r\sigma^{2}}{2}\operatorname{diag}(\mathbf{T}'\hat{\boldsymbol{\alpha}})\mathbf{T}'\hat{\boldsymbol{\alpha}}
\ge\mathbf{0}, \tag{IR}
\end{align*}
where $\boldsymbol{\psi}\big(\mathbf{C}\mathbf{T}'\hat{\boldsymbol{\alpha}}\big)$ is the vector of effort costs. The Lagrangian is given by:
\begin{equation*}
    \begin{aligned}
\mathcal{\mathbf{L}}( \hat{\boldsymbol{\beta}},\hat{\boldsymbol{\alpha}}, \boldsymbol{\mu})  = \, \,& \mathbf{1}'\mathbf{C}\mathbf{T}'\hat{\boldsymbol{\alpha}}
-\big(\mathbf{1}'\mathbf{C}\mathbf{T}'\hat{\boldsymbol{\alpha}}\big)\big(\mathbf{1}'\mathbf{T}'\hat{\boldsymbol{\alpha}}\big)
-\mathbf{1}'\mathbf{T}'\hat{\boldsymbol{\beta}} \\
& + \boldsymbol{\mu}'\left( \mathbf{T}'\hat{\boldsymbol{\beta}}
+\big(\mathbf{1}'\mathbf{C}\mathbf{T}'\hat{\boldsymbol{\alpha}}\big)\mathbf{T}'\hat{\boldsymbol{\alpha}}
-\boldsymbol{\psi}\!\big(\mathbf{C}\mathbf{T}'\hat{\boldsymbol{\alpha}}\big)
-\frac{r\sigma^{2}}{2}\operatorname{diag}(\mathbf{T}'\hat{\boldsymbol{\alpha}})\mathbf{T}'\hat{\boldsymbol{\alpha}}\right), 
    \end{aligned}
\end{equation*}
where $\boldsymbol{\mu}\in \mathbb{R}^{n}$ are the KKT multipliers on the (IR) constraints. Denote by $\mathbf{f}(\hat{\boldsymbol{\alpha}})$ the part of the individual rationality constraints that depends on incentives $\hat{\boldsymbol{\alpha}}$:
\begin{equation*}
    \begin{aligned}
        \mathbf{f}(\hat{\boldsymbol{\alpha}}) & = 
- \big(\mathbf{1}'\mathbf{C}\mathbf{T}'\hat{\boldsymbol{\alpha}}\big)\mathbf{T}'\hat{\boldsymbol{\alpha}}
+\boldsymbol{\psi}\!\big(\mathbf{C}\mathbf{T}'\hat{\boldsymbol{\alpha}}\big)
+\frac{r\sigma^{2}}{2}\operatorname{diag}(\mathbf{T}'\hat{\boldsymbol{\alpha}})\mathbf{T}'\hat{\boldsymbol{\alpha}} \\
& = - \big(\mathbf{1}'\mathbf{C}\mathbf{T}'\hat{\boldsymbol{\alpha}}\big)\mathbf{T}'\hat{\boldsymbol{\alpha}}
+ \frac{1}{2}\operatorname{diag}(\mathbf{CT}'\hat{\boldsymbol{\alpha}})\mathbf{CT}'\hat{\boldsymbol{\alpha}} 
-  \lambda \operatorname{diag}(\mathbf{CT}'\hat{\boldsymbol{\alpha}})\mathbf{GCT}'\hat{\boldsymbol{\alpha}}  +\frac{r\sigma^{2}}{2}\operatorname{diag}(\mathbf{T}'\hat{\boldsymbol{\alpha}})\mathbf{T}'\hat{\boldsymbol{\alpha}}.
    \end{aligned}
\end{equation*}
Then, the KKT conditions are given by: 
\begin{align}
    \frac{\partial \mathcal{\mathbf{L}}}{\partial \hat{\boldsymbol{\beta}}} & = -\mathbf{T1} + \mathbf{T}\boldsymbol{\mu} = \mathbf{0}, \label{eq:KKT_beta} \\
        \frac{\partial \mathcal{\mathbf{L}}}{\partial \hat{\boldsymbol{\alpha}}} & =  \mathbf{T}\mathbf{C}'\mathbf{1} - \Big[(\mathbf{1}'\mathbf{T}'\hat{\boldsymbol{\alpha}})\mathbf{T}\mathbf{C}'\mathbf{1} +(\mathbf{1}'\mathbf{C}\mathbf{T}'\hat{\boldsymbol{\alpha}})\mathbf{T}\mathbf{1}\Big] 
        - \left(\frac{\partial \mathbf{f}(\hat{\boldsymbol{\alpha}})}{\partial \hat{\boldsymbol{\alpha}}}\right)'\boldsymbol{\mu} = \mathbf{0}, \label{eq:KKT_alpha} \\
             \frac{\partial \mathcal{\mathbf{L}}}{\partial \boldsymbol{\mu}} & =     \mathbf{T}'\hat{\boldsymbol{\beta}} - \mathbf{f}(\hat{\boldsymbol{\alpha}}) \geq 0, \quad \boldsymbol{\mu} \geq 0, \quad \boldsymbol{\mu}' (\mathbf{T}'\hat{\boldsymbol{\beta}} - \mathbf{f}(\hat{\boldsymbol{\alpha}})) =  0 \label{eq:KKT_cs}.
\end{align}

Notice that from \eqref{eq:KKT_beta} we obtain that for each group $k$ we have:
\begin{equation*}
    \sum_{i\in k}\mu_{i} = n_{k}.
\end{equation*}
Moreover, for each worker in group $k$ the individual rationality constraint must hold, $\beta_{k} \geq f_{i}(\hat{\boldsymbol{\alpha}})$ for all $i\in k$. To ensure that $\beta_{k}$ satisfies all (IR) constraints in group $k$, the firm sets
\begin{equation*}
    \beta_k^{*}=\max_{i\in k}f_{i}(\hat{\boldsymbol{\alpha}})
=\max_{i\in k}\Big\{\psi_i\!\big(\mathbf{C}\mathbf{T}'\hat{\boldsymbol{\alpha}}\big)\Big\}
+\frac{r\sigma^{2}}{2}\hat{\alpha}_k^{2}
-\hat{\alpha}_k\mathbf{1}'\mathbf{C}\mathbf{T}'\hat{\boldsymbol{\alpha}},
\end{equation*}
where $\psi_i\!\big(\mathbf{C}\mathbf{T}'\hat{\boldsymbol{\alpha}}\big)=\psi_{i}(\mathbf{e}) = \frac{1}{2}(e_{i}^{*})^{2} - \lambda e_{i}^{*}\sum_{j}g_{ij}e_{j}^{*}$. 
This means that the KKT multipliers satisfy \eqref{eq:KKT_beta}, with $\mu_i=0$ whenever $\beta_k^{*}>f_i(\hat{\boldsymbol{\alpha}})$ and $\mu_i\geq 0$ whenever $\beta_k^{*}=f_i(\hat{\boldsymbol{\alpha}})$. Thus, workers whose effort cost is not maximal in their group have zero multipliers. If there is a unique worker with maximum effort cost, \eqref{eq:KKT_beta} implies $\mu_i=n_k$. If several workers tie, their nonnegative multipliers sum to $n_k$ and must additionally satisfy incentive stationarity \eqref{eq:KKT_alpha}. Which worker(s) bind is not fixed ex ante: it is precisely whichever $i\in k$ attain $\beta_k^{*}=\max_{j\in k}f_j(\hat{\boldsymbol{\alpha}})$. 

Next, notice that:
\begin{equation*}
    \begin{aligned}
        \left(\frac{\partial \mathbf{f}(\hat{\boldsymbol{\alpha}})}{\partial \hat{\boldsymbol{\alpha}}}\right) ' =&  
        - \left[\mathbf{TC}'\mathbf{1}\hat{\boldsymbol{\alpha}}'\mathbf{T} 
        + (\mathbf{1}'\mathbf{C}\mathbf{T}'\hat{\boldsymbol{\alpha}})\mathbf{T}\right]  + \mathbf{TC}'\operatorname{diag}(\mathbf{C}\mathbf{T}'\hat{\boldsymbol{\alpha}}) \\
        &
        - \lambda\mathbf{TC}'\Big[\operatorname{diag}(\mathbf{G}\mathbf{C}\mathbf{T}'\hat{\boldsymbol{\alpha}}) 
        + \mathbf{G}'\operatorname{diag}(\mathbf{C}\mathbf{T}'\hat{\boldsymbol{\alpha}})\Big] + r\sigma^{2} \mathbf{T}\operatorname{diag}(\mathbf{T}'\hat{\boldsymbol{\alpha}}).
    \end{aligned}
\end{equation*}
Thus, we can rewrite \eqref{eq:KKT_alpha} as follows:
\begin{equation}\label{eq:KKT_alpha2}
    \begin{aligned}
    & \mathbf{T}\mathbf{C}'\mathbf{1} 
    -\Big[(\mathbf{1}'\mathbf{T}'\hat{\boldsymbol{\alpha}})\mathbf{T}\mathbf{C}'\mathbf{1}
+(\mathbf{1}'\mathbf{C}\mathbf{T}'\hat{\boldsymbol{\alpha}})\mathbf{T}\mathbf{1}\Big] 
+ \left[\mathbf{TC}'\mathbf{1}\hat{\boldsymbol{\alpha}}'\mathbf{T} 
        + (\mathbf{1}'\mathbf{C}\mathbf{T}'\hat{\boldsymbol{\alpha}})\mathbf{T}\right]\boldsymbol{\mu}\\
&  
-  \mathbf{T}\mathbf{C}'\Big[\operatorname{diag}(\mathbf{C}\mathbf{T}'\hat{\boldsymbol{\alpha}})
-\lambda\operatorname{diag}(\mathbf{G}\mathbf{C}\mathbf{T}'\hat{\boldsymbol{\alpha}})
-\lambda\mathbf{G}'\operatorname{diag}(\mathbf{C}\mathbf{T}'\hat{\boldsymbol{\alpha}})\Big]\boldsymbol{\mu}
- r\sigma^{2} \mathbf{T}\operatorname{diag}(\mathbf{T}'\hat{\boldsymbol{\alpha}})\boldsymbol{\mu}=0.
    \end{aligned}
\end{equation}

Notice that $\mathbf{T}\boldsymbol{\mu} = \mathbf{T1}$ and $\mathbf{TC}'\mathbf{1}\hat{\boldsymbol{\alpha}}'\mathbf{T}\boldsymbol{\mu} 
+ (\mathbf{1}'\mathbf{C}\mathbf{T}'\hat{\boldsymbol{\alpha}})\mathbf{T}\boldsymbol{\mu} = 
\mathbf{TC}'\mathbf{1}(\hat{\boldsymbol{\alpha}}'\mathbf{T}\mathbf{1}) 
+ (\mathbf{1}'\mathbf{C}\mathbf{T}'\hat{\boldsymbol{\alpha}})\mathbf{T}\mathbf{1}$ so that the first two terms in brackets of \eqref{eq:KKT_alpha2} cancel out. Let $\mathbf{M} = \operatorname{diag}(\boldsymbol{\mu})$. We have $\operatorname{diag}(\mathbf{C}\mathbf{T}'\hat{\boldsymbol{\alpha}})\boldsymbol{\mu}  = \mathbf{C}\mathbf{T}'\hat{\boldsymbol{\alpha}}  \circ  \boldsymbol{\mu} = \boldsymbol{\mu}  \circ  \mathbf{C}\mathbf{T}'\hat{\boldsymbol{\alpha}} = \operatorname{diag}(\boldsymbol{\mu})\mathbf{C}\mathbf{T}'\hat{\boldsymbol{\alpha}} = \mathbf{M}\mathbf{C}\mathbf{T}'\hat{\boldsymbol{\alpha}}.$ Similarly, we have that $\operatorname{diag}(\mathbf{G}\mathbf{C}\mathbf{T}'\hat{\boldsymbol{\alpha}})\boldsymbol{\mu} = \mathbf{M}\mathbf{G}\mathbf{C}\mathbf{T}'\hat{\boldsymbol{\alpha}} $ and $ \mathbf{G}'\operatorname{diag}(\mathbf{C}\mathbf{T}'\hat{\boldsymbol{\alpha}})\boldsymbol{\mu}=\mathbf{G}'\mathbf{M}\mathbf{C}\mathbf{T}'\hat{\boldsymbol{\alpha}}$. Finally, notice that $ \mathbf{T}\operatorname{diag}(\mathbf{T}'\hat{\boldsymbol{\alpha}})\boldsymbol{\mu} =\mathbf{T} \operatorname{diag}(\boldsymbol{\mu})\mathbf{T}'\hat{\boldsymbol{\alpha}} = \mathbf{TMT}'\hat{\boldsymbol{\alpha}}=  \mathbf{TT}'\hat{\boldsymbol{\alpha}}$.\footnote{To understand the last equality, see that for any diagonal matrix $\mathbf{M}=\operatorname{diag}(\mu)$ with multipliers $\mu$ satisfying the common fixed-payment stationarity condition $\sum_{i\in k}\mu_i=n_k$, we have $(\mathbf{TMT}')_{k\ell}=\sum_{i\in k}\sum_{j\in \ell}T_{ki}M_{ij}T_{\ell j}= \sum_{i\in k} T_{ki}\mu_{i}T_{\ell i}$. Because a worker cannot be in two groups we have that $(\mathbf{TMT}')_{k\ell}=0$ for $k\neq \ell$. Moreover, the diagonal entries $k=\ell$ are $\sum_{i\in k} T_{ki}\mu_{i} = (\mathbf{T}\boldsymbol{\mu})_{k}=(\mathbf{T1})_{k} = \sum_{i\in k}T_{ki} = n_{k}$ by the KKT conditions. Hence $\mathbf{TMT}'=\operatorname{diag}(n_k)=\mathbf{TT}'$.} Thus, we can rewrite \eqref{eq:KKT_alpha2} as
\begin{equation*}
    \begin{aligned}
        \Big[r\sigma^{2}\mathbf{T}\mathbf{T}'
+\mathbf{T}\mathbf{C}'\big(\mathbf{M}-\lambda\mathbf{M}\mathbf{G}-\lambda\mathbf{G}'\mathbf{M}\big)\mathbf{C}\mathbf{T}'\Big]\hat{\boldsymbol{\alpha}}
=\mathbf{T}\mathbf{C}'\mathbf{1}.
    \end{aligned}
\end{equation*}

Notice that the Hessian matrix $\mathbf{H}=-(r\sigma^{2}\mathbf{T}\mathbf{T}'
+\mathbf{T}\mathbf{C}'\big(\mathbf{M}-\lambda\mathbf{M}\mathbf{G}-\lambda\mathbf{G}'\mathbf{M}\big)\mathbf{C}\mathbf{T}')$ is symmetric, invertible, and negative definite by Assumption~\ref{spectralradiusWB}.\footnote{To see this, notice that $(\mathbf{GC})'\mathbf{M}\mathbf{GC} = (\mathbf{M}^{1/2}\mathbf{GC})'(\mathbf{M}^{1/2}\mathbf{GC})$ where $\mathbf{M}^{1/2}= \operatorname{diag}(\sqrt{\mu_{i}})$. Let $\rho(\mathbf{A})$ denote the spectral radius of $\mathbf{A}$ and let $\left\lVert\cdot\right\rVert_{2}$ denote the spectral norm. We have $\rho((\mathbf{M}^{1/2}\mathbf{GC})'\mathbf{M}^{1/2}\mathbf{GC}) = \left\lVert \mathbf{M}^{1/2}\mathbf{GC}\right\rVert_{2}^{2}$ (see p.346 in \cite{hornjohnson2013}). Then,  $\rho((\mathbf{M}^{1/2}\mathbf{GC})'\mathbf{M}^{1/2}\mathbf{GC}) = \left\lVert \mathbf{M}^{1/2}\mathbf{GC}\right\rVert_{2}^{2}\leq \left\lVert \mathbf{M}^{1/2}\right\rVert_{2}^{2}\left\lVert\mathbf{GC}\right\rVert_{2}^{2}=(\max_{i}\mu_{i})\rho((\mathbf{\mathbf{GC}})'\mathbf{\mathbf{GC}})\leq \overline{n}\rho((\mathbf{\mathbf{GC}})'\mathbf{\mathbf{GC}})$, where the last inequality follows from \eqref{eq:KKT_beta}. Thus, for any $\mathbf{M}=\operatorname{diag}(\boldsymbol{\mu})$ we have that $\lambda^{2}\rho((\mathbf{GC})'\mathbf{M}\mathbf{GC})\leq \overline{n}\lambda^{2}\rho((\mathbf{\mathbf{GC}})'\mathbf{\mathbf{GC}})<(1+r\sigma^{2})$. Since $\mathbf{T}$ has full row rank, $\mathbf{H}$ is negative definite.} Thus, the optimal incentive rule under wage benchmarking is given by:
\begin{equation*}
    \begin{aligned}
        \hat{\boldsymbol{\alpha}}^{*} =\Big[ \mathbf{T}\left(\mathbf{C}'\big(\mathbf{M}-\lambda\mathbf{M}\mathbf{G}-\lambda\mathbf{G}'\mathbf{M}\big)\mathbf{C}+r\sigma^{2}\mathbf{I}
\right)\mathbf{T}'\Big]^{-1}\mathbf{T}\mathbf{C}'\mathbf{1} = \mathbf{W_{M}}\mathbf{T}\mathbf{C}'\mathbf{1},
    \end{aligned}
\end{equation*}
where $\mathbf{M}=\operatorname{diag}(\boldsymbol{\mu})$ and $\boldsymbol{\mu}$ is the KKT multiplier vector satisfying \eqref{eq:KKT_beta}-\eqref{eq:KKT_cs}.

To obtain the form of $\mathbf{W_{M}}$ given in the proposition, notice that 
\begin{equation*}
    \begin{aligned}
\mathbf{C}'(\mathbf{M}-\lambda \mathbf{M}\mathbf{G}-\lambda \mathbf{G}'\mathbf{M})\mathbf{C}
&= \mathbf{C}'\mathbf{M}\mathbf{C}-\lambda \mathbf{C}'\mathbf{M}\mathbf{G}\mathbf{C}-\lambda \mathbf{C}'\mathbf{G}'\mathbf{M}\mathbf{C}\\
&= \mathbf{C}'\mathbf{M}\mathbf{C}-\lambda \mathbf{C}'\mathbf{M}\mathbf{G}\mathbf{C}-(\mathbf{C}'-\mathbf{I})\mathbf{M}\mathbf{C}\\
&= \mathbf{M}\mathbf{C}-\lambda \mathbf{C}'\mathbf{M}\mathbf{G}\mathbf{C}\\
&= \mathbf{M}(\mathbf{C}-\mathbf{I})+\mathbf{M}-\mathbf{C}'\mathbf{M}(\mathbf{C}-\mathbf{I})\\
&= \mathbf{M}-(\mathbf{C}'-\mathbf{I})\mathbf{M}(\mathbf{C}-\mathbf{I})\\
&= \mathbf{M}-\lambda^{2}(\mathbf{G}\mathbf{C})'\mathbf{M}\mathbf{G}\mathbf{C},
    \end{aligned}
\end{equation*}
where in the second equality we have used the fact that $\lambda (\mathbf{GC})'=(\mathbf{C}'-\mathbf{I})$, in the fourth equality we add and subtract $\mathbf{M}$ and use the fact that $\lambda \mathbf{GC}=(\mathbf{C}-\mathbf{I})$, and in the last equality we use both facts again. Therefore, we have that: 
\begin{equation*}
    \mathbf{W_{M}}^{-1} =  \mathbf{T}\left(\mathbf{C}'\big(\mathbf{M}-\lambda\mathbf{M}\mathbf{G}-\lambda\mathbf{G}'\mathbf{M}\big)\mathbf{C}+r\sigma^{2}\mathbf{I}
\right)\mathbf{T}' =  \mathbf{T}\left(\mathbf{M}- \lambda^{2}(\mathbf{GC})'\mathbf{M}\mathbf{GC}+r\sigma^{2}\mathbf{I}
\right)\mathbf{T}'.
\end{equation*}
Using again the fact that $\mathbf{T}\mathbf{M}\mathbf{T}'= \operatorname{diag}(n_{k}) = \mathbf{TT}'$, we can finally write:
\begin{equation*}
    \begin{aligned}
        \mathbf{W_{M}}& =  \Big[\mathbf{T}\mathbf{T}'- \lambda^{2}\mathbf{T}(\mathbf{GC})'\mathbf{M}\mathbf{GC}\mathbf{T}'+r\sigma^{2}\mathbf{T}
\mathbf{T}' \Big]^{-1}
=   \Big[\mathbf{T}((1+r\sigma^{2})\mathbf{I}- (\lambda\mathbf{GC})'\mathbf{M}(\lambda\mathbf{GC}))\mathbf{T}'\Big]^{-1}.
    \end{aligned}
\end{equation*}

Finally, notice that $\alpha_{k}^{*}> 0$ for all $k$ and, thus, workers' efforts are interior. Under Assumptions~\ref{assumption1} and~\ref{spectralradiusWB}, $\mathbf{C}$ and $\mathbf{W_{M}}$ are entrywise nonnegative. Thus, we have $\mathbf{T}\mathbf{C}'\mathbf{1}\gg \mathbf{0}$ and so $\hat{\boldsymbol{\alpha}}^{*}=\mathbf{W_{M}}\mathbf{T}\mathbf{C}'\mathbf{1}\gg 0$. It follows that $\mathbf{e}^{*}=\mathbf{C}\mathbf{T}'\hat{\boldsymbol{\alpha}}^{*}\gg 0$. Because the workers' problem is strictly concave in $\mathbf{e}$, the constrained best response coincides with the FOC solution $\mathbf{e}=\mathbf{C}\boldsymbol{\alpha}$ at $\boldsymbol{\alpha}^{*}$.
\qed  

\subsection{Proof of Lemma \ref{profits_coarse}}

We can write the firm's profits at the optimal contract as: 
\begin{equation*}
    \begin{aligned}
         \mathbb{E}(\pi(\mathbf{e}^{*}\vert \boldsymbol{\alpha}^{*}, \boldsymbol{\beta}^{*})) & =  \mathbf{1}'\mathbf{C}\mathbf{T}'\hat{\boldsymbol{\alpha}}^{*}
-\big(\mathbf{1}'\mathbf{C}\mathbf{T}'\hat{\boldsymbol{\alpha}}^{*}\big)\big(\mathbf{1}'\mathbf{T}'\hat{\boldsymbol{\alpha}}^{*}\big)
-\mathbf{1}'\mathbf{T}'\hat{\boldsymbol{\beta}}^{*}.
    \end{aligned}
\end{equation*}

Using the complementary slackness condition \eqref{eq:KKT_cs} we obtain:
\begin{equation*}\
    \boldsymbol{\mu}'(\mathbf{T}'\hat{\boldsymbol{\beta}}- \mathbf{f}(\hat{\boldsymbol{\alpha}}^{*})) = \boldsymbol{\mu}'\mathbf{T}'\hat{\boldsymbol{\beta}}- \boldsymbol{\mu}'\mathbf{f}(\hat{\boldsymbol{\alpha}}^{*}) = \mathbf{1}'\mathbf{T}'\hat{\boldsymbol{\beta}} - \boldsymbol{\mu}'\mathbf{f}(\hat{\boldsymbol{\alpha}}^{*})=\mathbf{0} \,\, \implies \,\, \mathbf{1}'\mathbf{T}'\hat{\boldsymbol{\beta}} = \boldsymbol{\mu}'\mathbf{f}(\hat{\boldsymbol{\alpha}}^{*}),
\end{equation*}
where
\begin{equation}\label{eq:Lemma1_2}
    \begin{aligned}
        \boldsymbol{\mu}'\mathbf{f}(\hat{\boldsymbol{\alpha}}^{*})  = & -  \big(\mathbf{1}'\mathbf{C}\mathbf{T}'\hat{\boldsymbol{\alpha}}^{*}\big)\big(\mathbf{1}'\mathbf{T}'\hat{\boldsymbol{\alpha}}^{*}\big) + \frac{1}{2}\boldsymbol{\mu}'[\operatorname{diag}(\mathbf{CT}'\hat{\boldsymbol{\alpha}}^{*})\mathbf{CT}'\hat{\boldsymbol{\alpha}}^{*} - 2\lambda \operatorname{diag}(\mathbf{CT}'\hat{\boldsymbol{\alpha}}^{*})\mathbf{GCT}'\hat{\boldsymbol{\alpha}}^{*} \\
        & + r\sigma^{2} \operatorname{diag}(\mathbf{T}'\hat{\boldsymbol{\alpha}}^{*})\mathbf{T}'\hat{\boldsymbol{\alpha}}^{*}] \\
         = & -  \big(\mathbf{1}'\mathbf{C}\mathbf{T}'\hat{\boldsymbol{\alpha}}^{*}\big)\big(\mathbf{1}'\mathbf{T}'\hat{\boldsymbol{\alpha}}^{*}\big) + \frac{1}{2}(\hat{\boldsymbol{\alpha}}^{*})'[\mathbf{TC}'\mathbf{M}\mathbf{CT}' - 2\lambda \mathbf{TC}'\mathbf{MGCT}'+ r\sigma^{2} \mathbf{T}\mathbf{M}\mathbf{T}']\hat{\boldsymbol{\alpha}}^{*}. 
    \end{aligned}
\end{equation}

Using \eqref{eq:Lemma1_2} in the firm's expected profits at the optimal contract, we have: 
\begin{equation*}
    \begin{aligned}
         \mathbb{E}(\pi(\mathbf{e}^{*}|\boldsymbol{\alpha}^{*}, \boldsymbol{\beta}^{*}))  = \mathbf{1}'\mathbf{C}\mathbf{T}'\hat{\boldsymbol{\alpha}}^{*} - \frac{1}{2}(\hat{\boldsymbol{\alpha}}^{*})'[\mathbf{TC}'\mathbf{M}\mathbf{CT}' - 2\lambda \mathbf{TC}'\mathbf{MGCT}'+ r\sigma^{2} \mathbf{T}\mathbf{T}']\hat{\boldsymbol{\alpha}}^{*}.
    \end{aligned}
\end{equation*}

If $\mathbf{G}$ is not symmetric, the term $\mathbf{TC}'\mathbf{MGCT}'$ is also non-symmetric. However, the matrices $\mathbf{TC}'\mathbf{MCT}'$ and $\mathbf{TT}'$ are both symmetric. 
    Recall that the quadratic form for a non-symmetric matrix $\mathbf{P}$ only depends on the symmetric part of $\mathbf{P}$, i.e., $\mathbf{x}'\mathbf{Px} = \mathbf{x}'(\frac{\mathbf{P}+\mathbf{P}'}{2})\mathbf{x}$. Using this fact, we can rewrite the middle term in the brackets of the firm's profits as
    \begin{equation*}
        2\lambda\hat{\boldsymbol{\alpha}}'( \mathbf{TC}'\mathbf{MGCT}')\hat{\boldsymbol{\alpha}}  = \lambda\hat{\boldsymbol{\alpha}}'(\mathbf{TC}'\mathbf{MGCT}'+\mathbf{TC}'\mathbf{G}'\mathbf{MCT}')\hat{\boldsymbol{\alpha}},
    \end{equation*}
 to write firm's profits as
  \begin{equation*}
      \begin{aligned}
           \mathbb{E}(\pi(\mathbf{e}^{*}|\boldsymbol{\alpha}^{*}, \boldsymbol{\beta}^{*})) & =\mathbf{1}'\mathbf{C}\mathbf{T}'\hat{\boldsymbol{\alpha}} - \frac{1}{2}\hat{\boldsymbol{\alpha}}'[\mathbf{TC}'\mathbf{M}\mathbf{CT}' - 2\lambda \mathbf{TC}'\mathbf{MGCT}'+ r\sigma^{2} \mathbf{T}\mathbf{T}']\hat{\boldsymbol{\alpha}} \\
           & = \mathbf{1}'\mathbf{C}\mathbf{T}'\hat{\boldsymbol{\alpha}} - \frac{1}{2}\hat{\boldsymbol{\alpha}}'[\mathbf{TC}'\mathbf{M}\mathbf{CT}' - \lambda \mathbf{TC}'\mathbf{MGCT}'- \lambda \mathbf{TC}'\mathbf{G}'\mathbf{MCT}'+ r\sigma^{2} \mathbf{T}\mathbf{T}']\hat{\boldsymbol{\alpha}} \\
           & = \mathbf{1}'\mathbf{C}\mathbf{T}'\hat{\boldsymbol{\alpha}} - \frac{1}{2}\hat{\boldsymbol{\alpha}}'[\mathbf{TC}'(\mathbf{M} - \lambda \mathbf{MG}- \lambda \mathbf{G}'\mathbf{M})\mathbf{CT}'+ r\sigma^{2} \mathbf{T}\mathbf{T}']\hat{\boldsymbol{\alpha}} \\
           & = \mathbf{1}'\mathbf{C}\mathbf{T}'\hat{\boldsymbol{\alpha}} - \frac{1}{2}\hat{\boldsymbol{\alpha}}'\mathbf{TC}'\mathbf{1}  = \frac{1}{2}\mathbf{1}'\mathbf{C}\mathbf{T}'\hat{\boldsymbol{\alpha}} = \frac{1}{2}\sum_{i\in N}e_{i}^{*},
      \end{aligned}
  \end{equation*}
where in the second equality we use the symmetrization trick and in the fourth equality we use the optimal incentive rule \eqref{coarse_alpha}  from Proposition \ref{Coarse Contracts}.
\qed 

\subsection{Proof of Proposition \ref{profits_variance}}

We can write the optimal efforts for personalized contracts (P) and for the wage benchmarking case (WB) compactly as:
\begin{equation*}
    \mathbf{e}^{P} =\mathbf{C}\boldsymbol{\alpha}^{P} = \mathbf{C}\mathbf{W}\mathbf{C}'\mathbf{1},\quad \text{and} \quad \mathbf{e}^{WB} = \mathbf{C}\boldsymbol{\alpha}^{WB} =\mathbf{C}\mathbf{T}'\boldsymbol{\hat\alpha}^{WB} = \mathbf{C}\mathbf{T}'\mathbf{W}_{\mathbf{M}} \mathbf{T} \mathbf{C}'\mathbf{1},
\end{equation*}
where $ \mathbf{W}= [(1+r\sigma^{2})\mathbf{I}-(\lambda\mathbf{CG})'(\lambda\mathbf{CG})]^{-1}$ and $ \mathbf{W_{M}} = (\mathbf{T}[(1+r\sigma^{2})\mathbf{I}- (\lambda\mathbf{GC})'\mathbf{M}(\lambda\mathbf{GC})]\mathbf{T}')^{-1}$. Thus, the difference in output can be written as:
\begin{equation}
    \begin{aligned}
        \mathbf{1}' (\mathbf{e}^{P} - \mathbf{e}^{WB}) & =  \mathbf{1}'(\mathbf{C}\mathbf{W}\mathbf{C}'\mathbf{1} - \mathbf{C}\mathbf{T}'\mathbf{W}_{\mathbf{M}} \mathbf{T} \mathbf{C}'\mathbf{1})  = \mathbf{b}'[\mathbf{W}-\mathbf{T}'\mathbf{W}_{\mathbf{M}}\mathbf{T}]\mathbf{b},\label{A_P7_A}
    \end{aligned}
\end{equation}
where $\mathbf{b}=\mathbf{C'1}$ is the vector of outward Bonacich centralities.

We want to show that there exists a $\delta$ such that:
\[
\frac{1}{2} \sum_{i\in N} (\mathbf{e}_i^{P} - \mathbf{e}_i^{WB}) = \delta \frac{1}{n} \sum_k n_k \text{Var}(\mathbf{b}_k),
\]
where $\text{Var}(\mathbf{b}_k) = n_k^{-1}\sum_{i\in k}(b_i-\bar b_k)^2$ is the variance within group $k$ and $\sum_{k}$ sums over all $K$ groups. The average within-group variance, weighted by group size, is given by:
\begin{equation*}
    \begin{aligned}
        \frac{1}{n} \sum_k  \sum_{i\in k} (b_{i}-\bar{b}_{k})^{2} 
        = \frac{1}{n} \sum_k  \left(\sum_{i\in k} b_{i}^{2}-\frac{(\sum_{i\in k} b_{i})^{2}}{n_{k}}\right)
        = \frac{1}{n} \left[ \sum_{i\in N} b_{i}^{2} - \sum_{k}\frac{1}{n_{k}} (\sum_{i\in k}b_{i})^{2} \right],
    \end{aligned}
\end{equation*}
where the last equality follows from the fact that summing over workers in group $k$ and then over all groups $k$ is the same as summing over \emph{all} workers. Next, since the matrix $(\mathbf{T}\mathbf{T}')^{-1}$ is a $K \times K$ diagonal matrix with $(k,k)$ element equal one over group $k$'s size, we can write the expression above in matrix form as: 
\begin{equation}
    \begin{aligned}
         \frac{1}{n} \sum_k n_k \text{Var}(\mathbf{b}_k) & = \frac{1}{n} \left[ \mathbf{b}'\mathbf{b} - (\mathbf{T}\mathbf{b})'(\mathbf{T}\mathbf{T}')^{-1} \mathbf{T}\mathbf{b} \right] =  \frac{1}{n}\left(\mathbf{b}'\left[ \mathbf{I} - \mathbf{T'}(\mathbf{T}\mathbf{T}')^{-1}\mathbf{T}\right] \mathbf{b} \right).\label{A_P7_B}
    \end{aligned}
\end{equation}
Putting together \eqref{A_P7_A} and \eqref{A_P7_B} we have 
\[
 \frac{1}{2}\mathbf{b}' [\mathbf{W}-\mathbf{T}'\mathbf{W}_{\mathbf{M}}\mathbf{T}] \mathbf{b} = \delta \frac{1}{n} \mathbf{b}' (\mathbf{I} - \mathbf{T}'(\mathbf{T}\mathbf{T}')^{-1} \mathbf{T}) \mathbf{b}.
\]

Finally, notice that when $\lambda^{2}/r\sigma^{2} \to 0$ we have that  $\mathbf{W}\to  \frac{1}{1+r\sigma^{2}}\mathbf{I}$ and $\mathbf{W_{M}}\to  \frac{1}{1+r\sigma^{2}}(\mathbf{TT}')^{-1}$. Hence, in the limit, we have that
\begin{equation*}
    \frac{1}{2} \sum_{i\in N} (\mathbf{e}_i^{P} - \mathbf{e}_i^{WB}) 
    = \frac{1}{2(1+r\sigma^{2})} \mathbf{b}' [\mathbf{I} - \mathbf{T}'(\mathbf{T}\mathbf{T}')^{-1} \mathbf{T}] \mathbf{b}
    = \delta \frac{1}{n} \sum_k n_k \text{Var}(\mathbf{b}_k),
\end{equation*}
with $\delta = n/(2(1+r\sigma^{2}))$.\qed

\newpage

\renewcommand{\thepage}{Supp-\arabic{page}} 
\setcounter{page}{1}

\begin{center}
{\Large \textbf{Supplementary Appendix}}
\end{center}

\section{Optimal Incentive Contracts with Heterogeneous Workers}\label{SuppApp_General}

Consider the model described in Section \ref{baseline}, extended to allow for worker heterogeneity along four dimensions. First, workers may differ in their productivity per unit of effort, denoted by $\theta_i$. Second, they may exhibit heterogeneous risk aversion, indexed by $r_i$. Third, workers may differ in their reservation utilities, given by $-\exp[-r_i U_i]$. Finally, peer effects may vary across pairs of workers. 

We first solve the model in full generality, allowing for all four sources of heterogeneity simultaneously. We then examine each dimension of heterogeneity in isolation to clarify its distinct implications for the structure of optimal incentives.

The firm’s production function is now given by:
\[
X(\mathbf{e}) = \sum_{j\in N} \theta_{j} e_j + \varepsilon,
\]
where $\theta_i$ denotes worker $i$’s productivity per unit of effort. The certainty equivalent of worker $i$ is then given by
\[
\operatorname{CE}_i (\mathbf{e},\mathbf{G}) = \beta_i + \alpha_i  \sum_{j\in N} \theta_{j} e_j- \frac{1}{2} e_i^2 + \lambda \sum_{j\in N} g_{ij}e_i e_j - \alpha_i^2 \frac{ r_i \sigma^2}{2}.
\]

For any contract $(\alpha_{i},\beta_{i})$, the best-reply function of worker $i$ is given by:
\[
e_i^{*} = \theta_i\alpha_i +  \lambda\sum_{j\in N} g_{ij} e_j.
\]
Define $\mathbf{\Theta} = \operatorname{diag}(\boldsymbol{\theta})$, $\mathbf{R} = \operatorname{diag}(\mathbf{r})$, and $\mathbf{u} = ( U_{1}, U_{2}, \dots, U_{n})$. The vector of best responses is:
\[
\mathbf{e} =  \mathbf{\Theta}  \boldsymbol{\alpha} + \lambda \mathbf{G}\mathbf{e}.
\]
Under Assumption \ref{assumption1}, the unique Nash equilibrium effort profile $\mathbf e^{*}$ of the game can be characterized:
\[
\mathbf{e}^{*} = \left[ \mathbf{I} - \lambda \mathbf{G} \right]^{-1} \mathbf{\Theta}  \boldsymbol{\alpha}.
\]

As argued in the proof of Proposition~\ref{Optimal Contracts}, the firm can set fixed payments $\beta_{i}$ in order to extract all surplus from the workers, such that $\operatorname{CE}_{i}(\mathbf{e}) = U_{i}$. We can write the firm's problem as: 
\begin{align*}
 \max_{\alpha,\beta} \mathbb{E}[\pi(\mathbf{e} | \boldsymbol{\alpha , \beta})] & = \sum_{i}^{n} \theta_i e_i - \sum_{i}^{n} w_{i}(X) \\
 \text{subject to }\phantom{d} & \\
& \operatorname{CE}_i (\mathbf{e}, \mathbf{G}) = U_i, \, \quad  \forall i \in N\quad \tag{IR} \\
& \mathbf{e}^{*} = \left[ \mathbf{I} -\lambda \mathbf{G}  \right]^{-1}  \mathbf{\Theta}  \boldsymbol{\alpha}\quad  \tag{IC}
\end{align*}

The (IR) condition for worker $i\in N$ can be written as: 
\begin{equation}
    \beta_{i} + \alpha_{i}\sum_{j\in N} \theta_{j} e_{j} - \frac{1}{2}e_{i}^{2} +  \lambda \sum_{j\in N} g_{ij} e_{i} e_{j} - \alpha_{i}^{2} r_{i} \frac{\sigma^{2}}{2} = U_{i}.
\end{equation}

Substituting for $\beta_{i}$ into the objective function, ignoring the constant terms $\sum_{i\in N} U_{i}$, and writing the firm's expected profits in matrix form we have: 
\[
\begin{aligned}
\max_{\alpha} \mathbb{E}[\pi(\mathbf{e}|\boldsymbol{\alpha , \beta})] &= \{\mathbf{e}' \mathbf{\Theta} \boldsymbol{1} - \frac{1}{2} \mathbf{e}'\mathbf{e} - \frac{\sigma^2}{2} \boldsymbol{\alpha}' \mathbf{R} \boldsymbol{\alpha} +\lambda \mathbf{e}' \mathbf{G} \mathbf{e}\} \\
 \text{subject to }\phantom{d} & \\
\mathbf{e}^{*} &= \left[ \mathbf{I} - \lambda\mathbf{G}\right]^{-1}\mathbf{\Theta}  \boldsymbol{\alpha}. \quad &\text{(IC)}
\end{aligned}
\]
Taking $\mathbf{\tilde{C}} := \left[ \mathbf{I} -  \lambda\mathbf{G}  \right]^{-1} \mathbf{\Theta}$ and replacing $\mathbf{e}=\mathbf{\tilde{C}} \boldsymbol{\alpha}$ and $\mathbf{e}'=\boldsymbol{\alpha}' \mathbf{\tilde{C}}'$, the above maximization problem becomes:
\[
\max_{\boldsymbol{\alpha}} \mathbb{E}[\pi(\mathbf{e}|\boldsymbol{\alpha})] = \{\boldsymbol{\alpha}' \mathbf{\tilde{C}}' \mathbf{\Theta} \mathbf{1} - \frac{1}{2}\boldsymbol{\alpha}' \mathbf{\tilde{C}}' \mathbf{\tilde{C}} \boldsymbol{\alpha} - \frac{\sigma^2}{2} \boldsymbol{\alpha}' \mathbf{R} \boldsymbol{\alpha} + \lambda\boldsymbol{\alpha}' \mathbf{\tilde{C}}'\mathbf{G}\mathbf{\tilde{C}} \boldsymbol{\alpha}\}.
\]

This richer version of our model requires a revised version of Assumption \ref{spectralradius} in order to guarantee that the firm's problem is concave (i.e., the Hessian of the firm's problem is negative definite).

\renewcommand{\theassumption}{2'}
\begin{assumption}\label{spectralradius_het}
     The spectral radius of $\lambda^2(\boldsymbol{\Theta}^2+\sigma^{2}\mathbf{R})^{-1}(\mathbf{G}\mathbf{C}\boldsymbol{\Theta})^\prime\mathbf{G}\mathbf{C}\boldsymbol{\Theta}$ is less than 1.
 \end{assumption}

Taking the first order conditions with respect to $\boldsymbol{\alpha}$ and solving gives the optimal incentive rule under heterogeneity:
\begin{equation}\label{eq:GeneralIncentiveRule}
\boldsymbol{\alpha}^{*} = \left[ \left(\mathbf{\tilde{C}}' \left(\mathbf{I}-\lambda
(\mathbf{G}'+\mathbf{G})\right) \mathbf{\tilde{C}}\right) + \sigma^2 \mathbf{R} \right] ^{-1} \mathbf{\tilde{C}}' \boldsymbol{\theta}= \mathbf{\tilde{W}}\mathbf{\tilde{C}}'\boldsymbol{\theta}.
\end{equation}

The optimal induced effort in this case is given by $\mathbf{e}^{*}=\mathbf{\tilde{C}} \boldsymbol{\alpha}^{*}$. The vector of optimal fixed payments $\boldsymbol{\beta}^{*}$ can be recovered using (IR) in matrix form:
\[
\boldsymbol{\beta}^{*} (\boldsymbol{\alpha}^{*}, \mathbf{e}^{*}) = \frac{1}{2} (\mathbf{e}^{*} \circ \mathbf{e}^{*}) - (\mathbf{e}^{*} \circ \lambda\mathbf{G}\mathbf{e}^{*}) + \frac{\sigma^2}{2}(\boldsymbol{\alpha}^{*} \circ \mathbf{{R}} \boldsymbol{\alpha}^{*}) - \boldsymbol{\alpha}^{*} \circ \mathbf{1}(\mathbf{1}'\mathbf{\Theta}\mathbf{e}^{*}) + \mathbf{u},  
\]
where $\circ$ denotes the Hadamard product. We now specify equation \eqref{eq:GeneralIncentiveRule} to each form of heterogeneity separately, in order to better understand their implications.   


\subsection{Heterogeneous productivity.}

Assume the baseline model is extended to allow workers to differ in productivity. The general incentive rule in \eqref{eq:GeneralIncentiveRule} simplifies to:
\begin{equation*}
\boldsymbol{\alpha}^{*}
= \left[
\mathbf{\Theta}\mathbf{C}'
\left(\mathbf{I}-\lambda(\mathbf{G}'+\mathbf{G})\right)
\mathbf{C}\mathbf{\Theta}
+ r\sigma^2 \mathbf{I}
\right]^{-1}
\mathbf{\Theta}\mathbf{C}'\boldsymbol{\Theta}\mathbf{1}.
\end{equation*} In order to understand this incentive rule, consider worker $i$'s first-order condition:
$$
e_i=\alpha_i \theta_i+\lambda \sum_j g_{i j} e_j .
$$
Notice that the productivity parameter affects the return to exerting effort but leaves the strength of peer spillovers unchanged. Solving the system yields equilibrium effort:
$$
\mathbf{e}^{*}=\mathbf{C} \boldsymbol{\Theta} \boldsymbol{\alpha},
$$
where $\boldsymbol{\Theta}=\operatorname{diag}\left(\theta_i\right)$. The response of worker $j$'s effort to incentives assigned to worker $i$ is therefore:
$$
\frac{\partial e_j}{\partial \alpha_i}=(\mathbf{C} \boldsymbol{\Theta})_{j i}=\theta_i \mathbf{C}_{j i}=\theta_i \sum_{q=0}^{\infty} \lambda^q \mathbf{G}^q_{j i} .
$$
Thus, incentives directed at worker $i$ propagate through the network exactly as in the baseline model, but the magnitude of the amplification is scaled by $i$'s productivity. More productive workers therefore generate larger effort responses throughout the network.

Because output equals $\sum_j \theta_j e_j$, these effort responses translate to changes in output in proportion to $\theta_j$. Therefore, the total effect from raising $\alpha_i$ depends both on $i$'s productivity parameter and that of each one of $i$'s incentive targets:
$$
M B_{\alpha_i}=\frac{\partial}{\partial \alpha_i} \sum_j \theta_j e_j=\sum_j \theta_j \frac{\partial e_j}{\partial \alpha_i}=\sum_j \theta_i \theta_j \sum_{q=0}^{\infty} \lambda^q \mathbf{G}^q_{j i} .
$$
In matrix form the marginal benefit vector can be written as
$\boldsymbol{\Theta} \mathbf{C}^{\prime} \boldsymbol{\Theta} \boldsymbol{1}$. 
Productivity heterogeneity therefore enters the marginal benefit of incentives twice. First, more productive workers respond more strongly to their own incentives. Second, the effort they induce in others translates into more output when those peers are themselves productive.

The marginal cost component retains the same quadratic structure as in the homogeneous benchmark. However, the substitution and complementarity forces that determine the weighting matrix $\mathbf{W}$ are now scaled by the productivity of both the sender and the target of incentives, paralleling the same mechanisms discussed in the context of Figures  \ref{fig: substitutes} and \ref{fig:complements}.

\subsection{Heterogeneous risk aversion.}
When heterogeneity arises exclusively from differences in risk aversion, the general
incentive rule in \eqref{eq:GeneralIncentiveRule} simplifies to
\begin{equation*}
\boldsymbol{\alpha}^{*}
= \left[
\mathbf{C}'
\left(\mathbf{I}-\lambda(\mathbf{G}'+\mathbf{G})\right)
\mathbf{C}
+ \sigma^2 \mathbf{R}
\right]^{-1}
\mathbf{C}'\mathbf{1}.
\end{equation*}
In this case, unlike in the previous section on heterogeneous productivity, the firm's marginal benefit of raising $\boldsymbol{\alpha}$ remains unchanged. The reason is that risk aversion does not affect a worker's optimal contribution of effort, which remains $\mathbf{e}^{*}=\mathbf{C}  \boldsymbol{\alpha}$. Therefore, raising $\alpha_i$ has the same marginal effect on output as before ${MB}_{\alpha_i}=\sum_{j}\sum_{q=0}^{\infty} \lambda^q \mathbf{G}^q_{j i}$.

Risk aversion affects workers’ utility from a given level of incentives, independently of the effort they ultimately exert. For this reason, $\mathbf{R}$ enters the expression for the base salary $\boldsymbol{\beta}$ above and interacts with $\boldsymbol{\alpha}$ but not with equilibrium effort $\boldsymbol{e}^{*}$. In other words, risk aversion affects the marginal cost of providing incentives, but it does so only by shifting these costs. It does not interact with the complementarity and substitution forces discussed in Figures \ref{fig: substitutes} and \ref{fig:complements}. Consequently, it appears additively in the matrix $\mathbf{W}$ above.

\subsection{Heterogeneous reservation utilities.}

Relative to the preceding discussion of risk aversion, note that reservation utilities enter the model only through fixed payments $\boldsymbol{\beta}$. Unlike risk aversion, however, reservation utilities do not interact with any other component of the contract. They affect neither incentives $\boldsymbol{\alpha}$ nor equilibrium effort $\mathbf{e}$. Instead, they simply shift the base salary the firm must pay to each worker, since workers with higher reservation utilities require larger fixed transfers to satisfy their participation constraints. As a result, these parameters do not enter the incentive rule at all—not even as a shifter in $\mathbf{W}$. Their only effect is on the firm’s profits, given the optimal incentive rule $\boldsymbol{\alpha}$.

For completeness, note that because the firm maximizes profits subject to workers’ participation constraints, sufficiently high values of $U_i$ may force the firm to accept negative profits in order to induce participation.
We rule out this case by assuming that reservation utilities are heterogeneous but sufficiently close to zero. An alternative modeling approach would allow the firm to exclude workers with very high reservation utilities; we leave this extension for future work.

\subsection{Heterogeneous peer effects.}
Workers may also differ in how strongly they affect one another. There are several ways to introduce such heterogeneity at different levels of detail. For example, one could start from an unweighted network where $g_{ij}=1$ if workers $i$ and $j$ are connected and $0$ otherwise, but allow the strength of peer effects to vary across workers or across pairs of workers. Three natural specifications illustrate these possibilities.

\textbf{Model 1.} Spillovers enter worker $i$'s cost function as
$
\frac{1}{2}e_i^2 - e_i \sum_j g_{ij}\lambda_j e_j ,
$
so that spillovers depend on the characteristics of the \emph{sending} worker $j$.

\textbf{Model 2.} Spillovers enter worker $i$'s cost function as
$
\frac{1}{2}e_i^2 - e_i \lambda_i \sum_j g_{ij} e_j ,
$
so that the strength of peer effects depends on the \emph{receiving} worker $i$.

\textbf{Model 3.} Spillovers enter worker $i$'s cost function as
$
\frac{1}{2}e_i^2 - e_i \sum_j g_{ij}\lambda_{ij} e_j ,
$
allowing for fully pairwise-specific peer effects.

Because we allow the network $\mathbf{G}$ to be weighted, all of these specifications can be represented within our baseline formulation,
$
\frac{1}{2}e_i^2 - \lambda e_i \sum_j g_{ij} e_j ,
$
by absorbing the heterogeneous peer-effect parameters into the weights $g_{ij}$ of the network.

\section{Incentive Provision Under Individual Production} \label{SupApp_IndividualProduction}

Consider a version of the model in which each worker's effort results in a noisy individual production (IP) according to
\begin{equation*}
	q_{i} = e_{i} + \varepsilon_{i}.
\end{equation*}
The random variables $(\varepsilon_{i})_{i \in N}$ are assumed to be independently and normally distributed with mean zero and variance $\sigma^{2}$.\footnote{The model by \cite{holmstrom1987aggregation} has also been extended to situations with individual production and correlated outputs (see \cite{bolton2004contract}).} This means that worker $i$'s compensation is now conditional on individual output rather than the joint output $X(\mathbf{e})$. That is, 
\begin{equation*}
w_{i} = \beta_{i}^{IP} + \alpha_{i}^{IP}q_{i},	
\end{equation*}
where $\beta_{i}^{IP}$ is a fixed term of the compensation, and $\alpha_{i}^{IP}$ is a variable or performance-related compensation coefficient under individual production. In this case, worker $i$'s certainty equivalent is given by
\begin{equation} 
		CE_{i}^{IP}(\mathbf{e},\mathbf{G};\alpha_{i}^{IP},\beta_{i}^{IP}) = \beta_{i}^{IP} + {\alpha_{i}^{IP}e_{i}^{IP}} - \frac{1}{2}(e_{i}^{IP})^{2} + \lambda e_{i}^{IP}\sum_{j\in N}g_{ij}e_{j}^{IP} - (\alpha_{i}^{IP})^{2}\frac{r_{i}\sigma^{2}}{2}.\label{CE_IndividualProd}
	\end{equation}
Notice that the effort-provision problem of worker $i$ is the same when maximizing \eqref{CE} and \eqref{CE_IndividualProd}. However, the former expression has an extra term: $\alpha_{i}\sum_{j\neq i}e_{j}$. This implies that when the firm sets the fixed part of the compensation to guarantee that the participation constraint is satisfied for each worker, we have the following equivalence:
\[
\beta_{i}^{IP} = \beta_{i} + \alpha_{i}\sum_{j\neq i}e_{j}.
\]
However, looking at the firm's reduced maximization problem in both cases, we confirm $\alpha_{i}=\alpha_{i}^{IP}$. Under individual production, the firm maximizes:
\begin{equation*}
		\begin{aligned}
			\max_{\boldsymbol{\alpha}} && \sum_{i\in N} e_{i} - \sum_{i\in N} w_{i} = \sum_{i\in N} e_{i} - \sum_{i\in N}\beta_{i}^{IP} - \sum_{i\in N}\alpha_{i} e_{i}.
		\end{aligned}
	\end{equation*} 
Since $\beta_{i}^{IP} = \beta_{i} + \alpha_{i}\sum_{j\neq i}e_{j}$, this is the same problem under joint production:
\begin{equation*}
		\begin{aligned}
			\max_{\boldsymbol{\alpha}} && \sum_{i\in N} e_{i} - \sum_{i\in N}\beta_{i} - \sum_{i\in N}\alpha_{i}\sum_{j\neq i}e_{j} - \sum_{i\in N}\alpha_{i} e_{i} = \sum_{i\in N} e_{i} - \sum_{i\in N}\beta_{i} - \sum_{i\in N}\alpha_{i} \sum_{k\in N} e_{k}.
		\end{aligned}
	\end{equation*} 

\section{First-Best Efforts}\label{Sec: First-Best}

Surplus, $S$, corresponds to the profits collected by the firm and the consumption-equivalent units of expected utility received by workers: 
\begin{equation*}
    \begin{aligned}
          S &= \underbrace{ \mathbf{1}'\mathbf{e} - \mathbf{1}'\boldsymbol{\alpha}(\mathbf{1}'\mathbf{e}) - \mathbf{1}'\boldsymbol{\beta}}_{\mathbb{E}(\pi)} + \underbrace{\mathbf{1}' \boldsymbol{\beta} - \frac{1}{2}\mathbf{e}'\mathbf{e} +\lambda \mathbf{e}'\mathbf{Ge} + \mathbf{1}'\boldsymbol{\alpha}(\mathbf{1}'\mathbf{e}) -\frac{1}{2}r\sigma^{2}\boldsymbol{\alpha}'\boldsymbol{\alpha}}_{\sum_i \operatorname{CE}_i} 
          \\ & = \mathbf{1}'\mathbf{e}- \frac{1}{2}\mathbf{e}'\mathbf{e} +\lambda \mathbf{e}'\mathbf{Ge} -\frac{1}{2}r\sigma^{2}\boldsymbol{\alpha}'\boldsymbol{\alpha}.  
    \end{aligned}
\end{equation*}

\begin{proposition}[Surplus-maximizing Effort]\label{FirstBest}
    Assume that the spectral radius of $\lambda(\mathbf{G}+\mathbf{G}')$ is less than one. The effort vector that maximizes surplus is given by: 
    \begin{equation}\label{FB_efforts}
        \mathbf{e}^{0}=(\mathbf{I}-\lambda(\mathbf{G}+\mathbf{G}'))^{-1}\mathbf{1}. 
    \end{equation}
\end{proposition}

We now show that the optimal contract in Proposition~\ref{Optimal Contracts} yields the efficient level of efforts if and only if $\sigma^2=0$. To fix ideas, consider first the case without peer effects ($\lambda=0$). In this case, equation \eqref{FB_efforts} reduces to $e_i^0=1$ for all $i\in N$. In the main text we show that the optimal contract when $\lambda=0$ is $\alpha_i= 1/(1+r\sigma^2)$. So when $\sigma^2=0$ the firm makes workers full marginal claimants on output ($\alpha_i=1$) and thus $e_i=1$, which indeed corresponds to the efficient level as expected.

Intuitively, surplus is lost because the provision of incentives shifts risk from the risk-neutral principal to risk-averse agents, rather than because effort is unobservable per se. Indeed, when $\sigma^2 = 0$, the firm perfectly observes group output but cannot contract on individual effort directly. Nevertheless, the firm can still induce the surplus-maximizing effort profile $\mathbf{e}^{0}$ by setting $\alpha_i=1$ and extracting the residual surplus via $\beta_{i}$. Specifically, the firm charges each worker a participation fee $\beta_i = \tfrac{1}{2} - n < 0$. Given the firm’s budget-balance constraint,
\[ X \sum_i \alpha_i + \sum_i \beta_i \leq X,\]
it is easy to see that the first-best contract satisfies budget balance and, in fact, yields strictly positive profits, equal to $\pi = n/2$.

The setting just described (with $\lambda=0$ and $r\sigma^2=0$) coincides with the environment studied by \cite{holmstrom1982moral}, who shows that, in team production, no sharing rule can simultaneously satisfy budget balance and induce $\mathbf{e}^{FB}$ as a Nash equilibrium of the workers’ game. The key distinction---under which Holmström’s impossibility result does not apply to our analysis---is the contractual form we consider. We characterize linear contracts of the form $\omega_i=\beta_i+\alpha_i X$, whereas Holmström restricts attention to pure sharing rules $\omega_i=\alpha_i X$ (and, more generally, to  potentially nonlinear rules $s_i(X)\geq 0$). Under Holmström’s formulation, budget balance requires $\sum_i \alpha_i=1$, which indeed precludes efficiency unless the firm runs a deficit.\footnote{Holmström then shows that piecewise-linear sharing rules satisfying $\sum_i s_i(X)\leq X$ can restore efficiency.} The preceding paragraph makes clear that linear sharing rules can also implement the first best, provided the firm can extract residual surplus through fixed transfers $\beta_i$.

We now extend this argument and show that linear contracts are sufficiently flexible to guarantee  $\mathbf{e}^{*} = \mathbf{e}^0$ and budget balance, even in the presence of peer effects ($\lambda \neq 0$), provided that $\sigma^2=0$. To see this, take the expression for optimal effort from our benchmark model: 
\begin{equation*}
    \mathbf{e}^{*} = \mathbf{C}\boldsymbol{\alpha}^{*} =\mathbf{C}\mathbf{W\mathbf{C}'\mathbf{1}} =  \mathbf{C}\left[\mathbf{C}'(\mathbf{I}-\lambda(\mathbf{G}+\mathbf{G}'))\mathbf{C} +r\sigma^{2}\mathbf{I} \right]^{-1}\mathbf{C}'\mathbf{1}.
\end{equation*}
When $r\sigma^{2}=0$, this expression becomes: 
\begin{equation*}
    \begin{aligned}
        \mathbf{e}^{*}(r\sigma^{2}=0) = \mathbf{C}\left[\mathbf{C}'(\mathbf{I}-\lambda(\mathbf{G}+\mathbf{G}'))\mathbf{C} \right]^{-1}\mathbf{C}'\mathbf{1}
        = \left[\mathbf{I}-\lambda(\mathbf{G}+\mathbf{G}')\right]^{-1}\mathbf{1}
         = \mathbf{e}^{FB}.
    \end{aligned}
    \end{equation*}
Thus, even when peer effects are present, the firm can induce the optimal level of efforts when there is no risk. Moreover, the firm's budget balance condition is satisfied  and profits are strictly positive: 
\begin{equation*}
    X - X\sum_{i}\alpha_{i} - \sum_{i}\beta_{i} = \mathbb{E}[\pi^{FB}] = \frac{1 }{2}(\mathbf{e}^{FB})'\mathbf{1} = \frac{1}{2}\mathbf{1}'\left[ \mathbf{I} - \lambda (\mathbf{G}+\mathbf{G}') \right]^{-1}\mathbf{1} > 0.
\end{equation*}

\subsection{Proofs of Section~\ref{Sec: First-Best}}

\begin{proof}[\textbf{Proof of Proposition~\ref{FirstBest}}]
We maximize surplus over $\mathbf{e}\geq 0$. As we did for Proposition \ref{Optimal Contracts}, we first consider the unconstrained problem and verify that the nonnegativity constraint is slack. The first order conditions are given by:
\begin{equation*}
    \mathbf{1} - \mathbf{e} + \lambda(\mathbf{G}+\mathbf{G}')\mathbf{e} = 0 \quad \Rightarrow \quad (\mathbf{I} - \lambda(\mathbf{G}+\mathbf{G}'))\mathbf{e} = \mathbf{1}.
\end{equation*}
Since the spectral radius of $\lambda(\mathbf{G}+\mathbf{G}')$ is less than one, the Hessian matrix $\mathbf{H}=-(\mathbf{I} - \lambda(\mathbf{G}+\mathbf{G}'))$ is negative definite, while $\lambda\geq 0$ implies $(\mathbf{I} - \lambda (\mathbf{G}+\mathbf{G}'))^{-1}\geq 0$ by the Neumann series. Hence, the unique surplus-maximizing vector of efforts is given by:
\begin{equation*}
    \mathbf{e}^{0} = (\mathbf{I} - \lambda (\mathbf{G}+\mathbf{G}'))^{-1} \mathbf{1} \gg 0. \qedhere
\end{equation*}
\end{proof}

\section{Optimal Incentives and Bonacich Centrality in a Line Network}\label{sec: example supermarket}

In this section, we consider the 7-worker directed line network of Figure~\ref{fig:chain7workers} and show that, when risk is low, the firm optimally allocates larger incentives to worker 3 than to worker 1. In fact, plugging the line network into the characterization in Proposition~\ref{Optimal Contracts}, with $\sigma^{2}=0$, we obtain:
\begin{equation*}
	\alpha_{3}^{*} - \alpha_{1}^{*} = \frac{\lambda^{3}(1+2\lambda)}{1-4\lambda^{2}+2\lambda^{4}}. 
\end{equation*} 
For small $\lambda\geq0 $, we have $\alpha_{3}^{*}>\alpha_{1}^{*}$. However, we show below that $b_{1}(\lambda)>b_{3}(\lambda)$ when $\lambda>0$. Hence, the larger incentive assigned to worker 3 cannot be explained by output amplification alone. Instead, it reflects the second channel in equation~\eqref{systemprofits}: worker 3's incentives contribute more than worker 1's to lowering the marginal cost of incentivizing the rest of the firm.

\begin{figure}[t]
\centering
     \begin{tikzpicture}[node distance=2cm, every node/.style={circle, draw, minimum size=6mm}]
    \tikzset{
            blueNode/.style={circle, draw=blue!50, fill=blue!10, thick},
            redNode/.style={circle, draw=red!50, fill=red!10, thick},
        }
        \node[blueNode] (1) {1};
        \node[blueNode] (2) [right of=1] {2};
        \node[blueNode] (3) [right of=2] {3};
        \node[blueNode] (4) [right of=3] {4};
        \node[blueNode] (5) [right of=4] {5};
        \node[blueNode] (6) [right of=5] {6};
        \node[blueNode] (7) [right of=6] {7};

        \draw[->, -{Triangle},thick] (1) -- (2);
        \draw[->, -{Triangle},thick] (2) -- (3);
        \draw[->, -{Triangle},thick] (3) -- (4);
        \draw[->, -{Triangle},thick] (4) -- (5);
        \draw[->, -{Triangle},thick] (5) -- (6);
        \draw[->, -{Triangle},thick] (6) -- (7);
        
    \end{tikzpicture}
    \caption{A line network with 7 workers.}
    \label{fig:chain7workers}
\end{figure}

We first compare the amplification potential of both workers. First, recall that the firm's marginal benefit of providing incentives to worker $i$ is her Bonacich centrality: $b_{i}(\lambda) = \sum_{j=1}^{7}C_{ji}$. Then, notice that for any directed line, there is a unique directed path from $i$ to $j$ if and only if $j\ge i$, of length $j-i$. Thus, $C_{ji}=\lambda^{j-i}$ for $j\geq i$ and $C_{ij}=0$ otherwise. Therefore, the difference between the marginal benefits of $\alpha_1$ and $\alpha_3$ is given by:
\begin{equation*}\label{eq:MB13_newapp}
MB_{\alpha_1}-MB_{\alpha_3} = b_1(\lambda)-b_3(\lambda) = \lambda^5+\lambda^6 >0.
\end{equation*}
Notice that worker 3 is closer to all her targets than worker 1 is to those same targets. The additional amplification effect that worker 3 has with her targets is more than compensated by  the fact that worker 1 has additional targets: namely workers 1 and 2. Netting out these effects, we find that the marginal benefit from raising $\alpha_1$ is greater than the marginal benefit from raising $\alpha_3$.\footnote{Notice that this difference shrinks as $n$ increases since it only accounts for paths of length at least $n-2$.}

Next, we turn to the incentive cost-reduction channel. Following \eqref{systemprofits} with $\sigma^2=0$, the firm's marginal cost of providing incentives to worker $i$ is given by:  
\begin{equation*}\label{eq:foc_appE_new}
MC_{\alpha_{i}}=\alpha_i
-
\sum_{j=1}^7 \alpha_j K_{ij}(\lambda), \quad \text{where} \quad K_{ij}(\lambda)
:=
\sum_{s=1}^7
\left(\lambda\sum_{\ell=1}^7 g_{s\ell}\frac{\partial e_\ell}{\partial \alpha_j}\right)
\left(\lambda\sum_{m=1}^7 g_{sm}\frac{\partial e_m}{\partial \alpha_i}\right).
\end{equation*}
The entry $K_{ij}(\lambda)$ measures how much increasing $\alpha_j$ reduces the marginal cost of incentivizing worker $i$. Therefore, $\sum_i K_{ij}(\lambda)$ measures the aggregate reduction in marginal compensation costs generated by increasing worker $j$'s incentives. 

On the directed line, $g_{s\ell}=1$ only when $\ell=s-1$. Hence, we have:
\begin{equation*}
	K_{ij}(\lambda)
=
\sum_{s=2}^7
\left(\lambda\frac{\partial e_{s-1}}{\partial \alpha_j}\right)
\left(\lambda\frac{\partial e_{s-1}}{\partial \alpha_i}\right).
\end{equation*} 
Using $\frac{\partial e_t}{\partial \alpha_i}=\lambda^{t-i}$ when $t\ge i$ and zero otherwise, we can compute the total cost-reduction contributions of the incentives allocated to workers 1 and 3:
\begin{align*}
\sum_{i=1}^7 K_{i1}(\lambda)
&=
\lambda^2+\lambda^3+2\lambda^4+2\lambda^5+3\lambda^6+3\lambda^7+3\lambda^8 +2\lambda^9+2\lambda^{10}+\lambda^{11}+\lambda^{12},\\
\sum_{i=1}^7 K_{i3}(\lambda)
&=
\lambda^2+2\lambda^3+3\lambda^4+3\lambda^5+3\lambda^6+2\lambda^7+2\lambda^8+\lambda^9+\lambda^{10}.
\end{align*}
Subtracting, we obtain:
\begin{equation*}
\sum_{i=1}^7 K_{i3}(\lambda)-\sum_{i=1}^7 K_{i1}(\lambda)
=
\lambda^3+\lambda^4+\lambda^5-\lambda^7-\lambda^8-\lambda^9-\lambda^{10}-\lambda^{11}-\lambda^{12}>0,
\end{equation*}
for $\lambda$ small enough. Notice that there are positive and negative terms. This is because while the aggregate effect favors $\alpha_{3}$, increasing $\alpha_{1}$ has a larger effect on the marginal compensation cost of incentivizing some workers (workers 1 and 2) which worker 3 simply does not target. 
In fact, this can be seen by associating each reduction effect with the corresponding marginal cost it affects:\footnote{Notice that we retain the $\lambda^{2}$ effects that cancel out in the aggregate difference so that each bracket isolates the difference between the
effects of $\alpha_3$ and $\alpha_1$ on a particular marginal
compensation-cost term.}
\begin{equation*}
\sum_{i=1}^7 K_{i3}(\lambda)-\sum_{i=1}^7 K_{i1}(\lambda)
 =
\underbrace{(-\lambda^{2}-\lambda^{12})}_{MC_{\alpha_1}}
+
\underbrace{(-\lambda^{11})}_{MC_{\alpha_2}}
+
\underbrace{(\lambda^2-\lambda^{10})}_{MC_{\alpha_3}}
+
\underbrace{(\lambda^3-\lambda^9)}_{MC_{\alpha_4}}
+
\underbrace{(\lambda^4-\lambda^8)}_{MC_{\alpha_5}}
+
\underbrace{(\lambda^5-\lambda^7)}_{MC_{\alpha_6}}.\\
\end{equation*}
Although increasing $\alpha_1$ reduces marginal compensation costs for the targets that are exclusive to worker 1, increasing $\alpha_3$ has a stronger effect on those associated with all remaining workers, because worker 3 is closer than worker 1 to any common target they share downstream. Thus, if the line were shorter, worker 3 would lose part of this advantage and worker 1 would continue to receive stronger incentives. Conversely, if the line were even longer, then workers $4$ or $5$, etc. would gain this advantage and receive the steepest incentives.


Finally, notice that the same argument extends to sufficiently small positive $\sigma^2$. Indeed, the equilibrium incentive vector is continuous in $(\lambda,\sigma^2)$ on the admissible parameter region. Since $\alpha_3^\ast(\lambda,0)-\alpha_1^\ast(\lambda,0)>0$ for all sufficiently small $\lambda>0$, continuity implies that there exists $\bar{\sigma}^2(\lambda)>0$ such that:
$$
\alpha_3^\ast(\lambda,\sigma^2)-\alpha_1^\ast(\lambda,\sigma^2)>0
\qquad
\text{for all }\sigma^2\in[0,\bar{\sigma}^2(\lambda)).
$$
Thus the ranking $\alpha_3^{*}>\alpha_1^{*}$ is robust to small but positive output risk. Conversely, as shown in Proposition~\ref{monotonic}, for sufficiently high risk the
influence of the second channel becomes weak enough that optimal incentives preserve the strict ordering of Bonacich centrality.

\section{Optimal Incentives with Negative Spillovers}\label{sec: App Negative Spillovers}

In this section, we deal with the case of negative spillovers, i.e., $\lambda<0$. First of all, notice that under Assumption \ref{assumption1}, the Neumann expansion for $\mathbf{C}=(\mathbf{I}-\lambda\mathbf{G})^{-1}$ remains valid: 
\begin{equation*}
    \mathbf{C}(\lambda) = (\mathbf{I}-\lambda\mathbf{G})^{-1} = \sum_{q=0}^{\infty}(-1)^{q}(\vert\lambda\vert\mathbf{G})^{q} = \mathbf{I} - \vert\lambda\vert\mathbf{G} + \lambda^{2}\mathbf{G}^{2} - \vert\lambda^{3}\vert\mathbf{G}^{3} + \cdots. 
\end{equation*}
However, because the Neumann series alternates in sign, $\mathbf{C}(\lambda)$ need not be entrywise nonnegative. Therefore, the linear system $\mathbf{e}=\mathbf{C}(\lambda)\boldsymbol{\alpha}$ is no longer guaranteed to satisfy $\mathbf{e}\geq0$, even for $\boldsymbol{\alpha}\gg0$. The best-reply function of worker $i$ is given by:
\begin{equation*}
    e_{i} = \max\left\lbrace 0, \alpha_{i}- \vert\lambda\vert \sum_{j} g_{ij} e_{j} \right\rbrace.
\end{equation*}
Under strategic substitutes, a worker receiving low incentives may optimally exert zero effort if she is surrounded by sufficiently active neighbors. In other words, sufficiently unequal incentives (i.e., low incentives to a worker and high incentives to her connections) can generate free-riding and push towards corners.

Let the set of incentive vectors that induce an interior equilibrium effort profile be
\begin{equation*}
    \mathcal{A}_{\mathrm{int}}(\lambda)
    :=\Big\{\boldsymbol\alpha\in\mathbb R^n:\ 
    \mathbf{C}(\lambda)\boldsymbol{\alpha}\gg 0\Big\}.
\end{equation*}
For any $\boldsymbol{\alpha}\in\mathcal{A}_{\mathrm{int}}(\lambda)$, 
$\mathbf{e}^{*}:=\mathbf{C}(\lambda)\boldsymbol{\alpha}\gg0$ satisfies
$(\mathbf{I}+|\lambda|\mathbf{G})\mathbf{e}^{*}=\boldsymbol{\alpha}$,
and hence the best-response system. Moreover,
$\boldsymbol{\alpha}=(\mathbf{I}+|\lambda|\mathbf{G})\mathbf{e}^{*}\gg0$.
Thus, $\mathcal{A}_{\mathrm{int}}(\lambda)\subset\mathbb{R}_{++}^{n}$. Define the firm's maximal profits over contracts that induce an interior equilibrium as:
\begin{equation*}
    V_{int}(\lambda) := \sup_{\boldsymbol{\alpha} \in \mathcal{A}_{int}(\lambda)} 
    \Pi(\boldsymbol{\alpha}, \lambda) 
    = \max_{\boldsymbol{\alpha}:\, \mathbf{C}(\lambda)\boldsymbol{\alpha}\geq 0} 
    \Pi(\boldsymbol{\alpha}, \lambda),
\end{equation*}
where the equality follows because every weakly feasible contract can be
approximated by contracts in $\mathcal{A}_{int}(\lambda)$.\footnote{Since
$\mathcal{A}_{\mathrm{int}}(\lambda)\subset
\{\boldsymbol{\alpha}\geq0:\mathbf{C}(\lambda)\boldsymbol{\alpha}\geq0\}$,
one inequality is immediate. For the reverse, set
$\boldsymbol{\alpha}^{0}=(\mathbf{I}-\lambda\mathbf{G})\mathbf{1}\gg0$,
so $\mathbf{C}(\lambda)\boldsymbol{\alpha}^{0}=\mathbf{1}$.
For any weakly feasible $\boldsymbol{\alpha}$,
$(1-\varepsilon)\boldsymbol{\alpha}+\varepsilon\boldsymbol{\alpha}^{0}
\in\mathcal{A}_{\mathrm{int}}(\lambda)$ for every $\varepsilon\in(0,1)$
and converges to $\boldsymbol{\alpha}$ as $\varepsilon\to0$.
The result follows by continuity of $\Pi$.}

Consider any corner equilibrium and let $A=\operatorname{supp}(\mathbf e^*)$. For any $i\notin A$, replacing $\alpha_i$ by zero leaves the equilibrium unchanged, since $-|\lambda|\sum_jg_{ij}e_j^*\leq 0$, and weakly reduces risk compensation. Hence, without loss of generality,  the firm can set $\alpha_i =\beta_{i} =0$ for all $i\notin A$, essentially turning them off. 
The remaining workers interact through the
subnetwork induced by $A$. Let $\mathbf{G}_{A}$ denote its adjacency matrix and
$\mathbf{C}_{A}(\lambda):=(\mathbf{I}-\lambda\mathbf{G}_{A})^{-1}$. We define
\begin{equation*}
    V_A(\lambda)
    :=\max_{\boldsymbol{\alpha}_{A}:\,
    \mathbf{C}_{A}(\lambda)\boldsymbol{\alpha}_{A}\geq0}
    \Pi_A(\boldsymbol{\alpha}_{A},\lambda)
\end{equation*}
as the closed problem in which the equilibrium effort support is contained in $A$, and the firm's maximal profits over contracts that induce a corner equilibrium as
\begin{equation*}
    V_{\mathrm{corner}}(\lambda)
    :=\max_{ A\subsetneq N }V_A(\lambda).
\end{equation*}
Thus, $V_{\mathrm{corner}}(\lambda)$ is the \textit{upper envelope} of maximal profits over all strict induced subnetworks.  In other words, optimal contracts still coincide with our interior solutions, only restricted to an appropriate sub-network.

Finally, define $\Delta(\lambda) := V_{\mathrm{int}}(\lambda) - V_{\mathrm{corner}}(\lambda)$. In the absence of peer effects ($\lambda=0$), the firm's problem is separable across workers and the unique optimum, given in Corollary \ref{no peer effects}, yields $\alpha_{i}^{*} = e_{i}^{*}=1/(1+r\sigma^{2})$ for all $i\in N$.  The unique
optimal contract lies in $\mathcal A_{\mathrm{int}}(0)$ and strictly dominates every corner allocation, implying $\Delta(0)>0$. Hence, the incentive allocation characterized in Proposition \ref{Optimal Contracts} is valid.\footnote{In fact, by Proposition \ref{Profits and Spectrum} in Supplementary Appendix~\ref{Design}, profits at the optimal contract are equal to half the sum of efforts so that
\begin{equation*}
    \Delta(0) = V_{\operatorname{int}}(0)- V_{\operatorname{corner}}(0) = \frac{n}{2(1+r\sigma^{2})} - \frac{n-1}{2(1+r\sigma^{2})} = \frac{1}{2(1+r\sigma^{2})}>0.
\end{equation*}}

To show that workers play an interior equilibrium for sufficiently small $\lambda$ we just need to argue that these value functions are continuous in $\lambda$.
\begin{lemma}\label{Lemma:continuousVF}
    Under Assumptions~\ref{assumption1} and~\ref{spectralradius}, the value functions $V_{\operatorname{int}}(\lambda)$ and $V_{\mathrm{corner}}(\lambda)$ are continuous. 
\end{lemma}
This lemma implies that $\Delta(\lambda)$ is also continuous and, since we argued above that $\Delta(0)>0$, then  $\Delta(\lambda)>0$ for $\lambda$ sufficiently close to $0$. This leads to the following result.

\begin{proposition}\label{prop:existance}
    There exists a threshold value of spillovers, $\tilde{\lambda}$, such that the firm's optimal contract lies in $\mathcal{A}_{\operatorname{int}}(\lambda)$ for all $\vert \lambda\vert < \vert\tilde{\lambda}\vert$. 
\end{proposition}

Proposition \ref{prop:existance} ensures that if spillovers are small enough, the firm’s optimal contract continues to induce an interior effort profile, so the characterization in Propositions \ref{Optimal Contracts} and \ref{monotonic} applies. The remaining challenge is therefore one of selection: determining which sub-network attains $V_{\mathrm{corner}}(\lambda)$. This can be a daunting problem in general, given the large number of possible sub-networks. The examples below illustrate how the relevant threshold can be computed when it lies in the admissible region, and how the optimal corner sub-network may change as negative spillovers become stronger.


\begin{figure}[t]
\centering
\raisebox{0.15\height}{
    
\begin{tikzpicture}[scale=0.3]

    \node[circle, draw=black, fill=black, thick, minimum size=1mm, inner sep=0.5mm] (o1) at (12.5,18) {};
    \node[circle, draw=black, fill=black, thick, minimum size=1mm, inner sep=0.5mm] (o2) at (15.5,15) {};
    \node[circle, draw=black, fill=black, thick, minimum size=1mm, inner sep=0.5mm] (o3) at (9.5,15) {};
    \node[circle, draw=black, fill=black, thick, minimum size=1mm, inner sep=0.5mm] (o4) at (12.5,12) {};

    \draw[black] (o1) -- (o2);
    \draw[black] (o2) -- (o4);
    \draw[black] (o4) -- (o3);
    \draw[black] (o3) -- (o1);

\node[circle, draw=blue!80, fill=blue!80, thick, minimum size=1mm, inner sep=0.5mm] (a1) at (5.6,8.5) {};
\node[circle, draw=blue!80, fill=blue!80, thick, minimum size=1mm, inner sep=0.5mm] (a2) at (8,10.9) {};
\node[circle, draw=blue!80, fill=blue!80, thick, minimum size=1mm, inner sep=0.5mm] (a3) at (10.4,8.5) {};
\node[circle, draw=gray,    fill=white,   thick, minimum size=1mm, inner sep=0.5mm] (a4) at (8,6.1) {};

\draw[blue!80] (a1) -- (a2);
\draw[blue!80] (a2) -- (a3);
\draw[gray]    (a3) -- (a4);
\draw[gray]    (a4) -- (a1);

\node[text=blue!80] at (8,8.5) {\Large A};

\node[circle, draw=ForestGreen, fill=ForestGreen, thick, minimum size=1mm, inner sep=0.5mm] (b1) at (17,10.9) {};
\node[circle, draw=ForestGreen, fill=ForestGreen, thick, minimum size=1mm, inner sep=0.5mm] (b2) at (19.4,8.5) {};
\node[circle, draw=gray,        fill=white,       thick, minimum size=1mm, inner sep=0.5mm] (b3) at (17,6.1) {};
\node[circle, draw=gray,        fill=white,       thick, minimum size=1mm, inner sep=0.5mm] (b4) at (14.6,8.5) {};

\draw[ForestGreen] (b1) -- (b2);
\draw[gray]        (b2) -- (b3);
\draw[gray]        (b3) -- (b4);
\draw[gray]        (b4) -- (b1);

\node[text=ForestGreen] at (17,8.5) {\Large B};

\node[circle, draw=red!80, fill=red!80, thick, minimum size=1mm, inner sep=0.5mm] (c1) at (5.6,0.5) {};
\node[circle, draw=gray,   fill=white,  thick, minimum size=1mm, inner sep=0.5mm] (c2) at (8,2.9) {};
\node[circle, draw=red!80, fill=red!80, thick, minimum size=1mm, inner sep=0.5mm] (c3) at (10.4,0.5) {};
\node[circle, draw=gray,   fill=white,  thick, minimum size=1mm, inner sep=0.5mm] (c4) at (8,-1.9) {};

\draw[gray] (c1) -- (c2);
\draw[gray] (c2) -- (c3);
\draw[gray] (c3) -- (c4);
\draw[gray] (c4) -- (c1);

\node[text=red!80] at (8,0.5) {\Large C};

\node[circle, draw=brown!90, fill=brown!90, thick, minimum size=1mm, inner sep=0.5mm] (d1) at (17,2.9) {};
\node[circle, draw=gray,     fill=white,   thick, minimum size=1mm, inner sep=0.5mm] (d2) at (19.4,0.5) {};
\node[circle, draw=gray,     fill=white,   thick, minimum size=1mm, inner sep=0.5mm] (d3) at (17,-1.9) {};
\node[circle, draw=gray,     fill=white,   thick, minimum size=1mm, inner sep=0.5mm] (d4) at (14.6,0.5) {};

\draw[gray] (d1) -- (d2);
\draw[gray] (d2) -- (d3);
\draw[gray] (d3) -- (d4);
\draw[gray] (d4) -- (d1);

\node[text=brown!90] at (17,0.5) {\Large D};

\end{tikzpicture}
} 
    \hspace{0.5cm}
    \includegraphics[width=280pt]{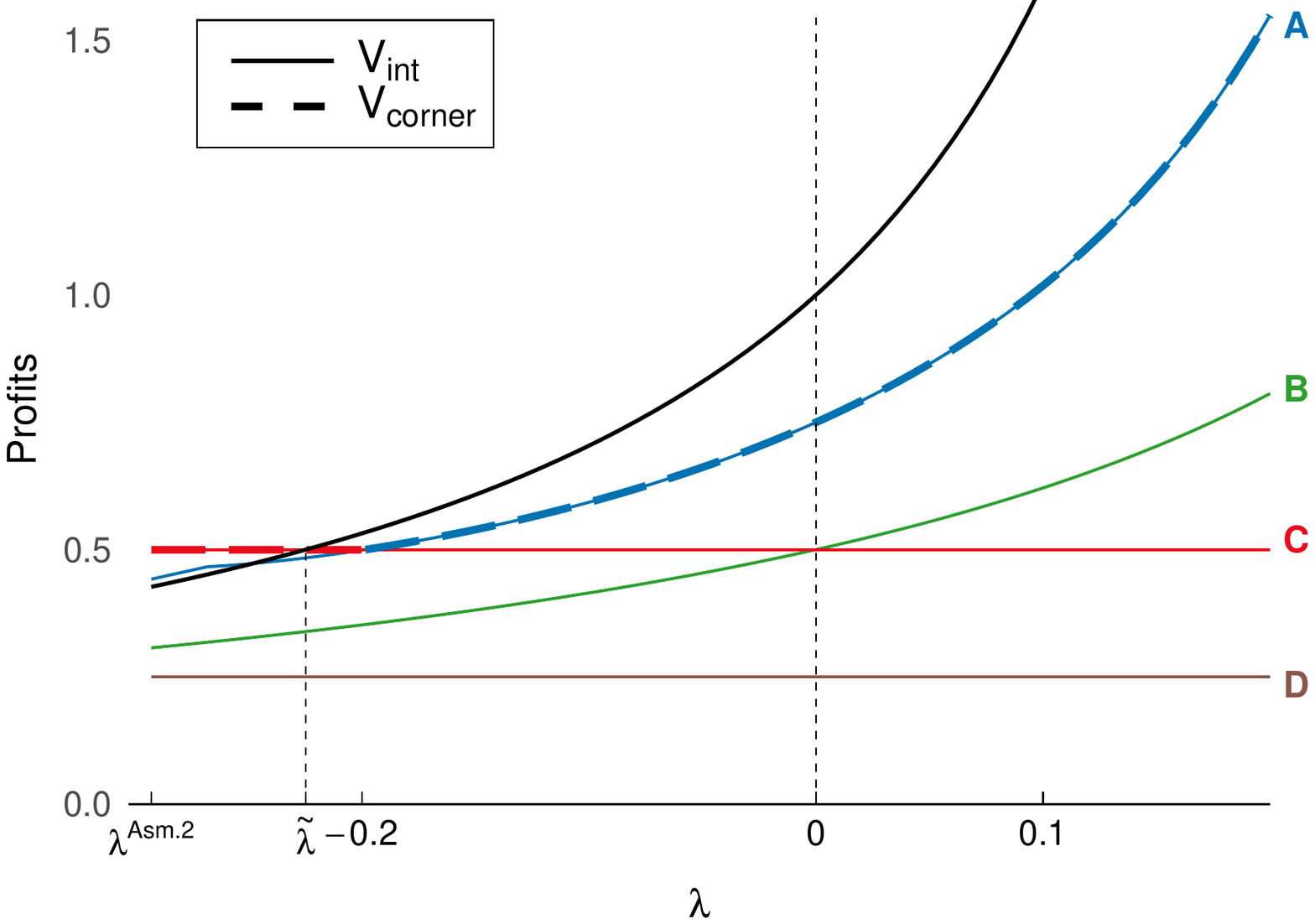}
    \caption{Firm’s maximal profits under interior and corner Nash equilibria in a 4-worker ring network. Corner solutions correspond to interior solutions on strict subnetworks: subnetwork A has 3 active workers arranged in a line, subnetworks B and C each have 2 active workers, and subnetwork D has a single active worker.}
    \label{fig:Ring_IntvsCorner}
\end{figure}

\subsection{Examples}

Consider the simple ring network with $n=4$ in Figure \ref{fig:Ring_IntvsCorner}. Set $r\sigma^2=1$. When $\lambda=0$, the optimal contract satisfies $\boldsymbol{\alpha}^{*}=\mathbf{e}^{*}=0.5*\mathbf{1}$ and, since $\Pi=\frac{1}{2}\sum_i e_i^{*}$, we obtain $\Pi=\frac{1}{2}\cdot n \cdot 0.5 = 1$, corresponding to the $y$-intercept of the black curve. Any strict sub-network contains fewer active workers and therefore yields strictly lower profits at $\lambda=0$, which is reflected by the lower y-intercepts for the profits of sub-networks A, B, C, and D.\footnote{Notice that, starting from a ring network with 4 workers, not all sub-networks with 3 workers are feasible. There are only 4 possible sub-networks to be considered. For example, excluding a worker from a ring network can only generate a sub-network of 3 workers in a line network. However, a further exclusion may generate a sub-network of two workers connected or isolated.}  As $\lambda$ becomes negative, profits decline faster in the full network due to its higher connectivity, while some sub-networks (being disconnected) generate profits that are independent of $\lambda$. Taking the upper envelope of profits across all sub-networks yields the dashed curve $V_{\operatorname{corner}}(\lambda)$, which corresponds to sub-network A for large $\lambda$ and sub-network C for low lambdas. It intersects $V_{\operatorname{int}}(\lambda)$ at $\tilde{\lambda}$. In this case, $\tilde{\lambda}$ lies in the value function of sub-network C, indicating that once interior equilibria cease to be optimal, the firm prefers to incentivize only two workers and shut down the other two, leaving the active nodes effectively disconnected. Notice also that $\tilde{\lambda}$ is around $-0.225$ while the $\lambda$ in Assumption 1 is about $-0.5$. That is, in general, we need a more stringent condition than Assumption 1 to guarantee that the firm's optimal contract induces an interior equilibrium.

\begin{figure}[t]
\centering
\raisebox{0.3\height}{
    \begin{tikzpicture}[scale=0.35]

        \node[circle, draw=black, fill=black, thick, minimum size=1mm, inner sep=0.5mm] (1) at (2,13) {};
        \node[circle, draw=black, fill=black, thick, minimum size=1mm, inner sep=0.5mm] (2) at (5,13) {};
        \node[circle, draw=black, fill=black, thick, minimum size=1mm, inner sep=0.5mm] (3) at (8,13) {};
        \node[circle, draw=black, fill=black, thick, minimum size=1mm, inner sep=0.5mm] (4) at (11,13) {};

        \draw[] (1) -- (2);
        \draw[] (2) -- (3);
        \draw[] (3) -- (4);

        \node[circle, draw=blue!80, fill=blue!80, thick, minimum size=1mm, inner sep=0.5mm] (5) at (2,10) {};
        \node[circle, draw=blue!80, fill=blue!80, thick, minimum size=1mm, inner sep=0.5mm] (6) at (5,10) {};
        \node[circle, draw=blue!80, fill=blue!80, thick, minimum size=1mm, inner sep=0.5mm] (7) at (8,10) {};
        \node[circle, draw=gray, thick, minimum size=1mm, inner sep=0.5mm] (8) at (11,10) {};
        

        \node[circle, draw=Orange, fill=Orange, thick, minimum size=1mm, inner sep=0.5mm] (9) at (2,8) {};
        \node[circle, draw=Orange, fill=Orange, thick, minimum size=1mm, inner sep=0.5mm] (10) at (5,8) {};
        \node[circle, draw=gray, thick, minimum size=1mm, inner sep=0.5mm] (11) at (8,8) {};
        \node[circle, draw=Orange, fill=Orange, thick, minimum size=1mm, inner sep=0.5mm] (12) at (11,8) {};


        \node[circle, draw=ForestGreen, fill=ForestGreen, thick, minimum size=1mm, inner sep=0.5mm] (13) at (2,6) {};
        \node[circle, draw=ForestGreen, fill=ForestGreen, thick, minimum size=1mm, inner sep=0.5mm] (14) at (5,6) {};
        \node[circle, draw=gray, thick, minimum size=1mm, inner sep=0.5mm] (15) at (8,6) {};
        \node[circle, draw=gray, thick, minimum size=1mm, inner sep=0.5mm] (16) at (11,6) {};


        \node[circle, draw=red!80, fill=red!80, thick, minimum size=1mm, inner sep=0.5mm] (17) at (2,4) {};
        \node[circle, draw=gray, thick, minimum size=1mm, inner sep=0.5mm] (18) at (5,4) {};
        \node[circle, draw=red!80, fill=red!80, thick, minimum size=1mm, inner sep=0.5mm] (19) at (8,4) {};
        \node[circle, draw=gray, thick, minimum size=1mm, inner sep=0.5mm] (20) at (11,4) {};


        \node[circle, draw=brown, fill=brown, thick, minimum size=1mm, inner sep=0.5mm] (21) at (2,2) {};
        \node[circle, draw=gray, thick, minimum size=1mm, inner sep=0.5mm] (22) at (5,2) {};
        \node[circle, draw=gray, thick, minimum size=1mm, inner sep=0.5mm] (23) at (8,2) {};
        \node[circle, draw=gray, thick, minimum size=1mm, inner sep=0.5mm] (24) at (11,2) {};


        \draw[blue!80] (5) -- (6);
        \draw[blue!80] (6) -- (7);
        \draw[gray] (7) -- (8);
        
        \draw[Orange] (9) -- (10);
        \draw[gray] (10) -- (11);
        \draw[gray] (11) -- (12);
        
        \draw[ForestGreen] (13) -- (14);
        \draw[gray] (14) -- (15);
        \draw[gray] (15) -- (16);

        \draw[gray] (17) -- (18);
        \draw[gray] (18) -- (19);
        \draw[gray] (19) -- (20);
        \draw[gray] (21) -- (22);
        \draw[gray] (22) -- (23);
        \draw[gray] (23) -- (24);

        \node[left, text=blue!80]     at (0.5,10) {\Large A};
        \node[left, text=Orange]      at (0.5,8)  {\Large B};
        \node[left, text=ForestGreen] at (0.5,6)  {\Large C};
        \node[left, text=red!80]      at (0.5,4)  {\Large D};
        \node[left, text=brown]       at (0.5,2)  {\Large E};

    \end{tikzpicture}}
    \hspace{0.5cm}
    \includegraphics[width=280pt]{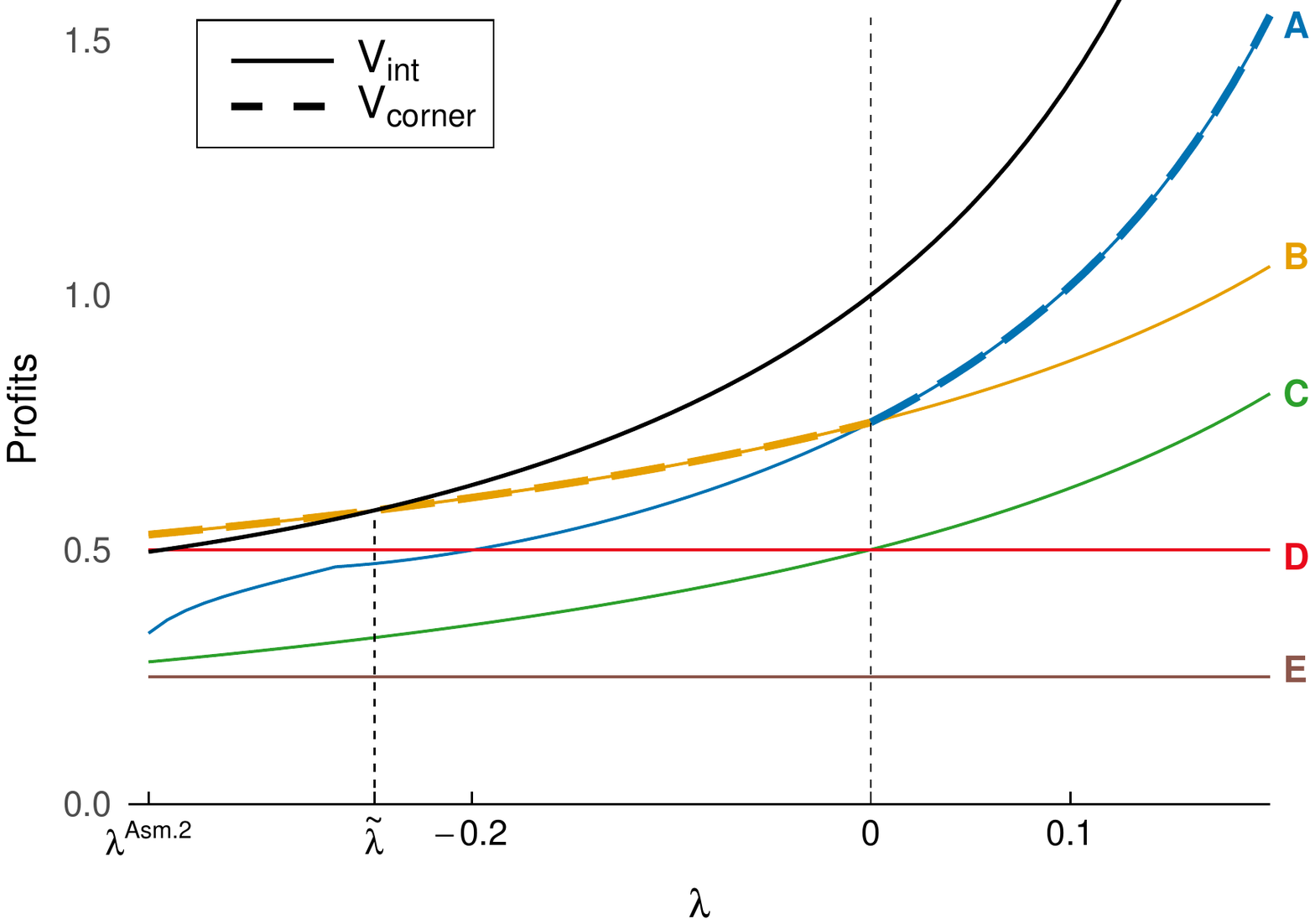}
    \caption{Firm’s maximal profits under interior and corner Nash equilibria in a 4-worker line network. Corner solutions correspond to interior solutions on strict subnetworks: A has 3 active workers arranged in a line, B has 2 active workers arranged in a line together with 1 isolated active worker, C has 2 active workers arranged in a line, D has 2 isolated active workers, and E has a single active worker.}
    \label{fig:Line_IntvsCorner}
\end{figure}

In Figure~\ref{fig:Line_IntvsCorner} we present $\tilde{\lambda}$ for a line network with 4 workers. In this example, the upper envelope of profits across all sub-networks $V_{corner}(\lambda)$ includes sub-networks A, B, and D. It intersects $V_{int}(\lambda)$ at $\tilde{\lambda}$. In this case, when interior equilibria cease to be optimal, the firm prefers to initially incentivize 3 workers and to shut down a worker in the middle of the line (sub-network B). However, as $\lambda$ keeps falling, the firm ends up incentivizing only two workers (sub-network D), as in the ring example.

\begin{figure}
    \centering
    \includegraphics[width=0.55\linewidth]{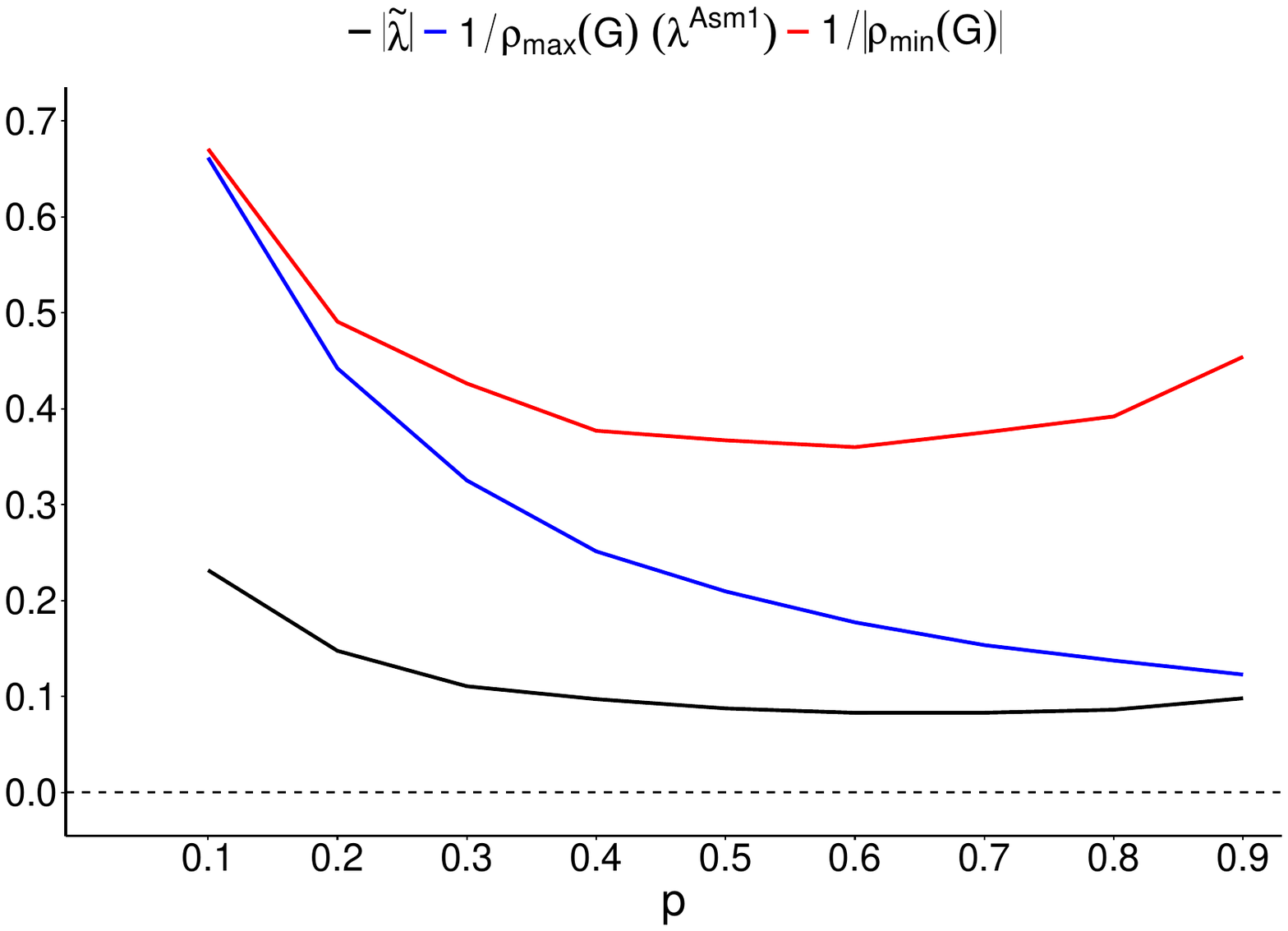}
    \caption{Bounds on the strength of peer effects for different levels of connectivity in Erd\H{o}s--R\'enyi networks.}
    \label{fig:BoundsLambda}
\end{figure}

To conclude, we conduct a Monte Carlo study based on $100$ Erd\H{o}s--R\'enyi random graphs with $n=10$ and $r\sigma^{2}=1$, varying the connectivity parameter $p$. The results, shown in figure \ref{fig:BoundsLambda}, indicate that as $p$ increases, the range of $\lambda$ for which the firm prefers an interior Nash equilibrium in the workers' effort--provision game, i.e., $\vert\lambda\vert<\vert\tilde\lambda\vert$, falls at first, then flattens, with a slight increase as $p$ approaches 1. This behavior suggests that as $p$ increases from low levels, higher connectivity both strengthens negative spillovers in the interior solution and expands the menu of feasible ``corner'' sub-networks, making the latter more attractive and pushing $\tilde\lambda$ toward 0. By contrast, as connectivity reaches high levels, the network becomes redundant, and exclusion of a few workers barely reduces negative spillovers, worsening the corner envelope relative to the interior and pushing $\vert \tilde\lambda \vert$ upwards. 

Comparing this with the lambda bound from Assumption \ref{assumption1} \citep{ballester2006}, which decreases monotonically with greater connectivity, we see that, in general, $\tilde{\lambda}$ is a more stringent condition. Finally, notice that the behavior of the red curve, i.e., the condition for uniqueness in games of strategic substitutes given in \cite{bramoulle2014strategic}, is U-shaped. That is because in an Erd\H{o}s--R\'enyi random graph, the magnitude of the smallest eigenvalue is mainly driven by random noise in the adjacency matrix. This noise has entrywise variance $p(1-p)$, which is smallest at the extremes and largest in the middle.

\subsection{Proofs of Section \ref{sec: App Negative Spillovers}}

\begin{proof}[\textbf{Proof of Lemma \ref{Lemma:continuousVF}}]
We first establish continuity of $V_{\mathrm{int}}(\lambda)$. Under Assumption~\ref{assumption1} and for $\lambda\leq0$, the change of variables $\mathbf e=\mathbf C(\lambda)\boldsymbol{\alpha}$ maps the feasible set $\{\boldsymbol{\alpha}\geq0:\mathbf C (\lambda)\boldsymbol{\alpha}\geq0\}$
onto $\mathbb R_+^n$. Indeed, for any $\mathbf e\geq0$, $\boldsymbol{\alpha}
    =
    (\mathbf I-\lambda\mathbf G)\mathbf e
    =
    \mathbf e+|\lambda|\mathbf G\mathbf e
    \geq0.$
Hence, we can use the closed representation of
$V_{\mathrm{int}}(\lambda)$ in terms of $\mathbf{e}$:
\[
V_{\mathrm{int}}(\lambda)
=
\max_{\mathbf e\geq0}
\left\{
\mathbf 1'\mathbf e
-\frac{1}{2}\|\mathbf e\|^2
+\lambda\mathbf e'\mathbf G\mathbf e
-\frac{r\sigma^2}{2}
\|(\mathbf I-\lambda\mathbf G)\mathbf e\|^2
\right\}.
\]
For $\lambda\leq0$, $\mathbf G\geq0$, and $\mathbf e\geq0$, the last two
terms are nonpositive. Therefore,
\[
    \Pi(\mathbf e,\lambda)
    \leq
    \mathbf 1'\mathbf e-\frac{1}{2}\|\mathbf e\|^2
    \leq
    \sqrt{n}\|\mathbf e\|-\frac{1}{2}\|\mathbf e\|^2,
\]
where the second inequality follows from Cauchy--Schwarz. Since
$\mathbf e=\mathbf0$ is feasible and yields zero profits, no maximizer can
satisfy $\|\mathbf e\|>2\sqrt n$. Thus, all maximizers lie in the common
compact set $\left\{\mathbf e\geq0:\|\mathbf e\|\leq2\sqrt n\right\}.$ The 
objective is continuous in $(\mathbf e,\lambda)$, so the maximum
theorem implies that $V_{\mathrm{int}}(\lambda)$ is continuous.

The same argument applies to every nonempty strict subnetwork
$A\subsetneq N$. Replacing $\mathbf G$ by $\mathbf G_A$ and $n$ by
$|A|$, we obtain
\[
    V_A(\lambda)
    =
    \max_{\mathbf e_A\geq0}
    \left\{
    \mathbf 1'\mathbf e_A
    -\frac{1}{2}\|\mathbf e_A\|^2
    +\lambda\mathbf e_A'\mathbf G_A\mathbf e_A
    -\frac{r\sigma^2}{2}
    \|(\mathbf I-\lambda\mathbf G_A)\mathbf e_A\|^2
    \right\},
\]
and every maximizer satisfies
$\|\mathbf e_A\|\leq2\sqrt{|A|}\leq2\sqrt n$.
Hence, each $V_A(\lambda)$ is continuous. Since there are finitely many nonempty strict subsets $A\subsetneq N$, $V_{\mathrm{corner}}(\lambda)$ is continuous as the maximum of finitely many continuous functions.
    
\end{proof}

\begin{proof}[\textbf{Proof of Proposition \ref{prop:existance}}]
By Lemma \ref{Lemma:continuousVF}, we know that $V_{\operatorname{int}}(\lambda)$ and $V_{\operatorname{corner}}(\lambda)$ are both continuous at $\lambda=0$. Thus, $\Delta(\lambda)$ is also continuous at $\lambda=0$. Because at $\lambda=0$ we know that $\Delta(0)>0$, it must be that there exists a threshold $\tilde\lambda$ such that $\Delta(\lambda)>0$ for all $\vert \lambda\vert <\vert \tilde{\lambda}\vert$. Depending on the example, $\Delta(\lambda)>0$ for all $\lambda\in [0,\lambda^{Asm.2}]$. Define $\tilde{\lambda} = \operatorname{max}\{\lambda^{Asm.2}, \bar{\lambda}\}$ where $\Delta(\bar{\lambda})=0$. The result follows. 
\end{proof}

\section{More Flexible Externalities}\label{SuppApp:MoreFlexibleExternalities}

This section considers the baseline model from Section~\ref{baseline} but with a richer specification of spillovers and derives the optimal incentive rule under these more flexible externalities. This extension allows for situations in which the overall utility effect runs counter to the strategic effect---for instance, I may exert more effort because a hardworking neighbor observes me, even though this lowers my overall welfare. We can capture this margin by adding a level effect (modulated by $\omega\geq0$) into the worker's cost structure: 
\begin{equation*}
\psi_i(\mathbf e;\mathbf G)=\frac{1}{2}e_i^2-\lambda e_i\sum_{j\in N}g_{ij}e_j+\omega\sum_{j\in N}g_{ij}e_j.
\label{eq:flexcost}
\end{equation*}
The last term is new and represents a level effect: it shifts utility holding own effort fixed. Thus, unlike the strategic effect from peer effects (modulated by $\lambda$), it does not alter the marginal cost of effort and, therefore, workers' best responses are still given by \eqref{eq:BestResponse}. 

The new term operates instead through the participation (IR) constraints, because higher effort by a worker's peers raises the fixed compensation required for the worker to accept the contract. Because it does not alter workers' marginal incentives, their best responses remain unchanged, and therefore the marginal-benefit side of the firm's problem is unaffected. The level shift does, however, influence the compensation workers require to participate in the contract, operating through the fixed salary term $\boldsymbol{\beta}$. In turn, this changes the firm's marginal cost of providing incentives. Specifically, \eqref{MCalphalong} is now given by:
\begin{equation*}
	M C_{\alpha_i}=\sum_s e_s \frac{\partial e_s}{\partial \alpha_i}-\lambda \sum_s \sum_{\ell} g_{s \ell}\left(e_s \frac{\partial e_{\ell}}{\partial \alpha_i}+e_{\ell} \frac{\partial e_s}{\partial \alpha_i}\right) + \omega \sum_s\sum_\ell g_{s\ell} \frac{\partial e_\ell}{\partial \alpha_i}+r \sigma^2 \alpha_i .
\end{equation*}
This implies that we can now write the firm's main optimality condition \eqref{systemprofits} for each $i\in N$ as:
\begin{equation*}
	b_i(\lambda)=\left(1+r \sigma^2\right) \alpha_i-\sum_j \alpha_j \sum_s\left(\lambda\sum_{\ell} \frac{\partial e_\ell}{\partial \alpha_{j}} g_{s\ell}\right)\left(\lambda\sum_m \frac{\partial e_m}{\partial \alpha_{i}} g_{sm}\right)+ \omega \sum_s\sum_\ell g_{s\ell} \frac{\partial e_\ell}{\partial \alpha_i}.
\end{equation*}
Notice that if $\omega=0$ we retrieve the baseline model in our main specification. Stacking up the $n$ equations and solving for incentives, we get the modified incentive rule.

\begin{proposition}[Optimal Incentives with More Flexible Externalities]
\label{prop:flexible_externalities}
Under Assumptions~\ref{assumption1},~\ref{spectralradius} and~\ref{omegabound}, there is a unique profit-maximizing incentive rule for any peer network $\mathbf{G}$ given by:
\begin{equation*}
\boldsymbol{\alpha}^*=\mathbf W\mathbf C'(\mathbf I-\omega \mathbf G')\mathbf 1.
\label{eq:flex_alpha}
\end{equation*}
\end{proposition}

Proposition~\ref{prop:flexible_externalities} shows that the structure of the optimal contract is the same as in Proposition~\ref{Optimal Contracts}, but with a modified centrality measure. Although the marginal-benefit term of the firm's problem remains $\mathbf{C}'\mathbf{1}$, it is helpful for interpretation to group the new $(\mathbf{I}-\omega\mathbf{G}')$ term with $\mathbf{C}'\mathbf{1}$ rather than with $\mathbf{W}$. This corresponds to viewing the firm as facing a ``net marginal benefit'' that is adjusted for the level costs workers impose on one another, while $\mathbf{W}$ continues to capture the cross-dependencies in the marginal cost of incentives. Under this interpretation, the incentive rule retains the same structure as in Proposition~\ref{Optimal Contracts} but with a modified centrality measure:
\begin{equation}\label{eq:bonacichomega}
\tilde{b}_i(\lambda, \omega)=1+\left(1-\frac{\omega}{\lambda}\right) \sum_s \sum_{q=1}^{\infty} \lambda^q \mathbf{G}_{s i}^q .
\end{equation}
The interpretation is straightforward. Workers' marginal incentives continue to be governed by $\lambda$, so---as in the baseline model---the $\lambda$-weighted paths from $i$ to $s$ determine how much influence worker $i$ has on output through $s$. However, these same effort spillovers now also generate level costs for the firm, captured by $\omega$. As $\omega$ increases relative to $\lambda$, the amplification generated by these paths becomes less valuable for the firm, and the effective centrality of worker $i$ correspondingly declines.

When $\omega=\lambda$, the two opposing forces exactly offset each other. The amplification generated by network interactions is therefore neutral from the firm's perspective, and any heterogeneity in incentives comes only from the marginal-cost interactions summarized by $\mathbf W$. The optimal contract therefore coincides with the incentive rule in Proposition~\ref{Optimal Contracts} except that it treats all workers as if they were equally central:
\begin{equation*}
	\boldsymbol{\alpha}=\mathbf{W1}.
\end{equation*}

When $\omega>\lambda$, however, workers with more outgoing paths generate a lower net marginal benefit for the firm. In particular, using \eqref{eq:bonacichomega}, it is easy to see that if 
\begin{equation*}
\omega>\lambda \frac{b_i}{b_i-1},
\end{equation*}
then worker $i$, with Bonacich centrality $b_i$, has a negative net marginal benefit. Importantly, this does not necessarily imply that the worker receives negative incentives. Since the optimal incentive vector $\boldsymbol{\alpha}$ is a linear combination of all workers' net marginal benefits, worker $i$ may still receive positive incentives $\left(\alpha_i>0\right)$ if they exert sufficient influence over workers with positive marginal benefits. Nevertheless, if $\omega$ becomes sufficiently large, workers may choose to exert zero effort and the firm optimally withdraws incentives altogether. Notice that efforts are now given by: 
\begin{equation*}
	\mathbf{e}^{*} = \mathbf{C}\boldsymbol{\alpha}^* = \mathbf{C}\mathbf W\mathbf C'(\mathbf I-\omega \mathbf G')\mathbf 1.
\end{equation*}

To rule out this degenerate case and focus on interior effort provision, we impose a bound on the size of the level effect. 

\renewcommand{\theassumption}{3}
\begin{assumption}\label{omegabound}
$ \mathbf{CWC}^{\prime}(\mathbf{I} - \omega\mathbf{G}') \mathbf{1} \gg 0$. 
 \end{assumption}
Assumption~\ref{omegabound} guarantees that the optimal contract induces strictly positive effort from all workers.

\medskip
A final implication of this extension concerns firm profits under the maximal contract (Proposition~\ref{Profits and Spectrum} in Section~\ref{Design}). In this case profits are given by:
$$
\Pi^*\left(\boldsymbol{\alpha}^*\right)=\frac{1}{2} \mathbf{e}^{\prime} \mathbf{1}-\omega \frac{1}{2} \mathbf{e}^{\prime} \mathbf{G}^{\prime} \mathbf{1}\leq \frac{1}{2} \mathbf{e}^{\prime} \mathbf{1},
$$
where the inequality is strict when $\omega>0$ and $\mathbf{G}$ is nonempty. Thus, unlike in the baseline model, profits are no longer equal to one half of total effort.
The intuition is straightforward. Workers' behavioral response to incentives is unchanged: effort continues to satisfy $\mathbf{e}=\mathbf{C} \boldsymbol{\alpha}$. By contrast, the firm's marginal cost of providing incentives is now affected by additional spillovers captured by $\omega$. As a result, the negative externalities created by effort interactions across workers are not fully internalized. These externalities therefore generate a wedge between total effort and firm profits, implying that the additional cost cannot be fully passed through to workers via incentives.

\subsection{Proofs of Section~\ref{SuppApp:MoreFlexibleExternalities}}

\begin{proof}[\textbf{Proof of Proposition~\ref{prop:flexible_externalities}}]
	We can write the participation (IR) constraints in matrix form as: 
	\begin{equation*}
		 \mathbf{1}' \boldsymbol{\beta} = \frac{1}{2}\mathbf{e}'\mathbf{e} 
         - \lambda \mathbf{e}'\mathbf{Ge} 
         - \mathbf{1}'\boldsymbol{\alpha}(\mathbf{1}'\mathbf{e}) 
         +\frac{1}{2}r\sigma^{2}\boldsymbol{\alpha}'\boldsymbol{\alpha} 
         +\omega (\mathbf{Ge})'\mathbf{1}.
	\end{equation*}
Substituting in the (IR) and (IC) constraints, the firm's problem is given by:
\begin{align*}
\max_{\boldsymbol{\alpha}} \: \mathbb{E}[\pi(\mathbf{e}\mid \boldsymbol{\alpha}, \boldsymbol{\beta})]  = \boldsymbol{\alpha}' \mathbf{C}'\mathbf{1} - \frac{1}{2}\boldsymbol{\alpha}'\mathbf{C}' \left(\mathbf{I} - 2\lambda\mathbf{G} \right) \mathbf{C}\boldsymbol{\alpha} - \frac{1}{2}\sigma^2 r\boldsymbol{\alpha}'\boldsymbol{\alpha}-\omega (\mathbf{GC}\boldsymbol{\alpha})'\mathbf{1}.
\end{align*}
Taking the first order condition and solving for $\boldsymbol{\alpha}$ we get:
\begin{align*}
 (\mathbf{C}'-\omega\mathbf{C}'\mathbf{G}')\boldsymbol{1} - \mathbf{C} '(\mathbf{I} - \lambda (\mathbf{G}+\mathbf{G}') )\mathbf{C}\boldsymbol{\alpha} - \sigma^2 r\boldsymbol{\alpha} = 0. 
\end{align*}
Solving for the optimal incentive vector we get: 
\begin{equation*}
	\boldsymbol{\alpha}^{*} = [\mathbf{C} '(\mathbf{I} - \lambda (\mathbf{G}+\mathbf{G}') )\mathbf{C} + \sigma^2 r\mathbf{I}]^{-1} (\mathbf{C}'-\omega\mathbf{C}'\mathbf{G}')\boldsymbol{1} = \mathbf{W} \mathbf{C}'(\mathbf{I}-\omega\mathbf{G}')\boldsymbol{1}. 
\end{equation*}
Finally, Assumption~\ref{omegabound} implies that $\mathbf{e}^{*} =\mathbf{C}\boldsymbol{\alpha}^{*} =\mathbf{CWC}'(\mathbf{I}-\omega\mathbf{G}')\mathbf{1} \gg 0.$ Thus, the interior best-response system $\mathbf{e}= \mathbf{C}\boldsymbol{\alpha}$ used above is valid at the optimum.
\end{proof}

\section{Incentives with Random Networks}\label{sec: random_networks}

In this section, we derive optimal incentive rules as a function of the parameters of a statistical process that generates a random network. First, we give the optimal allocation of incentives under the linear technology from the baseline model (Section \ref{baseline}) and, second, under the modular production from Section \ref{sec:Modular}.

\subsubsection{Random Networks and Linear Technology}

We now proceed by relaxing the full information assumption on the network structure. Imagine that, instead of knowing the realized network, the firm only knows the parameters of a data-generating process of random networks. Examples of such a statistical model include the stochastic block model (SBM) of random graphs and its special cases like the planted partition and  Erd\H{o}s-R\'{e}nyi models. We incorporate this possibility by considering that the optimal contract maps from a parameterized model of linking probabilities to a function of aggregate output. Employing a mean-field approximation, the firm could anticipate workers' behaviors as a function of the expected adjacency matrix $\mathbf{\bar{G}}$, rather than the realized network, and compute optimal contracts.\footnote{In fact, recent work using graphons as the underlying stochastic generative process of a network shows that Nash equilibria in finite sampled network games converge to Nash equilibria of the graphon game with high probability (\cite{parise2023graphon}). In such large and dense networks, it has also been shown that measures of centrality, such as Katz-Bonacich, of the expected adjacency matrix are very close to those of a realized network for stochastic block models (\cite{dasaratha2020random}) and, more generally, for graphons (\cite{avella2018random}) as stochastic generative models of network formation. Because large (and dense) firms are precisely the firms that are more likely to not have precise information about the peer network, recent work on random graph theory suggests that designing optimal contracts using a mean-field approximation could prove a valid approach. Moreover, as shown in \cite{parise2023graphon}, such an intervention in the space of the graphon is asymptotically optimal and tractable.} 

 A flexible family of random networks is the IRN model proposed by \cite{bollobas2007phase} where each worker has a specific ``type'' from a finite set and a worker of type $i$ is linked to a worker of type $j$ with independent probability $p_{ij}$. Although solving the optimal contract for a general class of models is beyond the scope of this paper, we consider a very natural special case that has been used extensively in the literature to capture \textit{homophily} in a parsimonious framework \citep{jackson2008social}. 

\begin{proposition}\label{pq_model}
     Consider a special case of the IRN model with two types in equally-sized groups (planted partition model). Let $p$ represent the within-type linking probability while $q$ represents the across-type linking probability. Then the optimal allocation of incentives is given by
     \[\alpha_i^{*}(p,q)= \frac{1-n\lambda \frac{p+q}{2}}{(1+r\sigma^{2})(1-n\lambda  \frac{p+q}{2} )^{2} - (n\lambda  \frac{p+q}{2})^{2}}, \quad \forall i \in N. \]
\end{proposition}

Consider how homophily affects optimal incentives in this generative \textit{planted partition random graph model} with two groups. Notice that incentive allocations in Proposition \ref{pq_model} depend on the expected degree of the network (parametrized by $p+q$), but not on the level of expected homophily (parameterized by $p/q$).

In Proposition \ref{pq_model} we only restrict the information available to the firm. We continue to assume that worker best-reply on the realized structure as in previous sections.\footnote{Previous work has analyzed the equilibrium of general network games when agents only have partial information on the interaction structure \citep{galeotti2010network, sundararajan2008local, fainmesser2016pricing}. In these models agents know some sufficient statistics of the network rather than the entire structure.} Notice that by comparing across different random network models we are able to vary the firm's level of informativeness. A very exciting line of future research follows the approach of \cite{fainmesser2016pricing} and considers how raising the information content (i.e., higher certainty on the true relationships within the firm) of the firm may affect the optimal contract, and how this may also depend on the information workers have on the network.\footnote{\cite{fainmesser2016pricing} consider a monopolist pricing a good with network effects and analyze how varying the information available to the monopolist affects the pricing strategy.} 

\subsubsection{Random Networks and Modular Production}

We now turn to investigate how homophily alters workers' optimal incentives as well as each module's productivity shares $\mu_{k}$ under the modular production in Section \ref{sec:Modular} when the network is generated according to the planted partition random graph model. 

\begin{figure}[t]
    \centering
    \includegraphics[width=0.45\linewidth]{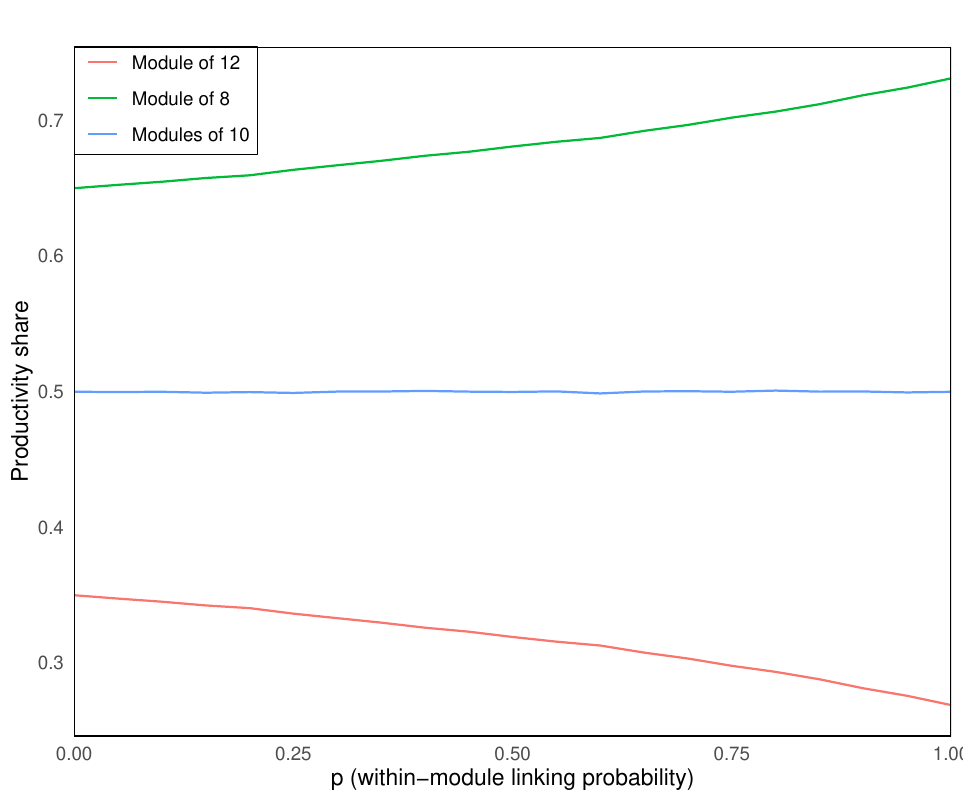}
    \includegraphics[width=0.45\linewidth]{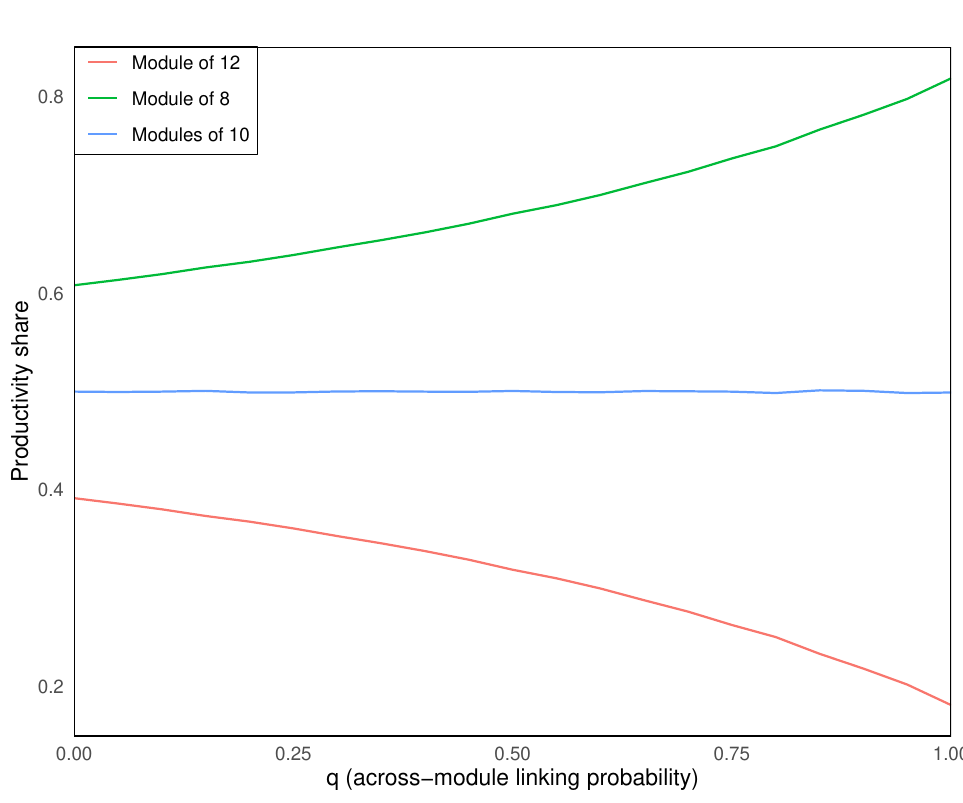}
    \caption{Modular share, $\mu_{k}$, for the planted partition model with linking parameters $p$ and $q$. Two equal-sized modules (blue line) and the asymmetric case with two modules of sizes 8 workers (green line) and 12 workers (red line) (parameters: $\lambda=0.025$, $\sigma^{2}=1$ and $r=1$).}
    \label{fig: pq_mus}
\end{figure}

We consider the example of a firm with $n=20$ workers and $K=2$ modules and look at two different cases: equal module sizes (i.e., 10 workers in each module) and asymmetric module sizes with 8 workers in one module and 12 workers in the other. In Figure \ref{fig: pq_mus}, for the two cases, we plot the mean productivity share of each module, $\mu_{k}$, across many random networks generated by fixing across-module linking probability $q=0.5$ and letting within-module linking probability $p$ increase (left) and by fixing $p=0.5$ and letting $q$ increase (right). This allows us to isolate the impact of each type of link on the productivity share. 

Consider first the case with symmetric modules. Because modules are of equal size, the module-specific productivity share is equal to $1/2$ for both modules. In this case, adding more links, no matter if they are within or across modules has no impact on these productivity shares, as illustrated by the flat lines in Figure \ref{fig: pq_mus}. 

Next, consider the case with asymmetric modules of 8 and 12 workers respectively. First, notice that even for $p=0$ or $q=0$, workers in the small module are assigned a larger productivity share. That is because these workers have higher costs of effort relative to the workers in the larger module. In fact, without peer effects, i.e., $\lambda=0$, the productivity shares are exactly 0.6 for the small module and 0.4 for the large one. However, as $p$ or $q$ increases, this gap widens. This is because the additional links always generate higher complementarities for workers in the large module, no matter if the links are within or across modules, and, hence, workers in the small module have even higher effort costs relative to those in the large module. Lastly, notice that this effect is larger for across-modules links than for within-modules links. 

\subsection{Proofs of Section~\ref{sec: random_networks}}

\begin{proof}[\textbf{Proof of Proposition \ref{pq_model}}]

Consider two groups of size $m$ such that $n=2m$. Let $\mathbf{J}_{m}$ be an $m\times m$ matrix of ones. Then, the expected adjacency matrix for the IRN model in this case is given by:
\begin{equation*}
    \mathbf{\bar{G}} = \begin{pmatrix}
        p\mathbf{J}_{m} & q \mathbf{J}_{m}\\
        q \mathbf{J}_{m} & p\mathbf{J}_{m}
    \end{pmatrix}.
\end{equation*}
Notice that the row sum of $\mathbf{\bar{G}}$ is equal to $m(p+q)$. Moreover, the row sum of $\mathbf{I} - \lambda\mathbf{\bar{G}}$ is $1-\lambda m(p+q)$. Because the row and column sums of $\mathbf{I} - \lambda\mathbf{\bar{G}}$ are all the same, we have that the row and column sums of $\mathbf{\bar{C}} = (\mathbf{I} - \lambda\mathbf{\bar{G}})^{-1}$ are equal to the reciprocal, i.e., $1/(1-\lambda m(p+q))$. This means we can write:
\begin{equation*}
\mathbf{\bar{C}}'\mathbf{1} = \left( \frac{1}{1-n\lambda \frac{p+q}{2}} \right) \mathbf{1}.
\end{equation*}
The firm anticipates workers' behaviors as a function of the expected adjacency matrix $\mathbf{\bar{G}}$, rather than the realized network, and computes optimal contracts. Thus, we have that $\boldsymbol{\alpha} = \alpha_{i} \mathbf{1} $, i.e., workers are symmetric. Then, the firm solves:
\begin{align*}
\max_{\alpha_i} \quad & \alpha_i \mathbf{1}' \mathbf{\bar{C}}' \mathbf{1} - \frac{1}{2} \alpha_i^2 \mathbf{1}' \mathbf{\bar{C}}' \mathbf{\bar{C}} \mathbf{1} + \lambda \alpha_i^2 \mathbf{1}' \mathbf{\bar{C}}' \mathbf{\bar{G}} \mathbf{\bar{C}} \mathbf{1} - \frac{r\sigma^2}{2} \alpha_i^2 n.
\end{align*}
Defining $\chi = 1/(1-n\lambda ((p+q)/2))$, substituting in for $\mathbf{\bar{C}}$, and using the fact that $\mathbf{1}' \mathbf{\bar{G}} \mathbf{1} = \frac{p+q}{2} n^2$, the firm's problem can be rewritten as:
\begin{align*}
\max_{\alpha_i} \quad & \alpha_{i}\chi - \frac{1}{2}\alpha_{i}^{2}(\chi^{2}(1-\lambda n (p+q))+r\sigma^{2}).
\end{align*}
Taking the first order condition we get:
\begin{align*}
\chi = \alpha_{i}(\chi^{2}(1-\lambda n (p+q))+r\sigma^{2}) \implies \alpha_{i} = \frac{\chi}{(\chi^{2}(1-\lambda n (p+q))+r\sigma^{2})}
\end{align*}
Finally, going back to the original notation and rearranging we get:
\begin{align*}
\alpha_i^{*} (p,q) &= \frac{1-\lambda n \frac{p+q}{2}}{(1+r\sigma^{2})(1-\lambda n \frac{p+q}{2})^{2} - (\lambda n \frac{p+q}{2})^{2}}. 
\end{align*}
\end{proof}

\section{Optimal Incentives under Modular  Production}\label{App:Modular_Production}

In this section, we extend Proposition~\ref{Modular Production} beyond the parameter range in which its unconstrained solution is implementable as a Nash equilibrium. We first allow for different modules to produce different levels of output and characterize optimal incentives when the set of modules overproducing is endogenous. We then impose equal module outputs again and characterize the firm's problem when workers' marginal costs must be nonnegative. 


\subsection{Unequal Module Outputs}\label{App:Modular_Production_Uneq}

In the main text, we argue that if peer effects are too strong, some Nash equilibria might feature modules generating different levels of total effort. The exact range over which this property holds depends on the network and the assignment of workers to modules. A simple and convenient sufficient condition is that every Nash equilibrium will feature all modules producing the same level $\hat{e}$ if
\begin{equation}
    \lambda
    \max_{k}
    \left\{
        \sum_{i\in k}\sum_j g_{ij}
    \right\}
    <1.
    \label{SuffCond_eqoutput}
\end{equation}
To see why, suppose instead that an equilibrium has unequal module outputs and let $k$ be a module with maximal total effort. Since $k$ is strictly non-bottleneck, each worker $i\in k$ must exert her cost-minimizing effort, $e_i=\lambda\sum_j g_{ij}e_j$. Summing across workers in $k$ and using that $e_j$ cannot exceed the total effort of the maximal module yields $1 \leq  \lambda\sum_{i\in k}\sum_j g_{ij},$ contradicting \eqref{SuffCond_eqoutput}. This is only a sufficient condition and, in some cases, even for $\lambda$ well above this bound, one can prove that the only Nash equilibria feature all modules producing $\hat{e}$. We now solve the firm's problem in the remaining cases.

Call the weakest-performing modules bottleneck modules. Let \(S\) denote an arbitrary candidate set of bottleneck modules and let \(B=|S|\). Conditional on \(S\), we can relabel modules and workers without loss of generality so that modules \(1,\ldots,B\) are the candidate bottleneck modules and modules \(B+1,\ldots,K\) are non-bottlenecks, with workers belonging to bottleneck modules ordered first. Let \(n_B\) denote the number of such workers. We can then write
\begin{equation*}
    \mathbf M=
    \begin{pmatrix}
        \mathbf M_B\\
        \mathbf M_{NB}
    \end{pmatrix},
\end{equation*}
where \(\mathbf M_B\) contains the rows corresponding to bottleneck modules. By definition, $\mathbf{M}\mathbf{e} \geq \hat{e}\mathbf{1}$. Conditional on S, this requires: $\mathbf{M}_{B}\mathbf{e} = \hat{e} \mathbf{1}_{B}$ and $\mathbf{M}_{NB}\mathbf{e} >  \hat{e} \mathbf{1}_{K-B}$. The inequality is an active-set consistency condition and can be verified ex post. As discussed in the main text, every Nash equilibrium must satisfy
\begin{equation*}
    \boldsymbol{\alpha} =  (\mathbf I-\lambda\mathbf G)\mathbf e\geq 0, \quad \mathbf{e}\geq 0.
\end{equation*}
Moreover, recall that workers in non-bottleneck modules receive zero incentives at the firm's optimum. Thus, conditional on $S$, the $n\times 1$ vector of incentives can be written as $\boldsymbol{\alpha}' =
\left( \boldsymbol{\alpha}_B',\,\, \mathbf 0_{n-n_B}' \right)
$, where $\boldsymbol{\alpha}_{B}\in \mathbb{R}^{n_{B}}$. Since \(\mathbf e=\mathbf C\boldsymbol{\alpha}\), let \(\mathbf C_B\) denote the \(n\times n_B\) matrix formed by the columns of \(\mathbf C\) corresponding to workers in bottleneck modules. It follows that $ \mathbf e=\mathbf C_B\boldsymbol{\alpha}_B$.

The firm's problem, conditional on a candidate configuration with $B$ bottleneck modules, can be written as: 
\begin{align*}
        \max_{\boldsymbol{\alpha}_{B}, \hat{e}} \: \mathbb{E}[\pi(\mathbf{e}\mid \boldsymbol{\alpha}, \boldsymbol{\beta})] &= \left( \hat{e} - \frac{1}{2} \mathbf{e}' (\mathbf{I}-2\lambda \mathbf{G})\mathbf{e} - \frac{\sigma^2 r}{2} 
        \boldsymbol{\alpha}_{B}' \boldsymbol{\alpha}_{B} \right) \\ 
         \text{subject to: \:}  \mathbf{e}&=\mathbf{C}_{B}\boldsymbol{\alpha}_{B}\geq 0, \quad \mathbf{e}\geq 0, \quad \mathbf{M_{B}}\mathbf{e}=\hat{e}\boldsymbol{1}_B, \quad \mathbf{M}_{NB}\mathbf{e} >  \hat{e} \mathbf{1}_{K-B}.
\end{align*}

\begin{proposition}[Modular Production with Unequal Outputs]
\label{propApp:Modular_uneqout}
Suppose Assumptions~\ref{assumption1} and~\ref{spectralradius} hold. Conditional on a candidate set $S$ of bottleneck modules, the optimal allocation of incentives is
\begin{equation*}
    \boldsymbol{\alpha}^{*} = \begin{pmatrix}
    \mathbf{W}_{B}\mathbf{C}_{B}'\boldsymbol{\mu}_{B} \\
    \mathbf{0}_{n-n_{B}}
    \end{pmatrix}
\end{equation*}
where $\mathbf{W}_{B} =  \left[\mathbf C_B' \bigl(\mathbf I-\lambda(\mathbf G+\mathbf G')\bigr) \mathbf C_B +r\sigma^2\mathbf I_{n_B}
    \right]^{-1}$, $ \boldsymbol{\mu}_{B} = \mathbf{M}_{B}'\mathbf{H}_{B}^{-1}\boldsymbol{1}_{B} / (\boldsymbol{1}_{B}'\mathbf{H}_{B}^{-1}\boldsymbol{1}_{B})$ is an $n\times 1$ vector of $B$ module-specific weights, with zero weight assigned to workers in non-bottlenecks modules, and $\mathbf{H}_{B}
    = \mathbf{M}_{B}\mathbf{C}_{B}\mathbf{W}_{B}\mathbf{C}_{B}' \mathbf{M}_{B}'$.
\end{proposition}

The incentive rule has the same structure as in Proposition~\ref{Modular Production}, but applies only to workers belonging to bottleneck modules. Workers in non-bottleneck modules receive no direct incentives: their effort does not affect output at the margin and is instead sustained by peers' efforts elsewhere in the network. Any positive incentives would only increase the risk premium paid by the firm. For everyone else, incentives are determined as in Proposition \ref{Modular Production}, but now the shadow value, $\mu_i$, must be $0$ for workers in non-bottleneck modules, and otherwise continues to reflect the marginal cost of raising (bottleneck) module $k$’s performance relative to other modules.

When all modules are bottlenecks, \(B=K\), \(\mathbf C_B=\mathbf C\), and \(\mathbf M_B=\mathbf M\), so the characterization collapses to Proposition~\ref{Modular Production}. When peer effects are sufficiently strong, however, a worker may optimally exert more effort than required for production even in the absence of direct incentives. The firm can then exploit this ``free'' effort by allowing that worker's module to overproduce and concentrating incentives on the bottleneck modules, as the following example shows.

\begin{example}
    Consider the firm with 3 workers in singleton modules in Figure~\ref{fig:3singletonmodules}. Worker 1 receives help from both workers 2 and 3. Fix $\lambda=3/4$ and $r\sigma^{2}=1$. In this example, condition \eqref{SuffCond_eqoutput} requires \(\lambda<1/2\), and is therefore violated. In fact, equal module outputs cannot constitute a positive-output Nash equilibrium. If \(e_1=e_2=e_3=\bar{e}\), worker \(1\)'s marginal cost is
\[
e_1-\lambda(e_2+e_3)
=
\bar e-\frac34(2\bar{e})
=
-\frac12\bar{e}<0,
\]
so worker \(1\) would profitably increase effort. Hence, any positive-output equilibrium must have unequal module outputs. Moreover, only module \(1\) can overproduce: workers \(2\) and \(3\) receive no peer assistance, so a strictly non-bottleneck worker in either module would optimally exert zero effort. Thus, modules \(2\) and \(3\) are bottlenecks, \(e_2=e_3=\hat e\), while worker \(1\), who receives no direct incentives, chooses
\[
e_1
=
\frac34(e_2+e_3)
=
\frac32\hat e.
\]

\begin{figure}[t]
    \centering
    \begin{tikzpicture}[
        scale=.8,
        every node/.style={circle, draw=blue, fill=lightblue, thick, minimum size=7mm, inner sep=0.2pt},
        every edge/.style={draw, thick, ->, >=stealth}
    ]

    \definecolor{lightblue}{rgb}{0.68, 0.85, 0.9}

    \fill[gray,opacity=0.1, rounded corners=0.4cm] (-.65,.65) rectangle (.65,-.65);
    \draw[thick, gray!50, rounded corners=0.4cm] (-.65,.65) rectangle (.65,-.65);

    \fill[gray,opacity=0.1, rounded corners=0.4cm] (-3.65,.65) rectangle (-2.35,-.65);
    \draw[thick, gray!50, rounded corners=0.4cm] (-3.65,.65) rectangle (-2.35,-.65);

    \fill[gray,opacity=0.1, rounded corners=0.4cm] (2.35,.65) rectangle (3.65,-.65);
    \draw[thick, gray!50, rounded corners=0.4cm] (2.35,.65) rectangle (3.65,-.65);


    \node (1) at (0,0) [draw=blue!50, fill=blue!10] {$1$};
    \node (2) at (-3,0) [draw=blue!50, fill=blue!10] {$2$};
    \node (3) at (3,0) [draw=blue!50, fill=blue!10] {$3$};

    \path (2) edge (1);
    \path (3) edge (1);

    \end{tikzpicture}
    \caption{An organization with three workers and individual modules.}
    \label{fig:3singletonmodules}
\end{figure}

The equilibrium effort profile is therefore
\[
\mathbf e
=
\left(
\frac32\hat e,\,
\hat e,\,
\hat e
\right),
\qquad
\boldsymbol{\alpha}
=
\left(
0,\,
\hat e,\,
\hat e
\right).
\]

Substituting these expressions into the firm's objective gives
\[
\mathbb E[\pi]
=
\hat e-\frac78\hat e^2,
\]
which is maximized at \(\hat e^*=4/7\). Hence,
\[
\mathbf e^*
=
\left(
\frac67,\,
\frac47,\,
\frac47
\right),
\qquad
\boldsymbol{\alpha}^*
=
\left(
0,\,
\frac47,\,
\frac47
\right).
\]
Worker \(1\) therefore produces strictly more than the bottleneck level despite receiving no direct incentives: her excess effort is sustained entirely by peer effects from workers \(2\) and \(3\).
\end{example}


\subsection{Generalized Modular Production}
\label{App:Modular_Production_eq}

Proposition~\ref{Modular Production} remains valid whenever the firm's preferred equal-output allocation satisfies $(\mathbf{I}-\lambda \mathbf{G})\mathbf{e}\geq 0$. That is, it remains valid when marginal effort costs at the allocation are nonnegative, so that no worker benefits from increasing effort beyond the prescribed level. This is always the case for sufficiently small peer effects: when \(\lambda=0\), the solution in Proposition~\ref{Modular Production} assigns strictly positive incentives and marginal costs, and the same remains true in a neighborhood of \(\lambda=0\) by continuity. In this region, the additional constraint $(\mathbf I-\lambda\mathbf G)\mathbf e\geq\mathbf 0$ is slack at the unconstrained optimum. Thus, the incentive rule in Proposition~\ref{Modular Production} and that of the constrained problem have the same solution. The exact range over which this condition holds depends on the network and the assignment of workers to modules.

As peer effects become stronger, this need no longer hold. The firm's unconstrained preferred allocation may assign some worker less effort than she would choose given the assistance received from her peers, so that $e_i-\lambda\sum_j g_{ij}e_j<0$. Such an allocation cannot be a Nash equilibrium, since worker \(i\) would profitably increase effort even though the increase does not raise output or compensation. The firm must therefore restrict attention to allocations for which workers' marginal costs are nonnegative.

Consider the firm's problem imposing the additional restriction that workers' marginal costs are nonnegative. For simplicity, we assume that peer effects are moderate enough that any Nash equilibrium must feature equal module totals (i.e. $\mathbf{M}\mathbf{e}=\hat{e}\mathbf{1}_K$). The firm's problem now is: 
\begin{align*}
    \max_{\mathbf e,\hat e}\quad
    & \hat e
    -\frac{1}{2}\mathbf e'
    \left(\mathbf I-2\lambda\mathbf G\right)\mathbf e
    -\frac{r\sigma^2}{2}
    \mathbf e'(\mathbf I-\lambda\mathbf G)'
    (\mathbf I-\lambda\mathbf G)\mathbf e
    \label{firm_problem_MC}\\
    \text{s.t.}\quad
    & \mathbf M\mathbf e=\hat e\mathbf 1_K, \quad  (\mathbf I-\lambda\mathbf G)\mathbf e\geq \mathbf 0, \quad  \mathbf e\geq\mathbf 0. \nonumber
\end{align*}
Let $\boldsymbol{\eta}$ be the $n\times 1$ vector of multipliers associated to the nonnegativity constraint $(\mathbf I-\lambda\mathbf G)\mathbf e\geq \mathbf 0$.  

\begin{proposition}[Generalized Modular Production]
\label{propApp:Modular_eqout_MCconstraint}
Suppose Assumptions~\ref{assumption1} and~\ref{spectralradius} hold. The optimal allocation of incentives is
\begin{equation*}
    \boldsymbol{\alpha}^{*}  =
     \boldsymbol{\alpha}^{Prop.4}+\mathbf{WC}'\boldsymbol{\Delta}(\boldsymbol{\eta}),
    \label{eq:modular_incentives}
\end{equation*}
where \(\boldsymbol{\Delta}(\boldsymbol{\eta})\) is defined in \eqref{A_DeltaEta} and  $\boldsymbol{\alpha}^{Prop.4}$ is the incentive rule from Proposition~\ref{Modular Production}.
\end{proposition}

The term \(\boldsymbol{\Delta}(\boldsymbol{\eta})\) captures the wedge between the incentive rule in Proposition~\ref{Modular Production} and the rule that respects these additional equilibrium constraints. This wedge reflects both the direct effect of the binding marginal-cost constraints and the associated adjustment in the module-specific shadow values. Since \(\boldsymbol{\Delta}(\mathbf 0)=\mathbf 0\), when the additional constraints are slack we recover \(\boldsymbol{\alpha}^{\operatorname{Prop\,4}}\), as expected. 

The incentive rule in Proposition~\ref{propApp:Modular_eqout_MCconstraint} generalizes the rule in Proposition~\ref{Modular Production} to peer effects large enough that, if unconstrained, the firm would select an effort allocation that is not a Nash equilibrium. The additional term $\mathbf{WC}'\boldsymbol{\Delta}(\boldsymbol{\eta})$ shows how the firm must adjust the allocation of incentives. Workers receiving strong peer influence must exert sufficient effort in equilibrium, or else they would be willing to raise effort for free. We now clarify this with an example.

\begin{example}
Consider the two-module organization in Figure~\ref{fig:2equaloutputmodules3workers}. Fix $\lambda=3/4$ and $r\sigma^{2}=1$. Here $\max_{k}\left\{\sum_{i\in k}\sum_{j} g_{ij} \right\}=1$, so condition~\eqref{SuffCond_eqoutput} is satisfied since \(\lambda<1 \), and every Nash equilibrium must feature equal module outputs. Nevertheless, the unconstrained solution in Proposition~\ref{Modular Production} is not implementable as a Nash equilibrium. Normalizing efforts by \(\hat e\), Proposition~\ref{Modular Production} yields
\[
\frac{\mathbf e^{\operatorname{Prop.\,4}}}{\hat e}
=
\left(
\frac{32}{73},
\frac{41}{73},
1
\right)
\approx
(0.44,0.56,1).
\]
Worker \(2\) receives assistance from worker \(3\), so her marginal cost is
\[
\frac{e_2-\lambda e_3}{\hat e}
=
\frac{41}{73}-\frac34
<0.
\]
Hence, worker \(2\) would profitably increase effort: the firm would like to assign her less effort than she is willing to provide given the assistance received from worker \(3\).

\begin{figure}[t]
    \centering
    \begin{tikzpicture}[
        scale=.8,
        every node/.style={circle, draw=blue, fill=lightblue, thick, minimum size=7mm, inner sep=0.2pt},
        every edge/.style={draw, thick, ->, >=stealth}
    ]

    \definecolor{lightblue}{rgb}{0.68, 0.85, 0.9}

    \fill[gray,opacity=0.1, rounded corners=0.4cm] (-.65,.65) rectangle (.65,-.65);
    \draw[thick, gray!50, rounded corners=0.4cm] (-.65,.65) rectangle (.65,-.65);

    \fill[gray,opacity=0.1, rounded corners=0.4cm] (-1.65,-1.35) rectangle (1.65,-2.65);
    \draw[thick, gray!50, rounded corners=0.4cm] (-1.65,-1.35) rectangle (1.65,-2.65);


    \node (1) at (0,0) [draw=blue!50, fill=blue!10] {$3$};
    \node (2) at (-1,-2) [draw=blue!50, fill=blue!10] {$1$};
    \node (3) at (1,-2) [draw=blue!50, fill=blue!10] {$2$};

    \path (1) edge (3);
    \path (2) edge (1);

    \end{tikzpicture}
    \caption{A two-module organization with three workers.}
    \label{fig:2equaloutputmodules3workers}
\end{figure}

Imposing the missing nonnegative marginal-cost constraint shifts effort within the first module. The constrained solution, again normalized by \(\hat e\), is
\[
\frac{\mathbf e^*}{\hat e}
=
\left(
\frac14,\frac34,1
\right).
\]
Worker \(2\)'s constraint now binds,
\[
e_2-\lambda e_3=0,
\]
so her optimal incentive is \(\alpha_2^*=0\). The corresponding normalized incentive vector is
\[
\frac{\boldsymbol{\alpha}^*}{\hat e}
=
\left(
\frac14,
0,
\frac{13}{16}
\right).
\]
Thus, relative to Proposition~\ref{Modular Production}, the binding constraint for worker \(2\), captured by \(\eta_2>0\), shifts effort toward worker \(2\) and away from worker \(1\). Since module \(1\) must continue to produce the same total output, the increase in \(e_2\) is accompanied by a reduction in \(e_1\). The resulting reallocation raises worker \(2\)'s effort to her cost-minimizing level, \(e_2=\lambda e_3\), eliminating her incentive to deviate upward and restoring a Nash equilibrium.
\end{example}



\subsection{Proofs of Section~\ref{App:Modular_Production}}

\begin{proof}[\textbf{Proof of Proposition \ref{propApp:Modular_uneqout}}]
We take the same approach as in Proposition~\ref{Modular Production}: we temporarily omit the nonnegativity constraints and verify ex-post that the solution satisfies them. Substituting in $\mathbf{e}=\mathbf{C}_{B}\boldsymbol{\alpha}_{B}$ and noting that $2\mathbf{e}'\mathbf{Ge} = \mathbf{e}'(\mathbf{G}+\mathbf{G}')\mathbf{e}$ we obtain: 
\begin{align*}
        \max_{\{\boldsymbol{\alpha}_{B}, \hat{e}\}} \:  & \left( \hat{e} - 
        \frac{1}{2} \boldsymbol{\alpha}_{B}' \overbrace{\left[  \mathbf{C}_{B}'(\mathbf{I}_{n}-\lambda( \mathbf{G}+\mathbf{G}'))\mathbf{C}_{B}+ \sigma^2 r \mathbf{I}_{n_{B}} \right]}^{\boldsymbol{\Sigma}_{B}} \boldsymbol{\alpha}_{B}
        \right)  = \hat{e} - \frac{1}{2}\boldsymbol{\alpha}_{B}'\boldsymbol{\Sigma}_{B}\boldsymbol{\alpha}_{B}\\ 
         & \text{subject to: \:}  \mathbf{M}_{B}\mathbf{C}_B\boldsymbol{\alpha}_B =\hat{e}\boldsymbol{1}_B. 
\end{align*}

The equalities \(\mathbf{M}_B\mathbf{C}_B\boldsymbol{\alpha}_B=\hat e\mathbf{1}_B\) correspond to the binding constraints of the underlying requirement \(\mathbf{M}\mathbf{e} \geq \hat e\mathbf{1}_K\). Hence, for \(S\) to constitute a valid active set, the associated multipliers must satisfy \(\boldsymbol{\mu}_B\geq\mathbf{0}\). We first solve the equality-constrained problem and verify this dual-feasibility condition ex post, together with the strict slackness of the constraints associated with non-bottleneck modules.

Considering the $B \times 1$ vector of Lagrangian multipliers $\tilde{\boldsymbol{\mu}}_{B}$, the Lagrangian of the above problem can be expressed as:
\[
\mathcal{L}(\boldsymbol{\alpha}_{B} , \hat{e}, \tilde{\boldsymbol{\mu}}_{B}) = \hat{e} - \frac{1}{2}\boldsymbol{\alpha}_{B}'\boldsymbol{\Sigma}_{B}\boldsymbol{\alpha}_{B} - \tilde{\boldsymbol{\mu}}_{B}' (\hat{e}\boldsymbol{1_B} - \mathbf{M}_{B}\mathbf{C}_B\boldsymbol{\alpha}_B).
\]
We have the following system of first order conditions:
\begin{align}
\frac{\partial \mathcal{L}}{\partial\boldsymbol{\alpha}_{B}} &= -\boldsymbol{\Sigma}_{B}\boldsymbol{\alpha}_{B} + \mathbf{C}_{B}'\mathbf{M}_{B}' \tilde{\boldsymbol{\mu}}_{B}  = \boldsymbol{0_{n_{B}}},  \label{A_ModularuneqFOC1} \\
\frac{\partial \mathcal{L}}{\partial \tilde{\boldsymbol{\mu}}_{B}} &= \mathbf{M}_{B}\mathbf{C}_B\boldsymbol{\alpha}_B - \hat{e}\boldsymbol{1_B}  = \boldsymbol{0_B}, \label{A_ModularuneqFOC2} \\
\frac{\partial \mathcal{L}}{\partial \hat{e}} &= 1-\boldsymbol{1_B}' \tilde{\boldsymbol{\mu}}_{B} = 0.\label{A_ModularuneqFOC3} 
\end{align}

Notice that the Hessian with respect to $\boldsymbol{\alpha}_{B}$ is $-\boldsymbol{\Sigma}_{B}$. Assumption 2 implies that $\boldsymbol{\Sigma}_{B}\succ 0$, and therefore the firm's problem is strictly concave in $\boldsymbol{\alpha}_{B}$.\footnote{Using $\mathbf{C}_{B}'(\mathbf{I}_{n}-\lambda\mathbf{G})'(\mathbf{I}_{n}-\lambda\mathbf{G})\mathbf{C}_{B}=\mathbf{I}_{n_{B}}$, we can write $\boldsymbol{\Sigma}_{B} = (1+r\sigma^{2})\mathbf{I}_{n_{B}} - \lambda^{2}(\mathbf{G}\mathbf{C}_{B})'(\mathbf{G}\mathbf{C}_{B}).$ Since $\mathbf{C}_{B}$ contains the columns of $\mathbf{C}$ corresponding to workers in bottleneck modules, $(\mathbf{G}\mathbf{C}_{B})'(\mathbf{G}\mathbf{C}_{B})$ is a principal submatrix of $(\mathbf{G}\mathbf{C})'(\mathbf{G}\mathbf{C})$. By the Poincar\'{e} separation theorem, its largest eigenvalue is weakly smaller. Hence $\boldsymbol{\Sigma}_{B}\succ 0$.}
We can thus define $\mathbf{W}_{B}:=\boldsymbol{\Sigma}_{B}^{-1}$ as the analogue of $\mathbf{W}$ in Proposition~\ref{Optimal Contracts}.

From \eqref{A_ModularuneqFOC1} we obtain:
\begin{equation}
    \boldsymbol{\alpha}_{B}
    =
    \mathbf{W}_{B}\mathbf{C}_{B}'\mathbf{M}_{B}'\tilde{\boldsymbol{\mu}}_{B}.
    \label{A_ModularAlpha}
\end{equation}

Let $\mathbf{H}_{B} := \mathbf{M}_{B}\mathbf{C}_{B}\mathbf{W}_{B}\mathbf{C}_{B}'\mathbf{M}_{B}'.$ Since $\mathbf{W}_{B}\succ 0$ and $\mathbf{M}_{B}\mathbf{C}_{B}$ has full row rank $B$, 
$\mathbf{H}_{B}\succ 0$ and therefore $\mathbf{H}_{B}^{-1}$ exists. Plugging 
\eqref{A_ModularAlpha} into \eqref{A_ModularuneqFOC2}, we obtain:
\begin{align*}
    \mathbf{M}_{B}\mathbf{C}_{B}
    \mathbf{W}_{B}\mathbf{C}_{B}'\mathbf{M}_{B}'
    \tilde{\boldsymbol{\mu}}_{B} =
    \hat{e}\boldsymbol{1}_{B} \quad 
    \Longrightarrow \qquad
    \tilde{\boldsymbol{\mu}}_{B} =
    \hat{e}\mathbf{H}_{B}^{-1}\boldsymbol{1}_{B}.
\end{align*}

Using \eqref{A_ModularuneqFOC3}, we can then solve for $\hat{e}$:
\begin{align*}
    \boldsymbol{1}_{B}'\tilde{\boldsymbol{\mu}}_{B}
    =
    \hat{e}\boldsymbol{1}_{B}'
    \mathbf{H}_{B}^{-1}\boldsymbol{1}_{B}
    =1 
    \Longrightarrow \qquad
    \hat{e}^{*}
    =
    \frac{1}
    {\boldsymbol{1}_{B}'\mathbf{H}_{B}^{-1}\boldsymbol{1}_{B}}.
\end{align*}

It follows that 
\begin{equation}
    \tilde{\boldsymbol{\mu}}_{B}
    =
    \frac{
        \mathbf{H}_{B}^{-1}\boldsymbol{1}_{B}
    }{
        \boldsymbol{1}_{B}'\mathbf{H}_{B}^{-1}\boldsymbol{1}_{B}
    }.
    \label{A_ModularMu}
\end{equation}

Finally, substituting \eqref{A_ModularMu} into \eqref{A_ModularAlpha} and defining the $n\times 1$ vector of worker-specific module multipliers as $\boldsymbol{\mu}_{B}=\mathbf{M}_{B}'\tilde{\boldsymbol{\mu}}_{B}$ we obtain:
\begin{equation}
    \boldsymbol{\alpha}_{B}^{*}
    =
        \mathbf{W}_{B}\mathbf{C}_{B}'\boldsymbol{\mu}_{B}.
    \label{A_ModularAlphaStar}
\end{equation}
Workers in non-bottleneck modules receive zero incentives, so the full incentive vector is
\begin{equation*}
    \boldsymbol{\alpha}^{*}
    =
    \begin{pmatrix}
        \mathbf{W}_{B}\mathbf{C}_{B}'\boldsymbol{\mu}_{B} \\
        \mathbf{0}_{n-n_{B}}
    \end{pmatrix},
    \qquad
    \mathbf e^*
    =
    \mathbf C_B\mathbf W_B\mathbf C_B'\boldsymbol{\mu}_{B}.
\end{equation*}

Next, for the candidate set \(S\) to be valid for the original problem, it must satisfy:
\begin{equation*}
    \mathbf M_{NB}\mathbf e^*  > \hat e^*\mathbf 1_{K-B},  \qquad  \tilde{\boldsymbol{\mu}}_B\geq0.
\end{equation*}
Under these conditions, and since by Assumption~\ref{spectralradius} \(\mathbf W_B\) and \(\mathbf C_B\) are entrywise nonnegative, the candidate solution also satisfies
\begin{equation*}
    \boldsymbol{\alpha}_B^*  =  \mathbf W_B\mathbf C_B'\boldsymbol{\mu}_B  \geq0,
    \qquad
    \mathbf e^*  = \mathbf C_B\boldsymbol{\alpha}_B^* \geq0.
\end{equation*}
Hence, the candidate solution is feasible for the original constrained problem.
\end{proof}


\begin{proof}[\textbf{Proof of Proposition \ref{propApp:Modular_eqout_MCconstraint}}]
We take the same approach as in Proposition~\ref{Optimal Contracts}: we temporarily omit the nonnegativity of effort condition and verify ex-post that the solution satisfies it. Let $\boldsymbol{\mu}_K$ denote the $K\times 1$ vector of multipliers associated with the equal-output constraint and let $\boldsymbol{\eta}\geq\mathbf 0$ denote the $n\times 1$ vector of multipliers associated with the workers' nonnegative marginal-cost constraints. The Lagrangian is given by
\begin{align*}
    \mathcal L = \, & \hat e 
    -\frac{1}{2}\mathbf e'
    (\mathbf I-2\lambda\mathbf G)\mathbf e
    -\frac{r\sigma^2}{2}
    \mathbf e'(\mathbf I-\lambda\mathbf G)'
    (\mathbf I-\lambda\mathbf G)\mathbf e
    +\boldsymbol{\mu}_K'
    \left(\mathbf M\mathbf e-\hat e\mathbf 1_K\right)
    +\boldsymbol{\eta}'
    (\mathbf I-\lambda\mathbf G)\mathbf e.
\end{align*}

Noting that \(2\mathbf e'\mathbf G\mathbf e =\mathbf e'(\mathbf G+\mathbf G')\mathbf e\), the KKT conditions are given by:
\begin{align}
\frac{\partial \mathcal{L}}{\partial\mathbf{e}} &= -\big[
    \mathbf I-\lambda(\mathbf G+\mathbf G')
    +r\sigma^2(\mathbf I-\lambda\mathbf G)'
    (\mathbf I-\lambda\mathbf G)
    \big]\mathbf e
    +
    \mathbf M'\boldsymbol{\mu}_K
    +(\mathbf I-\lambda\mathbf G)'\boldsymbol{\eta} =0,
    \notag\\
\frac{\partial \mathcal{L}}{\partial \hat{e}} &= \mathbf 1_K'\boldsymbol{\mu}_K - 1=0.
    \notag \\
\frac{\partial \mathcal{L}}{\partial \boldsymbol{\mu}_{K}} &= \mathbf M\mathbf e - \hat e\mathbf 1_K =0, \notag\\
    & (\mathbf I-\lambda\mathbf G)\mathbf e\geq\mathbf 0,\quad 
    \boldsymbol{\eta}\geq\mathbf 0,\quad 
    \eta_i
    \left[(\mathbf I-\lambda\mathbf G)\mathbf e\right]_i=0
    \qquad\text{for all }i.
    \label{CS_MC}
\end{align}

Notice that the Hessian with respect to $\mathbf{e}$ is $-[ \mathbf I-\lambda(\mathbf G+\mathbf G')
    +r\sigma^2(\mathbf I-\lambda\mathbf G)' (\mathbf I-\lambda\mathbf G)]$. Assumption 2 implies that this Hessian is negative definite and therefore the firm's objective is strictly concave in \(\mathbf e\) and these conditions characterize the solution.

Following the same steps as in the proof of Proposition~\ref{Modular Production}, the KKT conditions imply
\begin{align}
    \mathbf e
    &=\mathbf C\mathbf W\mathbf C'\mathbf M'\boldsymbol{\mu}_K
      +\mathbf C\mathbf W\boldsymbol{\eta}, \label{effort_MC}\\
    \boldsymbol{\mu}_K
    &=\mathbf H^{-1}
      \left(\hat e\mathbf1_K-\mathbf M\mathbf C\mathbf W\boldsymbol{\eta}\right),
      \label{mu_MC}\\
    \hat e
    &=
    \frac{1+\mathbf1_K'\mathbf H^{-1}\mathbf M\mathbf C\mathbf W\boldsymbol{\eta}}
    {\mathbf1_K'\mathbf H^{-1}\mathbf1_K}, \label{ehat_MC}\\
    \boldsymbol{\alpha}^*
    &=\mathbf W\mathbf C'\mathbf M'\boldsymbol{\mu}_K
      +\mathbf W\boldsymbol{\eta}. \label{MC_solution}
\end{align}


Together, equations~\eqref{effort_MC}--\eqref{MC_solution},
and the complementary-slackness conditions in \eqref{CS_MC} characterize the solution. Notice that since $\boldsymbol{\alpha}^{*}= (\mathbf{I}-\lambda\mathbf{G})\mathbf{e}^{*}\geq 0 $ and $\mathbf{C}\geq 0$, we have that $\mathbf{e}^{*}=\mathbf{C}\boldsymbol{\alpha}^{*}\geq 0$, verifying the omitted nonnegativity constraint. 

To obtain the representation stated in the proposition, add and subtract the incentive rule from Proposition~\ref{Modular Production}. Using
\(\mathbf C'(\mathbf I-\lambda\mathbf G)'=\mathbf I\), we can write
\begin{equation*}
\begin{split}
    \boldsymbol{\alpha}^*
    ={}&
    \underbrace{
    \mathbf W\mathbf C'\mathbf M'
    \frac{\mathbf H^{-1}\mathbf 1_K}
    {\mathbf 1_K'\mathbf H^{-1}\mathbf 1_K}
    }_{
    \boldsymbol{\alpha}^{\operatorname{Prop.\,4}}
    }
    +
    \mathbf W\mathbf C'
    \left[
        \mathbf M'
        \left(
            \boldsymbol{\mu}_K^*
            -
           \underbrace{\frac{\mathbf H^{-1}\mathbf 1_K}{(\mathbf 1_K'\mathbf H^{-1}\mathbf 1_K)}}_{\boldsymbol{\mu}^{Prop.4}}
        \right)
        +
        (\mathbf I-\lambda\mathbf G)'
        \boldsymbol{\eta}
    \right].
\end{split}
\end{equation*}
Defining
\begin{equation}
    \boldsymbol{\Delta}(\boldsymbol{\eta})
    :=
    \mathbf M'
    \left(
        \boldsymbol{\mu}_K^*
        -
        \frac{\mathbf H^{-1}\mathbf 1_K}
        {\mathbf 1_K'\mathbf H^{-1}\mathbf 1_K}
    \right)
    +
    (\mathbf I-\lambda\mathbf G)'
    \boldsymbol{\eta},
    \label{A_DeltaEta}
\end{equation}
we obtain
\[
    \boldsymbol{\alpha}^*
    =
    \boldsymbol{\alpha}^{\operatorname{Prop.\,4}}
    +
    \mathbf W\mathbf C'
    \boldsymbol{\Delta}(\boldsymbol{\eta}),
\]
as stated in the proposition.

If the additional constraints are slack,
then \(\boldsymbol{\eta}=\mathbf 0\), so
\[
    \hat e^*
    =
    \frac{1}{\mathbf1_K'\mathbf H^{-1}\mathbf1_K},
    \qquad
    \boldsymbol{\mu}_K^*
    =
    \frac{\mathbf H^{-1}\mathbf1_K}
    {\mathbf1_K'\mathbf H^{-1}\mathbf1_K},
\]
and hence \(\boldsymbol{\Delta}(\mathbf0)=\mathbf0\). Thus,
Proposition~\ref{Modular Production} is recovered.
\end{proof}


\section{Optimal Incentives under Wage Benchmarking}\label{App: WageBenchmarking}

In this section, we discuss the incentive rule from Proposition~\ref{Coarse Contracts} in more detail. We replicate the discussion following Proposition~\ref{Optimal Contracts} on the firm's marginal benefit and cost to clarify how wage benchmarking affects the allocation of incentives. We end the section with a couple of examples.

As discussed in the main text, wage benchmarking does not affect workers' amplification potential because, conditional on a particular level of effort chosen by worker $s$, her co-workers' choices are impacted exactly as in the baseline model. Therefore, the firm's marginal benefit of providing incentives is very similar to the baseline model, except that the firm now selects group-level incentives $\hat{\alpha}_r$ and therefore cares about the group-level centrality, $\mathbf{TC}^\prime\mathbf{1}$, which simply sums the centrality of all members in each group. At the margin, worker $s$ will respond to group $r$'s incentives in proportion to how many \emph{incentive paths} (of any length) lead to $s$ from all members of group $r$: 
\begin{equation*}
    \frac{\partial e_{s}}{\partial \hat{\alpha}_{r}}= \sum_{i\in r}\frac{\partial e_{s}}{\partial \alpha_{i}} = \sum_{i\in r} \sum_{q=0}^{\infty} \lambda^{q} \mathbf{G}_{si}^{q} \geq 0,
\end{equation*}
where the inequality follows from $\lambda \geq 0$ and $\mathbf{G}\geq 0$. The total marginal benefit of $\hat{\alpha}_{r}$ is therefore obtained by summing across all of group $r$'s targets. This equals the group-level Bonacich centrality, which captures the total \emph{amplification potential} of group $r$:
\begin{equation*}
    MB_{\hat{\alpha}_{r}} =  \sum_s \sum_{i\in r} \frac{\partial e_{s}}{\partial \alpha_{i}}   = \sum_{i\in r}b_{i}(\lambda). 
\end{equation*}

On the marginal cost side of the firm's problem, things look very different. Since the firm can only select group-level salaries, $\hat{\beta}_k$, not all members of group $k$ will have binding IR constraints. Intuitively, if worker $s$ in group $k(s)$ has positive rents (i.e. $\eta_s>0$) then the base salary of her group, $\hat{\beta}_{k(s)}$, is determined by someone else with strictly larger costs. Therefore, even if marginally raising some group $r$'s incentives leads to larger $e_s$, the firm cannot adjust worker $s$'s base salary as a response. Relative to our baseline model, $MC_{\hat{\alpha}_{r}}$ no longer depends on whether $i\in r$ has an incentive path to $s$, \textit{unless $s$'s IR constraint is binding}. 

Rewriting the firm's optimality condition but at the level of group $r$, we have that $MB_{\hat{\alpha}_r}=MC_{\hat{\alpha}_r}$ can be written as: 
\begin{equation}\label{systemprofits_benchmarking}
      \sum_{i\in r} b_{i}(\lambda) = (1+r\sigma^{2})n_{r}\hat{\alpha}_{r} - \sum_{k\in K}\hat{\alpha}_{k}\sum_{s\in N}\mu_{s}\left(\lambda \sum_{\ell \in N}\frac{\partial e_{\ell}}{\partial \hat{\alpha}_{k}}g_{s\ell} \right)\left( \lambda \sum_{m\in N}\frac{\partial e_{m}}{\partial \hat{\alpha}_{r}}g_{sm} \right).
\end{equation}
This is the group-level version of equation \eqref{systemprofits} under wage benchmarking, which essentially breaks down the incentive allocation rule in Proposition~\ref{Coarse Contracts} at the level of group $r$. It helps us see how workers with binding IR constraints shape the firm's marginal costs. Notice that common influence paths, which make up $MC_{\hat{\alpha}_r}$ (on the right-hand-side), are now weighted by $\mu_s\geq 0$: the multiplier associated to $s$'s IR constraint.  This is because raising incentives to some group $k$ will raise efforts of that group's targets. But this will only lower $MC_{\hat{\alpha}_r}$ if workers in group $r$ also share those targets, \textit{and these targets have binding IR constraints}. The last condition is the only substantial change with respect to our baseline model in Proposition \ref{Optimal Contracts}. In fact, stacking up equation \eqref{systemprofits_benchmarking} gives the optimal incentive allocation vector in Proposition~\ref{Coarse Contracts}, which only differs from the allocation vector in Proposition~\ref{Optimal Contracts} by the presence of the $n\times n$ diagonal matrix $\mathbf{M}=\operatorname{diag}(\boldsymbol{\mu})$ of multipliers, $\mu_i$, associated to workers' IR constraints. 

Thus, the optimal allocation rule under wage benchmarking now requires knowledge of the matrix of multipliers, $\mathbf{M}$. Only the highest-effort-cost worker(s) in each group have binding IR constraints and can therefore carry positive multipliers $\mu_s\geq 0$. Stationarity with respect to the common fixed payment implies that $\sum_{s\in k}\mu_s=n_k$, where $n_k$ is the number of workers in group $k$. In some situations the $\mu$'s are uniquely pinned down by the stationarity condition with respect to incentives. However, when multiple binders in the same team have equal network position vis-a-vis all groups---i.e. the rows of $\lambda \mathbf{GCT}'$ are equal---then any vector satisfying $\sum_{i\in k}\mu_i=n_k$ yields the same contract $\boldsymbol{\alpha}$.


\begin{figure}[t]
\centering
\begin{tikzpicture}[
    scale=1,
    every edge/.style={draw, thick, ->},
    lab/.style={draw=none, fill=none, font=\small}
]

    \definecolor{lightred}{rgb}{1.0, 0.82, 0.82}
    \definecolor{lightblue}{rgb}{0.80, 0.88, 1.0}

    \fill[gray,opacity=0.1, rounded corners=0.4cm] (-0.7,1.0) rectangle (2.7,2.2);
    \draw[thick, gray!50, rounded corners=0.4cm] (-0.7,1.0) rectangle (2.7,2.2);

    \fill[gray,opacity=0.1, rounded corners=0.4cm] (-0.7,-0.6) rectangle (2.7,0.6);
    \draw[thick, gray!50, rounded corners=0.4cm] (-0.7,-0.6) rectangle (2.7,0.6);

    \node[circle, draw=red!60, fill=red!45, thick, minimum size=6mm, inner sep=0pt] (1) at (0,1.6) {1};
    \node[circle, draw=red!60, fill=red!10, thick, minimum size=6mm, inner sep=0pt] (2) at (2,1.6) {2};

    \node[circle, draw=blue!60, fill=blue!10, thick, minimum size=6mm, inner sep=0pt] (3) at (0,0) {3};
    \node[circle, draw=blue!60, fill=blue!45, thick, minimum size=6mm, inner sep=0pt] (4) at (2,0) {4};

    \path (1) edge (2);
    \path (4) edge (2);
    \path (1) edge (3);
    \path (2) edge (3);
    \path (4) edge (3);
\end{tikzpicture}
\hspace{1cm}
\begin{tikzpicture}[
    scale=1,
    every edge/.style={draw, thick, ->},
    lab/.style={draw=none, fill=none, font=\small}
]

    \definecolor{lightred}{rgb}{1.0, 0.82, 0.82}
    \definecolor{lightblue}{rgb}{0.80, 0.88, 1.0}

    \fill[gray,opacity=0.1, rounded corners=0.4cm] (-0.7,1.0) rectangle (2.7,2.2);
    \draw[thick, gray!50, rounded corners=0.4cm] (-0.7,1.0) rectangle (2.7,2.2);

    \fill[gray,opacity=0.1, rounded corners=0.4cm] (-0.7,-0.6) rectangle (2.7,0.6);
    \draw[thick, gray!50, rounded corners=0.4cm] (-0.7,-0.6) rectangle (2.7,0.6);

    \node[circle, draw=red!60, fill=red!45, thick, minimum size=6mm, inner sep=0pt] (1) at (0,1.6) {1};
    \node[circle, draw=red!60, fill=red!10, thick, minimum size=6mm, inner sep=0pt] (2) at (2,1.6) {2};

    \node[circle, draw=blue!60, fill=blue!10, thick, minimum size=6mm, inner sep=0pt] (3) at (0,0) {3};
    \node[circle, draw=blue!60, fill=blue!45, thick, minimum size=6mm, inner sep=0pt] (4) at (2,0) {4};

    \path (4) edge (1);
    \path (1) edge (2);
    \path (4) edge (2);
    \path (1) edge (3);
    \path (2) edge (3);
    \path (4) edge (3);
\end{tikzpicture}
\hspace{1cm}
\begin{tikzpicture}[
    scale=1,
    every edge/.style={draw, thick, ->},
    lab/.style={draw=none, fill=none, font=\small}
]

    \definecolor{lightred}{rgb}{1.0, 0.82, 0.82}
    \definecolor{lightblue}{rgb}{0.80, 0.88, 1.0}

    \fill[gray,opacity=0.1, rounded corners=0.4cm] (-0.7,1.0) rectangle (2.7,2.2);
    \draw[thick, gray!50, rounded corners=0.4cm] (-0.7,1.0) rectangle (2.7,2.2);

    \fill[gray,opacity=0.1, rounded corners=0.4cm] (-0.7,-0.6) rectangle (2.7,0.6);
    \draw[thick, gray!50, rounded corners=0.4cm] (-0.7,-0.6) rectangle (2.7,0.6);

    \node[circle, draw=red!60, fill=red!45, thick, minimum size=6mm, inner sep=0pt] (1) at (0,1.6) {1};
    \node[circle, draw=red!60, fill=red!10, thick, minimum size=6mm, inner sep=0pt] (2) at (2,1.6) {2};

    \node[circle, draw=blue!60, fill=blue!10, thick, minimum size=6mm, inner sep=0pt] (3) at (0,0) {3};
    \node[circle, draw=blue!60, fill=blue!45, thick, minimum size=6mm, inner sep=0pt] (4) at (2,0) {4};

    \path (4) edge (1);
    \path (1) edge (2);
    \path (2) edge (4);
    \path (1) edge (3);
    \path (2) edge (3);
    \path (4) edge (3);

\end{tikzpicture}
\caption{\textbf{Panel A:} A firm with 2 equally-sized groups. \textbf{Panel B:} A firm with 2 equally-sized groups with one link to a binding worker. \textbf{Panel C:} A firm with 2 equally-sized groups with common influence on binding workers. (Darker nodes are binding workers.)}
\label{fig:nocommoninfluenceonbinders}
\end{figure}

\begin{ex}[Incentives under Wage Benchmarking]

Consider the networks shown in Figure~\ref{fig:nocommoninfluenceonbinders} with two groups: a red group with workers 1 and 2, and a blue group with workers 3 and 4. We vary the link structure across Panels A, B, and C to show how equation \eqref{systemprofits_benchmarking} works in different situations. In all three cases, worker $1$ is the binding worker of the red group and worker $4$ is the binding worker of the blue group. Therefore, the multipliers vector is given by $\boldsymbol{\mu}=(n_{\operatorname{red}},0,0,n_{\operatorname{blue}})=(2,0,0,2)$.

In Panel A, both groups share common targets (workers $2$ and $3$) but these workers don't have binding IR constraints. Therefore, raising incentives to one group does not make it cheaper to provide incentives to the other group. In fact, applying equation \eqref{systemprofits_benchmarking} yields $\sum_{i\in red}b_{i}= n_{red}(1+r\sigma^{2})\hat{\alpha}_{red}$ and $\sum_{i\in blue}b_{i} = n_{blue}(1+r\sigma^{2})\hat{\alpha}_{blue}$. Solving the system yields:
	\begin{equation*}
		\hat{\alpha}_{red} = \frac{\sum_{i\in red}b_{i}}{n_{\operatorname{red}}(1+r\sigma^{2})}, \quad \text{and} \quad \hat{\alpha}_{blue} =  \frac{\sum_{i\in blue}b_{i}}{n_{\operatorname{blue}}(1+r\sigma^{2})} . 
	\end{equation*}
In other words, despite the many cross-group links the two groups' optimal incentives are independent of each other. 

In Panel~B  we add an extra link: $g_{14}=1$. Now we can find paths of common influence that target a binding worker. Specifically, worker $1$ is the target of an incentive path from worker $4$. This means that raising incentives of the blue group will allow the firm to adjust $\hat{\beta}_{\operatorname{red}}$, and thus cheapen additional incentives to the blue group. The total size of this adjustment depends on $\mu_{1}=n_{\operatorname{red}}$. Thus applying equation \eqref{systemprofits_benchmarking} now yields $\sum_{i\in red}b_{i}= n_{red}(1+r\sigma^{2})\hat{\alpha}_{red}$ and $\sum_{i\in blue}b_{i} = n_{blue}(1+r\sigma^{2})\hat{\alpha}_{blue} - \mu_1 \lambda^{2}\hat{\alpha}_{blue}$. Solving the system yields:
	\begin{equation*}
		\hat{\alpha}_{red} = \frac{\sum_{i\in red}b_{i}}{n_{\operatorname{red}}(1+r\sigma^{2})}, \quad \text{and} \quad \hat{\alpha}_{blue} =  \frac{\sum_{i\in blue}b_{i}}{n_{\operatorname{blue}}(1+r\sigma^{2})-n_{\operatorname{red}}\lambda^{2}}. 
	\end{equation*}
 As before, both groups' incentives are independent of each other. This is because the new link ($g_{14}=1$) only generates common influence of the blue group with itself.\footnote{The own-group cost-reduction term makes the diagonal entries of $\mathbf{W_{M}}$ different. However, because the groups have no common influence on binding workers, the off-diagonal terms are still 0.} More importantly, the blue group's incentives are larger than the red group's because of this additional own-group cost-reduction term, and the larger the size of the red group the larger the difference in incentives. 

Finally, Panel C just reverses the direction of the link between workers $2$ and $4$. Now, worker $1$ is not only the target of an incentive path from worker $4$ (as before) but also from worker $2$ (via worker $4$). This means that raising incentives of the blue group cheapens the firm's cost of further incentivizing worker $1$, which makes the incentives on the red group cheaper. Now, paths of common influence make each group's incentives depend on the centrality of both groups because they share common influence paths with a binding worker: $
\hat{\alpha}_{\text {blue }}=(\mathbf{W}_{\mathbf{M}})_{21} \sum_{i \in \text { red }} b_i+ (\mathbf{W}_{\mathbf{M}})_{22}  \sum_{i \in \text { blue }} b_i$, where $(\mathbf{W_M})_{ij}$ is the $(i,j)$ element of the weighting matrix $\mathbf{W_M}$ in Proposition \ref{Coarse Contracts}.\footnote{We computed these examples numerically using a fixed-point procedure over incentives and binding workers. Starting from a candidate $\boldsymbol{\hat{\alpha}}$, we identify the implied binding workers, select a feasible multiplier vector $\boldsymbol{\mu}$ supported on those binding workers, update $\boldsymbol{\hat{\alpha}}$ via $\boldsymbol{\hat{\alpha}}=\mathbf{W_{M}TC}'\mathbf{1}$, and iterate until the binding worker set and incentives are self-consistent. The code is available upon request.}

Finally, keep in mind that groups might have multiple workers with binding IR constraints. Suppose we add links from worker 2 to worker 1, and from worker 4 to worker 2. This would imply that both workers in the red group receive the same influence in equilibrium. In this case, incentive stationarity depends on their multipliers only through their sum, so any nonnegative $(\mu_1,\mu_2)$ satisfying $\mu_1+\mu_2=n_{\mathrm{red}}$ is equivalent.

\end{ex}

\section{Organizational Design}\label{Design}

This section develops the model’s implications for organizational design, which are summarized in Section \ref{orgdesign}. First, we link firm profits—under the optimal contract from Proposition \ref{Optimal Contracts}—to the peer network’s structure via its principal components. Next, using spectral properties of specific graph families, we classify and compare networks based on their equilibrium profits. Finally, we identify for which networks firms prefer strengthening peer effects over equivalent investments in human capital.

\subsection{A Profit Decomposition Result}

How do profits under the optimal wage contract depend on network structure? For $\lambda>0$, additional links enhance incentives and boost firm profits, making the complete graph optimal.\footnote{For $\lambda<0$ the efficient network is empty.} When link formation is costly, prior work by \cite{belhaj2016efficient} and \cite{hiller2017peer} shows that constrained-efficient networks in games with strategic complements belong to the class of nested-split graphs.\footnote{These are graphs where each worker’s neighbors form a subset of the neighbors of any higher-degree worker. This result does not hold for strategic substitutes ($\lambda<0$).} However, existing research does not identify the most efficient network within this broad class. Moreover, nested-split graphs may be impractical to implement. Firms need to understand which networks yield equivalent profits and which outperform others.

To address this, we establish a direct connection between profits and the network’s structural (i.e., spectral) properties. The result relies on the spectral decomposition of the adjacency matrix $\mathbf{G}$. To obtain a clean characterization, we focus on the class of \emph{normal matrices}, for which the adjacency matrix admits a convenient spectral representation.

\begin{definition*}[Normal Matrix]
An adjacency matrix $\mathbf{G}$ is \emph{normal} if it commutes with its transpose, that is,
\[
\mathbf{G}\mathbf{G}' = \mathbf{G}'\mathbf{G}.
\]
\end{definition*}

This class includes many economically relevant network structures. In particular, all undirected networks ($\mathbf{G}=\mathbf{G}'$) are normal. It also includes several important families of directed graphs, such as unions of directed cycles, regular tournament graphs, and many other structured networks studied in graph theory. Focusing on normal adjacency matrices therefore allows us to characterize profits across a wide variety of organizational and interaction structures while maintaining analytical tractability.

The following proposition derives a spectral decomposition of expected profits for any normal adjacency matrix.

\begin{proposition}[Network Structure and Profits]\label{Profits and Spectrum}
        In expectation, a firm's profits are maximized at one-half of equilibrium output for any adjacency matrix $\mathbf{G}$, any level of peer effects $\lambda$, and any level of fundamental risk $\sigma^2$:
 \[\mathbb{E}(\pi^{*}(\mathbf{e}^{*}|\boldsymbol{\alpha}^{*}, \boldsymbol{\beta}^{*}))= \frac{1}{2}\mathbb{E} \left(X(\mathbf{e^{*}})\right).\]
        Moreover, assume $\mathbf{G}$ is a normal matrix and let $\mathbf{u}_{\ell}$ be the right unit-eigenvectors of  $\mathbf{G}$ associated to eigenvalues $\mu_{\ell}$. Expected profits are given by: 
        \begin{equation}\label{eigenprofits_directed}
    \mathbb{E}(\pi^{*}(\mathbf{e}^{*}|\boldsymbol{\alpha}^{*}, \boldsymbol{\beta}^{*})) = \frac{n}{2} \sum_{\ell} \frac{1-n\operatorname{Var}(\mathbf{u}_{\ell})}{1-2\lambda\operatorname{Re}(\mu_{\ell}) +\sigma^{2}r \vert 1-\lambda \mu_{\ell}\vert^{2}},
        \end{equation}
\end{proposition}
The first part of Proposition \ref{Profits and Spectrum} extends a well-known result from the team production literature to our setting with bilateral spillovers: optimized profits scale one-to-one with output (for any network). The second part of Proposition \ref{Profits and Spectrum} tells us how network structure  drives profits by decomposing the network effects into the \textit{principal components} of the underlying graph. 

In Section \ref{orgdesign} we approximate \eqref{eigenprofits_directed} by focusing on the leading term of the sum above.\footnote{This is a particularly good approximation when the spectral gap is large so that eigenvalues drop off quickly.} In this case, we can write:
\begin{equation}\label{directedprofits}
  \mathbb{E}\left(\pi^{*}\right) \approx \frac{n}{2} \frac{1-n \operatorname{Var}\left(\mathbf{u}_1\right)}{\left(1+r \sigma^2\right)\left(1-\lambda \mu_1\right)^2-\left(\lambda \mu_1\right)^2}  
\end{equation}

This simplified expression reveals the following important determinants of firms' profits:
\begin{enumerate}
    \item \underline{Profits decrease with risk}: higher $\sigma^2$ raises the transaction cost associated to moving incentives from firm to worker. Although the firm optimally responds with overall flatter incentives, profits unambiguously decline.
   
    \item \underline{Profits increase with leading eigenvalue $\mu_1$ if $\lambda>0$}: recall that workers' amplification potential is driven by $\mathbf{C}=(\mathbf{I}-\lambda \mathbf{G})^{-1}=\sum_{t=0}^{\infty}(\lambda \mathbf{G})^t$. Now, since $\mathbf{G}^t \approx \mu_1^t \mathbf{u}_1 \mathbf{u}_1^{\prime}$ for a normal adjacency matrix $\mathbf{G}$, we can approximate the Neumann series as $\mathbf{C}\approx \mathbf{u}_1 \mathbf{u}_1^{\prime}\sum_t\left(\lambda \mu_1\right)^t$ for a large spectral gap.   Intuitively, the leading eigenvalue captures walk expansions, so networks with more expansive link structures amplify incentives best and yield greater profits (the opposite is true if $\lambda<0$).
   
    \item \underline{Profits decrease with $\operatorname{Var}(\mathbf{u}_1)$}: profits are larger when centrality is evenly distributed across workers.  Even if the firm optimally allocates incentives away from centrality (case of low $\sigma^2$), the underlying network still transforms those incentives to profits through $\mathbf{e}=\mathbf{C}\boldsymbol{\alpha}$, which by the previous point is a geometry governed by $\mathbf{u}_1$.
    
    \item \underline{Size of the firm $n$}: there is no obvious comparative static on $n$, since most elements of \eqref{directedprofits} depend on $n$. However, it is helpful to see the numerator $n\left(1-n\operatorname{Var}(\mathbf{u}_1)\right)$ as capturing the \textit{number of effective workers} $N_{\mathrm{eff}}$: the size of a uniform network that would generate the same aggregate influence as the observed network. For instance, if $\mathbf{u}_1$ is uniform, $\mathbf{u}_1=\frac{1}{\sqrt{n}} 1$, then
$\operatorname{Var}\left(\mathbf{u}_1\right)=0$ and $N_{\text {eff }}=n$, while if $\mathbf{u}_1$ is concentrated on $k$ nodes, then $\operatorname{Var}\left(\mathbf{u}_1\right) \approx \frac{1}{n}-\frac{k}{n^2} $ and $ N_{\mathrm{eff}} =k$. As firms grow, profits increase only if this effective size grows with $n$; otherwise, adding workers has little impact on performance.

    
\end{enumerate}

\subsection{Comparing Network Structures}
The profit characterization above allows us to compare organizational structures based on their expected profits. We now provide additional insights for specific families of graphs for which we know a lot about their spectral properties.

\subsubsection{Bipartite Networks}

\begin{figure}[t]
    \centering
    
\begin{tikzpicture}[scale=0.6]
    \tikzset{
        blueNode/.style={circle, draw=blue!50, fill=blue!20, thick, minimum size=3mm},
        redNode/.style={circle, draw=red!50, fill=red!20, thick, minimum size=3mm},
    }

    \foreach \i in {1} {
        \node[blueNode] (B\i) at (\i+4,0) {};
    }

    \foreach \j in {1,...,9} {
        \node[redNode] (R\j) at (\j,-3) {};
    }

    \foreach \i in {1} {
        \foreach \j in {1,...,9} {
            \draw (B\i) -- (R\j);
        }
    }

\end{tikzpicture}
\hfill
\begin{tikzpicture}[scale=0.6]
    \tikzset{
        blueNode/.style={circle, draw=blue!50, fill=blue!20, thick, minimum size=3mm},
        redNode/.style={circle, draw=red!50, fill=red!20, thick, minimum size=3mm},
    }

    \foreach \i in {1,2,3} {
        \node[blueNode] (B\i) at (\i+2,0) {};
    }

    \foreach \j in {1,...,7} {
        \node[redNode] (R\j) at (\j,-3) {};
    }

    \foreach \i in {1,2,3} {
        \foreach \j in {1,...,7} {
            \draw (B\i) -- (R\j);
        }
    }
\end{tikzpicture}
\hfill
\begin{tikzpicture}[scale=0.6]
    \tikzset{
        blueNode/.style={circle, draw=blue!50, fill=blue!20, thick, minimum size=3mm},
        redNode/.style={circle, draw=red!50, fill=red!20, thick, minimum size=3mm},
    }

    \foreach \i in {1,...,5} {
        \node[blueNode] (B\i) at (\i+3,0) {};
    }

    \foreach \j in {1,...,5} {
        \node[redNode] (R\j) at (\j+3,-3) {};
    }

    \foreach \i in {1,...,5} {
        \foreach \j in {1,...,5} {
            \draw (B\i) -- (R\j);
        }
    }
\end{tikzpicture}
    \caption{Complete Bipartite graphs with $n=10$. An asymmetric bipartite graph (left panel) generates lower profits than a symmetric one (right panel).}
    \label{fig:bipartite}
\end{figure}

In what follows we look at classes of networks for which we can solve equation \eqref{eigenprofits_directed} directly and compute profits exactly. Consider a firm deciding how to delegate responsibilities, specifically determining the relative size of two interacting divisions. Suppose all relevant spillovers occur across divisions. The firm must then choose among all complete bipartite graphs of size $n$, where members are split into groups of size $m_{1}$ and \(m_{2}\) with $m_{1}+m_{2}=n$ (Figure \ref{fig:bipartite}). Using the spectral properties of these graphs,\footnote{For complete bipartite graphs, the eigenvalues are well known: \(\mu_1 = \sqrt{m_{1}m_{2}}\), \(\mu_2 = \mu_3 = \dots = \mu_{n-1} = 0\), and \(\mu_n = -\sqrt{m_{1}m_{2}}\). The corresponding unit eigenvectors satisfy:
\[
\mathbf{u}_{1,i} = 
\begin{cases} 
\frac{1}{\sqrt{2m_{1}}}, & \text{if } i \text{ is in group } m_{1}, \\
\frac{1}{\sqrt{2m_{2}}}, & \text{if } i \text{ is in group } m_{2}
\end{cases}, 
\quad
\mathbf{u}_{n,i} = 
\begin{cases} 
\frac{1}{\sqrt{2m_{1}}}, & \text{if } i \text{ is in group } m_{1}, \\
-\frac{1}{\sqrt{2m_{2}}}, & \text{if } i \text{ is in group } m_{2}
\end{cases}.
\]
Since \((\mathbf{u}_1^\prime \mathbf{1})^2 + (\mathbf{u}_n^\prime \mathbf{1})^2 = n\), only the first and last terms in Proposition \ref{Profits and Spectrum} contribute to profits. Lastly, we use the fact that $1-n\operatorname{Var}(\mathbf{u}_{\ell}) = (\mathbf{u}_{\ell}'\mathbf{1})^{2}/n$.} we can express expected profits in terms of $m_{1}$ and \(m_{2}\):  
\[
2 \mathbb{E}(\pi^{*})_{\mathrm{bipartite}} = \frac{\left(\sqrt{m_{1}}+\sqrt{m_{2}}\right)^2/2}{(1+r\sigma^2) (1-\lambda \sqrt{m_{1}m_{2}})^2 - (\lambda\sqrt{m_{1}m_{2}})^2} + \frac{\left(\sqrt{m_{1}}-\sqrt{m_{2}}\right)^2/2}{(1+r\sigma^2) (1+\lambda \sqrt{m_{1}m_{2}})^2 - (\lambda\sqrt{m_{1}m_{2}})^2}.
\]
From this, we identify the profit-maximizing structure among all complete bipartite graphs:

\begin{corollary}[\textbf{Complete Bipartite Networks}] \label{cor: bipartite}
    Among all complete bipartite graphs with $m_{1}$ nodes in group A and $m_{2}$ nodes in group B, expected profits are maximized when the two groups are of equal size, i.e., \( m_{1} = m_{2} \).
\end{corollary}

\begin{figure}[t]
    \centering
    
    \begin{tikzpicture}[scale=0.4]
        \tikzset{
            ringNode/.style={circle, draw=blue!50, fill=blue!20, thick, minimum size=2mm},
        }
    
        \foreach \i in {1,...,10} {
            \node[ringNode] (R\i) at ({\i*360/10}:4) {};
        }
    
        \foreach \i in {1,...,9} {
            \draw (R\i) -- (R\the\numexpr\i+1);
        }
        \draw (R10) -- (R1); 
    \end{tikzpicture}
    \hfill
    \begin{tikzpicture}[scale=0.4]
        \tikzset{
            ringNode/.style={circle, draw=red!50, fill=red!20, thick, minimum size=2mm},
        }
    
        \foreach \i in {1,...,5} {
            \node[ringNode] (R1\i) at ({\i*360/5+18}:4) {};
        }
    
        \foreach \i in {1,...,4} {
            \draw (R1\i) -- (R1\the\numexpr\i+1);
        }
        \draw (R15) -- (R11); 
    
        \foreach \i in {1,...,5} {
            \node[ringNode] (R2\i) at ({\i*360/5+18}:2) {};
        }
    
        \foreach \i in {1,...,4} {
            \draw (R2\i) -- (R2\the\numexpr\i+1);
        }
        \draw (R25) -- (R21); 
    \end{tikzpicture}
    \hfill
    \begin{tikzpicture}[scale=0.4]
        \tikzset{
            ringNode/.style={circle, draw=black!50, fill=black!20, thick, minimum size=2mm},
        }
    
        \foreach \i in {1,...,4} {
            \node[ringNode] (R1\i) at ({\i*360/4 + 45 }:1) {};
        }
    
        \foreach \i in {1,...,3} {
            \draw (R1\i) -- (R1\the\numexpr\i+1);
        }
        \draw (R14) -- (R11); 
    
        \foreach \i in {1,...,3} {
            \node[ringNode] (R2\i) at ({\i*360/3 + 210}:3.5) {};
        }
    
        \foreach \i in {1,...,2} {
            \draw (R2\i) -- (R2\the\numexpr\i+1);
        }
        \draw (R23) -- (R21); 
    
        \foreach \i in {1,...,3} {
            \node[ringNode] (R3\i) at ({\i*360/3 + 210 }:5) {};
        }
    
        \foreach \i in {1,...,2} {
            \draw (R3\i) -- (R3\the\numexpr\i+1);
        }
        \draw (R33) -- (R31); 
    \end{tikzpicture}
    
    \caption{All 2-regular graphs with $n=10$ give the same profits.}
    \label{fig: d-regular}
\end{figure}

\subsubsection{Regular Networks}

Now, consider a different scenario where a homogeneous organization---where each worker is influenced by the same number of peers---decides whether to split into $K$ divisions with sizes $C_1,\ldots,C_K$. Specifically, the CEO must choose between keeping all $n$ workers in a single \(d\)-regular structure or dividing them into smaller \(d\)-regular components (Figure \ref{fig: d-regular}). Again, Proposition \ref{Profits and Spectrum} helps solve this design problem. Given the spectral properties of \(d\)-regular graphs,\footnote{For a \(d\)-regular graph with \(k\) components of sizes \(C_1, C_2, \dots, C_k\), the eigenvalues are \(\mu_1 = \mu_2 = \dots = \mu_k = d\). The corresponding eigenvectors satisfy \(\mathbf{u}_i^\prime \mathbf{1} = C_i/\sqrt{C_i}\) for \(i = 1, \dots, k\), while all other eigenvectors are orthogonal to \(\mathbf{1}\). Lastly, we use the fact that $1-n\operatorname{Var}(\mathbf{u}_{\ell}) = (\mathbf{u}_{\ell}'\mathbf{1})^{2}/n$.} we express profits as:
\[
\mathbb{E}(\pi^{*})_{\mathrm{regular}} = \frac{1}{2} \sum_{i=1}^k \frac{C_i}{(1+r\sigma^2) (1-d\lambda)^2 - (d\lambda)^2} = \frac{1}{2} \frac{n}{(1+r\sigma^2) (1-d\lambda)^2 - (d\lambda)^2}.
\]
Expected profits depend only on local spillover structure, not on whether the organization is split into multiple divisions.

\begin{corollary}[\textbf{Regular Networks}] \label{cor: regular}
    All \(d\)-regular graphs of size $n$ yield the same expected profits.
\end{corollary}

\begin{figure}[t]
   \includegraphics[scale=0.3]{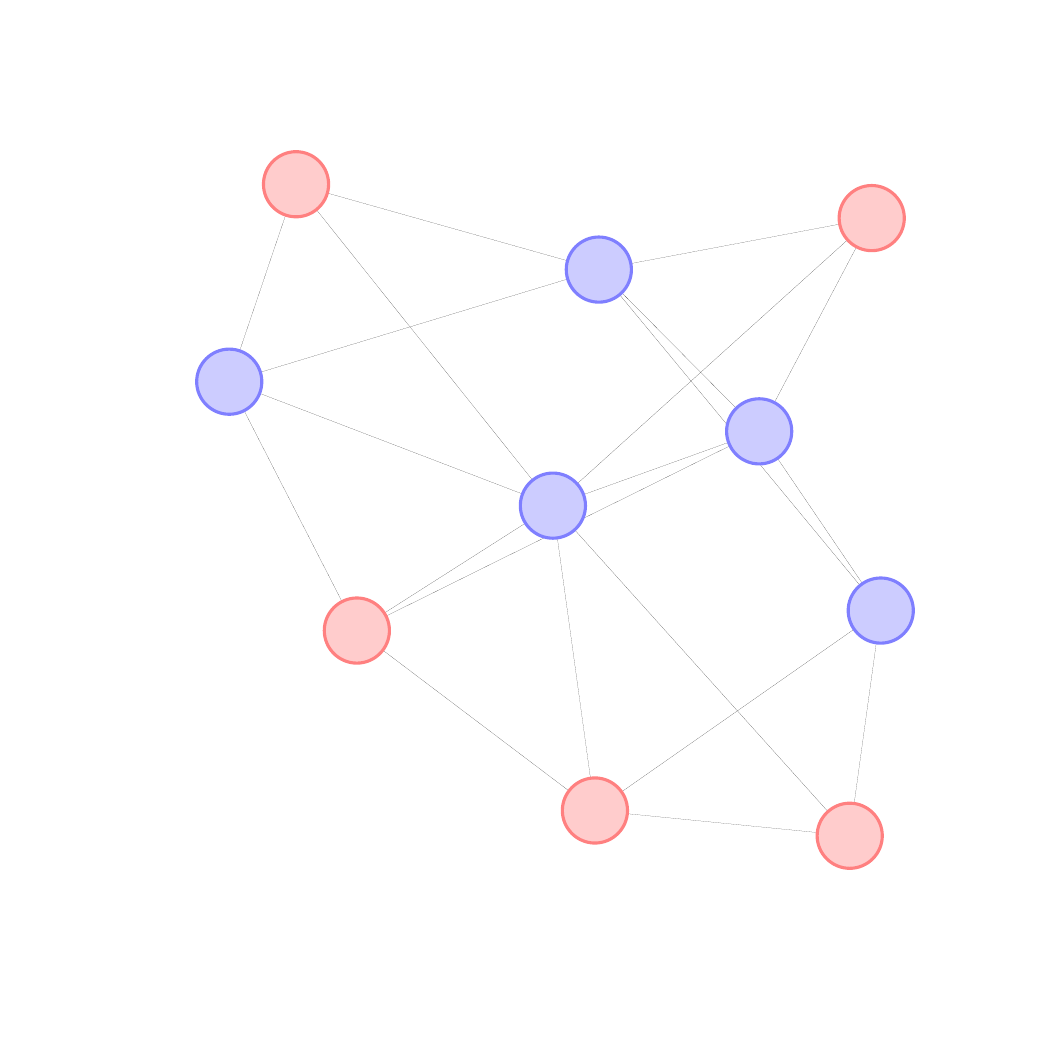}\hfill\includegraphics[scale=0.3]{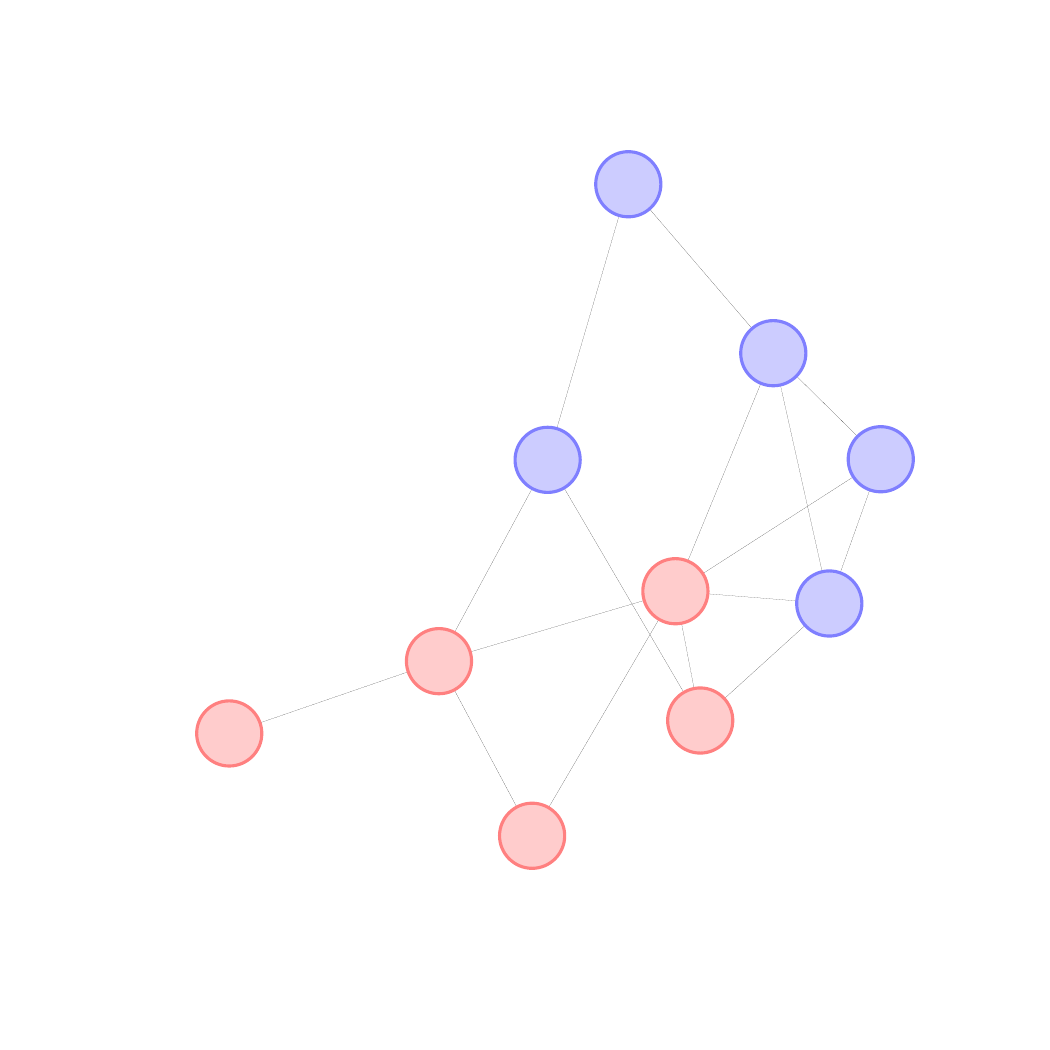}\hfill\includegraphics[scale=0.3]{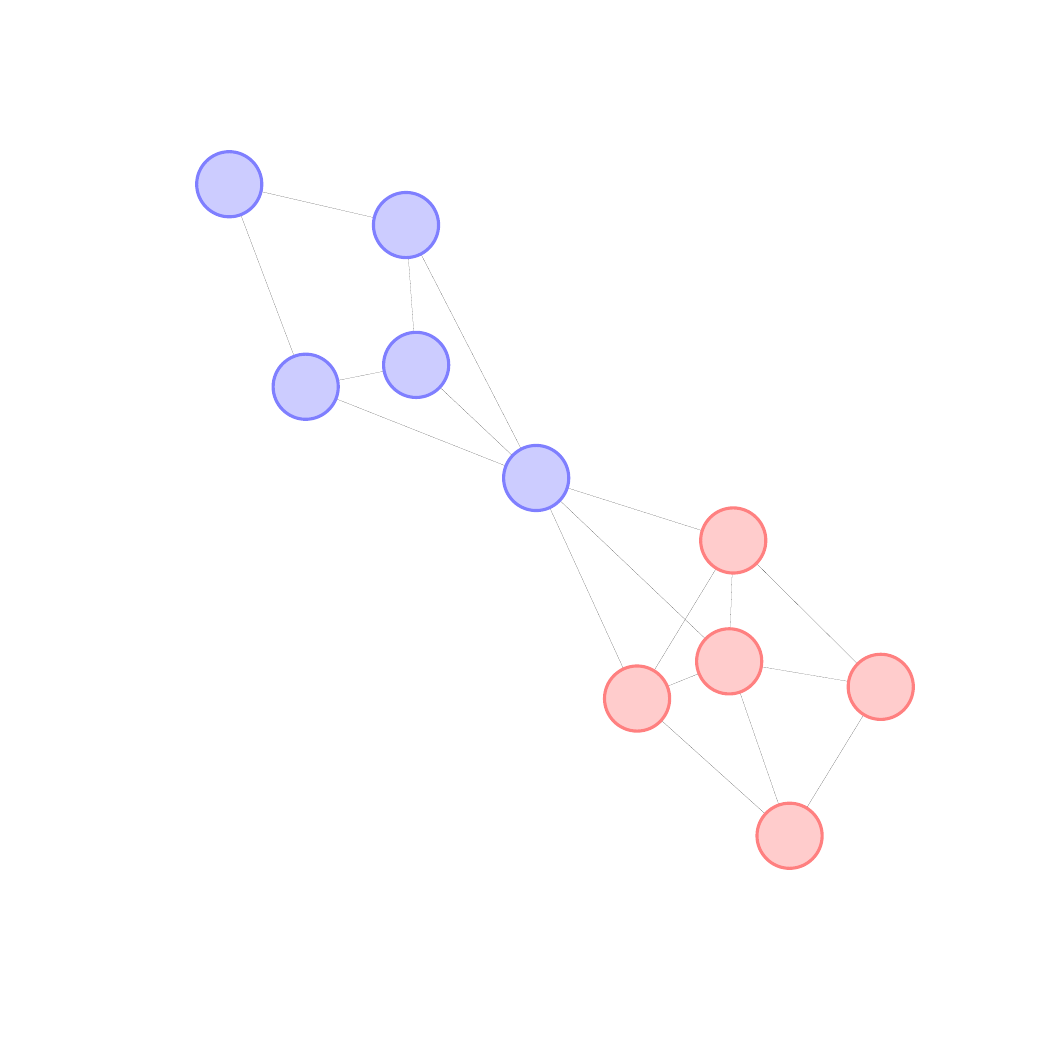}
    \caption{Planted Partition model with $n=10$ and $p+q=0.8$. \\
 \medskip
    \textit{Panel A:} $p=q=0.4$. \hfill \textit{Panel B:} $p=0.6, q=0.2$. \hfill \textit{Panel C:} $p=0.75, q=0.05$.}
\end{figure}

\subsubsection{Community Structure}

Next, consider the role of homophily---the tendency of individuals to form connections within their group. We analyze this using the planted partition random graph model from Proposition \ref{pq_model}, where \(p\) represents the probability of within-group connections and \(q\) the probability of cross-group connections. While \(p+q\) determines overall connectivity, homophily is captured by the ratio \(p/q\). Using Proposition \ref{Profits and Spectrum} and the expected adjacency matrix \(\Bar{\mathbf{G}} \in \{p,q\}^{n \times n}\) we obtain:\footnote{The only nonzero eigenvalues of \(\Bar{\mathbf{G}}\) are \(\mu_1 = n (p+q)/2 \) and \(\mu_2 = n (p-q)/2\), with \(\mathbf{u}_1^\prime \mathbf{1} = n/\sqrt{n}\) and \(\mathbf{u}_i^\prime \mathbf{1} = 0\) for \(i \geq 2\) and we use the fact that $1-n\operatorname{Var}(\mathbf{u}_{\ell}) = (\mathbf{u}_{\ell}'\mathbf{1})^{2}/n$.}
\[
\mathbb{E}(\pi^{*})_{\mathrm{planted\:partition}} = \frac{1}{2} \frac{n}{(1+r\sigma^2)(1 - \lambda n \frac{p+q}{2})^2 - (\lambda n \frac{p+q}{2})^2 }.
\]

\begin{corollary}[\textbf{Community Structure and Profits}] \label{RandomNetworks}
    In a planted partition model with connection probabilities \(p\) and \(q\), expected profits depend only on average degree, and not on expected homophily.
\end{corollary}

\subsubsection{Directed Networks}\label{App:Design_Directed}

We can also use Proposition \ref{Profits and Spectrum} to compute profits for directed networks. For instance, we could think of an assembly line production process in which each worker directly influences the next co-worker in a closed loop. Then, a \emph{directed cycle} of $n$ nodes (see the left network in Figure \ref{fig:directed_networks}) can be shown to have expected profits equal to:\footnote{For a directed cycle with homogeneous weights, the adjacency matrix is a permutation matrix. Its eigenvalues and eigenvectors are well-known. In particular, let $\omega=e^{2 \pi i / n}$. Then the eigenvalues are $\mu_k=\omega^k=e^{2 \pi i k / n}, \quad k=0,1, \ldots, n-1$ and a standard orthonormal basis of eigenvectors is $\mathbf{u}_k=\frac{1}{\sqrt{n}}\left(1, \omega^k, \omega^{2 k}, \ldots, \omega^{(n-1) k}\right)^{\top}
$. Now, for $k=0$ we have $\omega^0=1$ so $
\mathbf{u}_0^{\prime} \mathbf{1}=\frac{1}{\sqrt{n}} \sum_{j=0}^{n-1} 1=\frac{n}{\sqrt{n}}=\sqrt{n}
$ while for $k\neq 0$ it can be easily shown that $\mathbf{u}_k^{\prime} \mathbf{1}=0$.
} 
$$
\mathbb{E}\left(\pi^{*}\right)_{\mathrm{directed \:cycle}}=\frac{1}{2} \frac{n}{\left(1+r \sigma^2\right)(1-\lambda)^2-\lambda^2}$$
Here, the feedback loop means that when a worker takes an action, it has a spillover effect on the entire cycle. In other words, the presence of a cycle allows positive complementarities (when $\lambda>0$ ) to be amplified across the network. 

\begin{figure}[t]
    \centering
    \def\angleTwo{90}   
    \def\angleThree{162}
    \def\angleFour{234}  
    \def\angleFive{306}  
    \def\angleOne{18}    
    \def\angles{{\angleTwo,\angleThree,\angleFour,\angleFive,\angleOne}}
    
    \begin{subfigure}[b]{0.45\linewidth}
        \centering
        \begin{tikzpicture}[scale=.75]
            \foreach \i/\angle in {2/90, 3/162, 4/234, 5/306, 1/18} {
                \node[circle, draw=red!50, fill=red!20, thick, minimum size=3.5mm, inner sep=0.5mm] (C\i) at (\angle:2cm) {};
            }
            \draw[->, >=latex] (C2) -- (C3);
            \draw[->, >=latex] (C3) -- (C4);
            \draw[->, >=latex] (C4) -- (C5);
            \draw[->, >=latex] (C5) -- (C1);
            \draw[->, >=latex] (C1) -- (C2);
        \end{tikzpicture}
    \end{subfigure}
    \hfill
    \begin{subfigure}[b]{0.45\linewidth}
        \centering
        \begin{tikzpicture}[scale=.75]
            \foreach \i/\angle in {2/90, 3/162, 4/234, 5/306, 1/18} {
                \node[circle, draw=blue!50, fill=blue!20, thick, minimum size=3.5mm, inner sep=0.5mm] (T\i) at (\angle:2cm) {};
            }
            \draw[->, >=latex] (T1) -- (T2);
            \draw[->, >=latex] (T1) -- (T3);
            \draw[->, >=latex] (T2) -- (T3);
            \draw[->, >=latex] (T2) -- (T4);
            \draw[->, >=latex] (T3) -- (T4);
            \draw[->, >=latex] (T3) -- (T5);
            \draw[->, >=latex] (T4) -- (T5);
            \draw[->, >=latex] (T5) -- (T1);
            \draw[->, >=latex] (T4) -- (T1);
            \draw[->, >=latex] (T5) -- (T2);
        \end{tikzpicture}
    \end{subfigure}
    \caption{Two directed networks with 5 workers: (a) a directed cycle and (b) a regular tournament graph.}
    \label{fig:directed_networks}
\end{figure}

For another example, consider a \emph{regular tournament graph} (see the right network in Figure \ref{fig:directed_networks}).\footnote{A tournament graph is a complete directed graph in which every pair of nodes has exactly one directed link. A regular tournament graph is a tournament graph in which every node has the same in-degree and out-degree. Regular tournament graphs are guaranteed to exist when $n$ is odd.} While tournament graphs are often used to represent the outcome of a round-robin tournament, in our setting they can be interpreted as an organization in which, for each pair of workers, influence is one-sided. We can leverage Proposition \ref{Profits and Spectrum} to express expected maximized profits as:\footnote{For regular tournament graphs, each row adds up to the same constant $(n-1)/2$. This implies that $(n-1)/2$ is the leading eigenvalue with an eigenvector equal to $\mathbf{1}$.  Notice that the adjacency matrix of a regular tournament graph is normal, i.e., it is unitarily diagonalizable $\mathbf{G}=\mathbf{UMU}^{H}$, where $\mathbf{U}^{H}$ denotes the conjugate transpose of $\mathbf{U}$. We can choose an orthonormal basis of right eigenvectors $\lbrace \mathbf{u}_{\ell}\rbrace_{\ell=1}^{n}$ with $\mathbf{u}_{1} = \frac{1}{\sqrt{n}}\mathbf{1}$. It follows that $\mathbf{u}_{1}^{H}\mathbf{1}=\sqrt{n}$. Moreover, $\mathbf{u}_{\ell}^{H}\mathbf{1}=0$ for $\ell\geq 2$. Hence, only the leading eigenvalue contributes to profits. Lastly, we use the fact that $|\mathbf{u}_{\ell}^{H}\mathbf{1}|^{2}=n(1-n\operatorname{Var}(\mathbf{u_{\ell}}))$}
$$
\mathbb{E}\left(\pi^{*}\right)_{\mathrm{tournament}}=\frac{1}{2} \frac{n}{(1+r\sigma^{2})\left( 1-\lambda\frac{n-1}{2} \right)^{2}-\lambda^{2}\left(\frac{n-1}{2} \right)^{2}}.$$

In this case, each worker is, on average, similarly influential. Thus,  every worker's influence is aggregated over $\lambda \frac{n-1}{2}$ links, while in a directed cycle every node influences only one other node, so the effective amplification of a worker's action is governed by the factor $\lambda$. As a result, the denominator in the tournament case is smaller (for $\lambda>0$) than in the directed cycle case, leading to higher expected profits.

\subsubsection{Power Law Networks}

Consider networks whose degree distribution follows a power law $P(d) \propto d^{-\gamma}$. For $\gamma>3$, increasing $\gamma$ lightens the right tail of the distribution, leading to less degree heterogeneity. In standard configuration-type approximations, this reduces both the leading eigenvalue $\mu_1$ and the dispersion of the leading eigenvector $\operatorname{Var}\left(\mathbf{u}_1\right)$: with fewer extreme hubs, overall network amplification falls (lower $\mu_1$) and centrality becomes less concentrated (lower $\operatorname{Var}\left(\mathbf{u}_1\right)$). Since equation \eqref{profits_approx} implies that profits are increasing in $\mu_1$ but decreasing in $\operatorname{Var}\left(\mathbf{u}_1\right)$, a rise in $\gamma$ has two opposing effects on profits: it lowers $\mu_1$ (which decreases profits) but also lowers $\operatorname{Var}\left(\mathbf{u}_1\right)$ (which increases profits). Consequently, profits need not be monotone in $\gamma$, and may be maximized either at an interior value $\gamma^{*}$ (where the marginal effects offset) or at the boundary of the admissible region, depending on parameter values.

\medskip
A convenient approximation is $\mu_1 \approx\left\langle d^2\right\rangle /\langle d\rangle$ and $u_1(i) \approx d_i / \sqrt{\sum_j d_j^2}$, which implies $\operatorname{Var}\left(\mathbf{u}_1\right)= \frac{1}{n}\left(1-\frac{\langle d\rangle^2}{\left\langle d^2\right\rangle}\right)$. Using $\langle d\rangle=\frac{\gamma-1}{\gamma-2} k_{\min }$ and $\left\langle d^2\right\rangle=\frac{\gamma-1}{\gamma-3} k_{\min }^2$, we obtain $\frac{\langle d\rangle^2}{\left\langle d^2\right\rangle}=\frac{(\gamma-1)(\gamma-3)}{(\gamma-2)^2}$. Plugging these expressions into \eqref{profits_approx} and setting $k_{\text {min }}=1$ yields
$$
\mathbb{E}\left(\pi^{*}\right) \approx \frac{n}{2} \frac{1-\frac{1}{(\gamma-2)^2}}{\left(1+r \sigma^2\right)\left[1-\lambda\left(\frac{\gamma-2}{\gamma-3}\right)\right]^2-\left[\lambda\left(\frac{\gamma-2}{\gamma-3}\right)\right]^2} .
$$

\medskip
This expression characterizes how expected profits vary with the tail exponent over the parameter region in which the model's stability conditions hold. For fixed $\lambda>0$, this admissible region is bounded away from $\gamma=3$, since $\mu_1$ diverges as $\gamma \downarrow 3$. Within the admissible region, however, a rise in $\gamma$ has two opposing effects: it lowers $\mu_1$, reducing network amplification, but also lowers $\operatorname{Var}\left(\mathbf{u}_1\right)$, making centrality less concentrated. Consequently, the sign of the effect of $\gamma$ on profits is generally ambiguous and must be determined over the admissible parameter range.


\subsection{How to Invest in Workers}

Firms typically invest in their workforce through training programs that enhance either individual skills or teamwork. Should a firm focus on improving workers’ human capital or strengthening peer complementarities through team-building exercises? The profit decomposition in Proposition \ref{Profits and Spectrum} provides a framework to address such human resource decisions in large and complex organizations. Throughout this subsection, we focus on undirected networks, so that $\mathbf{G}=\mathbf{G}'$. 

Consider an extension of the baseline model where the marginal cost of effort is influenced by a human-capital parameter $\nu\geq 1 $. We can rewrite \eqref{psi} as
          \[\psi_i (\mathbf{e}) = \nu \frac{e_i^2}{2}  - \lambda \sum_{j\in N} g_{ji} e_i e_j.\]
Let $\mathbf{C}_{\nu}:=(\mathbf{I} - \frac{\lambda}{\nu}\mathbf{G})^{-1}$. For this extension, we maintain the following counterparts of Assumptions~\ref{assumption1} and~\ref{spectralradius}: 
\renewcommand{\theassumption}{1-$\nu$}
\begin{assumption}\label{assumption1nu}
The spectral radius of $\lambda\mathbf{G}$ is less than $\nu$.
 \end{assumption}
 \renewcommand{\theassumption}{2-$\nu$}
\begin{assumption}\label{spectralradiusnu}
The spectral radius of $\lambda^{2}/(1+\nu r \sigma^{2})(\mathbf{GC}_{\nu})'\mathbf{GC}_{\nu}$ is less than $\nu^{2}$. 
 \end{assumption}
          
Suppose a firm can invest one dollar to either \textit{decrease} $\nu$ (reducing effort costs) or \textit{increase} $\lambda$ (amplifying peer effects). To maintain comparability, we assume both investments have the same per-unit cost. Allowing for different costs would not alter the fundamental insight. The firm’s decision reduces to evaluating:
        \[\frac{\partial \mathbb{E}(\pi)}{\partial \lambda} \lessgtr  \Big\lvert \frac{\partial \mathbb{E}(\pi)}{\partial \nu}\Big\rvert.\]
Using Proposition \ref{Profits and Spectrum}, we can compute these marginal effects. We show that investing in team strength is preferable to investing in human capital only for those networks with:
\[\left(1-\mu_{\ell}\right)\left(1+r\sigma^2 \left(v-\lambda\mu_{\ell} \right)\right)<  \: 1/2, \:\:\ \forall \ell \text{ with } \mathbf{u}_\ell^\prime \mathbf{1}\neq 0.\]
The first thing to notice is that, as $r\sigma^2$ grows, it is less profitable to invest in team-building exercises, everything else equal. Intuitively, when the cost associated to providing risky incentives increases -- either because the firm is very risky or the workforce is very risk averse -- performance-based compensation is costly, so investing in peer effects has little impact. 

Network structure also plays a crucial role. For instance, the equation above tells us that the only \textit{regular network} for which investing in human capital dominates is the empty network.\footnote{To see this notice that Assumption 1 requires that $\nu > \lambda \mu_{\ell}$. Therefore since $\mu_{\ell}=d$ for all eigenvalues with $\mathbf{u}_{\ell}^\prime \mathbf{1}\neq 0$ the result follows.}  \textit{Scale-free networks}, which are characterized by a power-law degree distribution, and \textit{small-world networks} will typically favor investments in peer strength.\footnote{These arguments require checking how quickly the dominant eigenvalue exceeds $1$ for different parameter values.} When connections are sparse, investing in human capital is (obviously) preferable. However, as network density increases, the advantage shifts toward strengthening peer effects. One might wonder where this shift happens, and whether most networks of $n$ nodes favor one investment over the other. A structured way to increase network density is by raising the linking probability $p$ in an \textit{Erd\H{o}s-R\'{e}nyi random graph}. As $p$ increases from $0$ (isolated nodes) to $1$ (complete network), there is an exact threshold beyond which investing in human capital ceases to be optimal. Proposition~\ref{ER_model} provides a simpler sufficient condition that coincides with the familiar threshold for the emergence of a giant component.

\begin{proposition} \label{ER_model}
    In the Erd\H{o}s-R\'{e}nyi Random Graph model, investing in team strength outperforms investing in human capital if $np\geq 1$.
\end{proposition}

This sufficient condition implies that as long as each worker interacts with at least one other worker (in expectation), investing in team strength is superior to uniformly enhancing worker productivity. Therefore, investments in team-building exercises dominate investments in human capital for most real-world networks.

\subsection{Proofs of Section \ref{Design}}

\begin{proof}[\textbf{Proof of Proposition \ref{Profits and Spectrum}}]
Recall that the firm's problem can be written in matrix form as: 
\begin{align*}
\max_{\boldsymbol{\alpha}} \: \mathbb{E}[\pi(\mathbf{e}\mid \boldsymbol{\alpha}, \boldsymbol{\beta})]  = \boldsymbol{\alpha}' \mathbf{C}'\mathbf{1} - \frac{1}{2}\boldsymbol{\alpha}' \underbrace{\left[  \mathbf{C}' \mathbf{C} - 2\lambda  \mathbf{C}'\mathbf{G} \mathbf{C} +\sigma^2 r \mathbf{I}\right]}_{\mathbf{P}} \boldsymbol{\alpha}
\end{align*}

Notice that if $\mathbf{G}$ is not symmetric, then the term $\mathbf{C}'\mathbf{GC}$ is also non-symmetric. However, the matrices $\mathbf{C}'\mathbf{C}$ and $\mathbf{I}$ are both symmetric, even if $\mathbf{G}$ is not. Now, we use the fact that the quadratic form for a non-symmetric matrix $\mathbf{P}$ only depends on the symmetric part of $\mathbf{P}$, i.e., $\mathbf{x}'\mathbf{Px} = \mathbf{x}'\left( \frac{\mathbf{P}+\mathbf{P}'}{2}\right) \mathbf{x}$. This means we can replace $\mathbf{C}'\mathbf{GC}$ by its symmetric part $(1/2)(\mathbf{C}'\mathbf{GC}+\mathbf{C}'\mathbf{G}'\mathbf{C})$ without changing the value of the quadratic form $\boldsymbol{\alpha}'\mathbf{P}\boldsymbol{\alpha}$ and, hence, without changing the optimization problem, which we can write now as: 
\begin{align*}
\max_{\boldsymbol{\alpha}} \: \mathbb{E}[\pi(\mathbf{e}\mid \boldsymbol{\alpha}, \boldsymbol{\beta})]  = \boldsymbol{\alpha}' \mathbf{C}'\mathbf{1} - \frac{1}{2}\boldsymbol{\alpha}' \underbrace{\left[  \mathbf{C}'\left( \mathbf{I} - \lambda  (\mathbf{G}'+\mathbf{G})\right) \mathbf{C} +\sigma^{2}r\mathbf{I} \right]}_{\mathbf{P}} \boldsymbol{\alpha}.
\end{align*}
This problem yields our solution for optimal incentives in Proposition~\ref{Optimal Contracts}: 
\begin{equation*}
    \boldsymbol{\alpha}^{*} = \left( \mathbf{C} '(\mathbf{I} - \lambda (\mathbf{G}+\mathbf{G}') )\mathbf{C}+ \sigma^2 r\mathbf{I} \right)^{-1} \mathbf{C}'\mathbf{1} = \mathbf{P}^{-1}\mathbf{C}'\mathbf{1}.
\end{equation*}
Substituting this into the objective function we see that, at the optimum, maximum expected profits are:
\begin{equation*}
    \mathbb{E}[\pi^{*}] = (\mathbf{C}'\mathbf{1})'\mathbf{P}^{-1}\mathbf{C}'\mathbf{1} - \frac{1}{2}(\mathbf{C}'\mathbf{1})'\mathbf{P}^{-1}\mathbf{P}\mathbf{P}^{-1}\mathbf{C}'\mathbf{1} = \frac{1}{2}(\mathbf{C}'\mathbf{1})'\mathbf{P}^{-1}\mathbf{C}'\mathbf{1} = \frac{1}{2}\mathbf{1}'\mathbf{C}\boldsymbol{\alpha}^{*} = \frac{1}{2}\mathbf{1}'\mathbf{e}^{*}= \frac{1}{2}\mathbb{E}[X(\mathbf{e})],
\end{equation*}
where we have used the expression for equilibrium efforts $\mathbf{e} = \mathbf{C}\boldsymbol{\alpha}$. Thus, we have shown that, in expectation, the firm's profits are maximized at one-half of the equilibrium output for any network $\mathbf{G}$. 

We now prove the second part of the proposition. Because $\mathbf G$ is unitarily diagonalizable, there exists a unitary matrix $\mathbf U$ and a diagonal matrix $\mathbf{M}$ such that $\mathbf G = \mathbf U \mathbf{M} \mathbf U^{H}$ with $\mathbf U^{H}\mathbf U = \mathbf I$.\footnote{For a matrix $\mathbf{A}$, we denote by $\mathbf{A}^{H}$ its conjugate transpose.} 
The columns $\mathbf u_{\ell}$ of $\mathbf U$ are the orthonormal right eigenvectors of $\mathbf G$, satisfying $\mathbf G\mathbf u_{\ell}=\mu_{\ell}\mathbf u_{\ell}$, and the rows of $\mathbf U^{H}$ are their corresponding left eigenvectors $\mathbf u_{\ell}^{H}$. The diagonal entries $\mu_{\ell}$ of $\mathbf{M}$ may be complex.  

Recall the workers' equilibrium condition:
\begin{equation*}
    \mathbf{e}^{*} = \mathbf{C}\boldsymbol{\alpha} = \mathbf{C}\left( \mathbf{C}'(\mathbf{I} - \lambda (\mathbf{G}+\mathbf{G}') )\mathbf{C}+ \sigma^2 r\mathbf{I}\right)^{-1}\mathbf{C}'\mathbf{1}. 
\end{equation*}
Moving $\mathbf{C}=(\mathbf{I}-\lambda\mathbf{G})^{-1}$ and $\mathbf{C}'=(\mathbf{I}-\lambda\mathbf{G}')^{-1}$ inside the inverted matrix we can rewrite this condition as: 
\begin{equation*}
    \left[\left(\mathbf{I}-\lambda\left(\mathbf{G}+\mathbf{G}'\right)\right)+\sigma^2 r(\mathbf{I}-\lambda \mathbf{G}')\left(\mathbf{I}-\lambda \mathbf{G}\right)\right] \mathbf{e}^*=\mathbf{1}.
\end{equation*}


Pre-multiplying the equilibrium condition by $\mathbf{U}^{H}$ we obtain: 
\begin{equation*}
    \left[\left(\mathbf{I}-\lambda\left(\mathbf{M}+\mathbf{M}^{H}\right)\right)+\sigma^2 r(\mathbf{I}-\lambda \mathbf{M}^{H})\left(\mathbf{I}-\lambda \mathbf{M}\right)\right] \mathbf{U}^{H}\mathbf{e}^*=\mathbf{U}^{H}\mathbf{1}. 
\end{equation*}
Notice that all matrices inside the square brackets are diagonal. This means we can separate the system and treat each eigencomponent separately. Thus, for each $\ell$: 
\begin{equation*}
    \left[ 1 - \lambda (\mu_{\ell}+\overline{\mu_{\ell}}) + \sigma^{2} r (1-\lambda \overline{\mu_{\ell}})(1-\lambda \mu_{\ell})\right](\mathbf{U}^{H}\mathbf{e}^{*})_{\ell} = (\mathbf{U}^{H}\mathbf{1})_{\ell}. 
\end{equation*}
Because $\mu_{\ell}+\overline{\mu_{\ell}}= 2 \operatorname{Re}(\mu_{\ell})$ and $(1-\lambda\overline{\mu_{\ell}})(1-\lambda \mu_{\ell}) = \vert 1-\lambda \mu_{\ell}\vert^{2}$ we have: 
\begin{equation*}
    (\mathbf{U}^{H}\mathbf{e}^{*})_{\ell} = \frac{(\mathbf{u}_{\ell}^{H}\mathbf{1})}{1-2\lambda\operatorname{Re}(\mu_{\ell}) +\sigma^{2}r \vert 1-\lambda \mu_{\ell}\vert^{2}}.
\end{equation*}
Recovering $\mathbf{e}^{*} $ by multiplying each coefficient by its eigenvector $\mathbf{u}_{\ell}$ and summing for all $\ell$, we have:
\begin{equation*}
  \mathbf{1}' \mathbf{e}^{*} = \sum_{\ell}\frac{(\mathbf{u}_{\ell}^{H}\mathbf{1})(\mathbf{1}'\mathbf{u}_{\ell})}{1-2\lambda\operatorname{Re}(\mu_{\ell}) +\sigma^{2}r \vert 1-\lambda \mu_{\ell}\vert^{2}} = \sum_{\ell} \frac{\vert\mathbf{u}_{\ell}^{H}\mathbf{1}\vert^{2}}{1-2\lambda\operatorname{Re}(\mu_{\ell}) +\sigma^{2}r \vert 1-\lambda \mu_{\ell}\vert^{2}}. 
\end{equation*}
We can re-write the numerator since 
\begin{equation*}
    \operatorname{Var}(\mathbf{u}_{\ell}) = \frac{1}{n}\sum_{i=1}^{n}|u_{\ell i}|^{2} - |\bar{\mathbf{u}}_{\ell}|^{2}=\frac{1}{n} - |\bar{\mathbf{u}}_{\ell}|^{2},
\end{equation*}
where we have used the fact that $\mathbf{u}_{\ell}^{H}\mathbf{u}_{\ell}=1$ and $\bar{\mathbf{u}}_{\ell} = \frac{1}{n}\mathbf{1}'\mathbf{u}_{\ell}$. Multiplying both sides by $n$ and rearranging we obtain
\begin{equation*}
  n |\bar{\mathbf{u}}_{\ell}|^{2}  = 1 - n\operatorname{Var}(\mathbf{u}_{\ell}).
\end{equation*}
Moreover, notice that $\vert \mathbf{u}_{\ell}^{H}\mathbf{1} \vert = \vert \mathbf{1}'\mathbf{u}_{\ell}\vert$ since the two are complex conjugates. Then, we have:
\begin{equation*}
    |\mathbf{u}_{\ell}^{H}\mathbf{1}|^{2} = |\sum_{i=1}^{n}u_{\ell i}|^{2} = n^{2}|\bar{\mathbf{u}}_{\ell}|^{2}.
\end{equation*}
Putting everything together we can re-write expected profits at the optimal contract as: 
\begin{equation*}
    \mathbb{E}[\pi^{*}] =\frac{1}{2}\mathbf{1}'\mathbf{e^{*}} =\frac{1}{2} \sum_{\ell} \frac{n(1-n\operatorname{Var}(\mathbf{u}_{\ell}))}{1-2\lambda\operatorname{Re}(\mu_{\ell}) +\sigma^{2}r \vert 1-\lambda \mu_{\ell}\vert^{2}}.
\end{equation*}

When $\mathbf{G}$ is symmetric all eigenvalues are real and we can replace notation $^{H}$ with $'$ to obtain:
\begin{equation*}
    \mathbb{E}[\pi^{*}] =\frac{1}{2}\mathbf{1}'\mathbf{e^{*}} =\frac{1}{2} \sum_{\ell} \frac{(\mathbf{u}'_{\ell}\mathbf{1})(\mathbf{1}'\mathbf{u}_{\ell})}{1-2\lambda\mu_{\ell} +\sigma^{2}r (1-\lambda \mu_{\ell})^{2}}=\frac{1}{2} \sum_{\ell} \frac{n(1-n\operatorname{Var}(\mathbf{u}_{\ell}))}{1-2\lambda \mu_{\ell} +\sigma^{2}r (1-\lambda \mu_{\ell})^{2}}.
\end{equation*}

\end{proof}

\begin{proof}[\textbf{Proof of Proposition \ref{ER_model}}]
As in the baseline model, the firm can restrict attention without loss of
generality to contracts satisfying the linear system $(\nu \mathbf{I} - \lambda \mathbf{G})\mathbf{e}^{*}=\boldsymbol{\alpha}$ together with $\mathbf{e}^{*}\geq 0$. Then, under Assumption~\ref{assumption1nu}, equilibrium effort is unique and can be written as
\begin{equation*}
    \mathbf e=\frac{1}{\nu}\mathbf C_\nu\boldsymbol{\alpha},
\end{equation*}
where $\mathbf{C}_{\nu}$ is entrywise nonnegative. Substituting in for the (IR) and (IC) constraints, ignoring for a second $\mathbf{e}\geq 0$, the firm's problem is given by: 
\begin{equation*}
    \max_{\boldsymbol{\alpha}}
    \,\,\,\frac{1}{\nu}\boldsymbol{\alpha}'\mathbf{C}_{\nu}'\mathbf{1}
    -\frac{1}{2\nu}\boldsymbol{\alpha}'
    \left[
        (1+\nu r\sigma^{2})\mathbf{I}
        -\left(\frac{\lambda}{\nu}\right)^{2}
        (\mathbf{G}\mathbf{C}_{\nu})'\mathbf{G}\mathbf{C}_{\nu}
    \right]
    \boldsymbol{\alpha}.
\end{equation*}
Assumption~\ref{spectralradiusnu} guarantees that the matrix in brackets is positive definite. Thus, the Hessian is negative definite and the relaxed problem is strictly concave. Solving for the optimal incentive rule we obtain:
\begin{equation*}
    \boldsymbol{\alpha}^{*}_{\nu}
    =
    \left[
        (1+\nu r\sigma^{2})\mathbf{I}
        -\left(\frac{\lambda}{\nu}\right)^{2}
        (\mathbf{G}\mathbf{C}_{\nu})'\mathbf{G}\mathbf{C}_{\nu}
    \right]^{-1}
    \mathbf{C}_{\nu}'\mathbf{1} 
    = \mathbf{W}_{\nu}\mathbf{C}_{\nu}'\mathbf{1}.
\end{equation*}
As in the proof of Proposition~\ref{Optimal Contracts}, Assumptions~$1$-$\nu$ and~$2$-$\nu$, together with $\lambda\geq0$, imply that $\boldsymbol{\alpha}_{\nu}^{*}\gg0$ and hence
$\mathbf{e}^{*}_{\nu}=\nu^{-1}\mathbf{C}_{\nu}\boldsymbol{\alpha}_{\nu}^{*}\gg0$. Thus, the nonnegativity constraint on effort is slack, and the solution to the relaxed problem also solves the firm's constrained problem. 

Next, we can apply the same spectral decomposition from Proposition~\ref{Profits and Spectrum} to obtain:
$$
\mathbb{E}(\pi^{*})
=
\frac{1}{2\nu}
\sum_{\ell=1}^{n}
\frac{(\mathbf 1'\mathbf u_\ell)^2}
{(1+\nu r\sigma^2)\left(1-\frac{\lambda}{\nu}\mu_\ell\right)^2
-\left(\frac{\lambda}{\nu}\mu_\ell\right)^2},
$$

To compare the marginal effects of team strength $\lambda$ and human capital $\nu$, consider the contribution of a single eigenvalue $\mu_\ell$.
Since the denominator is positive, the sign of the marginal effect depends only on the numerator.\footnote{To see why the denominator is positive, notice that because $\mathbf{G}$ is symmetric, if $\mathbf{G}\mathbf{u}_{\ell} = \mu_{\ell} \mathbf{u}_{\ell}$ then $\mathbf{C}_{\nu}\mathbf{u}_{\ell} = \frac{1}{1-\frac{\lambda}{\nu}\mu_{\ell}}\mathbf{u}_{\ell}$. Hence $\mathbf{GC}_{\nu}$ has eigenvalue $\frac{\mu_{\ell}}{1-\frac{\lambda}{\nu}\mu_{\ell}}$. Plugging this into Assumption~\ref{spectralradiusnu} gives:
 \begin{equation*}
     \frac{(\lambda/\nu)^{2}}{1+\nu r \sigma^{2}}\frac{\mu_{\ell}^{2}}{(1-\frac{\lambda}{\nu}\mu_{\ell})^{2}}<1 \iff (1+\nu r\sigma^{2}) \left( 1- \frac{\lambda }{\nu}\mu_{\ell} \right)^{2} - \left(\frac{\lambda}{\nu}\mu_{\ell} \right)^{2} >0   \quad \text{for each } \ell.
 \end{equation*}} After simplification, a sufficient condition for team strength to have a larger marginal effect than human capital is given by:
\begin{equation}\label{nu_lambda}
2\mu_\ell\bigl(1+r\sigma^2(\nu-\lambda\mu_\ell)\bigr)
>
2r\sigma^2(\nu-\lambda\mu_\ell)+1  \iff
(1-\mu_\ell)\bigl(1+r\sigma^2(\nu-\lambda\mu_\ell)\bigr)
<
\frac{1}{2},
\end{equation}
for every $\ell$ such that $\mathbf{1}'\mathbf{u}_{\ell}\neq0$. When a single eigen-component satisfies $(\mathbf 1'\mathbf u_\ell)^2>0$, this condition is also necessary.

Now consider the balanced planted partition model with two groups of equal size. The expected adjacency matrix has two nonzero eigenvalues $\mu_1=\frac{n(p+q)}{2} $ and $\mu_2=\frac{n(p-q)}{2}$, and the corresponding eigenvectors satisfy $(\mathbf 1'\mathbf u_1)^2=n$ and $(\mathbf 1'\mathbf u_\ell)^2=0$ for all $\ell\ge2$. Hence,  only the principal eigenvalue contributes to the profit decomposition and condition \eqref{nu_lambda} becomes:
\begin{equation}\label{nu_lambda_pp}
(1-\mu_1)\bigl(1+r\sigma^2(\nu-\lambda\mu_1)\bigr)
<
\frac{1}{2}.
\end{equation}
Since only the principal eigencomponent contributes, condition~\eqref{nu_lambda_pp} is also necessary. For $r\sigma^2>0$ and $\lambda>0$, the equality associated with \eqref{nu_lambda_pp} has two roots,
\begin{equation*}
    \mu_{\pm}
    =
    \frac{
        1+r\sigma^2(\nu+\lambda)
        \pm
        \sqrt{\bigl(1+r\sigma^2(\nu-\lambda)\bigr)^2+2r\sigma^2\lambda}
    }
    {2r\sigma^2\lambda},
\end{equation*}
so that \eqref{nu_lambda_pp} holds for $\mu_1\in(\mu_-,\mu_+)$. Moreover, $\mu_-<1$ and $\mu_+>\nu/\lambda$. Since Assumption~\ref{assumption1nu} requires $\mu_1<\nu/\lambda$, condition~\eqref{nu_lambda_pp} is therefore equivalent, on the admissible parameter region, to
\begin{equation*}
    \mu_1>\mu_-.
\end{equation*}
Since $\mu_1=n(p+q)/2$ and $\mu_-<1$, the simpler condition
\begin{equation*}
    \frac{n(p+q)}{2}\geq1
\end{equation*}
is sufficient. In the Erd\H{o}s--R\'enyi case, $p=q$, and this reduces to $np\geq1$, proving the result.
\end{proof}



\end{document}